# Mathematics of Spacetime: A Guided Tour Through The Underlying Differential Topology and Differential Geometry

Johanna Barzen[0000-0001-8397-7973] and Frank Leymann[0000-0002-9123-259X]

University of Stuttgart, IAAS, Universitätsstr. 38, 70569 Stuttgart, Germany
{firstname.lastname}@iaas.uni-stuttgart.de

**Abstract.** Background in General Relativity (e.g. black holes, wormholes, or spacetime models in general) is needed to comprehend more recent efforts around understanding quantum phenomena like entanglement (e.g. "It from qubit" as well as "ER = EPR"). The former in turn requires a lot of knowledge from differential topology and differential geometry. While this knowledge is available in very good mathematics textbooks, it is scattered i.e. quite a bunch of sources need to be consulted to acquire it. The goal of this contribution is to provide the major background in a single place; in this sense, this contribution is some sort of guided tour through the corresponding literature.

## 1. Introduction

The ability to determine change or deviations is of paramount importance in many scientific disciplines. Change can be determined by observing differences at various points of that what is to be measured (points in time, points in location etc., in general points given by parameters of any kind). Very often deviations over very small ("infinitesimal") variations in parameters are important: at this stage domains that are often qualified by "differential" become relevant.

The scope at large of this contribution is change in physical quantities as given by functions, vector fields, and tensor fields, but also change in location within space and time (which are special kinds of parameters). Differential topology addresses especially generalized parameter spaces, i.e. manifolds, and the analysis of functions and fields on such manifolds. Differential geometry studies deviations of manifolds from a well-known parameter space, i.e. the Euclidian space. Our universe is such a manifold and its evolution (i.e. "change") in space and time as well as spacetime (i.e. as space and time interwoven) is investigated in general relativity theory. Understanding the universe as spacetime requires both, differential topology and differential geometry.

While there are many excellent text books on these subjects, more than one of them has to be studied to understand the mathematical underpinnings of general relativity. Our contributions aims to provide the necessary mathematical apparatus in a single source easing understanding of spacetime. In doing so, we focus on presenting and explaining the key concepts and the key facts and properties while leaving out proofs pointing to the literature for the detailed proofs; we present a few proofs that we did not find or that can be easily derived from what we presented before.

Furthermore, our overarching goal is to provide the mathematical basis for being able to follow the discussions around "ER=EPR" as well as "It from qubit": this discussion is about understanding the phenomenon of entanglement in quantum physics by means of gravitational effects and the importance of information in quantum gravitation. A presentation of these subjects in a separate contribution of the authors is ongoing work.

### 1.1. Structure of the Article

The article is structured as follows: Chapter 2 is mainly summarizing facts about differential calculus; readers will be familiar with the subject, thus, this chapter is brief and focuses on details needed in following chapters.

Chapter 3 first introduces differentiable manifolds with boundaries (all manifolds in this contribution are allowed to have a boundary if not said differently). Next, differentiable maps between such manifolds are defined. Tangent vectors are introduced as geometric entities and their interpretation as derivations is given. Tangent spaces and tangent bundles are defined. The pushforward (a.k.a. differentials) of a differential map as well as pullbacks are introduced.

Vector field on manifolds are defined in Chapter 4. Flows of vector fields give rise to local diffeomorphisms which are then used to define the Lie derivative of a vector field; a geometric interpretation of the Lie derivative is given. Also, the geometric meaning of the Lie bracket, which is also introduced, is provided. Finally, Lie derivatives of functions are defined.

Chapter 5 is extending the concept of a directional derivative to connections on manifolds (note that we omit the adjective "differentiable" because all manifolds are differentiable in this contribution). The definition of a connection is motivated by tangential directional derivatives in Euclidian space. Christoffel symbols are revealed as a means to perform computations with connections. The parallel transport between tangent spaces of a manifold is introduced. Torsions of connections are shown to indicate behavior of connections different from the behavior of directional derivatives in Euclidian space: a first indicator of the phenomenon of curvature. Geodesics are defined and geometrically interpreted.

The concept of Pseudo-Riemannian manifolds is subject of Chapter 6. First, pseudo-metrics and Riemannian metrics are discussed and the Lorentzian metric is shown as a special case of a pseudo-metric. The corresponding concept of a (pseudo-) Riemannian manifold is introduced (most manifolds are considered to be pseudo-Riemannian in this contribution, but because the existence of a pseudo-metric is not guaranteed we explicitly use the prefix "pseudo" to indicate results that are not only valid for Riemannian manifolds). Because spherical coordinates are important in many situations the metric of spheres in spherical coordinates is explicitly computed.

Finally, the gradient and Hessian of functions on (pseudo-) Riemannian manifolds is defined.

Many connections exist on a given manifold but only one symmetric and metric-compatible connection exists on a (pseudo-) Riemannian manifold: the Levi-Civita connection which is subject of Chapter 7. The Christoffel symbols of this connection are computed for a 3-sphere in spherical coordinates. The Riemann curvature tensor is introduced and geometrically illustrated. Holonomy as phenomenon of curvature is presented. Isometries are revealed as canonical "isomorphisms" of (pseudo-) Riemannian manifolds, Killing vector fields inducing symmetries by local isometries are presented. Finally, Ricci curvature, scalar curvature, and sectional curvature are defined and computed for the 2-sphere.

Chapter 8 presents spacetimes. At the beginning Minkowski spaces are explained and unit spheres for different metric signatures are shown. Manifolds of constant sectional curvature are classified by the Killing-Hopf theorem. Warped products as generic structures underlying spacetimes are introduced and their curvatures presented. These structures are specialized as Robertson-Walker spacetimes underlying the standard cosmological model. De Sitter spaces and anti-de Sitter spaces are presented as special Robertson-Walker spacetimes and as solutions of the vacuum Einstein Field Equation.

Exterior derivatives are the subject of Chapter 9. First, tensors and tensor fields are defined and their algebraic properties shown. Similarly, differential forms and corresponding computational laws are introduced. The definition of the exterior derivative, its properties as well as closeness and exactness of forms follows. De Rham cohomology is defined, and it is discussed how topological properties of an underlying manifold are revealed. Finally, antiderivation is explained.

The concept of an orientation is the focus of Chapter 10. Based on an understanding of orientations of vector spaces, oriented manifolds are defined via three equivalent approaches: equivalence classes of global frames, consistently oriented atlases, and orientation forms. The orientation of boundaries of oriented manifolds is derived. At the end of the chapter, the Riemann volume form is introduced.

Chapter 11 is discussing integration on oriented manifolds. Line integrals are defined, followed by a presentation of Stoke's theorem and its implications. Based on a discussion of line lengths and a geometric computation of the volume of parallelotopes, computing the volume of oriented manifolds is motivated. The chapter closes with a brief introduction of variational calculus, covering the Gâteaux derivative as well as the Euler-Langrange equations. In very sketchy manner it is indicated how Einstein-Hilbert actions are used to derive the Einstein Field Equation from the principle of least action.

## 2. Derivations in Euclidian Space

This section provides the fundamental definitions and concepts of differentiability and derivations in $\mathbb{R}^n$, i.e. in Euclidian space. We begin with the fundamental definition of differentiability of functions of a single variable:

**Definition 1**: Let $U \subseteq_{\text{open}} \mathbb{R}$; $f : U \to \mathbb{R}$ is called *differentiable* at $a \in U$ :⇔ the following limit exists:

$$\lim_{x \to 0} \frac{f(a+x) - f(a)}{x} =: f'(a) \tag{1}$$

$f'(a)$ is called the *derivative* or the *differential* of $f$ at $a$, also denoted by $\frac{df}{dx}(a)$. □

Thus, it is $f(a+x) = f(a) + f'(a)x + \varepsilon(x)$ where $\varepsilon : V \to \mathbb{R}$ is a function with $\lim_{x \to 0} \frac{\varepsilon(x)}{x} = 0$, i.e. $\varepsilon(x) = o(x)$. This means that $f$ is approximated locally (i.e. in a neighborhood of $a$) by the affine linear function $L(x) = f(a) + f'(a)x$. The derivative $f'(a)$ is the "slop" of this function $L$ (see Figure 1), and $\varepsilon(x)$ measures the corresponding approximation error. Thus, differentiability of a function means that it can be locally approximated by a linear function.

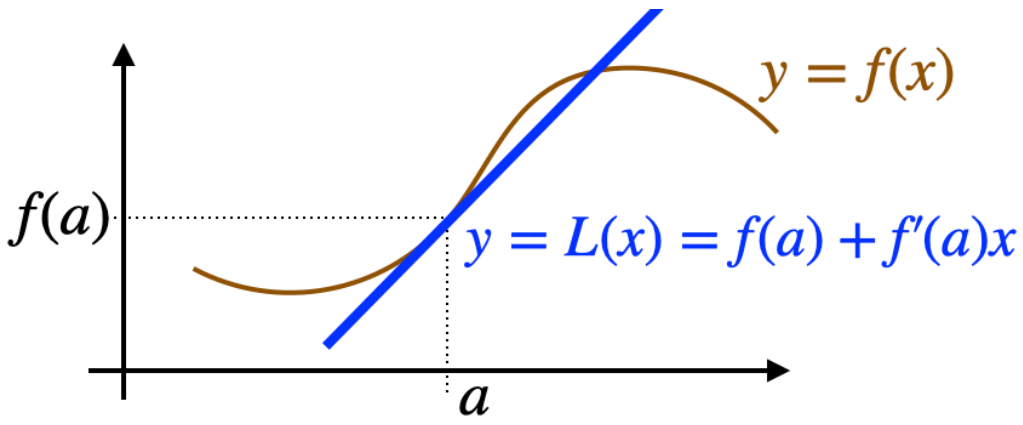

**Fig. 1**. Derivative as Slope of a Function

If a function has more than one variable derivatives w.r.t. any of these variables are of interest:

**Definition 2**: Let $U \subseteq_{\text{open}} \mathbb{R}^n$; $f : U \to \mathbb{R}$ is called *partially differentiable at* $a \in U$ w.r.t. $x_i$ :⇔

$$\lim_{h \to 0} \frac{f(a + he_i) - f(a)}{h} =: \partial_{x_i} f(a) \tag{2}$$

where $e_i$ is i-th unit vector in $\mathbb{R}^n$. $\partial_{x_i} f(a)$ is called the *partial derivative* w.r.t. $x_i$ at $a$, also denoted by $\frac{\partial f}{\partial x_i}(a)$.

A function $f : U \to \mathbb{R}$ is called *partially differentiable in* $U$ or simply *partially differentiable* :⇔ $f$ is partially differentiable at each point of $U$ and w.r.t. $x_1, \ldots, x_n$.

$f : U \to \mathbb{R}$ is called *continuously partially differentiable* :⇔ $f$ is partially differentiable and the partial derivatives are all continuous. □

The tuple of the derivatives w.r.t. all variables of a function is of utmost importance in many applications of calculus:

**Definition 3**: Let $U \subseteq_{\text{open}} \mathbb{R}^n$ and $f : U \to \mathbb{R}$ be partially differentiable. Then:

$$\operatorname{grad} f(x) := \left(\frac{\partial f}{\partial x_1}(x), \ldots, \frac{\partial f}{\partial x_n}(x)\right)^T \tag{3}$$

is called the *gradient* of $f$ in $x$. With

$$\nabla := \left(\frac{\partial}{\partial x_1}, \ldots, \frac{\partial}{\partial x_n}\right)^T \tag{4}$$

it is $\operatorname{grad} f(x) = \nabla f(x)$. $\nabla$ is a differential operator called *nabla*. □

Basically, a differential operator is a map with a domain of differentiable functions (of differentiable maps, in general) and a range of other functions (or maps) that computes the value of a function based on taking derivatives. The next definition introduces another differential operator; this operator applies to vector fields, i.e. functions that have vectors as values from the same Euclidian space that is the domain of the function.

**Definition 4**: Let $U \subseteq_{\text{open}} \mathbb{R}^n$; a map $f : U \to \mathbb{R}^n$ is called a *vector field* on $U$. A vector field $f = (f_1, \ldots, f_n) : U \to \mathbb{R}^n$ is called *partially differentiable* $:\Leftrightarrow$ $\forall i : f_i : U \to \mathbb{R}$ is *partially differentiable.* Then,

$$\operatorname{div} f := \langle \nabla, f \rangle = \sum_{i=1}^{n} \frac{\partial}{\partial x_i} f_i = \sum_{i=1}^{n} \frac{\partial f_i}{\partial x_i} \tag{5}$$

is called the *divergence* of the vector field $f$. □

In Equation 5, $\langle \cdot , \cdot \rangle$ is the standard scalar product of $\mathbb{R}^n$.

Let $U \subseteq_{\text{open}} \mathbb{R}^n$ and $f : U \to \mathbb{R}$. $\dfrac{\partial}{\partial x_i}$ is an operator that can be applied to functions $f$ that are partially differentiable w.r.t. $x_i$. Its application results in a function $\partial f / \partial x_i(x)$. If $\dfrac{\partial f}{\partial x_i}$ is partially differentiable w.r.t. $x_j$, the operator $\dfrac{\partial}{\partial x_j}$ can be applied:

$$\frac{\partial}{\partial x_j}\left(\frac{\partial f}{\partial x_i}\right) =: \frac{\partial^2 f}{\partial x_j \partial x_i} \tag{6}$$

**Definition 5**: $\dfrac{\partial^2 f}{\partial x_j \partial x_i}$ is called *2nd order partial derivative* of $f$ w.r.t. $x_i$ and $x_j$.

$f$ is called 2nd-order *continuously partially differentiabl*e (a.k.a. of differentiability class $C^2(U)$) $:\Leftrightarrow$ All 2nd-order partial derivatives of $f$ w.r.t. to all $x \in \{x_1, \ldots, x_n\}$ exists and are continuous.

By induction, the *differentiability class* $C^k(U)$ is defined. If $f$ is of class $C^k(U)$ (i.e. $f \in C^k(U)$) for all $k \in \mathbb{N}$, $f$ is called *smooth* or of class $C^\infty(U)$. □

The following theorem says that the order in which 2nd-order partial derivatives are taken is irrelevant([LT17], Th. 4.1):

**Theorem 1 (Schwarz's Theorem)**: Let $U \subseteq_{\text{open}} \mathbb{R}^n$ and $f \in C^\infty(U)$. Then:

$$\frac{\partial^2 f}{\partial x_i \partial x_j} = \frac{\partial^2 f}{\partial x_j \partial x_i} \tag{7}$$

for each $1 \leq i, j \leq n$. ∎

In fact, if a function is of class $C^k$ the order of up to any $k$ partial derivatives taken is irrelevant ([21], VI, §1). Since we need this fact only for two partial derivatives taken we formulated the theorem only accordingly.

**Definition 6**: Let $U \subseteq_{\text{open}} \mathbb{R}^n$; $f : U \to \mathbb{R}^m$ is called (*totally*) *differentiable* (or *Fréchet differentiable*) at $a \in U$ :⇔ There is a linear map $A : \mathbb{R}^n \to \mathbb{R}^m$ and a $V \subseteq_{\text{open}} \mathbb{R}^n$, $0 \in V$, such that for $x \in V$

$$f(a + x) = f(a) + Ax + \varepsilon(x) \tag{8}$$

where $\varepsilon : V \to \mathbb{R}^m$ is a function with $\lim\limits_{x \to 0} \dfrac{\varepsilon(x)}{\|x\|} = 0$ (i.e. $\varepsilon(x) = o(\|x\|)$).

$f$ is called (*totally*) *differentiable in U* or simply (*totally*) *differentiable* :⇔ $f$ is (totally) differentiable at each point of $U$.

The linear map $A$ is called *differential* (or *Fréchet derivative*) of $f$ at $a$, denoted by $df(a)$, $df_a$, or $Df(a)$; the corresponding matrix is called *Jacobian matrix* of $f$. □

Thus, the differential of $f$ at $a$ is the best linear approximation of $f$ in a neighborhood of $a$. It is $df(a) = \left(\frac{\partial f_i}{\partial x_j}(a)\right) =: J_f(a)$ the *Jacobian matrix* of $f$. Note, that the differential of a function is a row vector, while its gradient (see Definition 3) is a column vector (i.e. a vector field).

**Note 1**: It is

$$J_f(a) = \begin{pmatrix} \nabla f_1(a)^T \\ \vdots \\ \nabla f_m(a)^T \end{pmatrix} \tag{9}$$ ∎

While a partial derivative is concerned with the i-th unit vector $e_i$, derivatives w.r.t. any unit vector $v$ are of importance:

**Definition 7**: Let $U \subseteq_{\text{open}} \mathbb{R}^n$, $f : U \to \mathbb{R}$ and $v \in \mathbb{R}^n$ with $\|v\| = 1$. If the following limit exists,

$$\lim_{h \to 0} \frac{f(a + hv) - f(a)}{h} =: \partial_v f(a) \tag{10}$$

then $\partial_v f(a)$ is called the *directional derivative* of $f$ in $a$ along $v$. Instead of $\partial_v f(a)$, the following notations are also used: $D_v f(a)$, $\frac{\partial f}{\partial v}(a)$, $\nabla_v f(a)$. □

The notion of directional derivatives can be generalized to infinite dimensional vector spaces and is then called Gâteaux derivative (see Definition 96).

**Note 2**: Let $\gamma :]-\varepsilon,+\varepsilon[\to U$ be the (differentiable) curve defined by $t \mapsto a + tv$; then

$$\partial_v f(a) = (f \circ \gamma)'(a) = \frac{d}{dt} f(\gamma(t))\Big|_{t=0} \tag{11}$$ ∎

Note 2 is sometimes used as an equivalent definition of the directional derivative because it can be extended to differentiable manifolds (see later). Directional derivates of totally differentiable functions can be easily computed via the gradient of the function is shown by the next lemma:

**Lemma 1**: If $f : U \to \mathbb{R}$ is totally differentiable then for each $v \in \mathbb{R}^n$ with $\|v\| = 1$ it is

$$\partial_v f(a) = df(a)v = \langle \text{grad}\, f(a), v \rangle \tag{12}$$ ∎

Directional derivatives are also defined similarly for vector fields in Euclidian space, and their computation can be done as follows: Let $U \subseteq_{\text{open}} \mathbb{R}^n$, $f : U \to \mathbb{R}^n$ be a vector field and $v \in \mathbb{R}^n$ with $\|v\| = 1$. With $f(x) = (f_1(x), \ldots, f_n(x))$, the directional derivative of each component $f_i$ can be taken: $\partial_v f_i(a) = \langle \text{grad}\, f_i(a), v \rangle$. These directional derivatives are then combined ([18], Prop. 4.11):

$$D_v f(a) = \begin{pmatrix} \langle \text{grad}\, f_1(a), v \rangle \\ \vdots \\ \langle \text{grad}\, f_n(a), v \rangle \end{pmatrix} = \begin{pmatrix} \nabla f_1(a)^T \\ \vdots \\ \nabla f_m(a)^T \end{pmatrix} \cdot v = df(a)v \tag{13}$$

i.e. $D_v f(a)$ is the directional derivative of the vector field $f$ in $\mathbb{R}^n$.

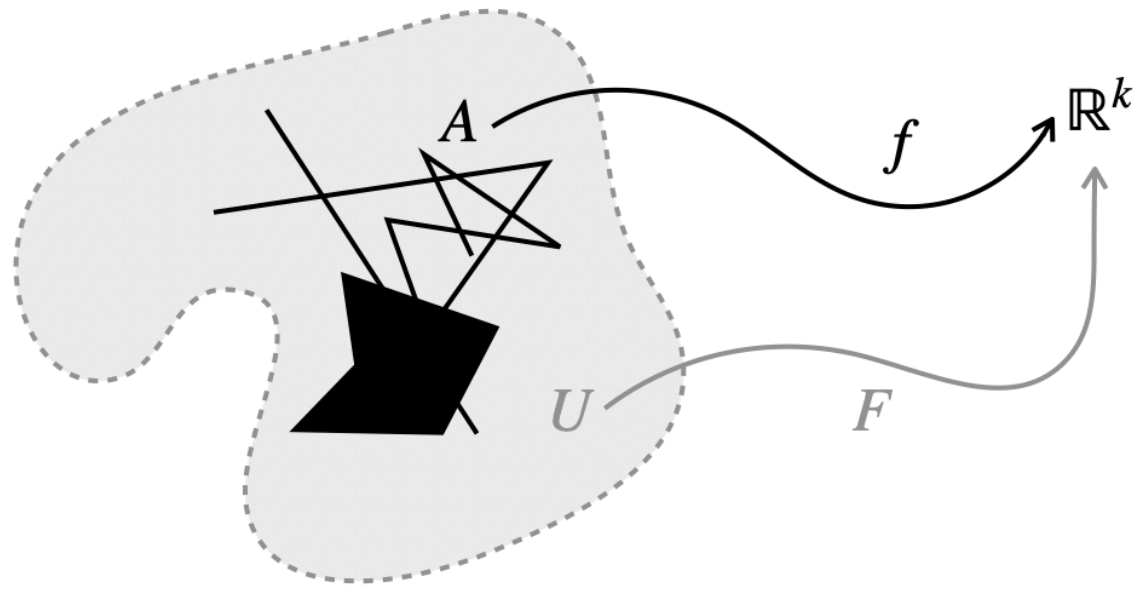

**Fig. 2**. Differentiability of Maps Defined on Arbitrary Sets

Often, the domain of a map is not an open subset of a Euclidian space but an arbitrary subset. In this situation differentiability is defined as follows (see Figure 2):

**Definition 8**: Let $A \subseteq \mathbb{R}^n$ be an arbitrary set, $f : A \to \mathbb{R}^k$. $f$ is called *differentiable of class $C^r$* :⇔ There is a map $F : U \to \mathbb{R}^k$, $A \subseteq U \subseteq_{\text{open}} \mathbb{R}^n$, $F$ is differentiable of class $C^r$, and $F|_A = f$. $F$ is called a *$C^r$-extension* of $f$. □

Figure 3 shows which property of a map implies which other property. Note, that the reverse directions are in general not true.

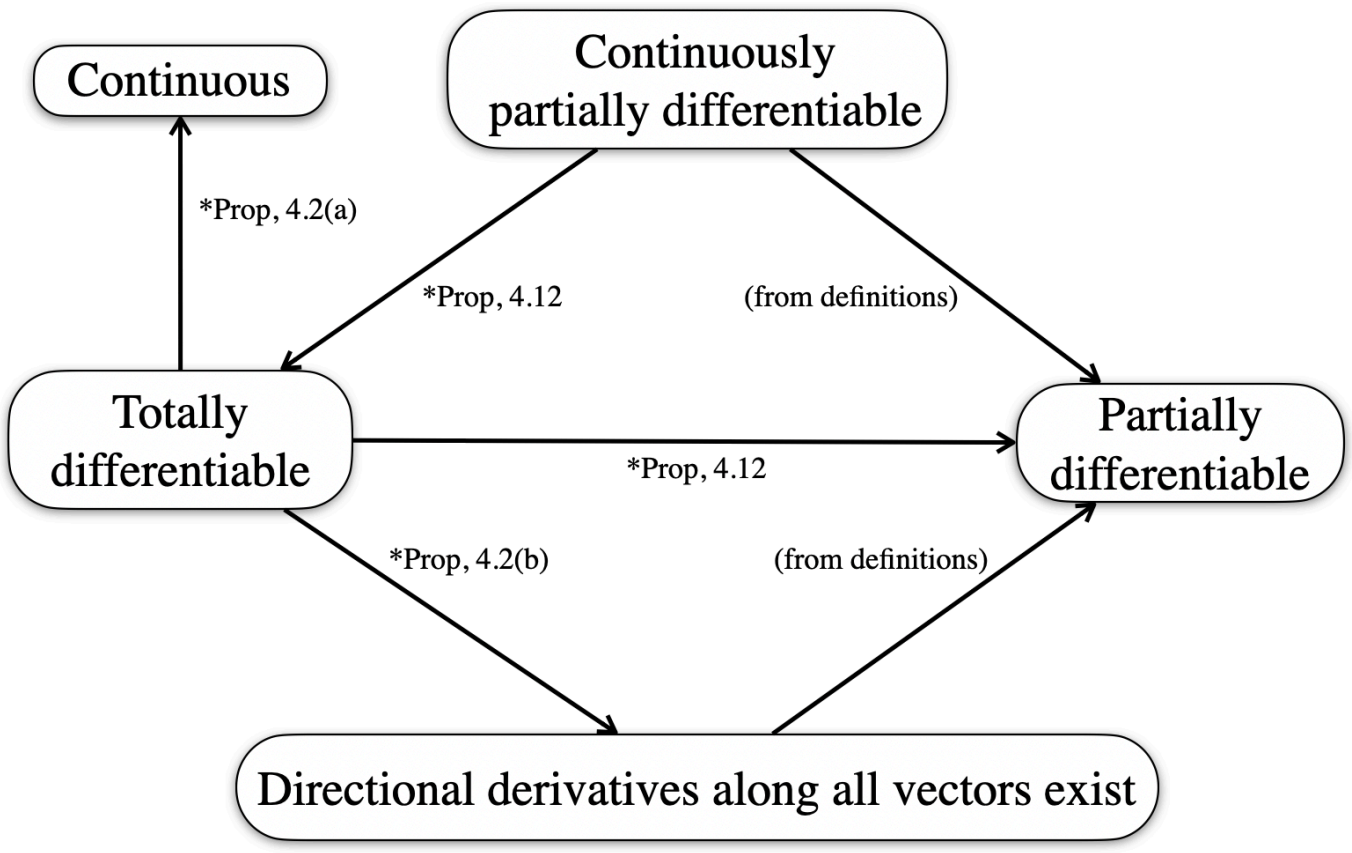

**Fig. 3**. Relations between the different notions of differentiability (*: for proofs see [18])

## 3. Differentiable Manifolds

In this section we provide the absolute basics about differentiable manifolds, tangent bundles and differentiable maps.

### 3.1. Manifolds with Boundaries

In many situations, spaces called differentiable manifolds with boundary are relevant. Such spaces are different from Euclidian space. The definition of these manifolds requires the use of the n-dimensional real half-space $\mathbb{R}^n_+ := \{x \in \mathbb{R}^n \mid x_n \geq 0\}$; it is $\partial\mathbb{R}^n_+ = \{x \in \mathbb{R}^n \mid x_n = 0\} \cong \mathbb{R}^{n-1}$ the *boundary* of $\mathbb{R}^n_+$.

Note, that in the context of differentiable manifolds, topological spaces are considered to be Hausdorff spaces and second countable. A *Hausdorff space* requires that any two different points of the space have disjoint neighborhoods; the set of all neighborhoods of a point $p$ is denoted by $\mathfrak{U}_p$. A space is *second countable* iff it has a countable basis, i.e. every open set of the space is the union of a subset of the countable basis. See [25] for the detailed definitions of these terms.

**Definition 9:** Let $M$ be a Hausdorff and second countable topological space. If each point $x \in M$ has a neighborhood $U \subseteq_{\text{open}} M$ and a homeomorphism $\varphi : U \to \varphi(U) \subseteq_{\text{open}} \mathbb{R}^n_+$, then $M$ is called a *topological manifold with boundary* of dimension $n$ (in symbol: $\dim M = n$). The pair $(U, \varphi)$ is called a *chart of M around x*. A set of charts $\mathfrak{A} = \{(U_i, \varphi_i) \,|\, i \in I\}$ with $\bigcup_{i \in I} U_i = M$ is called an *atlas of M*.

A point $x \in M$ with $\varphi(U) \subseteq_{\text{open}} \mathbb{R}^n$ is called *interior point*, and a point $x \in M$ with $\varphi(x) \in \partial\mathbb{R}^n_+$ is called *boundary point*. The set of boundary points of $M$ is called the *boundary of M*, denoted by $\partial M$.

For two intersecting charts $(U_i, \varphi_i), (U_j, \varphi_j) \in \mathfrak{A}$, i.e. charts with $U_i \cap U_j \neq \varnothing$ the map $\varphi_i \circ \varphi_j^{-1} : \varphi_j(U_i \cap U_j) \to \varphi_i(U_i \cap U_j)$ is called the *transition function* between the charts.

$M$ is called a *differentiable manifold of class* $C^k$ ($C^k$*-manifold* for short) if all transition functions are differentiable of class $C^k$; the corresponding atlas $\mathfrak{A}$ is called $C^k$-atlas. For $k = \infty$, the manifold (and the atlas) is called *smooth*. □

In Figure 4 $p$ and $q$ are interior points while $r$ is a boundary point.

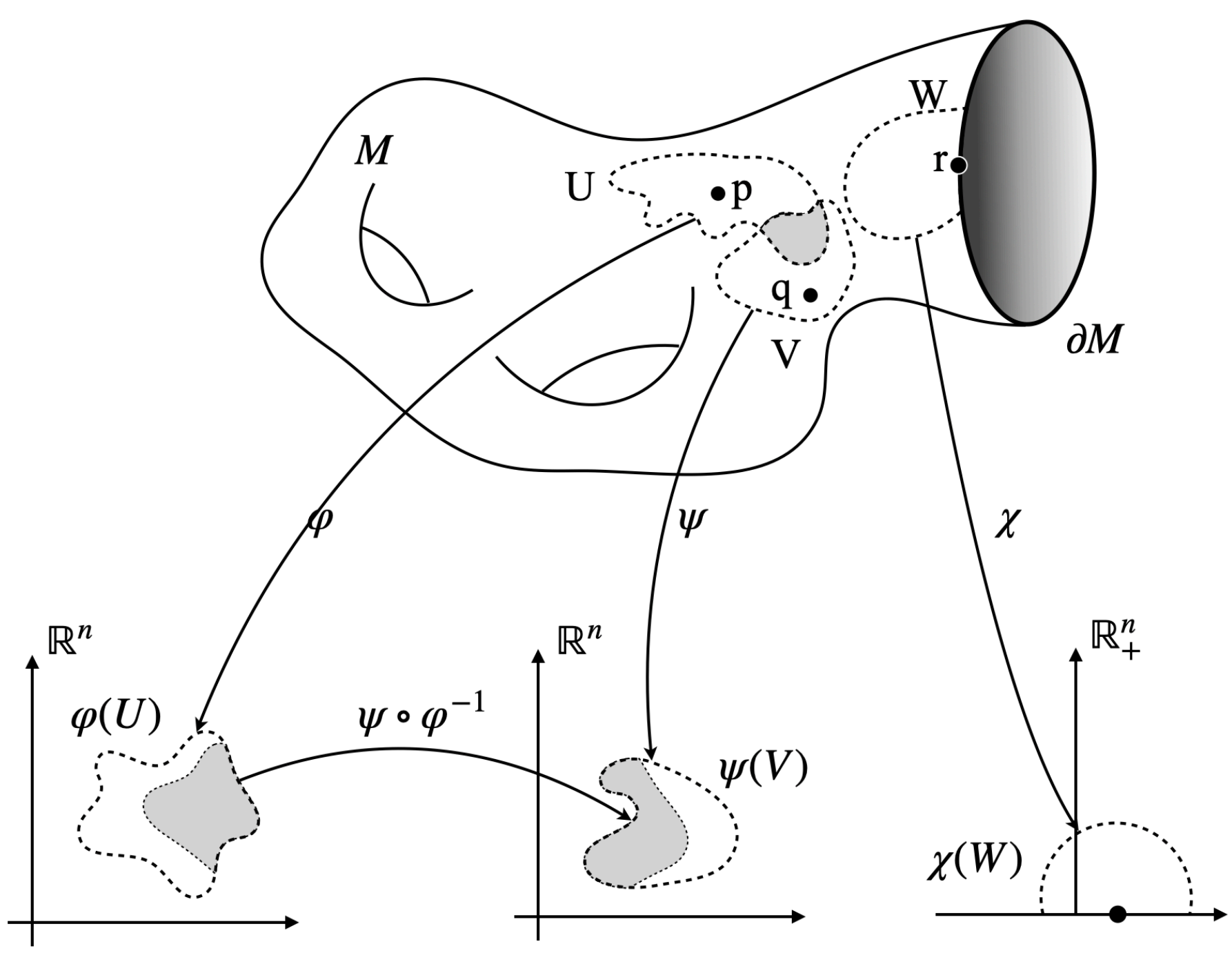

**Fig. 4**. Charts and Transition Functions of a Differentiable Manifold with Boundary

It is important to note that the definition of a manifold is independent of any ambient space like $\mathbb{R}^n$. Although figures of manifolds most often give the impression of them being "embedded" in Euclidian space, Definition 9 does not refer to the manifold $M$ being a subset of $\mathbb{R}^n$: manifolds are abstract topological spaces (but they are locally homeomorphic to open subsets of $\mathbb{R}^n_+$). This will require extra efforts to define familiar concepts known from Euclidian space (like tangent vectors - see Definition 11) which would otherwise be "inherited" from such an ambient space.

A $C^k$-manifold may have several $C^k$-atlases. For example, both, $\mathfrak{A}_1 = \{(\mathbb{R}^n, \mathrm{id})\}$ as well as $\mathfrak{A}_2 = \{(\mathbb{R}^n_{x_n > -1}, \mathrm{id}), (\mathbb{R}^n_{x_n < 1}, \mathrm{id})\}$ are $C^k$-atlases of $\mathbb{R}^n$ (for any $k$, and for any $n \geq 1$), and so is $\mathfrak{A}_1 \cup \mathfrak{A}_2$. In general, when adding a chart $(U, \varphi)$ to a given $C^k$-atlas $\mathfrak{A}$ and if the resulting atlas $\mathfrak{A} \cup \{(U, \varphi)\}$ is again a $C^k$-atlas, $(U, \varphi)$ is said to be *compatible* with $\mathfrak{A}$. Adding all compatible charts to $\mathfrak{A}$ results in the (unique) maximal atlas $\overline{\mathfrak{A}}$ (that contains $\mathfrak{A}$). A maximal $C^k$-atlas is called a *$C^k$-differentiable structure* of the corresponding manifold. Any $C^k$-atlas $\mathfrak{A}$ determines a unique $C^k$-differentiable structure (see [22], Prop. 1.17). Thus, we can assume that the atlases of a manifold are differentiable structures.

The next lemma is often used, and its proof can be found in [Lman, Prop. 1.38] and [T, Sec. 22.3].

**Lemma 2**: Let $M$ be a $C^k$-manifold with boundary, $\dim M = n$. Then, $\partial M \subseteq_{\text{closed}} M$ and $\partial M$ is a $C^k$-manifold without boundary, $\dim \partial M = n - 1$. ∎

Some results that we will present in the following are only valid for manifolds without boundary, i.e. manifolds $M$ with $\partial M = \emptyset$; we will point this out explicitly. In case a result holds for both, for manifolds with boundary as well as manifolds without boundary we will simply speak about "manifold" from now on.

### 3.2. Differentiable Maps

Next, we define differentiability of maps between differentiable manifolds.

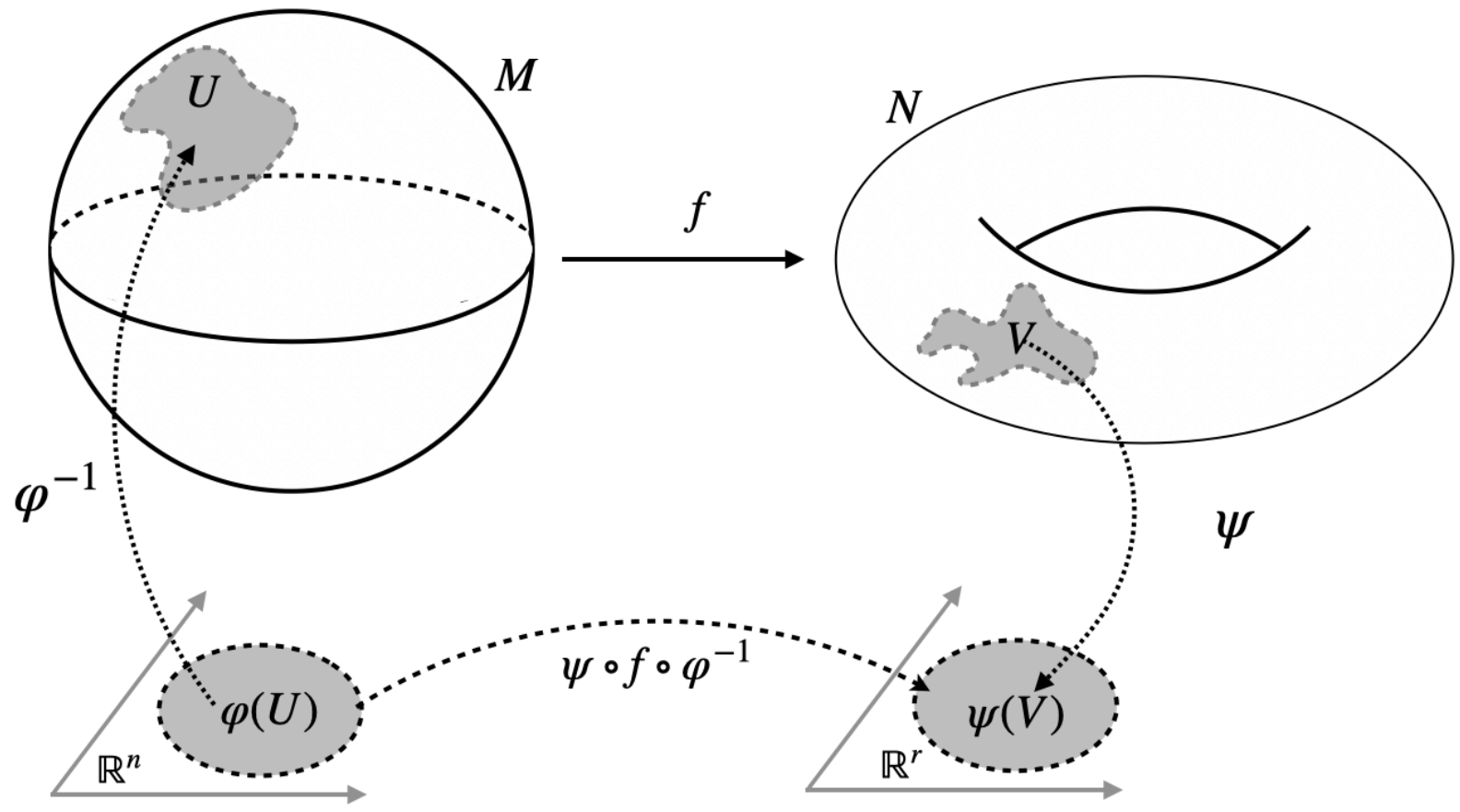

**Fig. 5**. Differentiability of a Map Between Differentiable Manifolds

**Definition 10:** Let $M$ an n-dimensional and $N$ be an r-dimensionale ($C^k$-) manifold. Let $f : M \to N$ continuous. $f$ is called ($C^k$-) *differentiable* :⇔ For each chart $(U, \varphi)$ of $M$ and each chart $(V, \psi)$ of $N$ with $f(U) \subseteq V$ the composed map $\psi \circ f \circ \varphi^{-1} : \varphi(U) \subseteq \mathbb{R}^n \to \psi(V) \subseteq \mathbb{R}^r$ is ($C^k$-) differentiable.

If $f$ is bijective and its inverse map $f^{-1}$ is also differentiable then $f$ is called a *diffeomorphism*. □

Figure 5 depicts this situation. Note, that we wrote the differentiability class $C^k$ of both, the manifold as well as of the map in parenthesis; this is because the actual differentiability class is often irrelevant (as long as $k \geq 1$). In this case, we will simply speak about "differentiable manifold", "differentiable map" etc., and we may even leaf out the adjective "differentiable" sometimes.

Diffeomorphisms are the maps that preserve the inherent structures of differentiable manifolds especially their differentiable structures. Diffeomorphic manifolds (or subsets thereof) are considered "the same".

### 3.3. Tangent Spaces

Next, the concept of a tangent vector at a point $p$ of a manifold is defined. First, a differentiable curve $\gamma : I \to M$ through this point is taken (i.e. $p = \gamma(t)$ for a $t \in I \subseteq_{\text{open}} \mathbb{R}$). The differentiability of $\gamma$ is defined according to Definition 10: $I \subseteq_{\text{open}} \mathbb{R}$ is a differentiable manifold with the atlas $\mathfrak{A}_I = \{(I, \text{id})\}$, i.e. $(I, \text{id})$ is a chart around $t$. For a chart $(U, \varphi)$ around $p$ the composed map $\varphi \circ \gamma \circ \text{id}^{-1} = \varphi \circ \gamma$ is then a differentiable map $\varphi \circ \gamma : I \to \mathbb{R}^n$ in the sense of Definition 6, i.e. its differential can be computed as Jacobi matrix $J_{\varphi \circ \gamma}$. In fact, $\varphi \circ \gamma$ is a curve in $\mathbb{R}^n$ thus, its differential

$$\gamma'(t) = \frac{d}{dt}\big(\varphi \circ \gamma\big)(t) \tag{14}$$

is a tangent vector at the curve $\varphi \circ \gamma$. This vector in $\mathbb{R}^n$ may be considered as a concrete representative of "the" tangent vector of the manifold at $p$. But there are many curves on $M$ through $p$ resulting in the same vector. In order to determine "the" tangent vector of the manifold at $p$ the equivalence of representatives has to be defined. For this purpose, we consider two curves $\gamma, \hat{\gamma} : ]-\varepsilon, \varepsilon[ \to M$ (for $\varepsilon > 0$) with $\gamma(0) = \hat{\gamma}(0) = p$. These curves are treated as equivalent (i.e. $\gamma \sim \hat{\gamma}$) if they result in the same representative of "the" tangent vector $v$, i.e.

$$\frac{d}{dt}\big(\varphi \circ \gamma\big)(0) = \frac{d}{dt}\big(\varphi \circ \hat{\gamma}\big)(0) \ \ (= v) \tag{15}$$

Figure 6 is a corresponding graphics. In summary:

**Definition 11:** Two curves $\gamma, \hat{\gamma} : ]-\varepsilon, \varepsilon[ \to M$ (for $\varepsilon > 0$) are equivalent (in symbol $\gamma \sim \hat{\gamma}$) $:\Leftrightarrow$

i. $\gamma(0) = \hat{\gamma}(0) = p$

ii. $\frac{d}{dt}\big(\varphi \circ \gamma\big)(0) = \frac{d}{dt}\big(\varphi \circ \hat{\gamma}\big)(0) \ \ (= v)$

The equivalence class of a curve γ is called *tangent vector* at $p$ and is denoted by $\gamma'(0)$. □

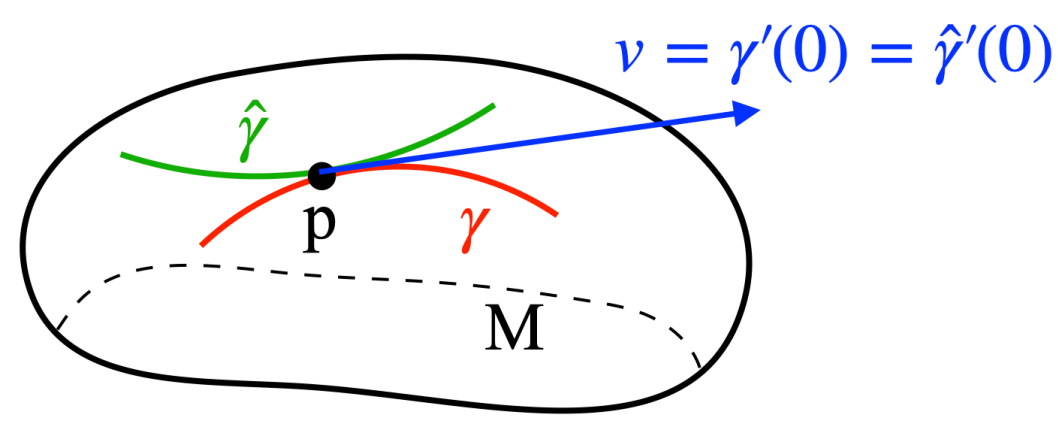

**Fig. 6**. Tangent Vector as Equivalence Class of Curves

I.e. it is the ∼-equivalence class of a curve $\gamma$ that is taken as a tangent vector of the manifold. This equivalence class is an abstract object, i.e. it is not an element of the Euclidian space $\mathbb{R}^n$, thus, it can be associated with the point $p$.

The set of tangent vectors at $p \in M$ is a vector space. We show only that the sum of two tangent vectors $v = \gamma'(0)$ and $w = \delta'(0)$ is again a tangent vector: for this purpose define the curve $\alpha(t) = \varphi^{-1}(\varphi(p) + t(v + w))$. Obviously, $\alpha(0) = p$ and $\alpha'(0) = \frac{d}{dt}\left(\varphi \circ \alpha\right)(0) = v + w$. Similarly, for $\lambda \in \mathbb{R}$ it is $\lambda v$ is a tangent vector. It can be shown that this construction is independent of the chart $(U, \varphi)$ used, i.e. the construction is well-defined. Also, the dimension of this vector space is equal to the dimension of the manifold. Thus:

**Lemma 3**: The set of all tangent vectors of $M$ at $p$ is a vector space called *tangent space* of $M$ at $p$ denoted as $T_pM$. It is $\dim T_pM = \dim M$. ∎

Next, the canonical basis of the tangent space is constructed (see Figure 7). Let $(U, \varphi)$ be a chart around $p$. Via $t \mapsto \varphi(p) + te_k$ a curve is given in direction of the basis vector $e_k$ through $\varphi(p)$. This is turned into curve $\gamma_k(t) := \varphi^{-1}\left(\varphi(p) + te_k\right)$ on $M$. Define

$$\partial_k := \gamma_k'(0) \in T_pM \tag{16}$$

The tangent vector $\partial_k$ corresponds to $e_k$ and is sometimes also denoted by $\partial/\partial x_k$.

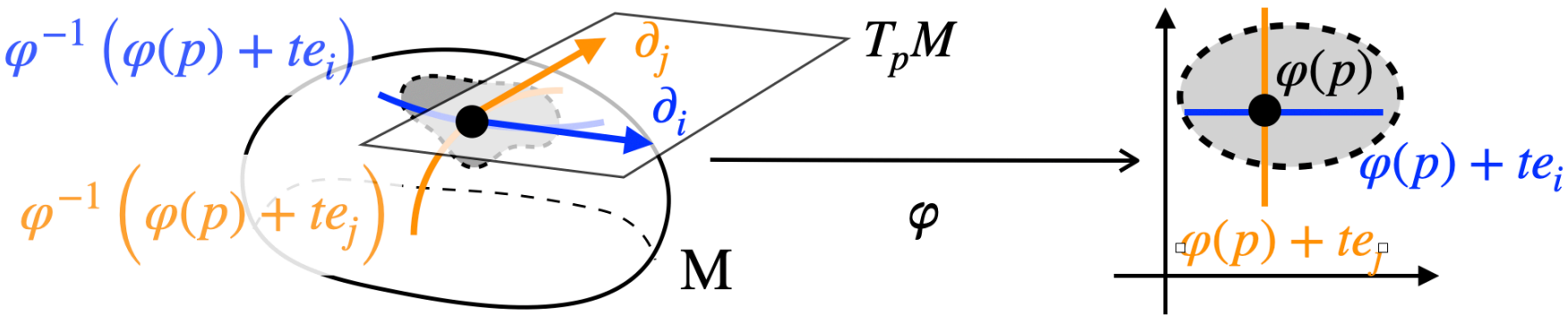

**Fig. 7**. Constructing the Canonical Basis $\{\partial_1, \ldots, \partial_n\}$ of the Tangent Space $T_pM$

**Lemma 4**: $\{\partial_1, \ldots, \partial_n\}$ is a basis of $T_pM$. ∎

It is important to note, that the basis vectors $\partial_k$ depends on the chart $(U, \varphi)$ chosen to derive the curve $\gamma_k(t) = \varphi^{-1}\left(\varphi(p) + te_k\right)$ whose derivate $\gamma_k'(0)$ result in $\partial_k$. To emphasize this dependency the phrase "in local coordinates" is used from time to time (see [22] pp. 60).

The tangent space defined before is sometimes referred to as "geometric tangent space". While it is quite illustrative it is hard to use in proofs and in concrete computations. Computations are supported by a different (but equivalent) point of view at tangent vectors. As seen in Note 2 each vector $v$ corresponds to a means to take the directional derivative of a function in $\mathbb{R}^n$ along it:

$$\partial_v f(p) = \frac{d}{dt} f(p + tv)\bigg|_{t=0}$$

This way each vector defines a map $C^\infty(\mathbb{R}^n) \to \mathbb{R}$ via $v(f) := \partial_v f(p)$, and this map is even a linear map (sum rule and constant factor rule - [33] Th. 2.3.2). The product rule ([33], Th. 2.3.4) further implies that $v(fg) = f(p)v(g) + g(p)v(f)$. Consequently, a vector corresponds to a corresponding differential operator. This is formally transferred to differentiable manifolds:

**Definition 12:** A linear map $v : C^\infty(M) \to \mathbb{R}$ is called *derivation* at $p \in M$ $:\Leftrightarrow$ $\forall f, g \in C^\infty(M) : v(fg) = f(p)v(g) + g(p)v(f)$ (Leibniz Rule). □

It can be proven (see [22], Prop. 3.2):

**Lemma 5**: The set of derivations at $p$ is a vector space isomorphic to $T_pM$. ∎

Because of this lemma, a derivation $v$ at $p$ is also called *tangent vector* at $p$, and the set of all derivations is referred to as tangent space of $M$ at $p$ also. In fact, it can be shown that both definitions of tangent spaces (the geometric one introduced above and the "algebraic" one introduced here) are equivalent ([22], pp.72).

Lemma 1 has shown that $v(f) = \partial_v f(p) = \langle \text{grad}\, f(a), v\rangle = \sum_i v_i \frac{\partial f}{\partial x_i}(p)$. This means that $v$ can be written as $v = \sum_i v_i \frac{\partial}{\partial x_i}$, i.e. $\{\frac{\partial}{\partial x_1}, \ldots, \frac{\partial}{\partial x_n}\}$ is a basis of the vector space of derivations. More precise: Let $(U, \varphi)$ be a chart around $p$, $\varphi : U \to \mathbb{R}^n$. Then, $\frac{\partial}{\partial x_i}\Big|_p : C^\infty(M) \to \mathbb{R}$, with $f \mapsto \frac{\partial}{\partial x^i}\Big|_{\varphi(p)} \left(f \circ \varphi^{-1}\right)$ is a derivation at $p$. With $\partial_i|_p := \frac{\partial}{\partial x_i}\Big|_p$ it is $\left\{\partial_1|_p, \ldots, \partial_n|_p\right\}$ a basis of $T_pM$.

The collection of tangent spaces at each point of a manifold is an important concept in differential topology and in differential geometry:

**Definition 13:** $TM := \bigcup_{x \in M} \{x\} \times T_xM$ is called *tangent bundle* of $M$. □

The nice situation is that the tangent bundle is again a manifold (see [22], Prop 3.18):

**Lemma 6**: $TM$ is a differentiable manifold with $\dim TM = 2 \dim M$. ∎

Figure 8 shows this concept: part (a) shows some tangent spaces at points of the differential manifold $\mathbb{S}^1$; each tangent space is isomorphic to $\mathbb{R}$. In part (b) the tangent spaces are "folded up", i.e. they are shown orthogonal to the basis $\mathbb{S}^1$. Finally, part (c) indicates the situation in which all tangent spaces are considered, and it is obvious that the tangent bundle $T\mathbb{S}^1$ is the manifold $\mathbb{S}^1 \times \mathbb{R}$.

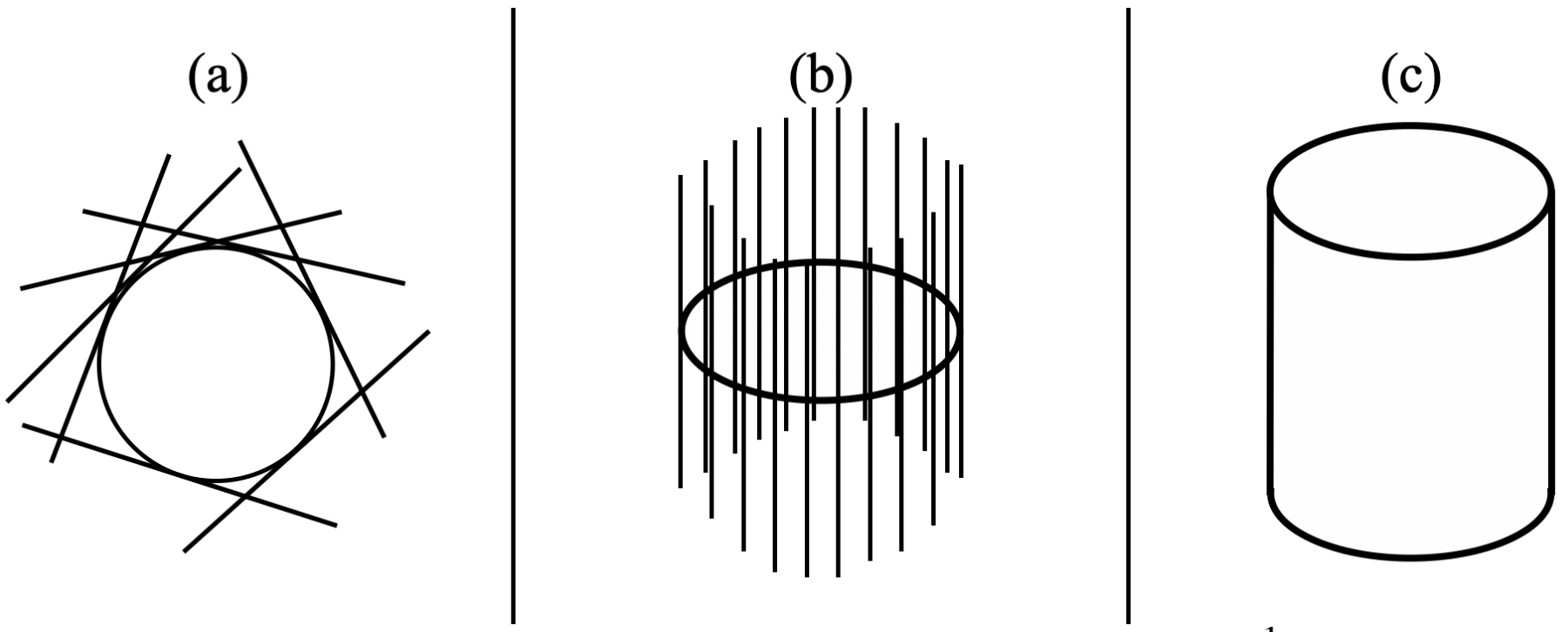

**Fig. 8**. (a) Collection of Tangent Spaces at the Manifold $\mathbb{S}^1$;
(b) Tangent Bundle as Tangent Spaces "Folded Up"; (c) Tangent Bundle as Manifold

It is important to note that in general it is <u>not</u> $TM \cong M \times \mathbb{R}^n$ (if $T_pM = \mathbb{R}^n$ for dim $M = n$). In case $TM \cong M \times \mathbb{R}^n$ the tangent bundle $TM$ is called *trivial*, and M is called *parallelizable.* A parallelizable manifold has a property that will become important later when we discuss integration on manifolds: a parallelizable manifold is always orientable (see [22], Prop 15.17). Also, if $M$ is a Lie-Group its tangent bundle is trivial (we do not discuss Lie-Groups in this article but it is worth mentioning this fact).

Tangent spaces are domain and codomain of the differential of differentiable maps between manifolds (see Figure 9) as defined next:

**Definition 14:** Let $M, N$ be differentiable manifolds and let $f : M \to N$ be a differentiable map. Then, $df_p : T_pM \to T_{f(p)}N$ defined by $df_p(\gamma'(0)) := (f \circ \gamma)'(0)$ is called *differential* or *pushforward* or *tangent map* of $f$. If pushforward or tangent map is used as name of $df_p$, the symbols $f_*$ or $T_pf$ instead $df_p$ are also used. □

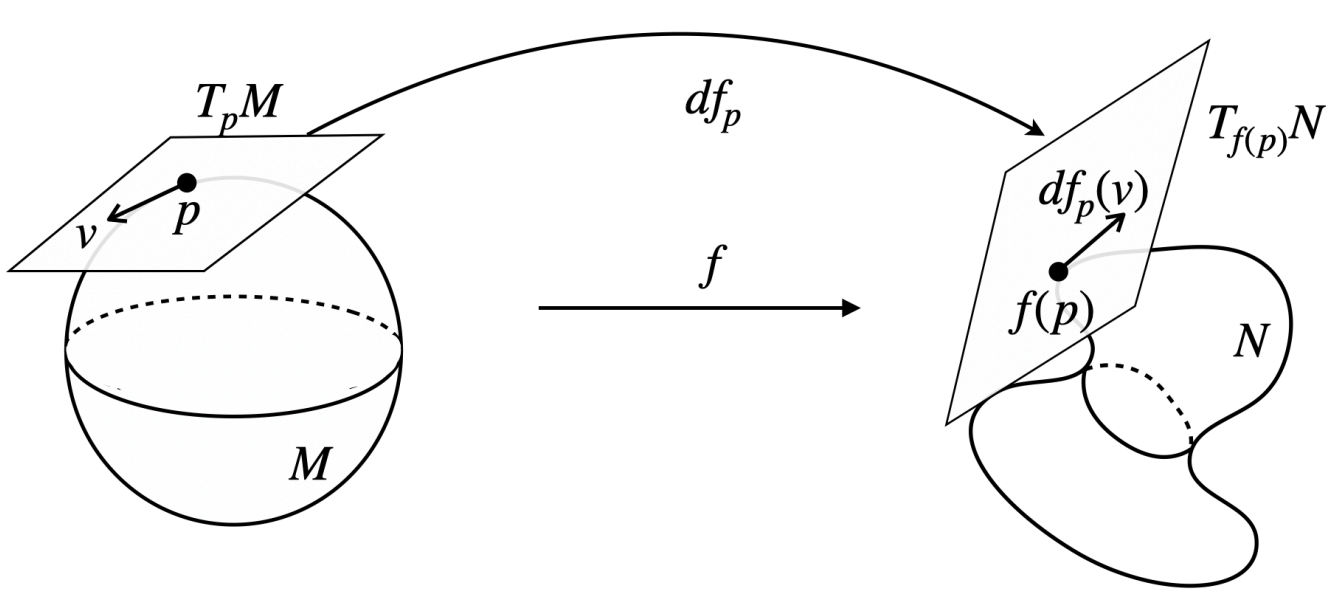

**Fig. 9**. The Differential of a Differentiable Map Between Manifolds

In Definition 14 the geometric interpretation of a tangent vector $v$ is taken, i.e. a tangent vector at $p \in M$ is represented by $\gamma'(0) = v$ with a curve $\gamma$ through $p$ on $M$. The map $f$ creates the curve $f \circ \gamma$ on the manifold $N$. The corresponding tangent vector at $f(p) \in N$ as represented by $(f \circ \gamma)'(0)$ is defined as the image of $v$ under the map $df_p$, i.e. the "differential" of $f$ at the point $p$. The name "pushforward" for the differential is now obvious: $f$ pushes the curve $\gamma$ forward to the curve $f \circ \gamma$ and the corresponding "velocity vector" $v = \gamma'(0) \in T_pM$ is pushed forward to the "velocity vector" $df_p = (f \circ \gamma)'(0) \in T_{f(p)}N$.

With the notion of a differential of a map between manifolds, $\gamma'$ can be interpreted accordingly: $I \subseteq_{\text{open}} \mathbb{R}$ is a differentiable manifold. Then, for a curve $\gamma : I \to M$ on $M$ and $\frac{d}{dt}|_{t_0} \in T_{t_o}I = \mathbb{R}$ being the base vector of $T_{t_o}I = \mathbb{R}$ it is

$$\gamma'(t_0) = d\gamma\Big|_{t_0}\left(\frac{d}{dt}\Big|_{t_0}\right) \in T_{\gamma(t_0)}M \tag{17}$$

and $\gamma'(t_0)$ is called the *velocity* of $\gamma$ at $t_0$ (sometimes also denoted by $\dot{\gamma}(t_0)$).

If the algebraic interpretation of a tangent vector as a derivative is taken, a tangent vector $v$ is mapped by $df_p$ to the derivation $df_p(v)$ as follows: for $\alpha \in C^\infty(N)$ it is $\alpha \circ f \in C^\infty(M)$. Thus,

$$\left(df_p(v)\right)(\alpha) = v\left(\alpha \circ f\right) \tag{18}$$

is a derivation at $f(p)$.

However, if charts $(U, \varphi)$ of $M$ around $p$ and $(V, \psi)$ of $N$ around $f(p)$ are given the Jacobian matrix $J_{\psi \circ f \circ \varphi^{-1}}(p)$ of $\psi \circ f \circ \varphi^{-1}$ as a map from $\mathbb{R}^m$ to $\mathbb{R}^n$ (with $\dim M = m$ and $\dim N = n$) represents the differential $df_p$: properties of $df_p$ like its rank correspond to the corresponding properties of $J_{\psi \circ f \circ \varphi^{-1}}(p)$ - these properties indicate key properties of the map $f$ (see [7], Section 4.6). It is said that "in local coordinates" $df_p$ is the Jacobian matrix $J_{\psi \circ f \circ \varphi^{-1}}(p)$.

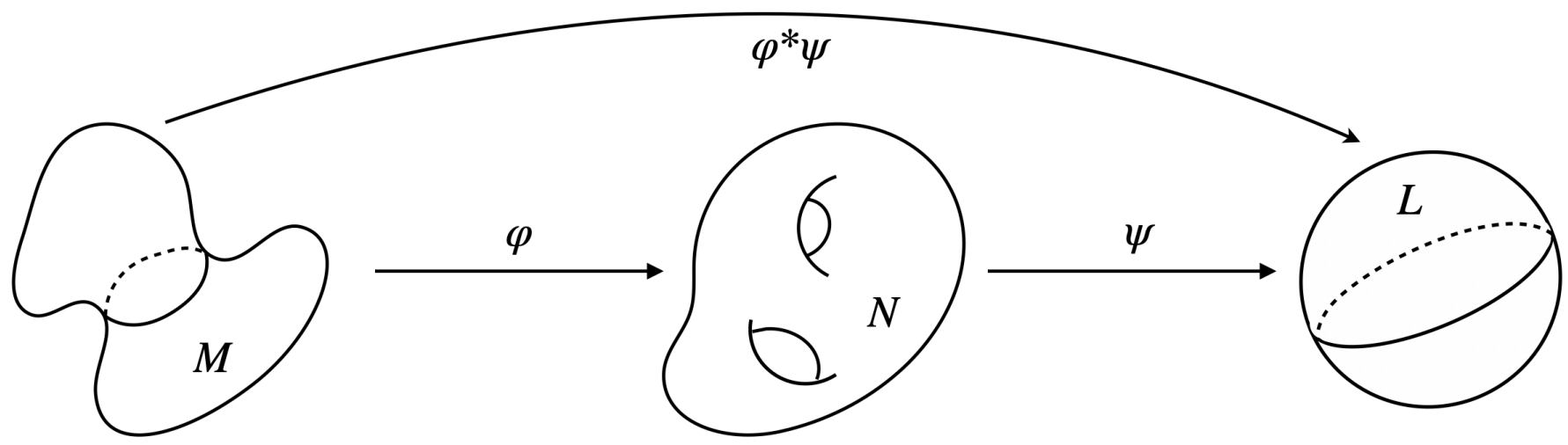

**Fig. 10**. The Differential of a Differentiable Map Between Manifolds

Similar to pushing tangent vectors forward, "pulling back" maps is often of interest (see Figure 10):

**Definition 15:** Let $L, M, N$ be differentiable manifolds and let $\varphi : M \to N$ and $\psi : N \to L$ be differentiable maps. $\varphi^*\psi : M \to L$ with $(\varphi^*\psi)(x) := \psi(\varphi(x))$ is called *pullback* of $\psi$ by $\varphi$. □

## 4. Lie Derivative of Vector Fields

Vector fields describe the dynamics of differentiable manifolds and are key to understand their geometric structure. Especially, this requires to be able to measure their change, i.e. their differential structure.

### 4.1. Vector Fields

Informally, a vector field on a manifold is a differentiable map that assigns to each point of the manifold a tangent vector. To be able to speak about differentiability of such a map Definition 10 is used which defines differentiability of maps between manifolds. Next, Lemma 6 guarantees that the tangent bundle $TM$ of a manifold $M$ is a manifold. Thus, differentiability of maps $M \to TM$ is defined: a vector field is a differentiable map $\mathscr{V} : M \to TM$. To ensure that a point $p \in M$ gets a tangent vector at $p$ assigned, the map $\pi : TM \to M$ is defined that assigns each point $(x, v) \in TM = \bigcup_x \{x\} \times T_xM$ to its "base point" $p$.

More formally:

**Definition 16:** The map $\pi : TM \to M$, $(x, v) \mapsto x$ is called the *canonical projection* of the tangent bundle. A differentiable map $\mathscr{V} : M \to TM$ with $\pi \circ \mathscr{V} = \mathrm{id}_M$ is called a *vector field* on $M$. □

Figure 11 depicts a vector field. It also makes it obvious why a map $\mathscr{V} : M \to TM$ with $\pi \circ \mathscr{V} = \mathrm{id}_M$ is called a *section* of the tangent bundle, in general.

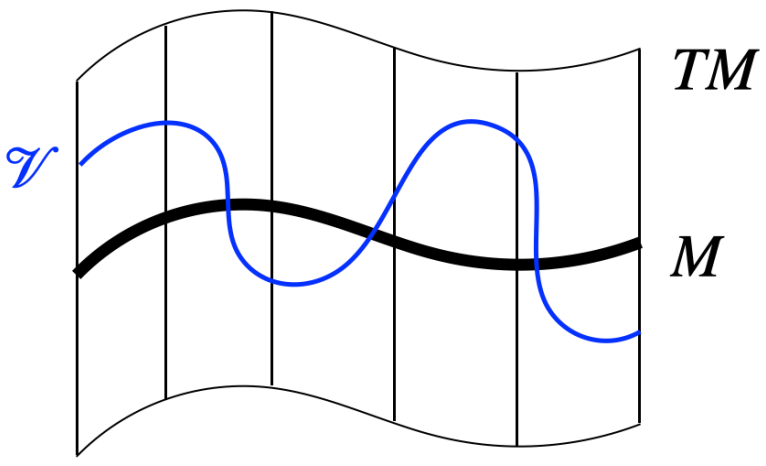

**Fig. 11**. A Vector Field on $M$

Vector fields can be added and multiplied by scalars, i.e. the sum of two vector fields is again a vector field (i.e. $(\mathscr{V} + \mathscr{W})(p) = \mathscr{V}(p) + \mathscr{W}(p)$) and a real multiple of a vector field is again vector field (i.e. $(a\mathscr{V})(p) = a\mathscr{V}(p)$). But even more, for any differentiable function $f \in C^\infty(M)$ defining

$$(f\mathscr{V})(p) := f(p)\mathscr{V}(p) \tag{19}$$

shows that $f\mathscr{V}$ is again a vector field on $M$. If $\mathfrak{X}(M)$ denotes the set of all vector fields on $M$ we get (see [22], Prop. 8.8):

**Lemma 7**: $\mathfrak{X}(M)$ is a module over the ring $C^\infty(M)$. ∎

Often, for a vector field $\mathscr{V} \in \mathfrak{X}(M)$ instead of $\mathscr{V}(p)$ we also may write

$$\mathscr{V}|_p := \mathscr{V}(p) \tag{19}$$

As an element of the tangent space, $\mathscr{X}(p) \in T_pM$ can be represented as a linear combination of the basis $\left\{\partial_1|_p, \ldots, \partial_n|_p\right\}$ of $T_pM$, i.e. $\mathscr{X}(p) = \sum_i X_i(p)\partial_i|_p$. Even more, since $\mathscr{X} \in \mathfrak{X}(M)$ is a differentiable map which turns the $X_i$ into differentiable functions around $p$ (see [22], Prop. 8.1):

**Lemma 8**: $\mathscr{X} \in \mathfrak{X}(M) \Leftrightarrow \mathscr{X}(p) = \sum_i X_i(p)\partial_i|_p$ with $X_i \in C^\infty(U)$ and a chart $(U, \varphi)$ around $p$. ∎

I.e. each vector field $\mathscr{X} \in \mathfrak{X}(M)$ is a linear combination of the vector fields $\left\{\partial_1, \ldots, \partial_n\right\}$. The functions $X_i \in C^\infty(U)$ are called *component functions* of $\mathscr{X}$.

Furthermore, each tangent vector $v \in T_pM$ can be extended to vector field on $M$ (see [22], Prop. 8.7):

**Lemma 9**: For each $v \in T_pM$ there exists a $\mathscr{V} \in \mathfrak{X}(M)$ such that $\mathscr{V}(p) = v$. ∎

Vector fields are generalizations of maps $\mathbb{R}^n \to \mathbb{R}^n$: $\mathbb{R}^n$ is a differentiable manifold (e.g. with the atlas $\mathfrak{A} = \{(\mathbb{R}^n, \mathrm{id})\}$). The tangent space at $p \in \mathbb{R}^n$ is $T_p\mathbb{R}^n \cong \mathbb{R}^n$. Thus, it is $T\mathbb{R}^n = \bigcup_{x\in M}\{x\} \times T_x\mathbb{R}^n = \bigcup_{x\in M}\{x\} \times \mathbb{R}^n \approx \mathbb{R}^n \times \mathbb{R}^n$. For a vector field $\mathscr{V} \in \mathfrak{X}(\mathbb{R}^n)$ it is $\mathscr{V}(x) = (x, F(x)) \in \mathbb{R}^n \times \mathbb{R}^n$ with a differentiable map $F : \mathbb{R}^n \to \mathbb{R}^n$. Consequently, a vector field $\mathscr{V} \in \mathfrak{X}(\mathbb{R}^n)$ can be considered as a map $F : \mathbb{R}^n \to \mathbb{R}^n$ with the value $F(x)$ being "glued" to the point $x$ (see Figure 12).

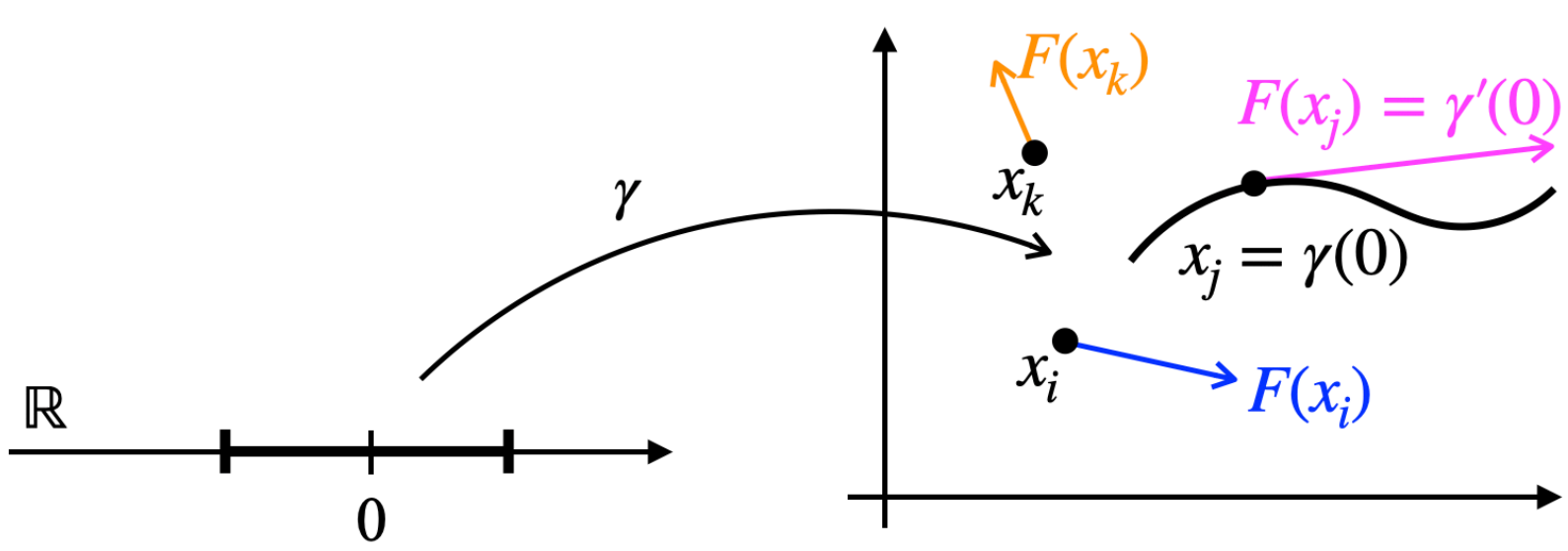

**Fig. 12**. A Map $F : \mathbb{R}^n \to \mathbb{R}^n$ as a Vector Field, and an Integral Curve $\gamma$

### 4.2. Flows

Vector fields give rise to curves whose tangent vectors are the values of the corresponding vector field. More precisely: Let $\mathscr{V} \in \mathfrak{X}(M)$; for each $p \in M$ there exists a maximal interval $I^{(p)} \subseteq_{\text{open}} \mathbb{R}$, $0 \in I^{(p)}$ and a unique curve $\gamma : I^{(p)} \to M$ with $\gamma(0) = p$ and $\gamma'(t) = \mathscr{V}(\gamma(t))$ — this is a result from the theory of ordinary differential equations (see, for example, [1], Th. 4.1.2, Th. 4.1.3). In Figure 12 the curve $\gamma$ satisfies $\gamma(0) = x_j$ and $\gamma'(t) = F(\gamma(t))$ for $F \in \mathfrak{X}(\mathbb{R}^n)$.

**Definition 17:** Let $\mathscr{V}$, $p$ and $\gamma$ be as before. $\gamma$ is called *integral curve* of $\mathscr{V}$ with *starting point* $p$. □

Collecting all the intervals of the integral curves of a vector field results in a set $D := \bigcup_{p \in M} I^{(p)} \times \{p\}$ with $D \subseteq_{\text{open}} \mathbb{R} \times M$ (see [22], p. 213) - see Figure 13.

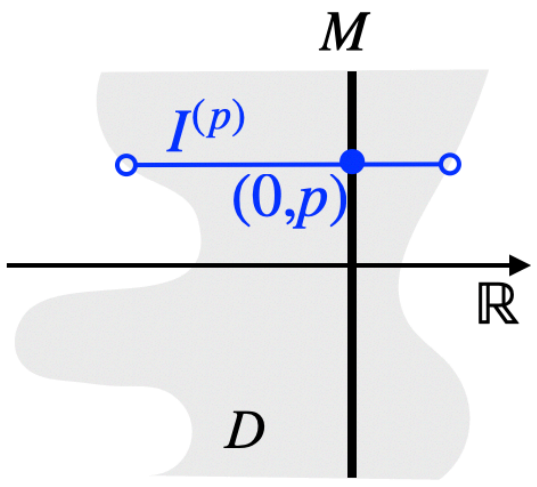

**Fig. 13**. The Domain of a Flow

The set $D$ is the domain of a map that moves point across the corresponding manifold in special manner:

**Definition 18:** A differentiable map $\Phi : D \to M$ is called a *flow* on M :⇔
i. $\Phi(0,p) = p$ and
ii. for $s \in I^{(p)}$ and $t \in I^{(\Phi(s,p))}$ with $s + t \in I^{(p)}$: $\Phi(t, \Phi(s,p)) = \Phi(t + s, p)$. □

The name "flow" indicates that $\Phi$ describes the movement of a particle at location $p$ during the time $t$. Condition (i) says that at the beginning (i.e. at $t = 0$) the observation of the movement starts at $p$. Condition (ii) means that moving the point $p$ for a time $s$ and after that moving the point $\Phi(s,p)$ reached by another time $t$ results in the point $\Phi(t, \Phi(s,p))$ which is the same point that is reached when moving $p$ from the beginning for the time $t + s$, i.e. $\Phi(t + s, p)$ is reached.

Vector fields result in such flows when particles are moved along the integral curves of the vector field (see [22], Th. 9.12):

**Theorem 2 (Fundamental Theorem on Flows)**: Let $\mathscr{V} \in \mathfrak{X}(M)$ be a vector field. There exists a unique maximal flow $\Phi_{\mathscr{V}} : D \to M$ with $\Phi_{\mathscr{V}}^{(p)} : I^{(p)} \to M$ and $\Phi_{\mathscr{V}}^{(p)}(t) := \Phi_{\mathscr{V}}(t,p)$ such that $\Phi_{\mathscr{V}}^{(p)}$ is integral curve of $\mathscr{V}$ with starting point $p$. $\Phi_{\mathscr{V}}$ is called the *flow of* $\mathscr{V}$. ∎

Descriptively, the flow $\Phi_{\mathscr{V}}$ of a vector field $\mathscr{V}$ deforms a subset $S$ of a manifold smoothly along the integral curves of the vector field: in Figure 14 the pink set $S$ is deformed after time $t$ into the blue set $\Phi_{\mathscr{V}}^{(S)}(t)$. This is because each point of the set $S$, e.g. $p$ and $q$ shown, is moved after time $t$ to the points $\Phi_{\mathscr{V}}^{(p)}(t)$ and $\Phi_{\mathscr{V}}^{(q)}(t)$ along the integral curves $\Phi_{\mathscr{V}}^{(p)}$ and $\Phi_{\mathscr{V}}^{(q)}$ starting at $p$ and $q$.

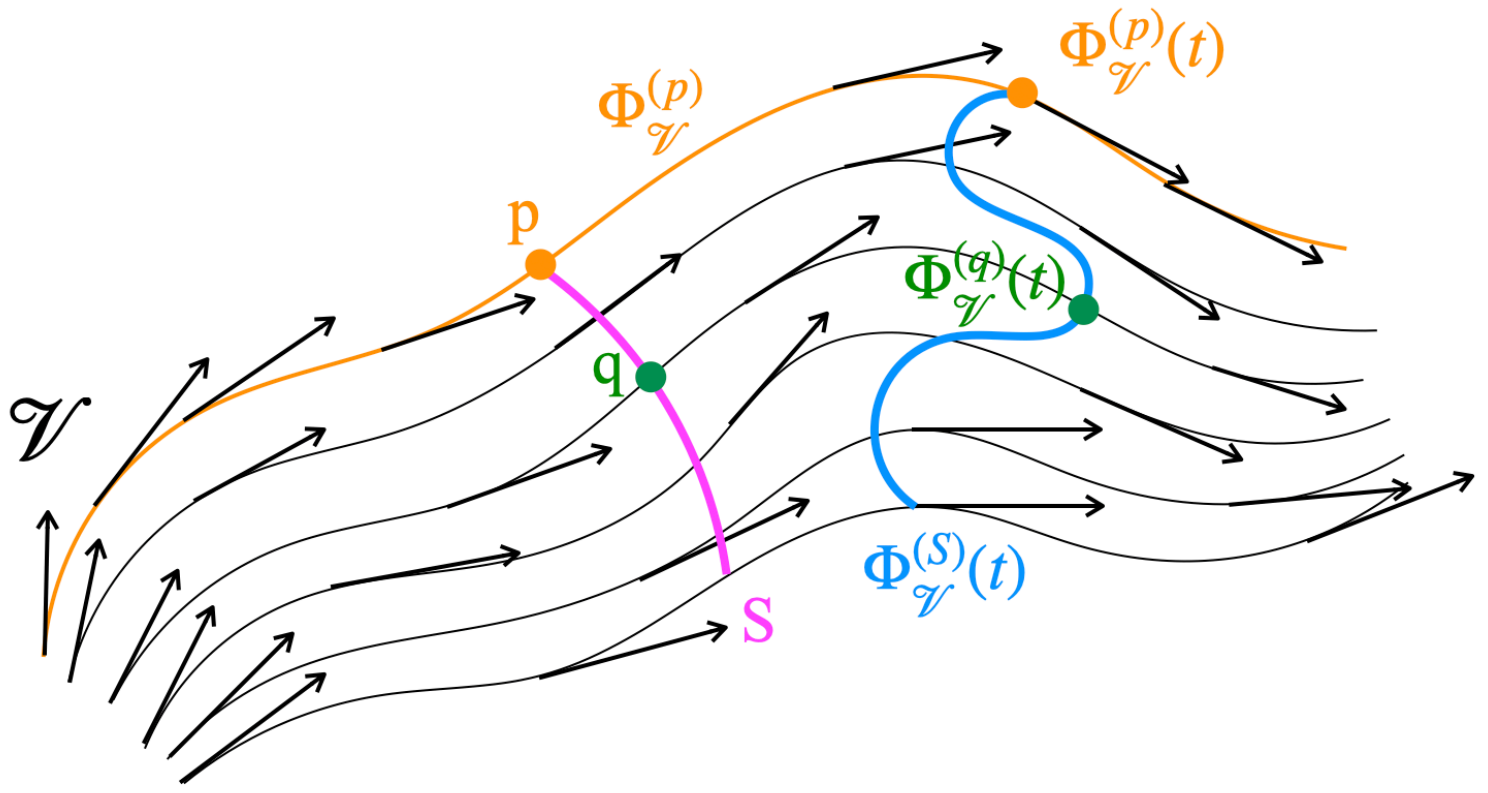

**Fig. 14**. Effects of the Flow of a Vector Field

Flows induce local diffeomorphisms. More precisely, we define the set $M_t := \{p \in M \mid (t,p) \in D\}$, i.e. all points of the manifold the can be moved by the flow for a time $t$ (Figure 15); it can be proven that this is an open set: $M_t \subseteq_{\text{open}} M$ (see [22], Th. 9.12).

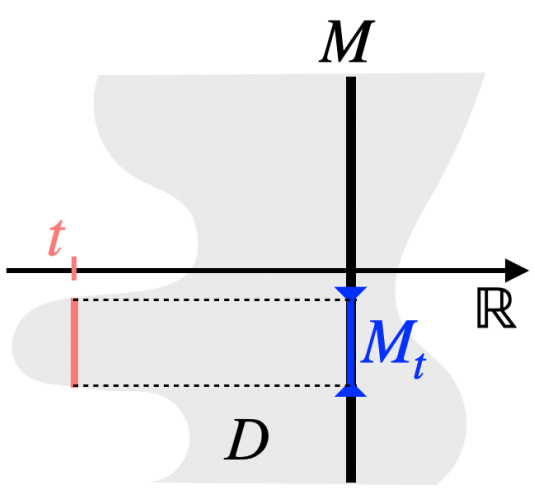

**Fig. 15**. Time-Dependent Domains of Flows

Furthermore, define $\Phi_{\mathscr{V},t} : M_t \to M_{-t}$ via $\Phi_{\mathscr{V},\,t}(p) := \Phi_{\mathscr{V}}(t,p)$. Then (see [22], Th. 9.12):

**Lemma 10**: Let $\mathscr{V} \in \mathfrak{X}(M)$ be a vector field. Then $\Phi_{\mathscr{V},t} : M_t \to M_{-t}$ is a diffeomorphism, with $(\Phi_{\mathscr{V},t})^{-1} = \Phi_{\mathscr{V},-t}$. ∎

Flows are of interest whose integral curves are define for all of $\mathbb{R}$.

**Definition 19:** A flow is called *global* :⇔ $\forall\, p \in M : I^{(p)} = \mathbb{R}$ (i.e. $D = \mathbb{R} \times M$, thus $M_t = M$) □

If the flow of a vector field $\Phi_{\mathscr{V}} : D \to M$ is global it is $M_t = M$, i.e. $\Phi_{\mathscr{V},t} : M \to M$ is for each $t \in \mathbb{R}$ a diffeomorphism of the whole manifold. Such vector fields have a special name:

**Definition 20:** A vector field $\mathscr{V} \in \mathfrak{X}(M)$ is called *complete* :⇔ Its flow $\Phi_{\mathscr{V}}$ is global. □

Vector fields on compact manifolds are always complete (see [22], Th. 9.16):

**Lemma 11**: Let $M$ be compact, $\mathscr{V} \in \mathfrak{X}(M) \Rightarrow \mathscr{V}$ is complete, i.e. $\Phi_{\mathscr{V}}$ is global. ∎

### 4.3. Lie Derivative

The directional derivative $D_v F(p)$ of a vector field $F \in \mathfrak{X}(\mathbb{R}^n)$ in direction of $v \in \mathbb{R}^n$ is defined in analogy to Definition 7 (see also Equation 13) as

$$D_v F(p) = \lim_{t \to 0} \frac{F(p + t\,v) - F(p)}{t} \tag{20}$$

The denominator of the fraction is the difference of the two vectors $F(p + t\,v) \in \mathbb{R}^n$ and $F(p) \in \mathbb{R}^n$. Especially, $F$ is evaluated along the path $\delta(t) = p + t\,v$ for small $t$, reaching $p$ at $t = 0$ and having the tangent vector $\delta'(0) = v$.

When trying to transfer this definition naively to arbitrary manifolds $M$ the following problem immediately occurs: For a vector field $\mathscr{W} \in \mathfrak{X}(M)$ and for two points $p \neq q \in M$ it is $\mathscr{W}(p) \in T_pM$ and $\mathscr{W}(q) \in T_qM$. But $T_pM \neq T_qM$, i.e. the vectors are elements of different vector spaces, thus, the difference of the vectors $\mathscr{W}(p), \mathscr{W}(q)$ can not be computed at all. Consequently, the expression $\mathscr{W}(p) - \mathscr{W}(q)$ is meaningless for $p \neq q$, i.e. the denominator and, thus, the whole fraction of the equation cannot be build.

Towards solving this problem, a vector field $\mathscr{V} \in \mathfrak{X}(M)$ is chosen with $\mathscr{V}(p) = v$ (Lemma 9) where $v \in T_pM$ is the vector in which direction the derivative of $\mathscr{W}$ should be taken. Next, the curve $\delta(t) = p + t\,v$ is substituted by the integral curve $\Phi^{(p)}_{\mathscr{V}}$ of $\mathscr{V}$ starting at $p$ (Theorem 2); denoting this curve as $\gamma := \Phi^{(p)}_{\mathscr{V}}$ it is $\gamma'(0) = v$ and $\gamma(0) = p$. Furthermore, $\gamma_t := \Phi_{\mathscr{V},t}$ is a diffeomorphism $\gamma_t : M_t \to M_{-t}$ with $(\gamma_t)^{-1} = (\Phi_{\mathscr{V},t})^{-1} = \Phi_{\mathscr{V},-t} = \gamma_{-t}$ (Lemma 10). Thus, the differential of $\gamma_{-t}$ is a map $d(\gamma_{-t}) : T_{\gamma_t(p)}M \to T_{\gamma_{-t}(\gamma_t(p))}M = T_pM$, pushing the vector $\mathscr{W}(\gamma_t(p)) \in T_{\gamma_t(p)}M$ to a vector in $T_pM$ where the difference $d(\gamma_{-t})\mathscr{W}(\gamma_t(p)) - \mathscr{W}(p)$ is defined and can be computed. The situation is depicted in Figure 16.

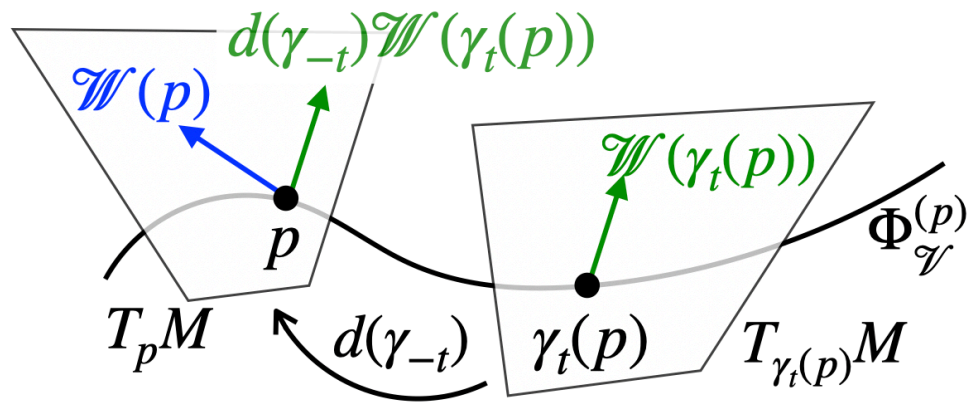

**Fig. 16**. Ingredients of Building the Lie-Derivative

Together, this allows the following generalization of a directional derivative:

**Definition 21:** The *Lie derivative* of $\mathscr{W} \in \mathfrak{X}(M)$ w.r.t. $\mathscr{V} \in \mathfrak{X}(M)$ at $p$ is the limit

$$(\mathscr{L}_{\mathscr{V}}\mathscr{W})_p = \lim_{t\to 0}\frac{d(\gamma_{-t})\mathscr{W}(\gamma_t(p)) - \mathscr{W}(p)}{t} \left( = \frac{d}{dt}\bigg|_{t=0} d(\gamma_{-t})_{\gamma_t(p)}(\mathscr{W}(\gamma_t(p))\right) \quad (21)\ \square$$

$(\mathscr{L}_{\mathscr{V}}\mathscr{W})_p$ measures how $\mathscr{W}$ changes when being moved “infinitesimally” from $p$ along the integral curve $\Phi^{(p)}_{\mathscr{V}}$.

When comparing Equation 21 and Equation 20, the question comes up why the differential $d(\gamma_{-t}) : T_{\gamma_t(p)}M \to T_pM$ is considered to move $\mathscr{W}(\gamma_t(p))$ "unmodified" from $T_{\gamma_t(p)}M$ back to $T_pM$? This can be understood very informative, very illustrative as follows (see Figure 17): Locally around $p \in M$, $M$ is approximated by the tangent space $T_pM$, i.e. there is $V \subseteq_{\text{open}} T_pM$ and $U \subseteq_{\text{open}} M$ with $p \in U$ and $0 \in V$ such that $U$ and $V$ are diffeomorphic (see [27], Prop. 4.4.4). Also, a differentiable map $f : M \to M$ is approximated locally around $p$ by its differential $df_p : T_pM \to T_{f(p)}M$, i.e. the effects of $f$ in $U$ correspond nearly to the effects of the linear map $df_p$. Consequently, for small $t$ the point $\gamma_t(p)$ is close to $p = \gamma(0)$, thus, $\gamma_t$ can be approximated by $d(\gamma_t)$. Since $\gamma_0 = \text{id}$ it is $d(\gamma_{-t}) \approx \text{id}$ for small $t$, thus, $d(\gamma_{-t})\mathscr{W}(\gamma_t(p)) \approx \mathscr{W}(\gamma_t(p))$ for small $t$: thus, $\mathscr{W}(\gamma_t(p))$ is moved by $d(\gamma_{-t})$ nearly unmodified.

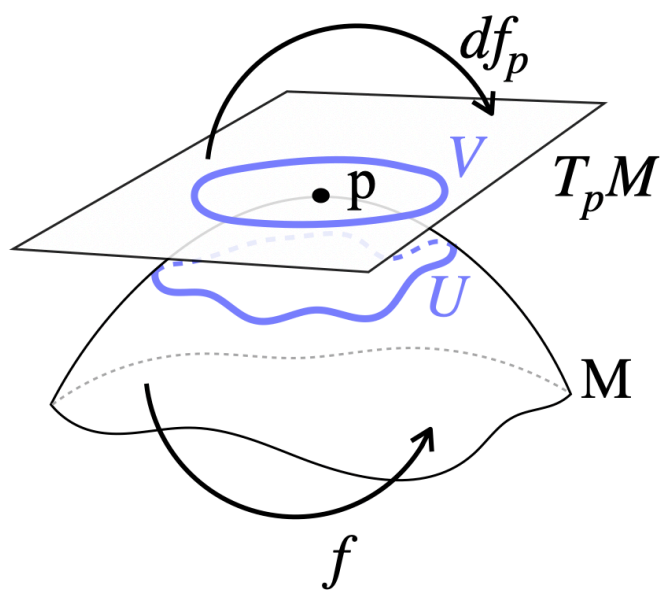

**Fig. 17**. Linear Approximations of Differentiable Maps and Manifolds

The Lie derivative results in a vector field (see [22], Th. 9.36):

**Lemma 12**: $\mathscr{L}_{\mathscr{V}}\mathscr{W} \in \mathfrak{X}(M)$. ∎

It is important to note that the Lie derivative $(\mathscr{L}_{\mathscr{V}}\mathscr{W})_p$ of $\mathscr{W}$ w.r.t. $\mathscr{V}$ at $p$ requires the knowledge about the vector <u>field</u> $\mathscr{V}$ in a neighborhood of $p$, i.e. just the vector $\mathscr{V}(p)$ (i.e. the direction of $\mathscr{V}$ at $p$) is <u>not</u> enough. In $\mathbb{R}^n$, the directional derivative $D_vF$ requires a vector $v$ only, not a vector field. If only a tangent <u>vector</u> $v \in T_pM$ should suffice to determine the change of $\mathscr{W}$ in direction of $v$ an additional structure on the manifold is needed — that is discussed in chapter 5 “Connections”.

### 4.4. Lie Bracket

According to Lemma 8, for $\mathscr{X} \in \mathfrak{X}(M)$ there exist smooth functions $X_i \in C^\infty(M)$ such that

$$\mathcal{X} = \sum_i X_i \frac{\partial}{\partial x_i} \tag{22}$$

As a tangent vector, $\mathcal{X}(p) \in T_pM$ is a derivation (Lemma 5), i.e. for $f \in C^\infty(M)$ it is

$$\mathcal{X}(p)f = \sum_i X_i(p) \frac{\partial f}{\partial x_i}(p) \tag{23}$$

Thus, when defining

$$(\mathcal{X}f)(p) := \mathcal{X}(p)f = \sum_i X_i(p) \frac{\partial f}{\partial x_i}(p) \tag{24}$$

$\mathcal{X}f$ becomes a smooth function: $\mathcal{X}f \in C^\infty(M)$.This means that $\mathcal{X}f$ can again be used as argument of a derivation.

Thus, let $\mathcal{V}, \mathcal{W} \in \mathfrak{X}(M)$ and let $f \in C^\infty(M)$. Then it is $\mathcal{W}f \in C^\infty(M)$, thus, it is $\mathcal{V}\mathcal{W}f \in C^\infty(M)$, and similarly $\mathcal{W}\mathcal{V}f \in C^\infty(M)$, which finally this means that $\mathcal{V}\mathcal{W}f - \mathcal{W}\mathcal{V}f \in C^\infty(M)$.

**Definition 22:** With $f \in C^\infty(M)$ define $[\mathcal{V}, \mathcal{W}]f := \mathcal{V}\mathcal{W}f - \mathcal{W}\mathcal{V}f$. This defines a map $[\mathcal{V}, \mathcal{W}] : C^\infty(M) \to C^\infty(M)$ the so-called *Lie bracket* of $\mathcal{V}$ and $\mathcal{W}$. □

The Lie bracket results in a vector field (see [22], Th. 8.25):

**Lemma 13**: $\mathcal{V}, \mathcal{W} \in \mathfrak{X}(M) \Rightarrow [\mathcal{V}, \mathcal{W}] \in \mathfrak{X}(M)$. ∎

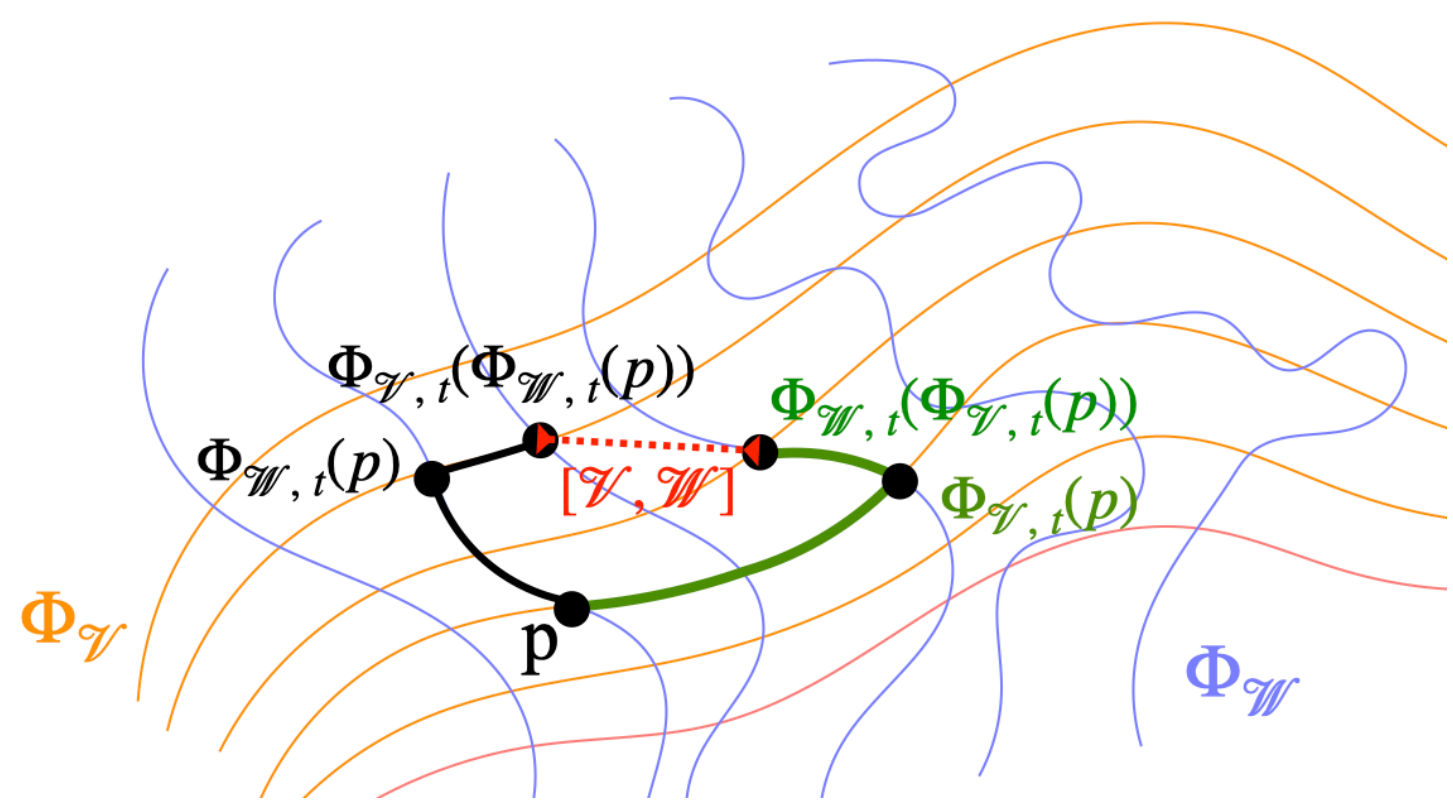

**Fig. 18**. Two Non-Commutative Flows and The Meaning of the Lie Bracket

The Lie bracket has a nice geometric interpretation (see Figure 18): Let $p \in M$ be a point that flows for a time $t$ via $\Phi_{\mathcal{V}}$ to $\Phi_{\mathcal{V},t}(p)$, and after this time $\Phi_{\mathcal{V},t}(p)$ flows via $\Phi_{\mathcal{W}}$ to $\Phi_{\mathcal{W},t}(\Phi_{\mathcal{V},t}(p))$. Similarly, $p \in M$ flows for the time $t$ via $\Phi_{\mathcal{W}}$ to $\Phi_{\mathcal{W},t}(p)$, then this point flows via $\Phi_{\mathcal{V}}$ to $\Phi_{\mathcal{V},t}(\Phi_{\mathcal{W},t}(p))$. I.e. the attempt is to create some sort of parallelogram by following integral curves. But in general, it is $\Phi_{\mathcal{W},t}(\Phi_{\mathcal{V},t}(p)) \neq \Phi_{\mathcal{V},t}(\Phi_{\mathcal{W},t}(p))$. It is said that the flows $\Phi_{\mathcal{V}}$ and $\Phi_{\mathcal{W}}$ are not commutative. I.e. in general no parallelogram can be built this way, a gap between the two endpoints reached remains: the distance of these two endpoints corresponds to the Lie bracket (see Lemma 14).

**Definition 23:** Two flows $\Phi_{\mathscr{V}}$ and $\Phi_{\mathscr{W}}$ are said to *commute* :⇔ For properly chosen (small) $s, t$ it is $\Phi_{\mathscr{V},\,t} \circ \Phi_{\mathscr{W},\,s} = \Phi_{\mathscr{W},\,s} \circ \Phi_{\mathscr{V},\,t}$ . □

The proper selection of $s$ and $t$ is important for non-global flows to ensure that the composition of the flows is defined. Similarly:

**Definition 24:** $\mathscr{V}, \mathscr{W} \in \mathfrak{X}(M)$ *commute* :⇔ $[\mathscr{V}, \mathscr{W}] = 0$ . □

The following lemma shows that two vector fields commute if and only if their corresponding flows commute (see [22], Th. 9.44):

**Lemma 14**: $\mathscr{V}, \mathscr{W}$ commute ⇔ $\Phi_{\mathscr{V}}$ and $\Phi_{\mathscr{W}}$ commute ∎

This means that at an infinitesimal scale, $[\mathscr{V}, \mathscr{W}]_p$ is some sort of measure of the difference between $\Phi_{\mathscr{V},\,t}(\Phi_{\mathscr{W},\,t}(p))$ and $\Phi_{\mathscr{W},\,t}(\Phi_{\mathscr{V},\,t}(p))$. Thus, the Lie bracket of two vector fields measures the gap when trying to construct a parallelogram out of the corresponding integral curves of the two vector fields (see Figure 18).

There is also a nice relation between the Lie bracket and the Lie derivative. According to Definition 21, in order to compute the Lie derivative $\mathscr{L}_{\mathscr{V}}\mathscr{W}$ for $\mathscr{V}, \mathscr{W} \in \mathfrak{X}(M)$ the flow $\Phi_{\mathscr{V}}$ must be known. But a flow, i.e. an analytical expressions of it, is often hard to determine. Luckily, computing the flows is not necessary (see [22], Th. 9.38):

**Lemma 15**: Let $\mathscr{V}, \mathscr{W} \in \mathfrak{X}(M)$. Then: $\mathscr{L}_{\mathscr{V}}\mathscr{W} = [\mathscr{V}, \mathscr{W}]$. ∎

Computing Lie brackets is somehow “easier”: Similar to Equation 22, let $\mathscr{V} = \sum_i V_i \frac{\partial}{\partial x_i}$ and $\mathscr{W} = \sum_i W_i \frac{\partial}{\partial x_i}$. Then (see [22], Th. 8.26):

**Lemma 16**: Let $\mathscr{V}, \mathscr{W} \in \mathfrak{X}(M)$. Then:

$$[\mathscr{V}, \mathscr{W}] = \sum_{i=1}^{n} \left( \sum_{j=1}^{n} \left( V_j \frac{\partial W_i}{\partial x_j} - W_j \frac{\partial V_i}{\partial x_j} \right) \right) \frac{\partial}{\partial x_i} \tag{25}$$ ∎

Lemma 16 and Lemma 15 together imply that

$$\mathscr{L}_{\mathscr{V}}\mathscr{W} = \sum_{i=1}^{n} \left( \sum_{j=1}^{n} \left( V_j \frac{\partial W_i}{\partial x_j} - W_j \frac{\partial V_i}{\partial x_j} \right) \right) \frac{\partial}{\partial x_i} \tag{26}$$

### 4.5. Lie Derivative of a Function

Finally, the Lie derivative of a function is defined. As before, let $\mathscr{V} \in \mathfrak{X}(M)$ be a vector field on $M$, let $\gamma_t := \Phi_{\mathscr{V},t}$ be the induced diffeomorphism, and let $f \in C^{\infty}(M)$.

**Definition 25:** $(\mathscr{L}_{\mathscr{V}} f)_p := \left.\frac{d}{dt}\right|_{t=0} (\gamma_t^* f)_p$ is called *Lie derivative* of $f$ w.r.t. $\mathscr{V}$. □

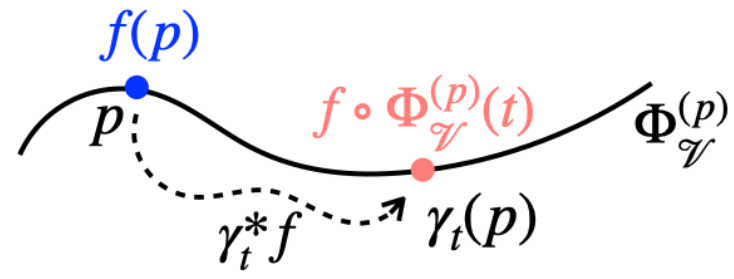

**Fig. 19**. Ingredients of the Lie Derivative of a Function

In this definition $\gamma_t^* f$ denotes the pullback of $f$ by $\gamma_t$ as introduced by Definition 15, i.e. $\gamma_t^* f = f \circ \gamma_t$ (see Figure 19). As before this requires knowledge of the flow $\Phi_{\mathscr{V}}$ and an analytical expressions of it which is typically difficult to compute. But corresponding computations can be simplified:

With $(\gamma_t^* f)_p = \gamma_t^* f(p) = f(\gamma_t(p)) = f \circ \Phi_{\mathscr{V},t}(p) = f \circ \Phi_{\mathscr{V}}^{(p)}(t)$, it is

$$
\begin{aligned}
(\mathscr{L}_{\mathscr{V}} f)_p &= \left.\frac{d}{dt}\right|_{t=0} \left(f \circ \Phi_{\mathscr{V}}^{(p)}(t)\right) \\
&= \left(f \circ \Phi_{\mathscr{V}}^{(p)}\right)'(0) \\
&\overset{(1)}{=} d\left(f \circ \Phi_{\mathscr{V}}^{(p)}\right)\Big|_{t=0} \left(\left.\frac{d}{dt}\right|_{t=0}\right) \\
&= df \circ d\Phi_{\mathscr{V}}^{(p)}\Big|_{t=0} \left(\left.\frac{d}{dt}\right|_{t=0}\right) \\
&= df \circ d\gamma_t\Big|_{t=0} \left(\left.\frac{d}{dt}\right|_{t=0}\right) \\
&= df(\gamma'(0)) = df(\mathscr{V}(p)) = \mathscr{V} f(p)
\end{aligned}
$$

where (1) is valid because of Equation 17. Thus it is (see also Equation 24):

**Lemma 17**: $\mathscr{L}_{\mathscr{V}} f = \mathscr{V} f$. ∎

## 5. Connections

Lie derivatives are directional derivatives in direction of a vector field. Its definition requires the flow defined by this vector field. Also, when taking the derivative at a point the field providing the direction must be known in a neighborhood of this point. Often, just a vector (not a field) should suffice to compute a directional derivative like taking directional derivatives in Euclidian space. This requires an additional structure on the manifold, a so-called “connection”.

### 5.1. Tangential Directional Derivative

Let $M \subset \mathbb{R}^n$ be an embedded manifold, i.e. a manifold with $\mathbb{R}^n$ as “ambient” space. Let $\mathcal{X} \in \mathfrak{X}(M)$, $v \in T_pM$ for $p \in M$. We want to define a directional derivative of $\mathcal{X}$ w.r.t. a vector $v$. As seen before (section 4.3), this can't be done by means of differences of values of $\mathcal{X}$ at different points because these values are elements in different tangent spaces of the manifold $M$.

Let $F : \mathbb{R}^n \to \mathbb{R}^n$ be a vector field and $F|_M = \mathcal{X}$ (i.e. $F$ is an extension of $\mathcal{X}$). With $F = (F_1, \ldots, F_n)$ we set $\overline{\nabla}_v F|_p := (\partial_v F_1(p), \ldots, \partial_v F_n(p))$ as the directional derivative of $F$ w.r.t. $v$ (see Equation 13, but here a different notation is used). This is well-defined based of the "usual" directional derivative in $\mathbb{R}^n$, but the result is not necessarily tangential to $M$ (see Figure 20), i.e. we can not set $\overline{\nabla}_v \mathcal{X}|_p := \overline{\nabla}_v F|_p$. Thus, the result $\overline{\nabla}_v F|_p$ is orthogonally projected onto the tangent space of $M$ at $p$ — and this is then considered to be the directional derivative of $\mathcal{X}$. I.e. we define $\nabla_v^{\perp} \mathcal{X}|_p := \pi_{\perp,p}(\overline{\nabla}_v F|_p)$ where $\pi_{\perp,p}$ is the orthogonal projection onto the tangent space $T_pM$.

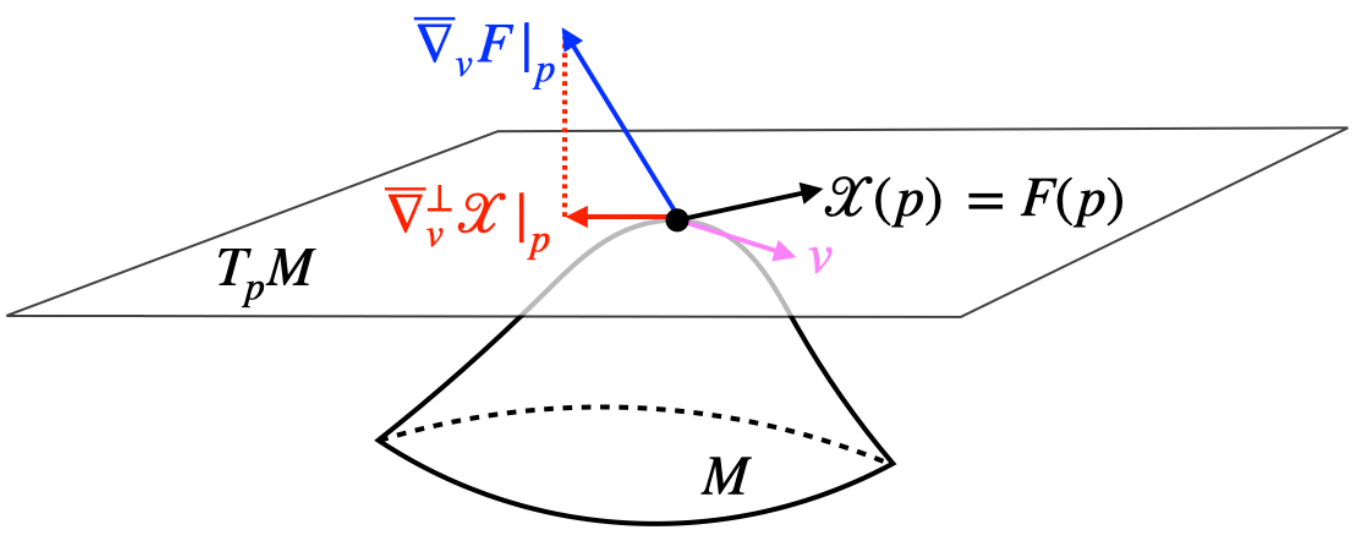

**Fig. 20**. Ingredients of the Lie Derivative of a Function

Straightforward computations prove:

**Note 3**: $\nabla^{\perp} : \mathfrak{X}(M) \times \mathfrak{X}(M) \to \mathfrak{X}(M)$ has the following properties:

1. $\nabla_{\mathcal{X}}^{\perp} \mathcal{Y}$ is linear over $C^{\infty}(M)$ in $\mathcal{X}$, i.e. for $\mathcal{X}_1, \mathcal{X}_2, \mathcal{Y} \in \mathcal{X}(M)$ and $f_1, f_2 \in C^{\infty}(M)$ it is

$$\nabla_{f_1\mathcal{X}_1 + f_2\mathcal{X}_2}^{\perp} \mathcal{Y} = f_1 \nabla_{\mathcal{X}_1}^{\perp} \mathcal{Y} + f_2 \nabla_{\mathcal{X}_2}^{\perp} \mathcal{Y}$$

2. $\nabla_{\mathcal{X}}^{\perp} \mathcal{Y}$ is linear over $\mathbb{R}$ in $\mathcal{Y}$, i.e. for $\mathcal{X}, \mathcal{Y}_1, \mathcal{Y}_2 \in \mathcal{X}(M)$ and $a_1, a_2 \in \mathbb{R}$ it is

$$\nabla_{\mathcal{X}}^{\perp}(a_1\mathcal{Y}_1 + a_2\mathcal{Y}_2) = a_1 \nabla_{\mathcal{X}}^{\perp} \mathcal{Y}_1 + a_2 \nabla_{\mathcal{X}}^{\perp} \mathcal{Y}_2$$

3. $\nabla^{\perp}$ satisfies the product rule, i.e for $\mathcal{X}, \mathcal{Y} \in \mathcal{X}(M)$ and $f \in C^{\infty}(M)$ it is

$$\nabla_{\mathcal{X}}^{\perp}(f\mathcal{Y}) = f \nabla_{\mathcal{X}}^{\perp} \mathcal{Y} + (\mathcal{X}f)\mathcal{Y}$$

∎

### 5.2. Connections

A map satisfying these three properties before is considered a meaningful directional derivative as an intrinsic feature (i.e. a feature without requiring references to an ambient space) of a differentiable manifold:

**Definition 26:** A map $\nabla : \mathfrak{X}(M) \times \mathfrak{X}(M) \to \mathfrak{X}(M)$ satisfying the following properties is called a *connection* (or *covariant derivative*) of $\mathcal{Y}$ in the direction of $\mathcal{X}$ on $M$:

1. $\nabla_{f_1\mathcal{X}_1+f_2\mathcal{X}_2}\mathcal{Y} = f_1\nabla_{\mathcal{X}_1}\mathcal{Y} + f_2\nabla_{\mathcal{X}_2}\mathcal{Y}$ for $\mathcal{X}_1, \mathcal{X}_2, \mathcal{Y} \in \mathfrak{X}(M)$ and $f_1, f_2 \in C^\infty(M)$, i.e. $\nabla_{\mathcal{X}}\mathcal{Y}$ is linear over $C^\infty(M)$ in $\mathcal{X}$
2. $\nabla_{\mathcal{X}}(a_1\mathcal{Y}_1 + a_2\mathcal{Y}_2) = a_1\nabla_{\mathcal{X}}\mathcal{Y}_1 + a_2\nabla_{\mathcal{X}}\mathcal{Y}_2$ for $\mathcal{X}, \mathcal{Y}_1, \mathcal{Y}_2 \in \mathfrak{X}(M)$ and $a_1, a_2 \in \mathbb{R}$, i.e. $\nabla_{\mathcal{X}}\mathcal{Y}$ is linear over $\mathbb{R}$ in $\mathcal{Y}$
3. $\nabla_{\mathcal{X}}(f\mathcal{Y}) = (\mathcal{X}f)\mathcal{Y} + f\nabla_{\mathcal{X}}\mathcal{Y}$ for $\mathcal{X}, \mathcal{Y} \in \mathfrak{X}(M)$ and $f \in C^\infty(M)$, i.e. $\nabla$ satisfies the product rule aka Leibniz rule. □

The term "covariant" has historical origin. In the past, it should indicate that $\nabla$ follows transformation rules that makes it independent of coordinates, which is now by definition. Also, the terms Koszul connection, affine connection, or linear connection are used. The name “connection” will be justified a bit later (see the discussion following Lemma 20).

But still there seems to be a problem that we wanted to avoid when discussing Lie derivatives: $\nabla_{\mathcal{X}}\mathcal{Y}$ requires a vector <u>field</u> $\mathcal{X}$ instead of a vector only for setting the direction of the derivative. At least it can be proven that identical vector fields in a neighborhood of $p$ result in identical connections (see [23], Lemma 4.1):

**Note 4**: Let $\mathcal{X}, \mathcal{Y}, \hat{\mathcal{X}}, \hat{\mathcal{Y}} \in \mathfrak{X}(M)$ and $U \in \mathfrak{U}(p)$ with $\mathcal{X}|_U = \hat{\mathcal{X}}|_U$ and $\mathcal{Y}|_U = \hat{\mathcal{Y}}|_U$. Then, $\nabla_{\mathcal{X}}\mathcal{Y}|_p = \nabla_{\hat{\mathcal{X}}}\hat{\mathcal{Y}}|_p$. ∎

This can be restricted even further w.r.t. to the direction providing field $\mathcal{X}$ (see [23], Prop. 4.5):

**Lemma 18**: $\nabla_{\mathcal{X}}\mathcal{Y}|_p$ is uniquely determined by $\mathcal{Y}|_U$ and $\mathcal{X}(p)$. ∎

Thus, picking $v \in T_pM$ and choosing $\mathcal{X} \in \mathfrak{X}(M)$ with $\mathcal{X}(p) = v$ (Lemma 9) the former lemma shows that the following is well-defined:

$$\nabla_v\mathcal{Y} := \nabla_{\mathcal{X}}\mathcal{Y}|_p \tag{27}$$

Thus, the definition of connections only depends on a vector giving the direction of the derivative taken. Consequently, connections are generalizations of directional derivatives w.r.t. a vector.

A connection as defined above is a map with certain properties. But as of now, it is not clear from the outset that such maps exists. To prove this let $\{\partial_1, \ldots, \partial_n\}$ be a basis of $T_pM$; then the connection of $\partial_j$ w.r.t. $\partial_i$ can be expressed in this basis as

$$\nabla_{\partial_i}\partial_j|_p = \sum_k \Gamma_{ij}^k(p)\partial_k \tag{28}$$

with differentiable functions $\Gamma_{ij}^k : U \to \mathbb{R}$ (Lemma 8) called *connection coefficients* (or *Christoffel symbols*) of $\nabla$. With $\mathcal{X} = \sum X_i\partial_i$ and $\mathcal{Y} = \sum Y_i\partial_i$ a straightforward computation using the rules of Definition 26 results in

$$\nabla_{\mathcal{X}}\mathcal{Y} = \sum_k \left(\mathcal{X}(Y_k) + \sum_{i,j} X_iY_j\Gamma_{ij}^k\right)\partial_k \tag{29}$$

As a consequence this shows that any choice of $n^3$ smooth functions $\{\Gamma_{ij}^k\}$ on $M$ according to Equation 29 yields a connection $\nabla$ on $M$ ($n = \dim M$) - see also [23], Lemma 4.10 and Prop. 4.11:

**Lemma 19**: Any differentiable manifold admits a connection. ∎

In fact, any manifold admits infinitely many connections according to Equation 29. By choosing $\Gamma_{ij}^k = 0$ and $\mathcal{X}, \mathcal{Y} \in \mathfrak{X}(\mathbb{R}^n)$ we get the "standard" directional derivative in $\mathbb{R}^n$:

$$\nabla_{\mathcal{X}}\mathcal{Y}|_p = \sum_k \mathcal{X}(p)(Y_k) \overset{(*)}{=} d\mathcal{Y}(p)(\mathcal{X}(p)) \tag{30}$$

where (*) is valid because of Equation 13. This shows again that the notion of a connection is a suitable generalization of the directional derivative.

### 5.3. Parallel Transport

Let $F : \mathbb{R}^n \to \mathbb{R}^n$ be a vector field in $\mathbb{R}^n$. $F$ is constant if and only if there is a $v \in \mathbb{R}^n$ such that for all $x \in \mathbb{R}^n : F(x) = v$; especially, the differential vanishes constantly: $dF \equiv 0$. This implies that all directional derivatives of $F$ vanish (Equation 13): $\forall w \in \mathbb{R}^n : D_wF(a) \equiv 0$. Also, all values $F(x)$ are parallel (see Figure 21). This observation yields to the following definition:

**Definition 27:** A vector field $\mathcal{Y} \in \mathfrak{X}(M)$ is called *parallel* on $M$ :⇔ $\nabla_v\mathcal{Y} \equiv 0$ for each $v \in T_pM$ and $p \in M$ □

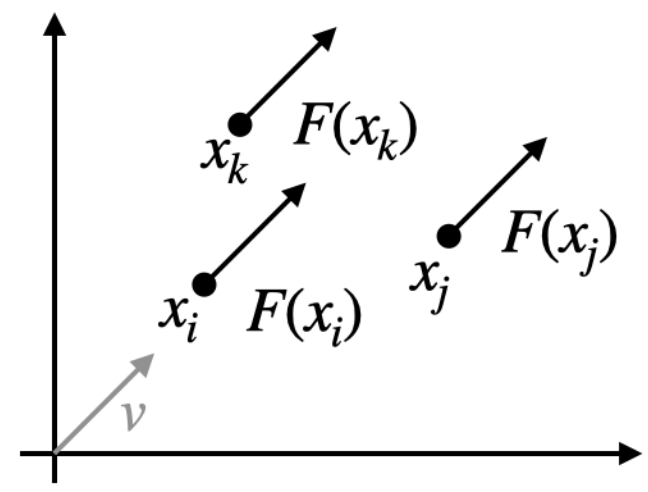

**Fig. 21**. A Parallel Vector Field in $\mathbb{R}^n$

Next, we are interested in vector fields along a curve, i.e. vector fields that are not necessarily defined on all of the manifold:

**Definition 28:** Let $\gamma : I \to M$ be a smooth curve on $M$. A smooth map $\mathscr{V} : I \to TM$ is called *vector field along* $\gamma$ $:\Leftrightarrow \forall t \in I : \mathscr{V}(t) \in T_{\gamma(t)}M$. $\mathfrak{X}(\gamma)$ denotes all vector fields along $\gamma$. □

Restricting the domain of a vector field defined on all of a manifold to the image of a curve yields a vector field along this curve: Let $\overline{\mathscr{V}} \in \mathfrak{X}(M)$ and define $\mathscr{V}(t) := \overline{\mathscr{V}}(\gamma(t))$; then $\mathscr{V} \in \mathfrak{X}(\gamma)$.

Especially, vector fields that do not change along a curve in direction of the tangent $\gamma'$ of the curve will become important:

**Definition 29:** $\mathscr{V} \in \mathfrak{X}(\gamma)$ is called *parallel along* $\gamma :\Leftrightarrow \nabla_{\gamma'(t)}\mathscr{V} \equiv 0$. □

Any vector on a curve gives raise to a unique vector field along the curve that is pointing towards that vector and that is parallel along the curve - Figure 22 depicts this situation (for a proof see [23], Th. 4.23):

**Theorem 3**: $\forall v \in T_{\gamma(t_0)}M \ \exists! \mathscr{V} \in \mathfrak{X}(\gamma) : \mathscr{V}(\gamma(t_0)) = v \wedge \mathscr{V}$ parallel along $\gamma$. ∎

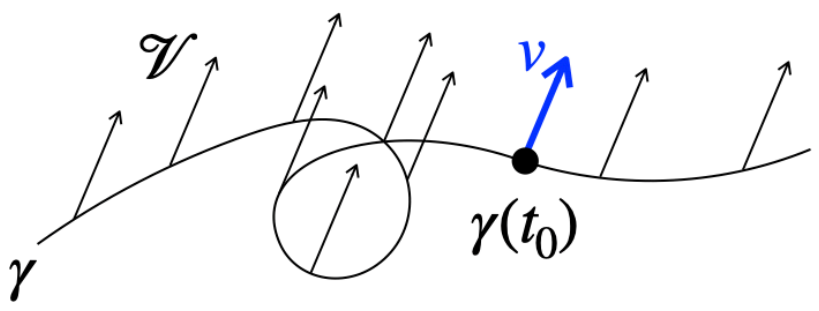

**Fig. 22**. A Parallel Vector Field Along a Curve With a Predefined Direction

This theorem is the basis for the following construction of parallel transport: Let $p, q \in M$ with $q$ "close" to $p$, and let $v \in T_pM$ an arbitrary tangent vector. $v$ should be transported to $T_qM$ in a "parallel" manner. First, we choose a curve $\gamma$ on $M$ with $\gamma(t_0) = p$ and $\gamma(t_1) = q$; such a $\gamma$ exist if $q$ is close to $p$ because of the local path connectedness of a manifold. Next, we chose $\mathscr{V} \in \mathfrak{X}(\gamma)$ with $\mathscr{V}(\gamma(t_0)) = v$ and $\mathscr{V}$ parallel along $\gamma$ (Theorem 3). $\mathscr{V}(\gamma(t_1))$ is then the result of $v$ being transported parallel to $T_{\gamma(t_0)}M = T_pM$. This way the following map is defined:

$$P^{\gamma}_{t_0,t_1} : T_{\gamma(t_0)}M \to T_{\gamma(t_1)}M \ , \ v \mapsto \mathscr{V}(\gamma(t_1)) \tag{31}$$

**Definition 30:** $P^{\gamma}_{t_0,t_1}$ is called *parallel transport* from $T_{\gamma(t_0)}M = T_pM$ to $T_{\gamma(t_1)}M = T_qM$. □

According to Definition 29 parallel transport only depends on the connection $\nabla$ chosen on the manifold $M$: each connection $\nabla$ defines a parallel transport between tangent spaces. Vice versa, a parallel transport determines “its” connection (see [23] Cor. 4.35):

**Lemma 20**: Let $\mathscr{X}, \mathscr{Y} \in \mathfrak{X}(M)$ and $p \in M$. Then

$$\nabla_{\mathcal{X}}\mathcal{Y}\,|_p = \lim_{t\to 0}\frac{P^{\gamma}_{t,0}\mathcal{Y}(\gamma(t)) - \mathcal{Y}(p)}{t} \tag{32}$$

for a curve $\gamma : I \to M$ with $\gamma(0) = p$ and $\gamma'(0) = \mathcal{X}(p)$. ∎

In this sense, connections and parallel transports are equivalent concepts. Via parallel transport two tangent spaces are connected, i.e. tangent vectors can be moved "unmodified" from one tangent space to another (if its base points are close to each other or can be connected by a curve). The equivalence of parallel transport and connections finally justifies the name of the latter.

Note: In contrast to Lie derivatives the curves used for parallel transport are typically not integral curves. Connections are independent of any flows.

### 5.4. Torsion

According to Equation 30 for two vector fields $\mathcal{X}, \mathcal{Y} \in \mathfrak{X}(\mathbb{R}^n)$ in Euclidian space the connection of $\mathcal{Y}$ in the direction of $\mathcal{X}$ is $\nabla_{\mathcal{X}}\mathcal{Y}\,|_p = d\mathcal{Y}(p)(\mathcal{X}(p))$. Thus, it is

$$\begin{aligned}
d\mathcal{Y}(p)(\mathcal{X}(p)) - d\mathcal{X}(p)(\mathcal{Y}(p)) &= \begin{pmatrix}\operatorname{grad}\mathcal{Y}_1(p)\\ \vdots\\ \operatorname{grad}\mathcal{Y}_n(p)\end{pmatrix}\begin{pmatrix}\mathcal{X}_1(p)\\ \vdots\\ \mathcal{X}_n(p)\end{pmatrix} - \begin{pmatrix}\operatorname{grad}\mathcal{X}_1(p)\\ \vdots\\ \operatorname{grad}\mathcal{X}_n(p)\end{pmatrix}\begin{pmatrix}\mathcal{Y}_1(p)\\ \vdots\\ \mathcal{Y}_n(p)\end{pmatrix}\\
&= \begin{pmatrix}\sum_i \frac{\partial \mathcal{Y}_1(p)}{\partial x_i}\mathcal{X}_i(p)\\ \vdots\\ \sum_i \frac{\partial \mathcal{Y}_n(p)}{\partial x_i}\mathcal{X}_i(p)\end{pmatrix} - \begin{pmatrix}\sum_i \frac{\partial \mathcal{X}_1(p)}{\partial x_i}\mathcal{Y}_i(p)\\ \vdots\\ \sum_i \frac{\partial \mathcal{X}_n(p)}{\partial x_i}\mathcal{Y}_i(p)\end{pmatrix}\\
&= \sum_j\left(\sum_i\left(\frac{\partial \mathcal{Y}_j(p)}{\partial x_i}\mathcal{X}_i(p) - \frac{\partial \mathcal{X}_j(p)}{\partial x_i}\mathcal{Y}_i(p)\right)\right)e_j\\
&\overset{(*)}{=} [\mathcal{X}, \mathcal{Y}]\Big|_p
\end{aligned}$$

where (*) is valid because of Lemma 16. This proves:

**Note 5**: For $\mathcal{X}, \mathcal{Y} \in \mathfrak{X}(\mathbb{R}^n)$ it is $\nabla_{\mathcal{X}}\mathcal{Y}\,|_p - \nabla_{\mathcal{Y}}\mathcal{X}\,|_p = [\mathcal{X}, \mathcal{Y}]\,|_p$. ∎

Because the goal is to mimic directional derivative in Euclidian space as good as possible via connections on differentiable manifolds the following definition is given:

**Definition 31:** Let $\nabla$ be a connection on $M$. $\nabla$ is called *symmetric* :⇔ $\forall\, \mathcal{X}, \mathcal{Y} \in \mathfrak{X}(M) : \nabla_{\mathcal{X}}\mathcal{Y} - \nabla_{\mathcal{Y}}\mathcal{X} = [\mathcal{X}, \mathcal{Y}]$. □

The deviation from symmetry is also defined:

**Definition 32:** Let $\nabla$ be a connection on $M$. The map $T : \mathfrak{X}(M) \times \mathfrak{X}(M) \to \mathfrak{X}(M)$ with $T(\mathcal{X}, \mathcal{Y}) := \nabla_{\mathcal{X}}\mathcal{Y} - \nabla_{\mathcal{Y}}\mathcal{X} - [\mathcal{X}, \mathcal{Y}]$ is called *torsion* of the connection $\nabla$. □

Using Christoffel symbols $\nabla_{\partial_i}\partial_j = \sum_k \Gamma_{ij}^k \partial_k$ it is

$$T(\partial_i, \partial_j) = \sum_k \left(\Gamma_{ij}^k - \Gamma_{ji}^k\right) \partial_k \tag{33}$$

Thus, $\forall\, i, j, k : \Gamma_{ij}^k = \Gamma_{ji}^k$ implies $T = 0$. I.e. symmetric Christoffel symbols imply a vanishing torsion and, thus, symmetry of the connection.

**Note 6**: A connection $\nabla$ is *symmetric* $\Leftrightarrow T = 0$. ∎

The torsion has a nice geometric interpretation: it measures the gap when trying to construct a parallelogram out of vectors that are parallel transported (see Figure 23, which is an extension and adaptation of [14], Figure 1). In contrast to this, the Lie-bracket measures the gap when trying to construct a parallelogram out of integral curves (see the discussion of Figure 18):

In Figure 23 part (a) the tangent vector $\mathscr{W}(p) \in T_pM$ is parallel transported from point $p$ to point $r$ along a curve starting at $p$ in direction of $\mathscr{V}(p)$ resulting in $\mathscr{W}(r)^{\parallel}$. Evaluating $\mathscr{W}$ in $r$ results in $\mathscr{W}(r)$ and the difference $\mathscr{W}(r) - \mathscr{W}(r)^{\parallel}$ is $\nabla_{\mathscr{V}}\mathscr{W}(r)$. Similarly, the tangent vector $\mathscr{V}(p) \in T_pM$ is parallel transported from $p$ to $q$ along a curve starting at $p$ in direction of $\mathscr{W}(p)$ resulting in $\mathscr{V}(q)^{\parallel}$. As before, the connection $\nabla_{\mathscr{W}}\mathscr{V}(q)$ is $\mathscr{V}(q) - \mathscr{V}(q)^{\parallel}$. Part (b) of Figure 23 depicts that $\nabla_{\mathscr{V}}\mathscr{W}(r)$ is parallel transported along a curve in direction of $\mathscr{W}(r)$ to point $r'$ and $\nabla_{\mathscr{W}}\mathscr{V}(q)$ is parallel transported along a curve in direction of $\mathscr{V}(q)$ to $q'$. The Lie bracket $[\mathscr{V}, \mathscr{W}]$ as the gap between the points $r'$ and $q'$ (see Figure 18) is shown. Vector algebra the results in the torsion $T(\mathscr{V}, \mathscr{W}) = \nabla_{\mathscr{V}}\mathscr{W} - \nabla_{\mathscr{W}}\mathscr{V} - [\mathscr{V}, \mathscr{W}]$ shown. In [28], Ch. 3.3.6, a different way resulting in the same interpretation of the torsion is provided.

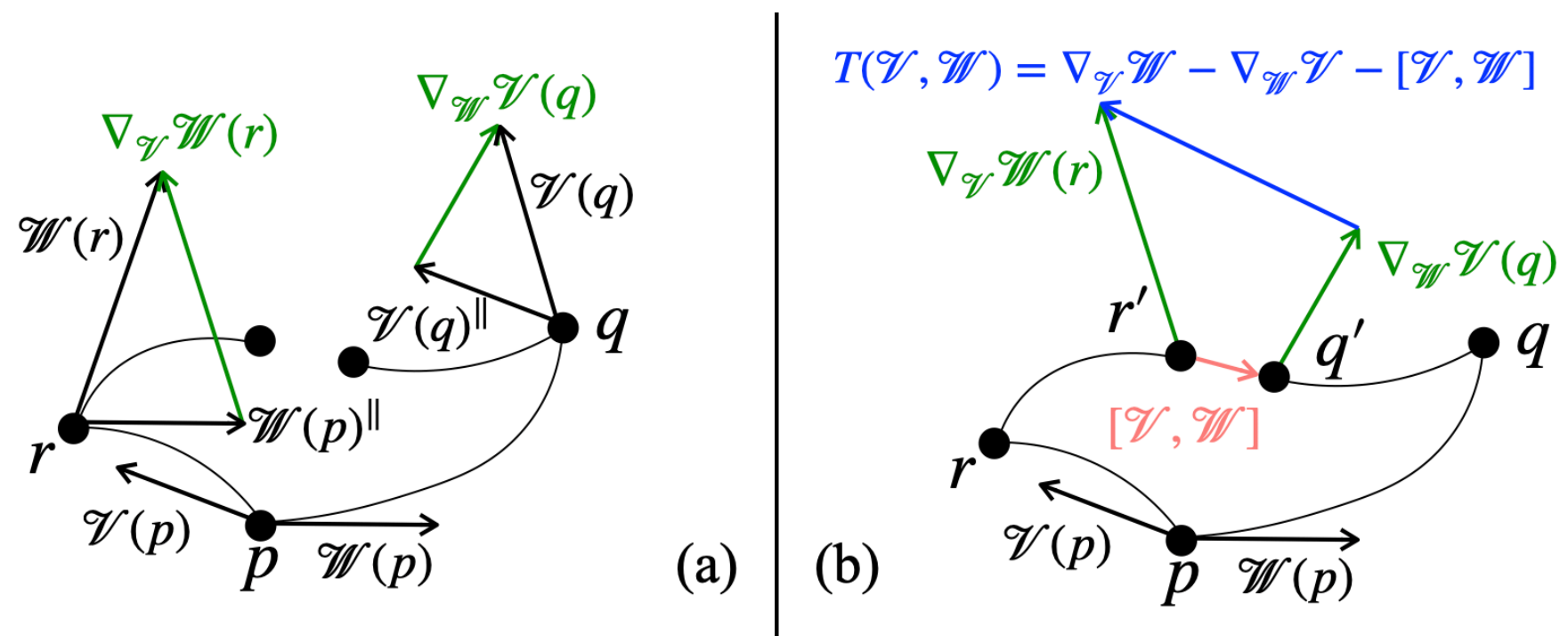

**Fig. 23**. Parallel Transportation of Vectors and The Meaning of The Torsion

## 5.5. Geodesics

Next, the concept of a straight line on a manifold is provided by analogy of a straight line in Euclidian space. There, a straight line $L$ connecting two points $p, q \in \mathbb{R}^n$ is the set $L = \{(1-t)p + tq \mid 0 \leq t \leq 1\}$. This set is the image of the smooth curve $\gamma : [0,1] \to \mathbb{R}^n$, $t \mapsto p + t(q-p)$ i.e. it is $\gamma([0,1]) = L$. This curve satisfies $\frac{d\gamma}{dt} = q - p$, i.e. $\dot{\gamma} = \text{const}$, and $\frac{d}{dt}\left(\frac{d\gamma}{dt}\right) = 0$, i.e. $\ddot{\gamma}(t) = 0$: the velocity along the curve is constant and the acceleration is vanishing.

This is generalized for a curve $\gamma : I \to M$ on a manifold: If the curve satisfies $\nabla_{\gamma'(t)}\gamma'(t) = 0$, it is considered to be "straight". Historically, such straight curves have been investigated for performing measurements on earth (i.e. $\mathbb{S}^2$) which resulted in the name "geodesic" for such curves.

**Definition 33:** A curve $\gamma : I \to M$ is called *geodesic* $:\Leftrightarrow \nabla_{\gamma'(t)}\gamma'(t) = 0$. □

Geodesics are the "straight lines" on manifolds. A curve being geodesic is equivalent to its tangent vector $\gamma'$ being parallel along $\gamma$ (compare Definition 29). As before, the velocity along a geodesic is constant and the acceleration is vanishing. In general relativity the geodesics are the paths followed by mass and light.

If $M$ is a Riemannian manifold (see Definition 35), then, a geodesic is locally minimizing the length of curves connecting two points (assuming $\|\gamma'\| = 1$, i.e. unit speed parameterization), i.e. a geodesic is locally the shortest paths between two points on a manifold. But "globally" a geodesic is not always the shortest path between two points: On a sphere $\mathbb{S}^2$ (see Figure 24) a geodesic is always a segment of a great circle. The red geodesic is the shortest path between p and q, but the green geodesic connects p and q too without being the shortest path between these points.

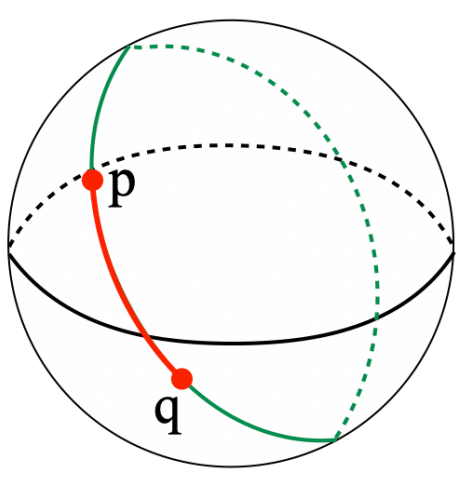

**Fig. 24**. A Great Circle as a Geodesic on $\mathbb{S}^2$

The following result states the existence and uniqueness of geodesics (see [23] Th. 4.27):

**Theorem 4 (Existence and Uniqueness of Geodesics)**: Let $\nabla$ be a connection on $M$, $p \in M$, $v \in T_pM$, and $t_0 \in \mathbb{R}$. Then there exists $]a,b[ \subseteq \mathbb{R}$ with $t_0 \in ]a,b[$ and a geodesic $\gamma : ]a,b[ \to M$ with $\gamma(t_0) = p$ and $\gamma'(t_0) = v$. If $\delta : ]c,d[ \to M$ is another such geodesic, then $\gamma = \delta$ on their common domain. ■

Using Equation 29 the derivative $\nabla_{\gamma'(t)}\gamma'(t)$ can be computed in local coordinates. This results in a system of ordinary differential equations of 2nd order - so-called *geodesic equation* - that allows to compute geodesics:

**Lemma 21**: $\gamma : I \to M$ is a geodesic $\Leftrightarrow$

$$\gamma_k'' + \sum_{i,j} \gamma_i'\gamma_j'\Gamma_{ij}^k = 0 \tag{34}$$

for all $1 \leq k \leq n$. ∎

Often, geodesics that are define on all of $\mathbb{R}$ are desirable:

**Definition 34:** $M$ is called (*geodesically*) *complete* :⇔ Every geodesic is define on all of $\mathbb{R}$. □

According to the Hopf-Rinow theorem any compact connected Riemannian manifold is geodesically complete. On a geodesically complete manifold any two points can be connected by a geodesic of minimal length (see [32] 18. Theorem).

# 6. (Pseudo-) Riemannian Manifolds

The ability to measure lengths, angles, or volumes etc on a manifold requires another intrinsic structure on a manifold, a so-called "metric" (see section 7.5 and section 11.3 for the role of metrics in such measurements); such a metric is a scalar product on each tangent space. While according to the comment after Lemma 19 a differentiable manifold has an infinite number of connections, on a manifold with a metric a unique connection can be singled out (see Theorem 6).

## 6.1. (Pseudo-) Metric

A metric associates with each tangent space a scalar product, and this assignment is done in a differentiable manner:

**Definition 35:** Let $M$ be a differentiable manifold, $g = \{g_p\}_{p \in M}$ be a set of scalar products $g_p : T_pM \times T_pM \to \mathbb{R}$. If for any two vector fields $\mathscr{V}, \mathscr{W} \in \mathfrak{X}(M)$ the map $M \to \mathbb{R}, p \mapsto g_p\left(\mathscr{V}(p), \mathscr{W}(p)\right)$ is differentiable $g$ is called *Riemannian metric* on $M$ and $\left(M, g\right)$ is called *Riemannian manifold.* □

The required differentiability of $g$ is a bit vague. After having introduced tensor fields, differentiability will be crisply defined (see Definition 69 and Lemma 51).

On every differentiable manifold, a Riemannian metric can be defined, i.e. every differentiable manifold is a Riemannian manifold (see [23] Prop. 2.4):

**Theorem 5 (Existence of Riemannian Metrics)**: Every smooth manifold admits a Riemannian metric. ∎

Many properties of Riemannian manifolds also hold when assuming a weaker notion of "metric", called a "pseudo-metric". While each $g_p$ is a positive-definite, symmetric bilinear form a pseudo-metric is a "non-degenerated" symmetric bilinear form:

**Definition 36:** Let $V$ be a vector space and $B : V \times V \to \mathbb{R}$ be a symmetric bilinear form. $B$ is called *non-degenerate* :⇔ If $B(v, w) = 0$ for each $w \in V$ then this implies $v = 0$. $(V, B)$ is called *pseudo-Euclidian* vector space. □

Let $b_1, \ldots, b_n$ be a basis of $V$, $(b_{ij}) := (B(b_i, b_j))$ be the corresponding so-called Gram-matrix. Then, $B(v, w) = v^T (b_{ij}) w$. The following holds (see [24] Th. 18.23, and [29] Th. 4.2):

**Lemma 22 (Sylvester's Law of Inertia)**: There exists a basis $b_1, \ldots, b_n$ of $V$ such that the corresponding Gram-matrix of $B$ is a diagonal matrix with diagonal elements $b_{ii} \in \{+1, -1\}$. With $n_+ := \text{card}\{i \mid b_{ii} = +1\}$ and $n_- := \text{card}\{i \mid b_{ii} = -1\}$ the number $n_+$ and $n_-$ are the same for each basis resulting in a diagonal Gram-matrix. ∎

The numbers $n_+$ and $n_-$ are key characteristics of a non-degenerated symmetric bilinear form:

**Definition 37:** $n_-$ is called *index* of $B$, the pair $(n_+, n_-)$ is called *signature* of $B$. Some authors call $n_+ - n_-$ the signature of $B$. □

According to the Spectral Theorem the following is true:

**Note 7**: $n_+ + n_- = \dim V$. Furthermore, $n_- = 0$ iff $B$ is positive-definite. ∎

Manifolds with pseudo-Euclidian tangent space play a key role in general relativity:

**Definition 38:** Let each $g_p : T_pM \times T_pM \to \mathbb{R}$ be a non-degenerate, symmetric, bilinear form for each $p \in M$. If the map $M \to \mathbb{R}$, $p \mapsto g_p\left(\mathscr{V}(p), \mathscr{W}(p)\right)$ is smooth ($\mathscr{V}, \mathscr{W} \in \mathfrak{X}(M)$ arbitrary) then $\left(M, g\right)$ is called a *pseudo-Riemannian manifold* and $g$ is called *pseudo-Riemannian metric* (i.e. $(T_pM, g_p)$ is a pseudo-Euclidian vector space). □

Sometimes pseudo-Riemannian manifolds are called semi-Riemannian manifolds (see [30]). While every differentiable manifold admits a Riemannian metric (Theorem 5) this is not true for a pseudo-Riemannian metric (see [23] Th. 2.69):

**Note 8**: Not every differentiable manifold admits a pseudo-Riemannian metric. ∎

If $M$ is connected, the signature of $g_p$ is the same for all $p \in M$ (called the *signature of* $M$). Pseudo-Riemannian manifolds of signature $(n-1,1)$ are fundamental in general relativity:

**Definition 39:** A pseudo-Riemannian manifold $(M, g)$ with signature $(n-1,1)$ is called *Lorentzian manifold*, and $g$ is called *Lorentzian metric.* □

The existence of a Lorentzian metric on a differentiable manifold $M$ is equivalent to (i) $M$ being non-compact manifolds or (ii) $M$ being compact with Euler number $\chi(M) = 0$ (see [30] Ch. 5 Prop. 37). Consequently, $\mathbb{S}^n$ is Lorentzian iff $n \geq 3$ is odd (see [30] p. 149).

For concrete computations, a (pseudo-) Riemannian metric $g$ is needed in local coordinates: Let $\mathscr{X}, \mathscr{Y} \in \mathfrak{X}(M)$ and let $\mathscr{X} = \sum_i X_i \partial_i$ and $\mathscr{Y} = \sum_j Y_j \partial_j$ for suitable smooth functions $X_i, Y_j \in C^\infty(M)$ (see Lemma 8). Bilinearity of a (pseudo-) Riemannian metric $g$ shows that $g(\mathcal{X}, \mathcal{Y}) = \sum_{i,j} X_i Y_j g(\partial_i, \partial_j)$.

**Definition 40:** $g_{ij} := g(\partial_i, \partial_j)$, $1 \le i, j \le n$ is called a *metric coefficient*. The matrix $(g_{ij})$ represents the (pseudo-) Riemannian metric in local coordinates. The inverse matrix $(g_{ij})^{-1}$ is denoted by $(g^{ij})$. □

Thus, in local coordinates it is

$$g(\mathcal{X}, \mathcal{Y}) = \sum_{i,j} g_{ij} X_i Y_j \tag{35}$$

By means of a basis transformation the matrix $(g_{ij})$ can be transformed into a diagonal matrix with diagonal elements in $\{+1, -1\}$ (Sylvester's Law of Inertia). Even more (see [24] Lemma. 18.22):

**Note 9**: In any basis, $(g_{ij})$ has only real eigenvalues, i.e. for the spectrum $\sigma(g_{ij})$ it is $\sigma(g_{ij}) \subseteq \mathbb{R}$. Furthermore, card $(\sigma(g_{ij}) \cap \mathbb{R}_{>0}) = n_+$ and card $(\sigma(g_{ij}) \cap \mathbb{R}_{<0}) = n_-$. ∎

If the corresponding signature of $g$ is $(n_+, n_-)$, this signature is written in more detail as a tuple of $+$ and $-$ signs with the sign in the i-th component of this tuple being the sign of $g_{ii}$ (since $g$ is diagonal for $i \neq j$ it is $g_{ij} = 0$). For example, $\text{diag}(-1, +1, +1, -1)$ corresponds to $(-, +, +, -)$, and $\text{diag}(-1, +1, +1, +1)$ corresponds to $(-, +, +, +)$. Note, that the latter signature indicates a Lorentzian metric (i.e. $n_- = 1$).

## 6.2. Intermezzo: Metric in Spherical Coordinates

Why performing computation in spherical symmetric situations (like black holes in general relativity, for example) Cartesian coordinates are not convenient to use. Instead spherical coordinates are much more appropriate.

Figure 25(a) depicts a sphere of radius $r$. The angle between $P = (x, y, z)^T$ and the z-axis is $\theta$. Thus, part (b) of the figure shows that $z = r\cos\theta$, where $r = \|P\|$. According to part (c) of the figure the angle between $P$ and the xy-plane is $\frac{\pi}{2} - \theta$, i.e. orthogonally projecting $P$ onto the xy-plane result in $N = r\cos\left(\frac{\pi}{2} - \theta\right) = r\sin\theta$. Finally (see part (d)), projecting $N$ orthogonally onto the x-axis results in $L = N\cos\rho$ which is $L = r\sin\theta\cos\rho$ the x-coordinate of $P$. Similarly, the orthogonal projection of $N$ onto the y-axis results in $M = N\sin\rho$ which is $M = r\sin\theta\sin\rho$ the y-coordinate of $P$. Together:

$$P = \begin{pmatrix} x \\ y \\ z \end{pmatrix} = r \begin{pmatrix} \cos\rho\sin\theta \\ \sin\rho\sin\theta \\ \cos\theta \end{pmatrix} \tag{36}$$

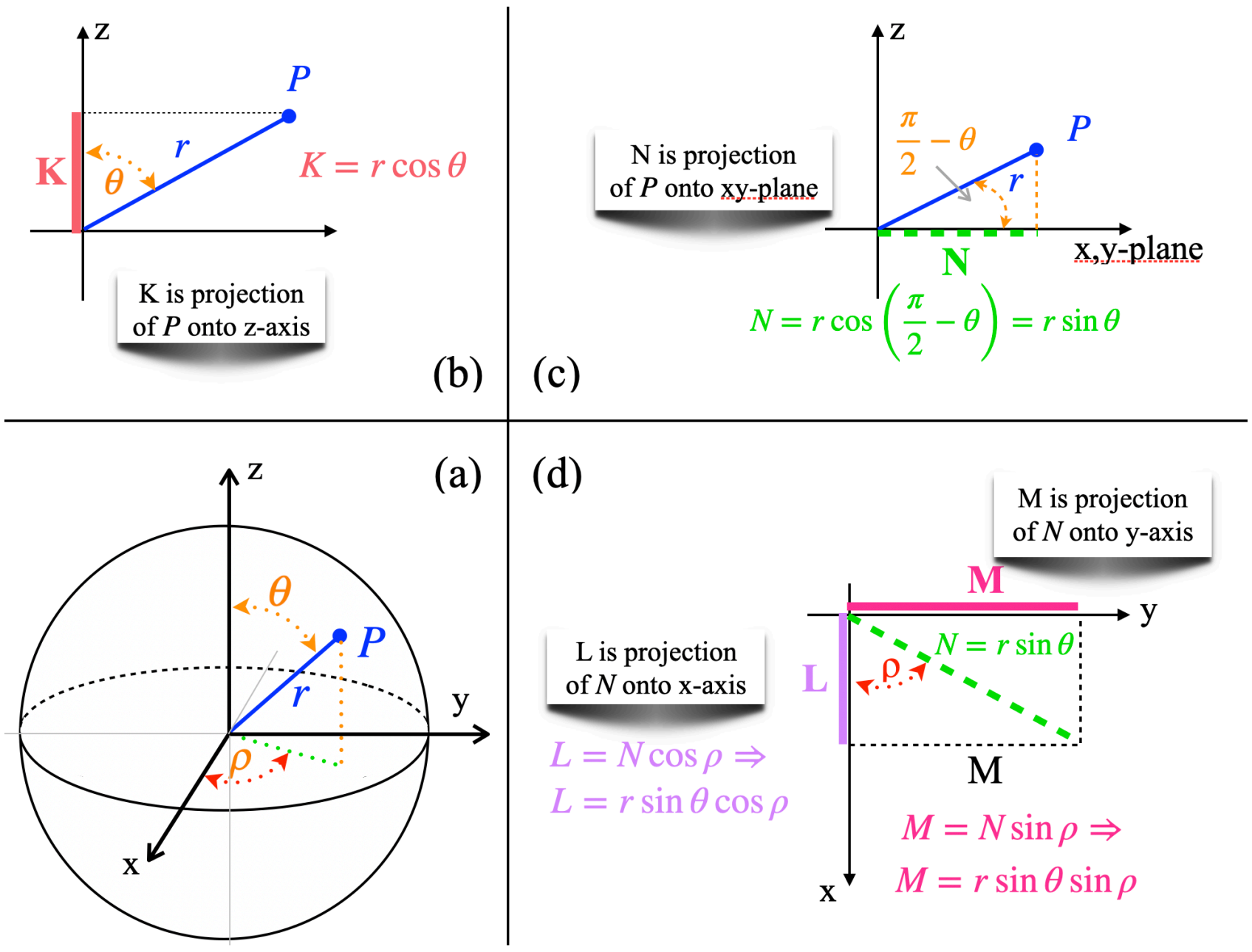

**Fig. 25**. Spherical Coordinates on $\mathbb{S}^2$

In Cartesian coordinates the distance $ds$ between two points $P_1, P_2 \in \mathbb{R}^3$ can be computed as (see Figure 26) $ds^2 = dx^2 + dy^2 + dz^2$. Substituting $x, y, z$ from Equation 36 results in:

$$ds^2 = \left[d\left(r\cos\rho\sin\theta\right)\right]^2 + \left[d\left(r\sin\rho\sin\theta\right)\right]^2 + \left[d\left(r\cos\theta\right)\right]^2 \qquad (37)$$

**Fig. 26**. The Distance Between Two Points in $\mathbb{R}^3$

In Definition 68 the computation of differentials of differentiable functions is introduced. Based in this the following results:

$$\left[d\left(r\cos\rho\sin\theta\right)\right]^2 = \left[\cos\rho\sin\theta dr - r\sin\rho\sin\theta d\rho + r\cos\rho\cos\theta d\theta\right]^2$$
$$\left[d\left(r\sin\rho\sin\theta\right)\right]^2 = \left[\sin\rho\sin\theta dr + r\cos\rho\sin\theta d\rho + r\sin\rho\cos\theta d\theta\right]^2 \quad (38)$$
$$\left[d\left(r\cos\theta\right)\right]^2 = \left[\cos\theta dr - r\sin\theta d\theta\right]^2$$

An exercise in algebraic and trigonometric computations result in:

$$ds^2 = dr^2 + r^2 d\theta^2 + r^2\sin^2\theta d\rho^2 \quad (39)$$

Restricting this on a 2-sphere i.e. for $r = \text{const} = R$ (i.e. $dr = 0$) finally results in the metric on a 2-sphere:

$$ds^2 = R^2(d\theta^2 + \sin^2\theta d\rho^2) \quad (40)$$

Spherical coordinates can be generalized to dimensions greater than three (see [5]). Let $P = (x_1, \ldots, x_n)^T$. Then for $\theta_1, \ldots, \theta_{n-2} \in [0,\pi]$ and $\theta_{n-1} \in [0,2\pi[$:

$$\begin{aligned}
x_1 &= r\cos\theta_1 \\
x_2 &= r\cos\theta_2\sin\theta_1 \\
&\vdots \\
x_i &= r\cos\theta_i\prod_{k=1}^{i-1}\sin\theta_k \\
&\vdots \\
x_{n-1} &= r\cos\theta_{n-1}\prod_{k=1}^{n-2}\sin\theta_k \\
x_n &= r\sin\theta_{n-1}\prod_{k=1}^{n-2}\sin\theta_k
\end{aligned} \quad (41)$$

Here, $x_i = \ldots$ is valid for $3 \le i \le n-1$. The formula for $x_{n-1}$ is covered by the formula for $x_i$ but it shows nicely that $\cos\theta_{n-1}$ is substituted by $\sin\theta_{n-1}$ in the formula for $x_n$. Finally, note that the order of coordinates in Equation 41 has been modified compared to [5]. The coordinates represented by Equation 41 are also referred to as *hyperspherical coordinates*.

In general relativity, the universe is represented as a four dimensional pseudo-Riemannian manifold. The pseudo-metric in four dimensions is derived by first using Equation 41 to rewrite Cartesian coordinates in hyperspherical coordinates thereby renaming $x_1, x_2, x_3, x_4$ as $x, y, z, w$ as well as renaming the angles:

$$
\begin{aligned}
x &= r\cos\chi \\
y &= r\cos\theta\sin\chi \\
z &= r\cos\phi\sin\chi\sin\theta \\
w &= r\sin\phi\sin\chi\sin\theta
\end{aligned} \tag{42}
$$

and then substituting these formulas in $ds^2 = dx^2 + dy^2 + dz^2 + dw^2$:

$$
\begin{aligned}
ds^2 &= dx^2 + dy^2 + dz^2 + dw^2 \\
&= \left[d\left(r\cos\chi\right)\right]^2 + \left[d\left(r\cos\theta\sin\chi\right)\right]^2 + \left[d\left(r\cos\phi\sin\chi\sin\theta\right)\right]^2 + \left[d\left(r\sin\phi\sin\chi\sin\theta\right)\right]^2
\end{aligned} \tag{43}
$$

Again, an exercise in algebraic and trigonometric computations result in:

$$
ds^2 = dr^2 + r^2\left(d\chi^2 + \sin^2\chi(d\theta^2 + \sin^2\theta d\phi^2\right) \tag{44}
$$

A 3-sphere has constant radius $r = R$, i.e. $dr = 0$. Based on the equation before, it is $ds^2 = R^2\left(d\chi^2 + \sin^2\chi(d\theta^2 + \sin^2\theta d\phi^2)\right)$. In this latter equation the expression $\sin^2\chi(d\theta^2 + \sin^2\theta d\phi^2)$ is the metric on a 2-sphere with constant radius $r = \sin\chi$ (i.e. $dr = 0$) - refer to Equation 40. This metric on a 2-sphere with constant radius $r = \sin\chi$ is denoted as $d\Omega_2$, i.e.

$$
d\Omega_2^2 = \sin^2\chi(d\theta^2 + \sin^2\theta d\phi^2) \tag{45}
$$

In summary, the metric on a 3-sphere in spherical coordinates is

$$
ds^2 = R^2\left(d\chi^2 + d\Omega_2^2\right) \tag{46}
$$

This metric on a 3-sphere with constant radius is often denoted by $d\Omega_3$. Substituting this in Equation 44 result in the following metric on $\mathbb{R}^4$:

$$
ds^2 = dr^2 + r^2 \overbrace{\left(d\chi^2 + \underbrace{\sin^2\chi(d\theta^2 + \sin^2\theta d\phi^2)}_{\text{metric } d\Omega_2^2 \text{ on 2-sphere}}\right)}^{\text{metric } d\Omega_3^2 \text{ on 3-sphere}} \tag{47}
$$

The second term of Equation 47 is the metric $d\Omega_3^2$ on a 3-sphere of radius *r*, while the last summand is the metric $d\Omega_2^2$ on a 2-sphere of radius $\sin\chi$. This somehow reflects the topological structure of $\mathbb{S}^3$ as a fibre space with base $\mathbb{S}^2$ and fibre $\mathbb{S}^1$ (so-called Hopf-fibration showing that $\mathbb{S}^3/\mathbb{S}^1 \cong \mathbb{S}^2$ - see [12] p. 135).

### 6.3. Gradient, Hessian, and Laplacian

Concrete computations on a (pseudo-) Riemannian manifold $(M, g)$ require generalizations of the gradient (see Definition 3), the Hessian, and the Laplacian of a function $f \in C^\infty(M)$.

The *gradient* grad $f$ is a vector field grad $f \in \mathfrak{X}(M)$, also denoted by $\nabla f$. It is uniquely determined by the property $g(\text{grad } f, \mathscr{V}) = \mathscr{V} f$ for every $\mathscr{V} \in \mathfrak{X}(M)$; in local coordinates it is (see [22] p. 343):

$$\text{grad } f = \sum_{i,j} g^{ij} \frac{\partial f}{\partial x_i} \frac{\partial}{\partial x_j} \tag{48}$$

Recall that $(g^{ij})$is the inverse matrix of $(g_{ij})$ (see Definition 40). On $\mathbb{R}^n$ with the Euclidian metric $g = \langle\, , \rangle$, it is $g_{ij} = \langle e_i, e_j \rangle = \delta_{ij}$, thus $g^{ij} = \delta_{ij}$. Thus, according to Equation 48 it is $\text{grad } f = \sum_i \frac{\partial f}{\partial x_i} \frac{\partial}{\partial x_i} = \left( \frac{\partial f}{\partial x_1}(x), \ldots, \frac{\partial f}{\partial x_n}(x) \right)^T$, which recovers the definition of the gradient of functions on $\mathbb{R}^n$.

$H(f) := \nabla \nabla f \in \mathscr{T}^2(M)$ is called the *Hessian* of $f$; $\mathscr{T}^2(M)$ denotes a tensor field which will be defined precisely in Definition 69 later. Here it suffice consider $H(f)$ as a matrix with differentiable coefficients. It is:

$$H(f)(\mathscr{X}, \mathscr{Y}) = \nabla \nabla f(\mathscr{X}, \mathscr{Y}) = \mathscr{X}(\mathscr{Y} f) - (\nabla_{\mathscr{X}} \mathscr{Y}) f \tag{49}$$

and in local coordinates (see [23] Ex. 4.22):

$$\nabla \nabla f = \sum_{i,j} \left( \frac{\partial^2 f}{\partial x_i \partial x_j} - \sum_k \Gamma_{ij}^k \frac{\partial f}{\partial x_k} \right) d x_i \otimes d x_j \tag{50}$$

The meaning of $d x_i \otimes d x_j$ is not important for now and will be explained by Definition 64.

Finally, $\Delta f := \text{tr } H(f)$ is called the *Laplacian* of $f$.

## 7. Connections on (Pseudo-) Riemannian Manifolds

In this section we define the fundamental concept of the Levi-Civita connection on a (pseudo-) Riemannian manifold. Based on this, the Riemann curvature tensor is defined and based on this various notions of curvature follow.

### 7.1. Levi-Civita Connection

Lemma 19 and the remark after it shows that a differentiable manifold admits an infinite number of connections. On (pseudo-) Riemannian manifold one of these connections can be distinguished which is called the Levi-Civita connection (see Definition 42).

**Definition 41:** Let $(M, g)$ be a (pseudo-) Riemannian manifold and $\nabla$ a connection of $M$. $\nabla$ is called *metric-compatible* or a *metric connection* :⇔ $\forall\, \mathscr{X}, \mathscr{Y}, \mathscr{Z} \in \mathfrak{X}(M)$ :

$$\mathscr{X} \left( g(\mathscr{Y}, \mathscr{Z}) \right) = g \left( \nabla_{\mathscr{X}} \mathscr{Y}, \mathscr{Z} \right) + g \left( \mathscr{Y}, \nabla_{\mathscr{X}} \mathscr{Z} \right) \tag{51}$$ □

I.e. the derivative of the metric of two vector fields of a metric-compatible connection can be computed by means of the metric of the connection of the corresponding vector fields. This easies the computation of derivatives of the metric. Another desirable property (which directional derivatives in Euclidian space have) is being symmetric (see Definition 31) which is equivalent to being torsion-free (Note 6). Connections with these two properties are of utmost importance:

**Definition 42:** A metric-compatible and symmetric connection $\nabla$ is called *Levi-Civita connection*. If $(M, g)$ is Riemannian, $\nabla$ is also called *Riemannian connection*. □

Existence of Levi-Civita connections, even its uniqueness is subject of the following (see [23] Th. 5.10):

**Theorem 6 (Fundamental Theorem of Riemannian Geometry)**: Let $(M, g)$ be a (pseudo-) Riemannian manifold. There exists a unique Levi-Civita connection $\nabla$ of $M$. ∎

Remember that each connection is determined by the Christoffel symbols (refer to Equation 29). Furthermore, the next lemma shows that the Christoffel symbols are completely determined by the (pseudo-) metric $g$ (i.e. the metric coefficients - see Definition 40). Together, this means that the Levi-Civita connection is uniquely determined by $g$ (see [23] Cor. 5.11):

**Lemma 23**: The Christoffel symbols of the Levi-Civita connection are determined by the (pseudo-) metric $g$:

$$\Gamma_{ij}^{k} = \frac{1}{2} \sum_{l} g^{kl} \left( \frac{\partial g_{lj}}{\partial x_i} + \frac{\partial g_{il}}{\partial x_j} - \frac{\partial g_{ij}}{\partial x_l} \right) \quad (52)$$ ∎

More precisely, the Christoffel symbols (and consequently the Levi-Civita connection) are determined by the metric coefficients of the (pseudo-) metric and their partial derivatives.

### 7.2. Intermezzo: Christoffel Symbols of a Levi-Civita Connection

Christoffel symbols of the Levi-Civita connection of $\mathbb{S}_R^2 \subset \mathbb{R}^3$, the two-dimensional sphere of radius $R$ in $\mathbb{R}^3$.

The metric on $\mathbb{S}_R^3$ in spherical coordinates is (Equation 40):

$$ds^2 = R^2 d\theta^2 + R^2 \sin^2 \theta d\rho^2 = g_{\theta\theta} d\theta^2 + g_{\rho\rho} d\rho^2$$

Thus, the metric coefficients are $g_{\theta\theta} = R^2$ and $g_{\rho\rho} = R^2 \sin^2 \theta$, and $g_{\theta\rho} = g_{\rho\theta} = 0$. Consequently,

$$(g_{ij}) = \begin{pmatrix} R^2 & 0 \\ 0 & R^2 \sin^2 \theta \end{pmatrix} = \begin{pmatrix} g_{\theta\theta} & g_{\theta\rho} \\ g_{\rho\theta} & g_{\rho\rho} \end{pmatrix}$$

This matrix is diagonal, i.e.

$$(g_{ij})^{-1} = \begin{pmatrix} \frac{1}{R^2} & 0 \\ 0 & \frac{1}{R^2 \sin^2 \theta} \end{pmatrix} = \begin{pmatrix} g^{\theta\theta} & g^{\theta\rho} \\ g^{\rho\theta} & g_{\rho\rho} \end{pmatrix} \tag{a}$$

According to Equation 52, the Christoffel symbols are computed as

$$\Gamma_{ij}^k = \frac{1}{2} \sum_l g^{kl} \left( \frac{\partial g_{lj}}{\partial x_i} + \frac{\partial g_{il}}{\partial x_j} - \frac{\partial g_{ij}}{\partial x_l} \right)$$

The Levi-Civita connection is symmetric, i.e. $\Gamma_{ij}^k = \Gamma_{ji}^k$ (see remarks after Definition 32). Furthermore, $(g^{ij}) = 0$ for $i \neq j$ as seen in Equation (a). Thus, the the formula for the Christoffel symbols reduces to

$$\Gamma_{ij}^k = \frac{1}{2} g^{kk} \left( \frac{\partial g_{kj}}{\partial x_i} + \frac{\partial g_{ik}}{\partial x_j} - \frac{\partial g_{ij}}{\partial x_k} \right) \tag{b}$$

The only non-constant metric coefficient is $g_{\rho\rho} = R^2 \sin^2 \theta$, i.e

$$\frac{\partial g_{\rho\rho}}{\partial \theta} = 2R^2 \cos\theta \sin\theta \tag{c}$$

and all other partial derivatives are zero, i.e.

$$\frac{\partial g_{\rho\rho}}{\partial \rho} = \frac{\partial g_{\theta\theta}}{\partial \theta} = \frac{\partial g_{\theta\theta}}{\partial \rho} = \frac{\partial g_{\theta\rho}}{\partial \rho} = \frac{\partial g_{\theta\rho}}{\partial \theta} = \frac{\partial g_{\rho\theta}}{\partial \theta} = \frac{\partial g_{\rho\theta}}{\partial \rho} = 0 \tag{d}$$

Using equation (b) with $k = \theta$ and $g^{\theta\theta} = \frac{1}{R^2}$:

$$\Gamma_{ij}^\theta = \frac{1}{2R^2} \left( \frac{\partial g_{\theta j}}{\partial x_i} + \frac{\partial g_{i\theta}}{\partial x_j} - \frac{\partial g_{ij}}{\partial \theta} \right) \tag{e}$$

From (e) and for $i = \theta$ and $j = \theta$ we get with (d)

$$\Gamma_{\theta\theta}^\theta = \frac{1}{2R^2} \left( \frac{\partial g_{\theta\theta}}{\partial \theta} + \frac{\partial g_{\theta\theta}}{\partial \theta} - \frac{\partial g_{\theta\theta}}{\partial \theta} \right) = \frac{1}{2R^2}(0 + 0 - 0) = 0$$

From (e) and for $i = \theta$ and $j = \rho$ we get with (d) and $\Gamma_{ij}^k = \Gamma_{ji}^k$

$$\Gamma_{\theta\rho}^\theta = \frac{1}{2R^2} \left( \frac{\partial g_{\theta\rho}}{\partial \theta} + \frac{\partial g_{\theta\theta}}{\partial \rho} - \frac{\partial g_{\theta\rho}}{\partial \theta} \right) = \frac{1}{2R^2}(0 + 0 - 0) = 0 = \Gamma_{\rho\theta}^\theta$$

From (e) and for $i = \rho$ and $j = \rho$ we get with (d)

$$\Gamma^{\theta}_{\rho\rho} = \frac{1}{2R^2}\left(\frac{\partial g_{\theta\rho}}{\partial\rho} + \frac{\partial g_{\rho\theta}}{\partial\rho} - \frac{\partial g_{\rho\rho}}{\partial\theta}\right)$$
$$= \frac{1}{2R^2}(0 + 0 - 2R^2\cos\theta\,\sin\theta) = -\cos\theta\,\sin\theta$$

Next, let $k = \rho$ and $g^{\rho\rho} = \frac{1}{R^2\sin^2\theta}$. Substituting this in (b):

$$\Gamma^{\rho}_{ij} = \frac{1}{2R^2\sin^2\theta}\left(\frac{\partial g_{\rho j}}{\partial x_i} + \frac{\partial g_{i\rho}}{\partial x_j} - \frac{\partial g_{ij}}{\partial\rho}\right) \tag{f}$$

From (f) and for $i = \theta$ and $j = \theta$ we get with (d)

$$\Gamma^{\rho}_{\theta\theta} = \frac{1}{2R^2\sin^2\theta}\left(\frac{\partial g_{\rho\theta}}{\partial\theta} + \frac{\partial g_{\theta\rho}}{\partial\theta} - \frac{\partial g_{\theta\theta}}{\partial\rho}\right) = 0$$

Similarly, with $i = \rho$ and $j = \rho$ we get with (d)

$$\Gamma^{\rho}_{\rho\rho} = \frac{1}{2R^2\sin^2\theta}\left(\frac{\partial g_{\rho\rho}}{\partial\rho} + \frac{\partial g_{\rho\rho}}{\partial\rho} - \frac{\partial g_{\rho\rho}}{\partial\rho}\right) = 0$$

Finally, with $i = \theta$ and $j = \rho$ we get with (d), (c) and $\Gamma^k_{ij} = \Gamma^k_{ji}$

$$\Gamma^{\rho}_{\theta\rho} = \frac{1}{2R^2\sin^2\theta}\left(\frac{\partial g_{\rho\rho}}{\partial\theta} + \frac{\partial g_{\theta\rho}}{\partial\rho} - \frac{\partial g_{\theta\rho}}{\partial\rho}\right)$$
$$= \frac{1}{2R^2\sin^2\theta}(2R^2\cos\theta\,\sin\theta + 0 - 0) = \frac{\cos\theta}{\sin\theta} = \cot\theta = \Gamma^{\rho}_{\rho\theta}$$

Putting this all together, the Christoffel symbols of the Levi-Civita connection of $\mathbb{S}^2_R \subset \mathbb{R}^3$ in spherical coordinates are:

$$\Gamma^{\theta}_{\theta\theta} = 0 \quad \Gamma^{\theta}_{\theta\rho} = 0 \quad \Gamma^{\theta}_{\rho\theta} = 0 \quad \Gamma^{\theta}_{\rho\rho} = -\cos\theta\,\sin\theta$$
$$\Gamma^{\rho}_{\theta\theta} = 0 \quad \Gamma^{\rho}_{\theta\rho} = \cot\theta \quad \Gamma^{\rho}_{\rho\theta} = \cot\theta \quad \Gamma^{\rho}_{\rho\rho} = 0$$

An interesting observation is that the Christoffel symbols are independent of the radius $R$ of the sphere.

### 7.3. Riemann Curvature

The Levi-Civita connection of a (pseudo-) Riemannian manifold $(M, g)$ allows to define the following map that supports to measuring curvature of a $M$:

**Definition 43:** The map $R : \mathfrak{X}(M) \times \mathfrak{X}(M) \times \mathfrak{X}(M) \to \mathfrak{X}(M)$ defined by

$$R(\mathcal{X},\mathcal{Y})\mathcal{Z} := \nabla_{\mathcal{X}}\nabla_{\mathcal{Y}}\mathcal{Z} - \nabla_{\mathcal{Y}}\nabla_{\mathcal{X}}\mathcal{Z} - \nabla_{[\mathcal{X},\mathcal{Y}]}\mathcal{Z} \tag{53}$$

is called *Riemann curvature tensor* (often *Riemann curvature* for short). □

Again, the concept of a tensor is not important for now but it will be introduced in section 9.1. As shown in [23], Prop. 7.3:

**Note 10**: The Riemann curvature tensor $R$ is multilinear over $C^\infty(M)$. ∎

Furthermore, the Riemann curvature tensor can be computed by means of the Christoffel symbols and their partial derivatives (see [23] Prop. 7.4):

**Lemma 24**: With $R(\partial_i,\partial_j)\partial_k = \sum_l R^l_{ijk}\partial_l$ it is

$$R^l_{ijk} = \frac{\partial\Gamma^l_{jk}}{\partial x_i} - \frac{\partial\Gamma^l_{ik}}{\partial x_j} + \sum_m \left(\Gamma^m_{jk}\Gamma^l_{im} - \Gamma^m_{ik}\Gamma^l_{jm}\right) \tag{54}$$ ∎

Thus, the components $R^l_{ijk}$ (and, thus, the Riemann curvature tensor) are determined by the (pseudo-) metric and its 1st and 2nd order partial derivatives because the Christoffel symbols $\Gamma^c_{ab}$ used in Equation 54 are determined by the (pseudo-) metric $g$ (Lemma 23).

Fixing two vector fields $\mathcal{X},\mathcal{Y} \in \mathfrak{X}(M)$ turns the Riemann curvature tensor into a map $R(\mathcal{X},\mathcal{Y})$ on the set of vector fields $\mathfrak{X}(M)$; this map will become important for defining further notions of curvature (see Definition 48 and Equation 65) later on:

**Definition 44:** Let $\mathcal{X},\mathcal{Y} \in \mathfrak{X}(M)$. The map $R(\mathcal{X},\mathcal{Y}) : \mathfrak{X}(M) \to \mathfrak{X}(M)$ with $\mathcal{Z} \mapsto R(\mathcal{X},\mathcal{Y})\mathcal{Z}$ is called *curvature endomorphism* (because it induces an endomorphism $T_pM \to T_pM$ for each $p \in M$). □

Next, we give an algebraic interpretation of the Riemann curvature tensor. Let $f \in C^\infty(M)$ be arbitrary; then $[\partial_i,\partial_j]$ is a vector field on $M$ (Lemma 13). Thus, $[\partial_i,\partial_j]|_p$ is a derivation (see Definition 12) i.e. $[\partial_i,\partial_j](f)$ can be computed (see Equation 23 and Definition 22):

$$[\partial_i,\partial_j](f) = \frac{\partial}{\partial x_i}\left(\frac{\partial f}{\partial x_j}\right) - \frac{\partial}{\partial x_j}\left(\frac{\partial f}{\partial x_i}\right) = \frac{\partial^2 f}{\partial x_i \partial x_j} - \frac{\partial^2 f}{\partial x_j \partial x_i} = 0 \tag{55}$$

The latter equal sign is valid because of Schwarz's Theorem (Theorem 1). Because $f$ was an arbitrary function we got

$$[\partial_i,\partial_j] = 0 \tag{56}$$

Consequently, $\nabla_{[\partial_i,\partial_j]}\mathcal{Z} = 0$ for each $\mathcal{Z} \in \mathfrak{X}(M)$, thus

$$R(\partial_i,\partial_j)\mathcal{Z} = \nabla_{\partial_i}\nabla_{\partial_j}\mathcal{Z} - \nabla_{\partial_j}\nabla_{\partial_i}\mathcal{Z} - \nabla_{[\partial_i,\partial_j]}\mathcal{Z} = \nabla_{\partial_i}\nabla_{\partial_j}\mathcal{Z} - \nabla_{\partial_j}\nabla_{\partial_i}\mathcal{Z} \tag{57}$$

which shows that

$$R(\partial_i, \partial_j) = \nabla_{\partial_i}\nabla_{\partial_j} - \nabla_{\partial_j}\nabla_{\partial_i} \tag{58}$$

Thus, the Riemann curvature tensor $R$ measures the extend to which the second covariant derivatives of the Levi-Civita connection commute.

There is also a geometric interpretation of the Riemann curvature tensor $R$ (see Figure 27): The left side of the figure shows a (small) "rectangle" formed by curves $\gamma_1, \ldots, \gamma_4$ on $M$ with $p \in M$ be the start of $\gamma_1$ and the end of $\gamma_4$. The right side of the figure depicts one possible way how this "rectangle" is constructed: take a chart $\varphi : U \to V \subset_{\text{open}} \mathbb{R}^n$ around $p$, take a parallelogram contained in $V$ with $\varphi(p)$ as a vertex, represent each edge of the parallelogram as a linear curve, and lift these curves via $\varphi^{-1}$ to $M$ resulting into $\gamma_1, \ldots, \gamma_4$. For $\mathcal{X}, \mathcal{Y}, \mathcal{Z} \in \mathfrak{X}(M)$, it can even be achieved such that $\mathcal{X}(p)$ is tangent to $\gamma_1$, $\mathcal{Y}(p)$ is tangent to $\gamma_4$. Choose $z = \mathcal{Z}(p) \in T_pM$. Then, parallel transport $z$ along $\gamma_1, \gamma_2, \gamma_3, \gamma_4$ (the "rectangle" $\gamma_4 \circ \gamma_3 \circ \gamma_2 \circ \gamma_1$ is a closed piecewise smooth curve). The resulting vector is $\hat{z}$. In general, it $\hat{z} \neq z$ - a phenomenon called *holonomy* (see next Section 7.3). For small "rectangles", this difference is $R(\mathcal{X}, \mathcal{Y})\mathcal{Z}$.

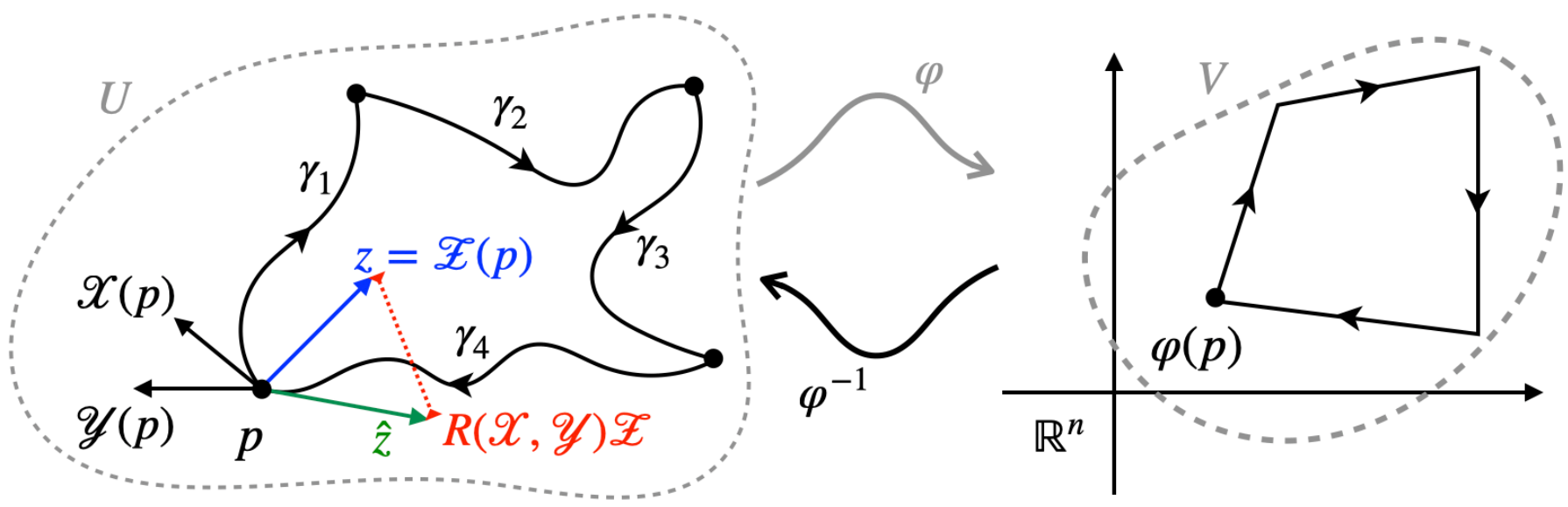

**Fig. 27**. Geometric Interpretation of the Riemann Curvature Tensor

### 7.4. Holonomy

Holonomy is an indicator of curvature. Figure 28 depicts two examples of holonomy. In part (a) a sphere is shown. $p$ and $q$ are two diametrically points, and $\alpha$ is half of the equator, i.e. a curve from $p$ to $q$. The parallel transport $P^\alpha$ along $\alpha$ transports the tangent vector $v$ at $p$ to a tangent vector $P^\alpha(v)$ at $q$. Similarly, $\beta$ is half of a great circle, another curve from $p$ to $q$. The parallel transport $P^\beta$ along $\beta$ transports the tangent vector $v$ at $p$ to a tangent vector $P^\beta(v)$ at $q$. Obviously, It is $P^\alpha(v) \neq P^\beta(v)$. This is an effect of the curvature of the sphere: parallel transport in the (flat) plane along different paths between two points will always result in the same vector.

In part (b) of the figure, the tangent vector $v \in T_pM$ at $p$ is parallel transported along $\gamma_1$. The resulting tangent vector at $q$ is parallel transported along curve $\gamma_2$. This resulting tangent vector at $r$ is parallel transported along curve $\gamma_3$. The resulting vector $w \in T_pM$ at $p$ is in general not the original vector $v$, i.e. $v \neq w$. The angle α between

$v$ and $w$ is a measure of the curvature of $M$ "close to $p$". Again, this it the phenomenon of *holonomy.*

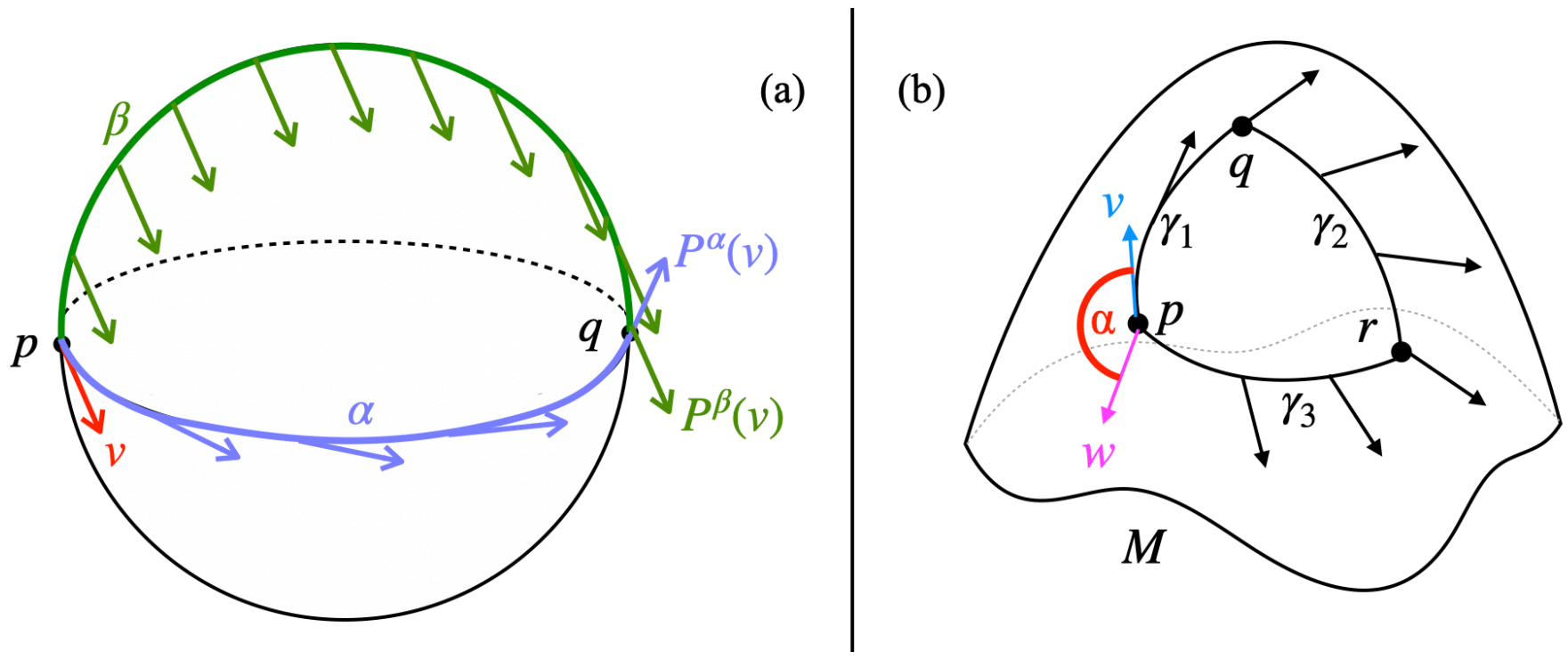

**Fig. 28**. Holonomy On a Sphere and On an Arbitrary Manifold

An important observation about parallel transport is preservation of lengths: Let $(M, g)$ be a (pseudo-) Riemannian manifold with Levi-Civita connection $\nabla$, and let $\mathscr{V} \in \mathfrak{X}(\gamma)$ be parallel along $\gamma$, i.e. $\nabla_{\gamma'(t)}\mathscr{V} \equiv 0$. The length of $\mathscr{V}(t)$ w.r.t. $g$ is $\|\mathscr{V}(t)\|_g = \sqrt{g\left(\mathscr{V}(\gamma(t)), \mathscr{V}(\gamma(t))\right)}$. With $\nabla_{\gamma'(t)}\mathscr{V} \equiv 0$ the following results:

$$\frac{d}{dt} g\left(\mathscr{V}(\gamma(t)), \mathscr{V}(\gamma(t))\right) \overset{(*)}{=} \gamma'(t)\Big(g\left(\mathscr{V}(\gamma(t)), \mathscr{V}(\gamma(t))\right)\Big)$$
$$\overset{(**)}{=} g\left(\nabla_{\gamma'(t)}V(\gamma(t)), V(\gamma(t)\right) + g\left(V(\gamma(t), \nabla_{\gamma'(t)}V(\gamma(t))\right) = 0$$

where (*) follows from Definition 14 and Equation 13, while (**) is the metric-compatibility of the Levi-Civita connection $\nabla$ (Equation 51); an alternative proof can be found in [12], Prop. 15.11. Thus, $g\left(\mathscr{V}(\gamma(t)), \mathscr{V}(\gamma(t))\right)$ is constant, i.e. the length of $\mathscr{V}(t)$ is constant along $\gamma$:

**Note 11**: The length of a vector that is parallel transported w.r.t. the Levi-Civita connection does not change. ∎

In analogy to Euclidian spaces the angle between two vectors in the tangent space of a (pseudo-) Riemannian manifold can be defined for $v, w \in T_pM$ with $g(v, v) \neq 0$ and $g(w, w) \neq 0$ based on the (pseudo-) metric $g$:

$$\sphericalangle(v, w) = \arccos \frac{g(v, w)}{\sqrt{|g(v, v)||g(w, w)|}} \tag{59}$$

Note that the absolute values $|g(v, v)|$ and $|g(w, w)|$ must be taken because $g(\cdot, \cdot)$ may be negative.

Let $\mathscr{V}, \mathscr{W} \in \mathfrak{X}(\gamma)$ be parallel along $\gamma$, i.e. $\nabla_{\gamma'(t)}\mathscr{X} = \nabla_{\gamma'(t)}\mathscr{Y} = 0$. As before it is

$$\frac{d}{dt} g\left(\mathscr{X}(\gamma(t)), \mathscr{Y}(\gamma(t))\right) = g\left(\nabla_{\gamma'(t)}\mathscr{X}(\gamma(t)), \mathscr{Y}(\gamma(t)\right) + g\left(\mathscr{X}(\gamma(t), \nabla_{\gamma'(t)}\mathscr{Y}(\gamma(t))\right) = 0$$

i.e. $g\left(\mathscr{X}(\gamma(t)), \mathscr{Y}(\gamma(t))\right)$ is constant along $\gamma$. Since $g\left(\mathscr{X}(\gamma(t)), \mathscr{X}(\gamma(t))\right)$ and $g\left(\mathscr{Y}(\gamma(t)), \mathscr{Y}(\gamma(t))\right)$ are constant, the angle $\sphericalangle\left(\mathscr{X}(\gamma(t)), \mathscr{Y}(\gamma(t))\right)$ is constant along $\gamma$:

**Note 12**: The angle between two vectors that are parallel transported w.r.t. the Levi-Civita connection along the same curve does not change. ∎

Let $\gamma : I \to M$ be a closed path, $p \in \gamma(I)$. Then, parallel transport along $\gamma$ induces a map $P^\gamma : T_pM \to T_pM$ called *holonomy map* (or *holonomy of* $\gamma$) which is even a linear isomorphism (see [12], Prop. 15.7). A proof that the set of all holonomy maps are a group can be found in [19] Chapter II, Prop. 3.3: first, it is shown that compositions of closed curves is again a closed curve, thus, holonomy maps compose. Next, the inverse of a holonomy map is the parallel transport along the curve in reverse direction.

Because of Note 11 holonomy maps are length preserving, i.e. they are orthogonal maps:

**Note 13**: The set of holonomy maps at $p$ is a group $\mathfrak{Hol}_p(M) \lhd O(n)$, $\dim M = n$. ∎

Furthermore, if $M$ is path-connected, it is $\mathfrak{Hol}_p(M) = \mathfrak{Hol}_q(M)$ for $p, q \in M$. Basically, this is because a path $\alpha$ from $p$ to $q$ and a closed path $\gamma$ at $q$ results in a closed path $\alpha^{-1} \circ \gamma \circ \alpha$ at $p$ which shifts holonomy maps from $\mathfrak{Hol}_q(M)$ to $\mathfrak{Hol}_p(M)$. This implies that the *holonomy group* $\mathfrak{Hol}(M) \lhd O(n)$ of $M$ is a well-defined subgroup of the orthogonal group $O(n)$ for $n = \dim M$.

### 7.5. Isometries

Each mathematical structure has its “isomorphisms”, i.e. definitions of equivalence of structures of the same kind: the “isomorphisms” of vector spaces are “linear isomorphisms”, that of topological spaces are “homeomorphisms”, that of differentiable manifolds are “diffeomorphisms”, and that of (pseudo-) Riemannian manifolds are “isometries”:

**Definition 45:** Let $(M, g)$ and $\left(\widehat{M}, \widehat{g}\right)$ be (pseudo-) Riemannian manifolds. A diffeomorphism $f : M \to \widehat{M}$ is called *isometry* $:\Leftrightarrow f^*\widehat{g} = g$. □

Hereby, $f^*\widehat{g}$ is called *pullback of the metric* $\widehat{g}$ and is defined as

$$(f^*\widehat{g})_p(v_1, v_2) := \widehat{g}_{f(p)}(df_p(v_1), df_p(v_2)) \tag{60}$$

Thus, by definition an isometry satisfies

$$g_p(v_1, v_2) = \widehat{g}_{f(p)}(df_p(v_1), df_p(v_2)) \tag{61}$$

Consequently, $f: M \to \widehat{M}$ is an isometry if and only if $df_p : T_pM \to T_{f(p)}\widehat{M}$ is a linear isometry for each $p \in M$ which means that $df_p$ is an orthogonal map.

Isometries also preserve the Levi-Civita connections on the manifolds (see [23] Prop. 5.13):

**Lemma 25**: Let $(M, g)$, $\left(\widehat{M}, \widehat{g}\right)$ be (pseudo-) Riemannian manifolds with Levi-Civita connections $\nabla$, $\widehat{\nabla}$ and let $f : M \to \widehat{M}$ be an isometry. Then $f^* \widehat{\nabla} = \nabla$. ∎

The lemma before is somehow intuitive because isometries preserve the metric and the metric determines uniquely the Levi-Civita connection. This in turn implies that the Riemann curvature tensor is preserved under isometries: it is

$$\nabla_{\mathcal{X}}\mathcal{Y} = (f^*\widehat{\nabla})_{\mathcal{X}}\mathcal{Y} = \widehat{\nabla}_{f^*\mathcal{X}}(f^*\mathcal{Y})$$

which implies $f^* \widehat{R} = R$:

**Note 14**: The Riemann curvature tensor is preserved under isometries. ∎

This last note shows that the Riemann curvature is invariant under isometries.

**Definition 46:** $(M, g)$ is called *flat* ($\dim M = n$) $:\Leftrightarrow$ $M$ is locally isometric to an open subset $U$ of $\mathbb{R}^n_+$. □

The Riemann curvature tensor indicates flatness of a manifold (see [23] Th. 7.10):

**Theorem 7**: $(M, g)$ is flat $\Leftrightarrow$ $R = 0$. ∎

Theorem 7 justifies the term "curvature" in "Riemann curvature tensor". It is important to note that the Riemann curvature $R$ is an intrinsic property of a manifold, i.e. $R$ is independent of any ambient space: namely $R$ is defined by means of the Levi-Civita connection which in turn is determined by the (pseudo-) metric (see comment after Lemma 24).

This has consequences which seem to be counterintuitive first place: the circle $\mathbb{S}^1$ is a flat Riemannian manifold, but as embedded in the Euclidian space (i.e. located in an ambient space) the circle $\mathbb{S}^1 \subset \mathbb{R}^2$ is of course curved. Similarly, the torus $\mathbb{S}^1 \times \mathbb{S}^1$ with a suitable metric is flat while the "donut-shaped" torus in $\mathbb{R}^3$ is obviously not flat. The reason is that Riemann curvature is a different measure than the intuitive curvature of curves or surfaces. More notions of curvature will be defined in the next Section 7.5.

Isometries represent symmetries of a (pseudo-) Riemannian manifold. Symmetries imposed by the flow of a vector field play an important role in several domains. Vector fields that keep the (pseudo-) metric unchanged along their flows turn out to result in such symmetries:

**Definition 47:** Let $(M, g)$ be a (pseudo-) Riemannian manifold. $\mathcal{V} \in \mathfrak{X}(M)$ is called a *Killing (vector) field* $:\Leftrightarrow$ $\mathcal{L}_{\mathcal{V}}g = 0$. □

While the Lie derivative of a vector field has been defined in Definition 21, and the Lie derivative of a function in Definition 25, $\mathcal{L}_{\mathcal{V}}g$ is defined as follows (we generalize Lie derivatives to tensor fields in Definition 70): Let $\Phi_{\mathscr{V}} : D \to M$ be the maximal flow of the vector field $\mathscr{V}$, and $(\gamma_{-t})^* g_p$ be the pull back of the metric along an integral curve; then:

$$(\mathscr{L}_{\mathscr{V}} g)_p = \lim_{t \to 0} \frac{(\gamma_{-t})^* g_p - g_p}{t} \tag{62}$$

Killing vector fields result in symmetries along their flows (see [15], Lemma 2.3.7):

**Lemma 26**: $\mathscr{V} \in \mathfrak{X}(M)$ is a Killing field $\Leftrightarrow$ $\Phi_{\mathscr{V},t} : M_t \to M_{-t}$ is isometric for each $t$. ∎

Note 14 and Lemma 26 immediately imply that the Riemann curvature tensor is preserved along the flow of a Killing field. Also, this lemma finally gives a nice geometric interpretation of Killing fields in terms of symmetries. The left part of Figure 29 shows a Killing vector field $\mathscr{V}$; since $\Phi_{\mathscr{V},t}$ is an isometry (Lemma 26) it preserves the (pseudo-) metric $g$, which in turn implies that the lengths and angles of the shape $\Delta$ do not change along the time period $t$ by the flow of $\mathscr{V}$ (Note 11 and Note 12): the manifold looks symmetric along the flow of $\mathscr{V}$. In contrast, the right part of Figure 29 depicts the flow of a vector field $\mathscr{W}$; this vector field is not a Killing vector field because its flow $\Phi_{\mathscr{W},t}$ changes the lengths and angles of he shape $\Delta$, i.e. the $\Phi_{\mathscr{W},t}$ cannot preserve $g$, i.e. it cannot be an isometry implying that $\mathscr{W}$ is not a Killing vector field.

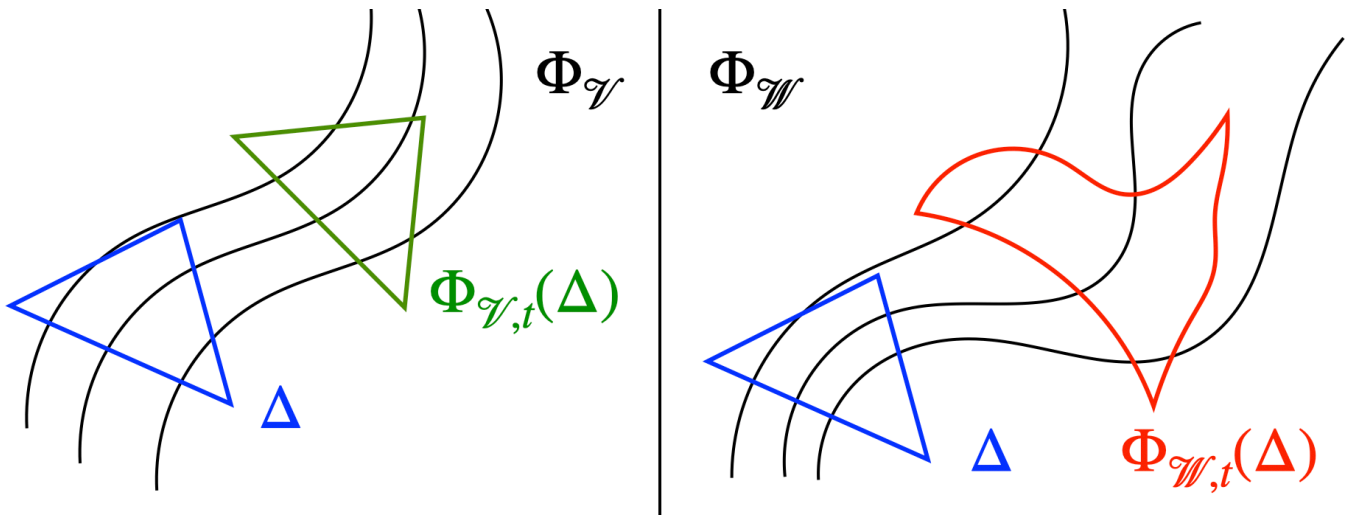

**Fig. 29**. Geometric Interpretation of Killing Vector Fields

In local coordinates, Killing vector fields can be determined via the following equation (see [22] Ex. 13.13(b)):

**Lemma 27 (Killing Equation)**: $\mathscr{V}$ is a Killing field $\Leftrightarrow$ in local coordinates, $\mathscr{V}$ satisfies for all $1 \le i, j \le k$

$$\sum_k \left( \mathscr{V}_k \frac{\partial g_{ij}}{\partial x^k} + g_{jk} \frac{\partial \mathscr{V}_k}{\partial x^i} + g_{ik} \frac{\partial \mathscr{V}_k}{\partial x^j} \right) = 0 \tag{63}$$ ∎

W.r.t. the Levi-Civita connection $\nabla$ of $(M, g)$ and $\mathscr{X}, \mathscr{Y} \in \mathfrak{X}(M)$ the following equivalence results (see [30] Ch. 9 Prop. 25):

**Lemma 28**: $\mathscr{V}$ is a Killing (vector) field $\Leftrightarrow$

$$g(\nabla_{\mathcal{X}}\mathcal{V},\mathcal{Y})+g(\mathcal{X},\nabla_{\mathcal{Y}}\mathcal{V})=0 \qquad (64)\ ∎$$

### 7.6. Curvatures

As a tensor field the value of the Riemann curvature tensor $R$ at a point $p$ only depends on the values $\mathcal{X}(p), \mathcal{Y}(p), \mathcal{Z}(p)$ of the argument vector fields $\mathcal{X}, \mathcal{Y}, \mathcal{Z}$ but not on the vector fields in a whole neighborhood of $p$ (see Note 31 later). Especially, fixing two vector fields $\mathcal{Y}, \mathcal{Z} \in \mathfrak{X}(M)$ induces a map

$$\varrho_p^{\mathcal{Y},\mathcal{Z}} : T_pM \to T_pM \text{ with } v \mapsto \big(R(\mathcal{X},\mathcal{Y})\mathcal{Z}\big)(p) \qquad (65)$$

for any $\mathcal{X} \in \mathfrak{X}(M)$ with $\mathcal{X}(p) = v$. Remember that such as vector field $\mathcal{X}$ exists according to Lemma 9, and the value of $\varrho_p^{\mathcal{Y},\mathcal{Z}}(v)$ is independent of a corresponding vector field $\mathcal{X}$ chosen because $R$ only depends on the values of the vector fields at point $p$. Also, $\varrho_p^{\mathcal{Y},\mathcal{Z}}$ is a linear map (Note 10).

Again, $\varrho_p^{\mathcal{Y},\mathcal{Z}}$ depends only on the values $\mathcal{Y}(p)$ and $\mathcal{Z}(p)$. Consequently, for $(y,z) \in T_pM \times T_pM$ and $\mathcal{Y}, \mathcal{Z} \in \mathfrak{X}(M)$ with $\mathcal{Y}(p) = y$ and $\mathcal{Z}(p) = z$ the following bilinear map can be defined:

**Definition 48:** The bilinear map

$$\mathrm{Ric}_p : T_pM \times T_pM \to \mathbb{R} \text{ with } (y,z) \mapsto \mathrm{Tr}\,\varrho_p^{\mathcal{Y},\mathcal{Z}} \qquad (66)$$

for any vector fields $\mathcal{Y}(p) = y, \mathcal{Z}(p) = z$ is called *Ricci tensor*. □

$\mathrm{Ric}_p$ is a symmetric bilinear map (see [30], Ch. 3, Lemma 52), and symmetric bilinear maps are determined by their values on unit vectors (see [23] Lemma 8.11):

**Lemma 29**: $\mathrm{Ric}_p$ is determined by its values on $\mathbb{S}^n \subset T_pM$, $n = \dim M$. ∎

Thus justifies the following definition:

**Definition 49:** Let $v \in T_pM$, $\|v\| = 1$. Then $\mathrm{Ric}_p(v,v)$ is called the *Ricci Curvature* at $p$ in direction of $v$. □

The following shows how the Ricci-curvature can be locally computed (see [28] Sec. 4.2.1 and the computation there following Def. 4.2.6):

**Lemma 30**: With $\mathrm{Ric} = (R_{ij})$ it is

$$R_{ij} = \sum_{a=1}^{n} \frac{\partial \Gamma_{ij}^a}{\partial x_a} - \sum_{a=1}^{n} \frac{\partial \Gamma_{ai}^a}{\partial x_j} + \sum_{a=1}^{n}\sum_{b=1}^{n} \left(\Gamma_{ab}^a \Gamma_{ij}^b - \Gamma_{ib}^a \Gamma_{aj}^b\right) \qquad (67)\ ∎$$

According to this lemma, $R_{ij}$ is determined by the Christoffel symbols and their derivatives, and because the Christoffel symbols are determined by the metric (Equation 50) Ric is determined by the metric. Also, Ric is determined by the

Riemann curvature tensor (which is somehow obvious from the definition of Ric) (see [23] Prop. 7.15):

**Note 15**: $R_{ij} = \sum_k R^k_{kij}$. ∎

A manifold is called *Ricci flat* iff $\mathrm{Ric} = 0$. Obviously, a flat manifold (see Definition 46) is Ricci flat (because $R = 0$, Theorem 7) but the converse is not true (see [30] p. 87). Thus, the Ricci curvature is less precise than the Riemann curvature: while the Riemann curvature contains all information about the curvature of a manifold the Ricci curvature is some sort of a summary of the Riemann curvature (it is its trace).

Even the Ricci curvature can be further “summarized” by taking its trace:

**Definition 50:** Let $g^{ij}$ denote the coefficients of $(g_{ij})^{-1}$. Then, $S := \sum_{i,j} g^{ij} R_{ij}$ is called *Ricci scalar* or *scalar curvature* (also denoted by Scal). □

The Ricci tensor and the Ricci scalar play a central role in general relativity - they are parameters of Einstein’s Field Equation (see Equation 123 as well as [10] Equation 4.44).

Let $v, w \in T_pM$ be linear independent tangent vectors, and let $\mathscr{V}, \mathscr{W} \in \mathfrak{X}(M)$ be vector fields with $\mathscr{V}(p) = v, \mathscr{W}(p) = w$. Then, $g_p(v,v)g_p(w,w) - g_p(v,w)$ is the square of the area of the parallelogram spanned by $v$ and $w$ (this is elementary geometry). Next, $g_p\left(R(v,w)w, v\right)$ is a measure of the square of the area of the parallelogram spanned by $v$ and the holonomy of $w$ (refer to the explanation of Figure 27). Thus, the quotient

$$K_p(v,w) := \frac{g_p\left(R(\mathscr{V},\mathscr{W})\mathscr{W},\mathscr{V}\right)}{g_p(v,v)g_p(w,w) - g_p(v,w)} \tag{68}$$

measures how much the first parallelogram in the tangent space is squeezed or stretched as a consequence of curvature: obviously, for a flat manifold (i.e. $R = 0$) it is $K_p(v,w) = 0$.

Furthermore, $K_p(v,w)$ only depends on the plane $\sigma \subseteq T_pM$ spanned by $v, w$ (see [30] Ch.3, Lemma 39 ), which justifies the notation $K_p(\sigma)$:

**Definition 51:** Let $\sigma \subseteq T_pM$ be a two dimensional subplane of $T_pM$. Then, $K_p(\sigma)$ is called *sectional curvature* of $\sigma$ at $p$. □

Note, that for $\dim M = 2$ and $M \subseteq \mathbb{R}^3$ the sectional curvature is the same as the Gauss curvature known from basic differential geometry.

While $R = 0$ implies by definition a vanishing sectional curvature $K_p(\sigma) = 0$ for each plane $\sigma$, the converse is non-trivial (see [30] Ch.3, Prop. 41) but true, i.e. a manifold is flat if and only if its sectional curvature vanishes:

**Note 16**: If $K_p(\sigma) = 0$ for each plane $\sigma \subseteq T_pM$ and each $p$, then $R = 0$. ∎

This note indicates that the sectional curvature influences the Riemann curvature: in fact, let $\partial_i$ and $\partial_j$ be two standard basis vectors, $\sigma$ be the span of these vectors; then

$$K_p(\sigma) = \frac{g_p\left(R(\partial_i, \partial_j)\partial_j, \partial_i\right)}{g_p(\partial_i, \partial_i)g_p(\partial_j, \partial_j) - g_p(\partial_i, \partial_j)} = \frac{R_{ijij}}{g_{ii}g_{jj} - g_{ij}} = \frac{g_{ii}R^i_{jij}}{g_{ii}g_{jj} - g_{ij}} \tag{69}$$

I.e. $R_{ijij} = K_p(\sigma)(g_{ii}g_{jj} - g_{ij})$ which means that the sectional curvature determines the Riemann curvature tensor (see [31] Th. 5.12).

The next lemma provides a geometric interpretations of the Ricci curvature and the scalar curvature (i.e. the Ricci scalar) which emphasizes the importance of the sectional curvature (see [23] Prop. 8.32):

**Lemma 31**: Let $\{b_1, \ldots, b_n\} \subseteq T_pM$ be any orthonormal basis and $v \in T_pM$ with $\|v\| = 1$. Then:

i. $S = \sum_{i=2}^n K_p(b_1, b_i)$
ii. $\mathrm{Ric}_p(v, v) = \sum_{i=2}^n K_p(v, b_i)$ (i.e. w.l.o.g. $b_1$ is replaced by $v$). ∎

Item (i) of this lemma shows that the scalar curvature is the sum of the sectional curvatures of the planes spanned by pairs of distinct basis vectors of an orthonormal basis and the scalar curvature is the same for any such basis $\{b_i\}$; this is interpreted as the scalar curvature being some sort of an average of the sectional curvature.

Item (ii) computes the Ricci curvature in direction of $v$ as sum of the sectional curvatures of planes having $v$ as a fixed joint direction. Since $\{b_i\}$ is arbitrary it also shows that the Ricci curvature is determined by the sectional curvature.

Because of the importance of sectional curvature the following definition is made:

**Definition 52:** A (pseudo-) Riemannian manifold has *constant sectional curvature* $K$ $:\Leftrightarrow$ $K_p(\sigma) = K$ for all $p$ and all $\sigma$. □

Certain Riemannian manifolds that have constant sectional curvature can be classified (see Theorem 9). The following theorem about manifolds of constant sectional curvature is useful in computations (see [23] Prop. 8.36, and [30] Ch.3, Cor. 43 & Ex. 5):

**Theorem 8**: Let $(M, g)$ be (pseudo-) Riemannian manifold of constant sectional curvature $K$, $\dim M = n$, and let $\mathcal{X}, \mathcal{Y}, \mathcal{Z} \in \mathfrak{X}(M)$. Then:

i. $R(\mathcal{X}, \mathcal{Y})\mathcal{Z} = K \cdot \left(g(\mathcal{Y}, \mathcal{Z})\mathcal{X} - g(\mathcal{X}, \mathcal{Z})\mathcal{Y}\right)$
ii. $\mathrm{Ric}(\mathcal{X}, \mathcal{Y}) = (n-1)K \cdot g(\mathcal{X}, \mathcal{Y})$
iii. $\mathrm{Scal} = n(n-1)K$ ∎

Manifolds whose Ricci tensor is proportional to its (pseudo-) metric (see item (ii) in the theorem before) are important in general relativity. Thus:

**Definition 53:** A (pseudo-) Riemannian manifold is called *Einstein manifold* :⇔ $\mathrm{Ric}_p = \lambda g_p$ for each $p$ and a fixed $\lambda \in \mathbb{R}$. The metric is then called *Einstein metric*. □

This definition has its origin in Einstein's vacuum field equation

$$\mathrm{Ric} - \frac{1}{2}\mathrm{Scal}\, g = 0 \tag{70}$$

i.e. an Einstein metric is a solution of this equation (setting $\lambda = \frac{1}{2}\mathrm{Scal}$).

A possible extension of the definition of an Eistein metric is $\mathrm{Ric}_p = f(p) \cdot g_p$ with $f : M \to \mathbb{R}$. But it turns out that for $\dim M \geq 3$ this is not needed (see [23] Prop. 7.19):

**Lemma 32 (Schur's Lemma)**: Let $f : M \to \mathbb{R}$ and $\mathrm{Ric}_p = f(p) g_p$. If $\dim M \geq 3$ and $M$ connected then $f$ is constant, i.e. $g$ is Einsteinsch. ∎

Not only cases are of interest where the Ricci tensor is proportional to its metric but also situations in which two metrics are proportional:

**Definition 54:** Let $(M, g)$ and $(\overline{M}, \bar{g})$ be Riemannian manifolds. $(M, g)$ and $(\overline{M}, \bar{g})$ are called *conformally equivalent* :⇔ There exists a diffeomorphism $\varphi : M \to \overline{M}$ with $\varphi^* \bar{g} = fg$ for a $f \in C^\infty(M), f > 0$. $\varphi$ is called *conformal diffeomorphism* or a *conformal transformation.* □

The angle between two tangent vectors is

$$\sphericalangle(v, w) = \frac{g(v, w)}{\sqrt{g(v, v) g(w, w)}} \,,$$

thus, conformal maps preserve angles between tangent vectors (and, consequently) curves. But conformal maps do not preserve lengths. It can be shown that any compact manifold allows to turn its metric conformally into one whose scalar curvature is constant (see [23] p. 211):

**Lemma 33**: Let $(M, g)$ be a compact Riemannian manifold. Then, there exists a metric $\bar{g}$ conformal to $g$ having constant scalar curvature. ∎

## 7.7. Intermezzo: Curvatures of the 2-Sphere

We want to compute the curvature of the two-dimensional sphere of radius $R$ in $\mathbb{R}^3$, i.e. $\mathbb{S}^2_R \subset \mathbb{R}^3$. The metric on $\mathbb{S}^3_R$ in spherical coordinates is used (see Section 6.2). Its Christoffel symbols have been computed in Section 7.2:

$$\Gamma^\theta_{\theta\theta} = 0 \quad \Gamma^\theta_{\theta\rho} = 0 \quad \Gamma^\theta_{\rho\theta} = 0 \quad \Gamma^\theta_{\rho\rho} = -\cos\theta \sin\theta$$
$$\Gamma^\rho_{\theta\theta} = 0 \quad \Gamma^\rho_{\theta\rho} = \cot\theta \quad \Gamma^\rho_{\rho\theta} = \cot\theta \quad \Gamma^\rho_{\rho\rho} = 0$$

The components of the Riemann curvature tensor are (according to Equation 54):

$$R^l_{ijk} = \frac{\partial \Gamma^l_{jk}}{\partial x_i} - \frac{\partial \Gamma^l_{ik}}{\partial x_j} + \sum_m \left( \Gamma^m_{jk}\Gamma^l_{im} - \Gamma^m_{ik}\Gamma^l_{jm} \right)$$

For all combinations of $l, i, j, k \in \{\theta, \rho\}$ the expressions for $R^l_{ijk}$ have to be computed. Note, that substituting the index $i$ in $x_i$ by $\theta$ or $\rho$ results in $\partial x_i = \partial\theta$ or $\partial x_i = \partial\rho$, respectively (similar for $\partial x_j$).

First, consider the sum: the summation index $m$ it is $m \in \{\theta, \rho\}$. Thus

$$\sum_m \left( \Gamma^m_{jk}\Gamma^l_{im} - \Gamma^m_{ik}\Gamma^l_{jm} \right) = \left( \Gamma^\theta_{jk}\Gamma^l_{i\theta} - \Gamma^\theta_{ik}\Gamma^l_{j\theta} \right) + \left( \Gamma^\rho_{jk}\Gamma^l_{i\rho} - \Gamma^\rho_{ik}\Gamma^l_{j\rho} \right)$$

Next, all combinations of $j$ and $k$ have to be considered, and within each pair $j, k \in \{\theta, \rho\}$ the index $l \in \{\theta, \rho\}$ is chosen, and within each such triple $j, k, l \in \{\theta, \rho\}$ the index $i \in \{\theta, \rho\}$ is chosen. This results in 16 combinations. As a sample, we perform the computations for 7 combinations; the other computations follow similarly.

*7.7.1 Let $j = \theta$, and $k = \theta$*

$$\sum_m \left( \Gamma^m_{jk}\Gamma^l_{im} - \Gamma^m_{ik}\Gamma^l_{jm} \right) = \left( \Gamma^\theta_{\theta\theta}\Gamma^l_{i\theta} - \Gamma^\theta_{i\theta}\Gamma^l_{\theta\theta} \right) + \left( \Gamma^\rho_{\theta\theta}\Gamma^l_{i\rho} - \Gamma^\rho_{i\theta}\Gamma^l_{\theta\rho} \right)$$

Then with $\Gamma^\theta_{\theta\theta} = 0$ and $\Gamma^\rho_{\theta\theta} = 0$:

$$\sum_m \left( \Gamma^m_{jk}\Gamma^l_{im} - \Gamma^m_{ik}\Gamma^l_{jm} \right) = -\Gamma^\theta_{i\theta}\Gamma^l_{\theta\theta} - \Gamma^\rho_{i\theta}\Gamma^l_{\theta\rho}$$

*a. Let $l = \theta$*

$$\sum_m \left( \Gamma^m_{jk}\Gamma^l_{im} - \Gamma^m_{ik}\Gamma^l_{jm} \right) = -\Gamma^\theta_{i\theta}\Gamma^\theta_{\theta\theta} - \Gamma^\rho_{i\theta}\Gamma^\theta_{\theta\rho}$$

With $\Gamma^\theta_{\theta\theta} = 0$ and $\Gamma^\theta_{\theta\rho} = 0$ we get:

$$j = \theta \wedge k = \theta \wedge l = \theta \Rightarrow \quad \sum_m \left( \Gamma^m_{jk}\Gamma^l_{im} - \Gamma^m_{ik}\Gamma^l_{jm} \right) = 0$$

For $i = \theta \wedge j = \theta \wedge k = \theta \wedge l = \theta$ it is $\Gamma^l_{jk} = \Gamma^\theta_{\theta\theta} = 0$, thus $\frac{\partial \Gamma^l_{jk}}{\partial x_i} = 0$; similarly $\Gamma^l_{ik} = \Gamma^\theta_{\theta\theta} = 0$, thus $\frac{\partial \Gamma^l_{ik}}{\partial x_j} = 0$

Substituting this for $R^l_{ijk}$ gives:

$$R^\theta_{\theta\theta\theta} = 0$$

For $i = \rho \wedge j = \theta \wedge k = \theta \wedge l = \theta$ it is $\Gamma^l_{jk} = \Gamma^\theta_{\theta\theta} = 0$, thus $\frac{\partial \Gamma^l_{jk}}{\partial x_i} = 0$; similarly $\Gamma^l_{ik} = \Gamma^\theta_{\rho\theta} = 0$, thus $\frac{\partial \Gamma^l_{ik}}{\partial x_j} = 0$. Thus:

$$R^{\theta}_{\rho\theta\theta} = 0$$

*b. Let $l = \rho$*

$\sum_m \left(\Gamma^m_{jk}\Gamma^l_{im} - \Gamma^m_{ik}\Gamma^l_{jm}\right) = -\Gamma^{\theta}_{i\theta}\Gamma^{\rho}_{\theta\theta} - \Gamma^{\rho}_{i\theta}\Gamma^{\rho}_{\theta\rho}$

With $\Gamma^{\rho}_{\theta\theta} = 0$ and $\Gamma^{\rho}_{\theta\rho} = \cot\theta$ ist is:

$$j = \theta \wedge k = \theta \wedge l = \rho \Rightarrow \sum_m \left(\Gamma^m_{jk}\Gamma^l_{im} - \Gamma^m_{ik}\Gamma^l_{jm}\right) = -\Gamma^{\rho}_{i\theta}\cot\theta$$

For $i = \theta$ it is $\Gamma^{\rho}_{\theta\theta} = 0$, thus we get

$$i = \theta \wedge j = \theta \wedge k = \theta \wedge l = \rho \Rightarrow \quad \sum_m \left(\Gamma^m_{jk}\Gamma^l_{im} - \Gamma^m_{ik}\Gamma^l_{jm}\right) = 0$$

For $i = \theta \;\wedge\; j = \theta \;\wedge\; k = \theta \;\wedge\; l = \rho$ it is $\Gamma^l_{jk} = \Gamma^{\rho}_{\theta\theta} = 0 = \Gamma^l_{ik}$, thus $\dfrac{\partial\Gamma^l_{jk}}{\partial x_i} = \dfrac{\partial\Gamma^{\rho}_{\theta\theta}}{\partial\theta} = 0 = \dfrac{\partial\Gamma^l_{ik}}{\partial x_j}$.

Substituting this for $R^l_{ijk}$ gives

$$R^{\rho}_{\theta\theta\theta} = 0$$

For $i = \rho$ it is $\Gamma^{\rho}_{\rho\theta} = \cot\theta$, thus we get

$$i = \rho \wedge j = \theta \wedge k = \theta \wedge l = \rho \Rightarrow \quad \sum_m \left(\Gamma^m_{jk}\Gamma^l_{im} - \Gamma^m_{ik}\Gamma^l_{jm}\right) = -\cot^2\theta$$

For $i = \rho \wedge j = \theta \wedge k = \theta \wedge l = \rho$ it is $\Gamma^l_{jk} = \Gamma^{\rho}_{\theta\theta} = 0$ and $\Gamma^l_{ik} = \Gamma^{\rho}_{\rho\theta} = \cot\theta$

thus $\dfrac{\partial\Gamma^l_{jk}}{\partial x_i} = \dfrac{\partial\Gamma^{\rho}_{\theta\theta}}{\partial\rho} = 0$ and $\dfrac{\partial\Gamma^l_{ik}}{\partial x_j} = \dfrac{\partial\Gamma^{\rho}_{\rho\theta}}{\partial\theta} = \dfrac{\partial}{\partial\theta}\cot\theta = -\dfrac{1}{\sin^2\theta}$ (i.e. $-\dfrac{\partial\Gamma^l_{ik}}{\partial x_j} = \dfrac{1}{\sin^2\theta}$).

Substituting this for $R^l_{ijk}$ gives

$$R^{\rho}_{\rho\theta\theta} = \frac{1}{\sin^2\theta} - \cot^2\theta = \frac{1}{\sin^2\theta} - \frac{\cos^2\theta}{\sin^2\theta} = \frac{1-\cos^2\theta}{\sin^2\theta} = \frac{\sin^2\theta}{\sin^2\theta} = 1,$$

i.e.

$$R^{\rho}_{\rho\theta\theta} = 1$$

*7.7.2. Let $j = \rho$, and $k = \rho$*

$\sum_m \left(\Gamma^m_{jk}\Gamma^l_{im} - \Gamma^m_{ik}\Gamma^l_{jm}\right) = \left(\Gamma^{\theta}_{\rho\rho}\Gamma^l_{i\theta} - \Gamma^{\theta}_{i\rho}\Gamma^l_{\rho\theta}\right) + \left(\Gamma^{\rho}_{\rho\rho}\Gamma^l_{i\rho} - \Gamma^{\rho}_{i\rho}\Gamma^l_{\rho\rho}\right)$

then with $\Gamma^{\theta}_{\rho\rho} = -\cos\theta\sin\theta$ and $\Gamma^{\rho}_{\rho\rho} = 0$:

$\sum_m \left(\Gamma^m_{jk}\Gamma^l_{im} - \Gamma^m_{ik}\Gamma^l_{jm}\right) = -\Gamma^l_{i\theta}\cos\theta\sin\theta - \Gamma^\theta_{i\rho}\Gamma^l_{\rho\theta} - \Gamma^\rho_{i\rho}\Gamma^l_{\rho\rho}$

*a. Let $l = \theta$*

$\sum_m \left(\Gamma^m_{jk}\Gamma^l_{im} - \Gamma^m_{ik}\Gamma^l_{jm}\right) = -\Gamma^\theta_{i\theta}\cos\theta\sin\theta - \Gamma^\theta_{i\rho}\Gamma^\theta_{\rho\theta} - \Gamma^\rho_{i\rho}\Gamma^\theta_{\rho\rho}$

With $\Gamma^\theta_{\rho\theta} = 0$ and $\Gamma^\theta_{\rho\rho} = -\cos\theta\sin\theta$

$\sum_m \left(\Gamma^m_{jk}\Gamma^l_{im} - \Gamma^m_{ik}\Gamma^l_{jm}\right) = -\Gamma^\theta_{i\theta}\cos\theta\sin\theta + \Gamma^\rho_{i\rho}\cos\theta\sin\theta$

Next, let $i = \theta$; then

$\sum_m \left(\Gamma^m_{jk}\Gamma^l_{im} - \Gamma^m_{ik}\Gamma^l_{jm}\right) = -\Gamma^\theta_{\theta\theta}\cos\theta\sin\theta + \Gamma^\rho_{\theta\rho}\cos\theta\sin\theta$

With $\Gamma^\theta_{\theta\theta} = 0$ and $\Gamma^\rho_{\theta\rho} = \cot\theta$ it is

$\sum_m \left(\Gamma^m_{jk}\Gamma^l_{im} - \Gamma^m_{ik}\Gamma^l_{jm}\right) = \cot\theta\cos\theta\sin\theta = \frac{\cos\theta}{\sin\theta}\cos\theta\sin\theta = \cos^2\theta$, thus

$$i = \theta \wedge j = \rho \wedge k = \rho \wedge l = \theta \Rightarrow \quad \sum_m \left(\Gamma^m_{jk}\Gamma^l_{im} - \Gamma^m_{ik}\Gamma^l_{jm}\right) = \cos^2\theta$$

For $i = \theta \ \wedge \ j = \rho \ \wedge \ k = \rho \ \wedge \ l = \theta$ it is $\Gamma^l_{jk} = \Gamma^\theta_{\rho\rho} = -\cos\theta\sin\theta$ and $\Gamma^l_{ik} = \Gamma^\theta_{\theta\rho} = 0$, thus

$$\frac{\partial\Gamma^l_{jk}}{\partial x_i} = \frac{\partial\Gamma^\theta_{\rho\rho}}{\partial\theta} = -\frac{\partial}{\partial\theta}\cos\theta\sin\theta = -(-\sin\theta\sin\theta + \cos\theta\cos\theta) = \sin^2\theta - \cos^2\theta$$

and $\dfrac{\partial\Gamma^l_{ik}}{\partial x_j} = \dfrac{\partial\Gamma^\theta_{\theta\rho}}{\partial\rho} = 0$ .

Substituting this for $R^l_{ijk}$ gives $R^\theta_{\theta\rho\rho} = \sin^2\theta - \cos^2\theta + \cos^2\theta = \sin^2\theta$, i.e.

$$R^\theta_{\theta\rho\rho} = \sin^2\theta$$

*b. Let $l = \rho$*

$\sum_m \left(\Gamma^m_{jk}\Gamma^l_{im} - \Gamma^m_{ik}\Gamma^l_{jm}\right) = -\Gamma^\rho_{i\theta}\cos\theta\sin\theta - \Gamma^\theta_{i\rho}\Gamma^\rho_{\rho\theta} - \Gamma^\rho_{i\rho}\Gamma^\rho_{\rho\rho}$

With $\Gamma^\rho_{\rho\theta} = \cot\theta$ and $\Gamma^\rho_{\rho\rho} = 0$

$\sum_m \left(\Gamma^m_{jk}\Gamma^l_{im} - \Gamma^m_{ik}\Gamma^l_{jm}\right) = -\Gamma^\rho_{i\theta}\cos\theta\sin\theta - \Gamma^\theta_{i\rho}\cot\theta$

Next, let $i = \theta$; then

$\sum_m \left(\Gamma^m_{jk}\Gamma^l_{im} - \Gamma^m_{ik}\Gamma^l_{jm}\right) = -\Gamma^\rho_{\theta\theta}\cos\theta\sin\theta - \Gamma^\theta_{\theta\rho}\cot\theta$

with $\Gamma^\rho_{\theta\theta} = 0$ and $\Gamma^\theta_{\theta\rho} = 0$ we get

$$i = \theta \wedge j = \rho \wedge k = \rho \wedge l = \rho \Rightarrow \sum_m \left( \Gamma^m_{jk}\Gamma^l_{im} - \Gamma^m_{ik}\Gamma^l_{jm} \right) = 0$$

For $i = \theta \wedge j = \rho \wedge k = \rho \wedge l = \rho$ it is $\Gamma^l_{jk} = \Gamma^\rho_{\rho\rho} = 0$ and $\Gamma^l_{ik} = \Gamma^\rho_{\theta\rho} = \cot\theta$, thus $\frac{\partial \Gamma^l_{jk}}{\partial x_i} = \frac{\partial \Gamma^\rho_{\rho\rho}}{\partial \theta} = 0$ and $\frac{\partial \Gamma^l_{ik}}{\partial x_j} = \frac{\partial \Gamma^\rho_{\theta\rho}}{\partial \rho} = \frac{\partial}{\partial \rho}\cot\theta = 0$

Substituting this for $R^l_{ijk}$ gives

$$R^\rho_{\theta\rho\rho} = 0$$

*7.7.3. Let $j = \theta$, and $k = \rho$*

$$\sum_m \left( \Gamma^m_{jk}\Gamma^l_{im} - \Gamma^m_{ik}\Gamma^l_{jm} \right) = \left( \Gamma^\theta_{\theta\rho}\Gamma^l_{i\theta} - \Gamma^\theta_{i\rho}\Gamma^l_{\theta\theta} \right) + \left( \Gamma^\rho_{\theta\rho}\Gamma^l_{i\rho} - \Gamma^\rho_{i\rho}\Gamma^l_{\theta\rho} \right)$$

Then with $\Gamma^\theta_{\theta\rho} = 0$ and $\Gamma^\rho_{\theta\rho} = \cot\theta$:

$$\sum_m \left( \Gamma^m_{jk}\Gamma^l_{im} - \Gamma^m_{ik}\Gamma^l_{jm} \right) = -\Gamma^\theta_{i\rho}\Gamma^l_{\theta\theta} + \Gamma^l_{i\rho}\cot\theta - \Gamma^\rho_{i\rho}\Gamma^l_{\theta\rho}$$

*a. Let $l = \theta$*

$$\sum_m \left( \Gamma^m_{jk}\Gamma^l_{im} - \Gamma^m_{ik}\Gamma^l_{jm} \right) = -\Gamma^\theta_{i\rho}\Gamma^\theta_{\theta\theta} + \Gamma^\theta_{i\rho}\cot\theta - \Gamma^\rho_{i\rho}\Gamma^\theta_{\theta\rho}$$

With $\Gamma^\theta_{\theta\theta} = 0$ and $\Gamma^\theta_{\theta\rho} = 0$:

$$\sum_m \left( \Gamma^m_{jk}\Gamma^l_{im} - \Gamma^m_{ik}\Gamma^l_{jm} \right) = \Gamma^\theta_{i\rho}\cot\theta$$

Let $i = \rho$; then $\Gamma^\theta_{\rho\rho} = -\cos\theta\sin\theta$ and $\sum_m \left( \Gamma^m_{jk}\Gamma^l_{im} - \Gamma^m_{ik}\Gamma^l_{jm} \right) = -\cos\theta\sin\theta\cot\theta = -\cos\theta\sin\theta\frac{\cos\theta}{\sin\theta} = -\cos^2\theta$, thus

$$i = \rho \wedge j = \theta \wedge k = \rho \wedge l = \theta \Rightarrow \sum_m \left( \Gamma^m_{jk}\Gamma^l_{im} - \Gamma^m_{ik}\Gamma^l_{jm} \right) = -\cos^2\theta$$

For $i = \rho \wedge j = \theta \wedge k = \rho \wedge l = \theta$ it is $\Gamma^l_{jk} = \Gamma^\theta_{\theta\rho} = 0$ and $\Gamma^l_{ik} = \Gamma^\theta_{\rho\rho} = -\cos\theta\sin\theta$, thus $\frac{\partial \Gamma^l_{jk}}{\partial x_i} = \frac{\partial \Gamma^\theta_{\theta\rho}}{\partial \rho} = 0$ and

$$-\frac{\partial \Gamma^l_{ik}}{\partial x_j} = -\frac{\partial \Gamma^\theta_{\rho\rho}}{\partial \theta} = -\frac{\partial}{\partial \theta}(-\cos\theta\sin\theta) = \frac{\partial}{\partial \theta}(\cos\theta\sin\theta) = -\sin^2\theta + \cos^2\theta$$

Substituting this for $R^l_{ijk}$ gives

$$R^\theta_{\rho\theta\rho} = -\sin^2\theta$$

All remaining combinations are computed the same way. The result is:

Summary (on Riemann Tensor): The components of the Riemann curvature tensor $R$ are:

$$R^{\rho}_{\rho\theta\theta} = 1, R^{\rho}_{\theta\rho\theta} = -1, R^{\theta}_{\theta\rho\rho} = \sin^2\theta, R^{\theta}_{\rho\theta\rho} = -\sin^2\theta$$

all other components vanish. ∎

According to Theorem 7, this proves that $\mathbb{S}^2_R \subset \mathbb{R}^3$ is not flat, i.e. it is no locally isometric to an open subset $U$ of $\mathbb{R}^n_+$.

Next, we compute the Ricci tensor. According to Note 15 it is $R_{ij} = \sum_k R^k_{kij}$. Thus:

$$R_{\theta\theta} = \sum_k R^k_{k\theta\theta} = R^{\theta}_{\theta\theta\theta} + R^{\rho}_{\rho\theta\theta} = 0 + 1 = 1$$

$$R_{\theta\rho} = \sum_k R^k_{k\theta\rho} = R^{\theta}_{\theta\theta\rho} + R^{\rho}_{\rho\theta\rho} = 0 + 0 = 0$$

$$R_{\rho\theta} = \sum_k R^k_{k\rho\theta} = R^{\theta}_{\theta\rho\theta} + R^{\rho}_{\rho\rho\theta} = 0 + 0 = 0$$

$$R_{\rho\rho} = \sum_k R^k_{k\rho\rho} = R^{\theta}_{\theta\rho\rho} + R^{\rho}_{\rho\rho\rho} = \sin^2\theta + 0 = \sin^2\theta$$

Summary (on Ricci Tensor): The components of the Ricci curvature tensor Ric are:

$$R_{\theta\theta} = 1, R_{\rho\rho} = \sin^2\theta, R_{\rho\theta} = 0, R_{\theta\rho} = 0 \quad ∎$$

Thus, $\mathbb{S}^2_R \subset \mathbb{R}^3$ is not Ricci flat. In Section 7.2 we have shown that $g_{\theta\theta} = R^2$, $g_{\rho\rho} = R^2 \sin^2\theta$, $g_{\theta\rho} = g_{\rho\theta} = 0$. Thus, $R_{ij} = \frac{1}{R^2} g_{ij}$ with a fixed $-\frac{1}{R^2} g_{ij} \in \mathbb{R}$. According to Definition 53, $\mathbb{S}^2_R \subset \mathbb{R}^3$ is an Einstein manifold.

Next, we compute the scalar curvature. It is $\mathrm{Scal} = \sum_{i,j} g^{ij} R_{ij}$ (Definition 50). As shown in Section 7.2 it is $g^{\theta\theta} = \frac{1}{R^2}$, $g^{\rho\rho} = \frac{1}{R^2 \sin^2\theta}$, and $g^{\theta\rho} = g^{\rho\theta} = 0$; thus

$$\mathrm{Scal} = g^{\theta\theta} R_{\theta\theta} + g^{\rho\rho} R_{\rho\rho} = \frac{1}{R^2} + \frac{1}{R^2 \sin^2\theta} \sin^2\theta = \frac{2}{R^2}$$

Finally, we compute the sectional curvature. According to Equation 69 it is $K_p(\sigma) = \frac{R_{ijij}}{g_{ii}g_{jj} - g_{ij}} = \frac{g_{ii} R^i_{ijj}}{g_{ii}g_{jj} - g_{ij}}$. Since $\dim \mathbb{S}^2_R = 2 = \dim T_p\mathbb{S}^2_R$ only a single two dimensional subplane in $T_p\mathbb{S}^2_R$, i.e. $i = \theta$ and $j = \rho$. Thus

$$K_p(\sigma) = \frac{g_{\theta\theta} R^{\theta}_{\theta\rho\rho}}{g_{\theta\theta} g_{\rho\rho} - g_{\theta\rho}} = \frac{R^2 \sin^2\theta}{R^2 R^2 \sin^2\theta - 0} = \frac{1}{R^2}$$

## 8. Spacetimes

In general relativity theory pseudo-Riemannian manifolds play a key role that are products of two pseudo-Riemannian manifolds where the metric on one component of the product is “disturbed”. In this section, the basic properties of such manifolds are described.

**8.1. Minkowski Spaces**

Tensors will be explained in much more detail in section 9.1. For the purpose of this section only very little about tensors is needed which we briefly provide now: a map $dx_i : \mathbb{R}^n \to \mathbb{R}^n$ is given via $dx_i(v) = v_i$, i.e. $dx_i$ is the projection of a vector onto its i-th component. The notion $dx_i \otimes dx_j$ then takes two vectors $v, w$ as input and results in the product of the two projections:

$$dx_i \otimes dx_j(v, w) = v_i \cdot w_j \tag{71}$$

Also, for $i = j$ the short-hand notation $dx_i \otimes dx_i = dx_i^2$ is used, i.e. $dx_i^2(v, w) = v_i \cdot w_i$. Thus, it is

$$\sum_{i=1}^n dx_i \otimes dx_i(v, w) = \sum_{i=1}^n dx_i^2(v, w) = \sum_{i=1}^n v_i w_i = \langle v, w \rangle \tag{72}$$

i.e. $\sum_{i=1}^n dx_i \otimes dx_i$ is just another notation for the canonical scalar product $\langle \, , \rangle$ in $\mathbb{R}^n$. Using this notation, each (pseudo-) metric $g$ can be written as

$$g = \sum_{i=1,j}^n g_{ij} dx_i \otimes dx_j \tag{73}$$

with metric coefficients $g_{ij}$ derived from the standard basis of the underlying tangent space (see Equation 35). The pseudo-metric $\langle\langle v, w \rangle\rangle := -v_1 w_1 + \sum_{i=2}^n v_i w_i$, i.e.

$$\langle\langle \cdot, \cdot \rangle\rangle = -dx_1 \otimes dx_1 + \sum_{i=2}^{n+1} dx_i \otimes dx_i \tag{74}$$

is a special Lorentzian metric (Definition 39) i.e. it has signature $(n-1,1)$ (Definition 37) with $g_{11} = -1$ and $g_{ii} = +1$ for $i \geq 2$.

**Definition 55:** $\mathbb{R}^{n+1}$ endowed with the metric of Equation (74) is called a *Minkowski space* and $\langle\langle \cdot, \cdot \rangle\rangle$ is called *Minkowski metric*. The corresponding manifold is denoted by $\mathbb{R}_{n,1}^{n+1} := (\mathbb{R}^n, \langle\langle \cdot, \cdot \rangle\rangle)$. □

In Minkowski space $\mathbb{R}_{n,1}^{n+1}$ the sphere $\mathbb{S}_{n,1}^n(r)$ of radius $r$ is the set of all points $x$ having a fixed positive distance ("radius" $r$) from the origin, i.e. $\|x\|^2 = \langle\langle x, x \rangle\rangle = r^2$:

$$\mathbb{S}_{n,1}^n(r) = \{(x_1, x_2, \ldots, x_{n+1} \mid \langle\langle x, x \rangle\rangle = -x_1^2 + \sum_{i=2}^{n+1} x_i^2 = r^2\} \tag{75}$$

Notably, a sphere in a Minkowski space is very different from a sphere in a Euclidian space. For example, Figure 30 depicts the unit sphere $\mathbb{S}_{2,1}^2(1)$ in the three dimensional Minkowski space $\mathbb{R}_{2,1}^3$. In general, $\mathbb{S}_{n,1}^n(r)$ is a one-sheeted hyperboloid in $\mathbb{R}^{n+1}$, diffeomorphic to $\mathbb{R} \times \mathbb{S}^{n-1}(r)$ (see [30] Ch. 4 Lemma 25):

**Note 17**: $\mathbb{S}_{n,1}^n(r) \approx \mathbb{R} \times \mathbb{S}^{n-1}(r)$. ∎

As a pseudo-Riemannian manifold the pseudo-sphere $\mathbb{S}^n_{n,1}(r)$ is endowed with the metric $ds^2 = -dx_1^2 + \sum_{i=2}^{n+1} dx_i^2$ induced from its ambient space, i.e. $ds^2 = \iota^*\langle\langle\cdot,\cdot\rangle\rangle$, where $\iota : \mathbb{S}^n_{n,1}(r) \to \mathbb{R}^{n+1}$ is the inclusion (see Equation 60 for the definition of a pullback of a metric). As usual in general relativity theory (see [10]) the symbol $ds^2$ is used instead of $g$ for a (pseudo-) metric.

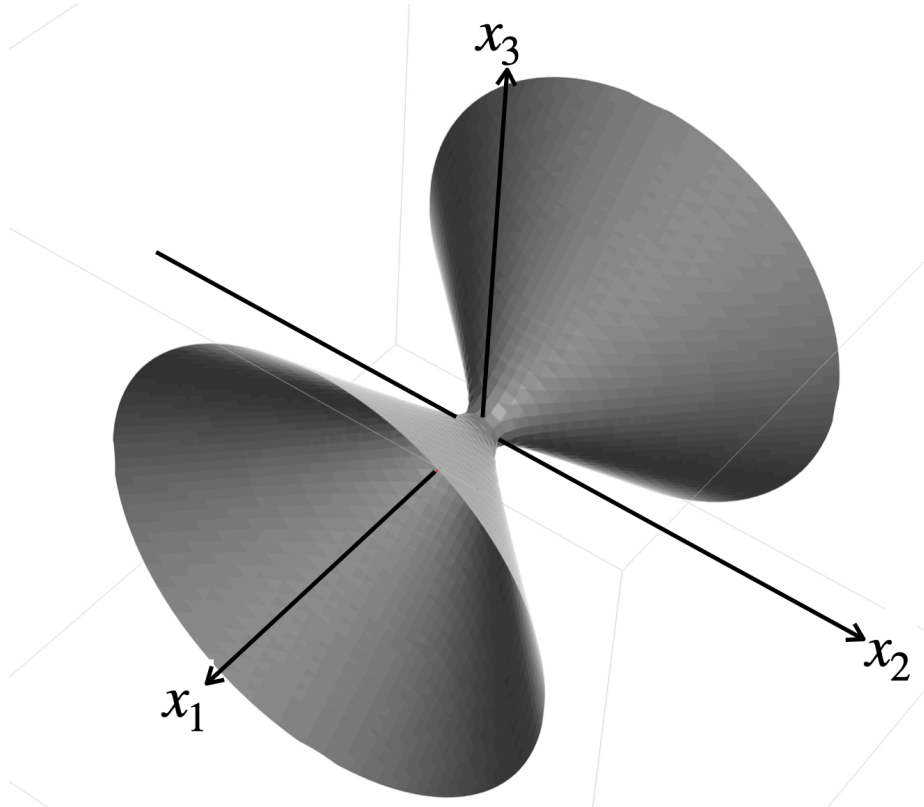

**Fig. 30**. The Unit Sphere $\mathbb{S}^2_{2,1}(1)$ in Minkowski Space $\mathbb{R}^3_{2,1}$

A *hyperboloid* of radius $r$ in the Minkowski space is defined as the set of all points having a fixed distance $r$ from the origin, where the distance is taken with negative sign:

$$\mathscr{H}^n_{n,1}(r) = \{(x_1, x_2, \dots, x_{n+1} \mid -x_1^2 + \textstyle\sum_{i=2}^{n+1} x_i^2 = -r^2\} \tag{76}$$

$\mathscr{H}^n_{n,1}(r)$ has two sheets. A *hyperbolic space* in $\mathbb{R}^{n+1}_{n,1}$ is the upper sheet (i.e. $x_1 > 0$) of a two-sheeted hyperboloid $\mathscr{H}^n_{n,1}(r) \subseteq \mathbb{R}^{n+1}$ denoted by $\mathbb{H}^n_{n,1}(r)$:

$$\mathbb{H}^n_{n,1}(r) = \{(x_1, x_2, \dots, x_{n+1} \mid -x_1^2 + \textstyle\sum_{i=2}^{n+1} x_i^2 = -r^2 \wedge x_1 > 0\} \tag{77}$$

Figure 31 shows the unit hyperboloid $\mathbb{H}^2_{2,1}(1)$ in Minkowski space $\mathbb{R}^3_{2,1}$. $\mathbb{H}^n_{n,1}(r)$ is diffeomorphic to $\mathbb{R}^n$ (see [30], paragraph before Ch. 4 Definition 26):

**Note 18**: $\mathbb{H}^n_{n,1}(r) \approx \mathbb{R}^n$. ∎

The hyperboloid is a pseudo-Riemannian manifold where the endowed metric on $\mathbb{H}^n_{n,1}(r)$ is induced from its ambient space, i.e. $ds^2 = -dx_1^2 + \sum_{i=2}^{n+1} dx_i^2$. In many computational situations, this metric is needed in spherical coordinates, especially for $\mathbb{H}^3_{3,1}(R)$ (i.e. in dimension $n = 3$ with $r = R = \text{const}$) which is relevant in general relativity (see [13] Equation B.16):

$$ds^2 = R^2 \left(d\chi + \sinh^2\chi(d\theta^2 + \sin^2\theta d\phi^2)\right) \tag{78}$$

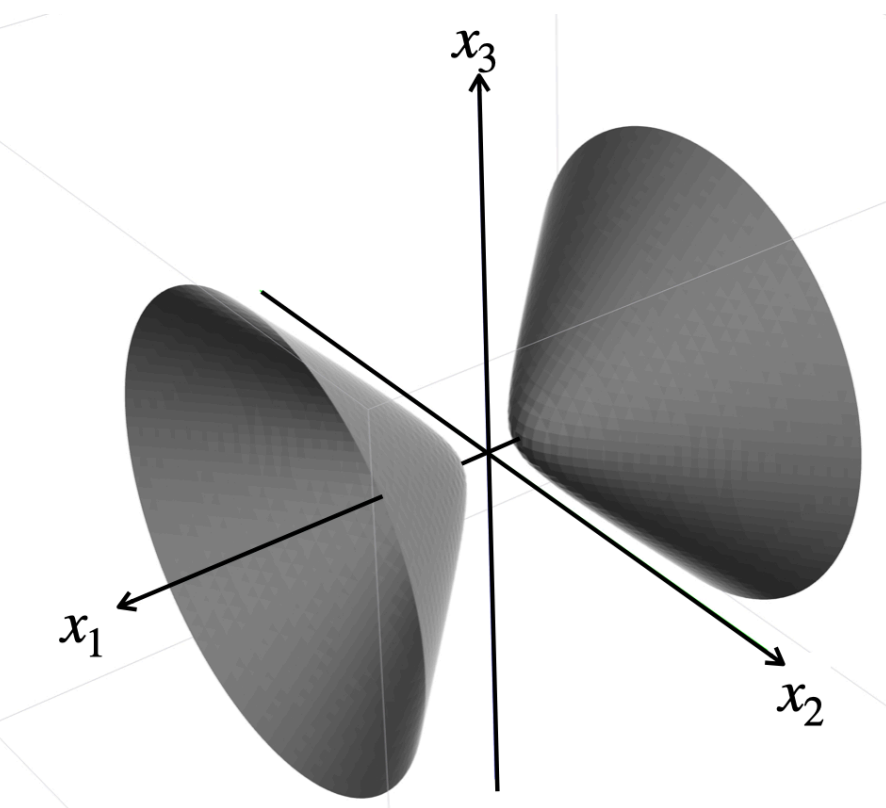

**Fig. 31**. The Unit Hyperboloid $\mathbb{H}^2_{2,1}(1)$ (Left Sheet $x_1 > 0$) in Minkowski Space $\mathbb{R}^3_{2,1}$

Minkowski metrics are generalized to metrics with signatures of more than one negative sign, i.e. $n_- > 1$:

$$ds^2 = -\sum_{i=1}^{n_-} dx_i^2 - \sum_{i=n_-+1}^{n_+ + n_-} dx_i^2 \quad (79)$$

Of special interest is $\mathbb{R}^n$ endowed with such a metric:

**Definition 56:** $\mathbb{R}^n_{n_+,n_-}$ denotes $\mathbb{R}^n$ endowed with the (pseudo-) metric $ds^2 = -\sum_{i=1}^{n_-} dx_i^2 - \sum_{i=n_-+1}^{n_+ + n_-} dx_i^2$. □

A hyperboloid with radius $r$ in $\mathbb{R}^{n+1}_{n-1,2}$ is

$$\mathbb{H}^n_{n-1,2}(r) = \{(x_1, x_2, \ldots, x_{n+1} \mid -x_1^2 - x_2^2 + \sum_{i=3}^{n+1} x_i^2 = -r^2\} \quad (80)$$

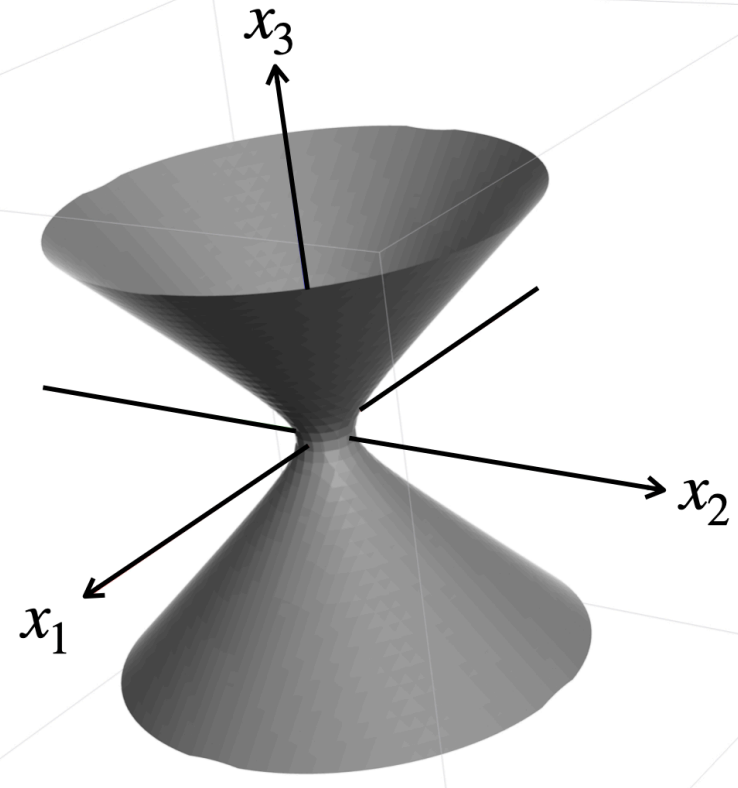

**Fig. 32**. The Unit Hyperboloid $\mathbb{H}^2_{1,2}(1)$ in $\mathbb{R}^3_{1,2}$ (note interchanged axis compared to $\mathbb{S}^2_{2,1}(1)$)

This hyperboloid is single sheeted (see Figure 32). In dimension $n = 2$ the unit hyperboloid $\mathbb{H}^2_{1,2}(1)$ is homeomorphic to $\mathbb{S}^2_{2,1}(1)$ (see Figure 30), but note that the $x_1$- and $x_3$-axis are interchanged. In general, $\mathbb{H}^n_{n-1,2}(r)$ is diffeomorphic to $\mathbb{R}^{n-1} \times \mathbb{S}^1$ (see [30] Ch. 4 Lemma 25):

**Note 19**: $\mathbb{H}^n_{n-1,2}(r) \approx \mathbb{R}^{n-1} \times \mathbb{S}^1$. ∎

This diffeomorphism is indicated by the rewrite

$$-x_1^2 - x_2^2 + \sum_{i=3}^{n+1} x_i^2 = -r^2 \Leftrightarrow x_1^2 + x_2^2 = r^2 + \sum_{i=3}^{n+1} x_i^2$$

where $x_1^2 + x_2^2$ corresponds to the circle $\mathbb{S}^1$ and $r^2 + \sum_{i=3}^{n+1} x_i^2$ corresponds to $\mathbb{R}^{n-1}$. As a pseudo-Riemannian manifold $ds^2 = -dx_1^2 - dx_2^2 + \sum_{i=3}^{n+1} dx_i^2$ is the pseudo-metric on $\mathbb{H}^n_{n-1,2}(r)$ induced by its ambient space.

The spaces introduced in this chapter allow for a classification of manifolds of constant sectional curvature:

**Theorem 9 (Killing-Hopf)**: Let $(M, g)$ be a complete, simply connected and Riemannian manifold ($\dim M = n \geq 2$). If $M$ has constant sectional curvature $K$, then:

i. $K = 0 \Rightarrow M$ is isometric to $\mathbb{R}^n$
ii. $K > 0 \Rightarrow M$ is isometric to $\mathbb{S}^n(r)$ for some $r \in \mathbb{R}_+$ ($K = \frac{1}{r^2}$)
iii. $K < 0 \Rightarrow M$ is isometric to $\mathbb{H}^n_{n,1}(r)$ for some $r \in \mathbb{R}_+$ ($K = -\frac{1}{r^2}$)

∎

$\mathbb{S}^n(r)$ is the (usual) sphere of radius $r$ in $\mathbb{R}^{n+1}$ with the Euclidian metric inherited from the induced metric. $\mathbb{H}^n_{n,1}(r)$ is the upper sheet of two-sheeted hyperboloid $\mathscr{H}^n_{n,1}(r) \subseteq \mathbb{R}^{n+1}$.

## 8.2. Warped Products

Let $(B, g_B)$ and $(F, g_F)$ be (pseudo-) Riemannian manifolds. The product manifold $B \times F$ is again a (pseudo-) Riemannian manifolds in a canonical manner: denote by $\pi : B \times F \to B$ and $\sigma : B \times F \to F$ the canonical projections, then with

$$(\pi^* g_B + \sigma^* g_F)(v, w) = g_B(\pi^* v, \pi^* w) + g_F(\sigma^* v, \sigma^* w) \tag{81}$$

it is $\pi^* g_B + \sigma^* g_F$ a (pseudo-) Riemannian metric on $B \times F$ (see Equation 60 for the pullback of a metric). Thus:

**Note 20**: If $(B, g_B)$ and $(F, g_F)$ are two (pseudo-) Riemannian manifolds then the product manifold $(B \times F, \pi^* g_B + \sigma^* g_F)$ is a (pseudo-) Riemannian manifold. ∎

This can be generalized as follows:

**Definition 57:** Let $(B, g_B)$ and $(F, g_F)$ be (pseudo-) Riemannian manifolds and $f : B \to \mathbb{R}_+$ be a smooth function. The *warped product* $M = B \times_f F$ is the product manifold $B \times F$ endowed with the (pseudo-) metric $\pi^* g_B + (f \circ \pi)^2 \sigma^* g_F$ where $\pi : B \times F \to B$ and $\sigma : B \times F \to F$ are the canonical projections. The function $f$ is called a *warping function*, $\pi^{-1}(p)$ (as well as $F$) is called a *fiber*, $\sigma^{-1}(q)$ is called a *leaf*, $B$ is called the *basis*. □

Figure 33 depicts a warped product: the warping function $f$ contracts or expands the product manifold $M = B \times F$ in the direction of the fibers $F$.

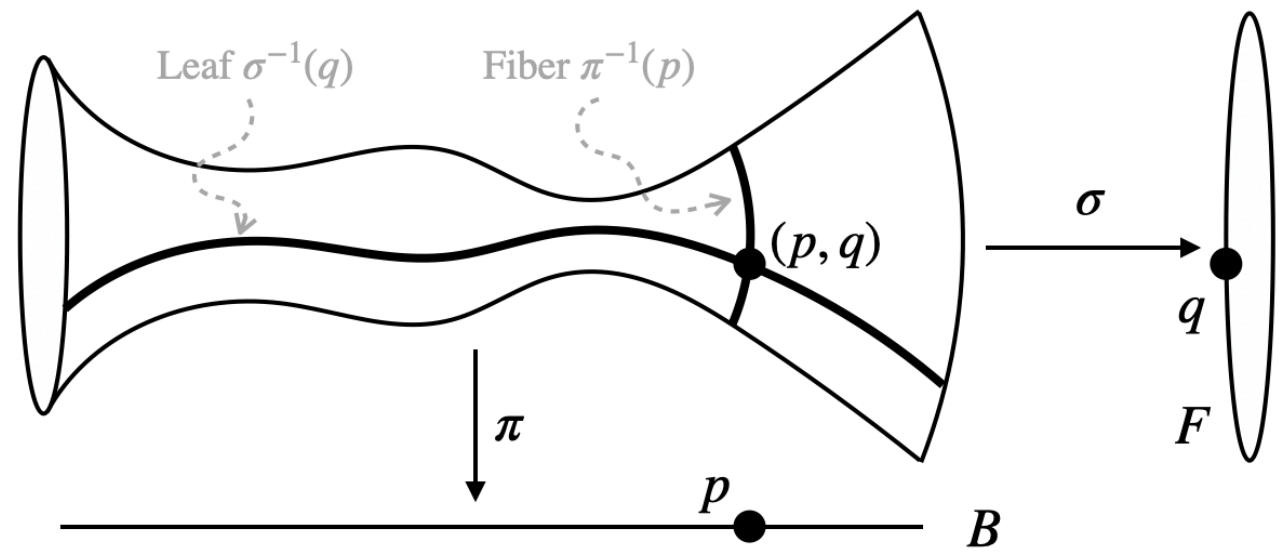

**Fig. 33**. Structure of a Warped Product

Just like the metric on a warped product is composed of the metrics of the fibre and the basis the same is true for its Levi-Civita connection. And while the warping function changes the metric on the fibre while moving along a leaf, the next lemma shows that similar is true for the Levi-Civita connection. The proofs of the lemma can be found in [8], in ([9] Lemma 7.3) and in ([30] Ch.7, Prop. 35):

**Lemma 34**: Let $\mathcal{X}, \mathcal{Y}, \mathcal{Z} \in \mathfrak{X}(B)$ and $\mathcal{U}, \mathcal{V}, \mathcal{W} \in \mathfrak{X}(F)$, and let $\nabla^B$, $\nabla^F$ and $\nabla$ be the Levi-Civita connections of $B$, $F$, and $M$, respectively. Then:

i. $\nabla_{\mathcal{X}} \mathcal{Y} = \nabla^B_{\mathcal{X}} \mathcal{Y}$

ii. $\nabla_{\mathcal{X}} \mathcal{V} = \nabla_{\mathcal{V}} \mathcal{X} = \dfrac{\mathcal{X} f}{f} \mathcal{V}$

iii. $\nabla_{\mathcal{V}} \mathcal{W} = \nabla^F_{\mathcal{V}} \mathcal{W} - \dfrac{g(\mathcal{V}, \mathcal{W})}{f} \operatorname{grad} f$ ∎

List item (i) of the lemma means that the Levi-Civita connection along a leaf is that of the basis unchanged. Item (ii) means that moving along a leaf vectors or vector fields of the fibre are scaled depending on the warping function. Interestingly, the second term of the right side of equation (iii) is normal to the fibre (i.e. it is tangental to the corresponding leave - see [9] Lemma 7.3(3)) and it shows that $f$ scales the geometry of the fibre.

The Riemann curvature tensor of the warped product is composed of the Riemann curvature tensors of the fibre and the basis and it is influenced by the warping function too. References ([30] Ch.7, Prop. 42) and ([9] Lemma 7.4) prove the next lemma:

**Lemma 35**: Let $\mathcal{X}, \mathcal{Y}, \mathcal{Z} \in \mathfrak{X}(B)$ and $\mathcal{U}, \mathcal{V}, \mathcal{W} \in \mathfrak{X}(F)$, and let $R^B$, $R^F$ and $R$ be the Riemann curvature tensor of $B$, $F$, and $M$, respectively. Then:

i. $R(\mathscr{X},\mathscr{Y})\mathscr{Z} = R^B(\mathscr{X},\mathscr{Y})\mathscr{Z}$

ii. $R(\mathscr{X},\mathscr{Y})\mathscr{V} = R(\mathscr{V},\mathscr{W})\mathscr{X} = 0$

iii. $R(\mathscr{V},\mathscr{W})\mathscr{U} = R^F(\mathscr{V},\mathscr{W})\mathscr{U} - \frac{\|\text{grad } f\|^2}{f^2}\left(g(\mathscr{V},\mathscr{U})\mathscr{W} - g(\mathscr{W},\mathscr{U})\mathscr{V}\right)$

iv. $R(\mathscr{V},\mathscr{X})\mathscr{Y} = \frac{H(f)(\mathscr{X},\mathscr{Y})}{f}\mathscr{V}$

v. $R(\mathscr{X},\mathscr{V})\mathscr{W} = \frac{g(\mathscr{V},\mathscr{W})}{f}\nabla_{\mathscr{X}}\text{grad } f$ ∎

List item (i) means that the Riemann curvature of the leafs is unchanged. Item (ii) means that the holonomy of a vector of a fibre when moved along a path within a leaf vanishes (and the same is true for moving a vector of a leaf along a fibre). The surprising result is item (iii): even if the fibre is flat (i.e. $R^F = 0$), a non-vanishing gradient of the warping function creates a curved overall manifold (i.e. even a flat space may result in a curved spacetime - see the section 8.3). Items (iv) and (v) show how the holonomy along paths partially contained in a leaf and partially contained in a fibre are determined by the warping function (see section 6.3 for the definition of the gradient and the Hessian).

Similarly, the warping function influences the Ricci tensor of $M$ which is composed of the Ricci tensor of the fibre and the basis (see [30] Ch. 7, Cor. 43):

**Lemma 36**: Let $\mathscr{X},\mathscr{Y},\mathscr{Z} \in \mathfrak{X}(B)$ and $\mathscr{U},\mathscr{V},\mathscr{W} \in \mathfrak{X}(F)$, and let $\text{Ric}^B$, $\text{Ric}^F$ and Ric be the Ricci tensors of $B$, $F$, and $M$ respectively. Then:

i. $\text{Ric}(\mathscr{X},\mathscr{Y}) = \text{Ric}^B(\mathscr{X},\mathscr{Y}) - \frac{\dim F}{f}H(f)(\mathscr{X},\mathscr{Y})$

ii. $\text{Ric}(\mathscr{V},\mathscr{W}) = \text{Ric}^F(\mathscr{V},\mathscr{W}) - \left(\frac{\Delta f}{f} + (\dim F - 1)\frac{\|\text{grad } f\|^2}{f^2}\right)g(\mathscr{V},\mathscr{W})$

iii. $\text{Ric}(\mathscr{X},\mathscr{V}) = 0$ ∎

Equation (i) means that the Ricci tensor along a leaf is the one of the basis corrected by a component influenced by the warping function; the higher the dimension of the fibre the higher this influence. Similar, the Ricci tensor along a fibre is the Ricci tensor of the fibre again corrected by a component influenced by the warping function (list item (ii)); as before for Riemann curvature, even if the fibre is Ricci flat (i.e. $\text{Ric}^F = 0$) a curved overall manifold may result. The Ricci tensor in mixed directions vanishes (list item (iii)). Obviously, the same results hold for the Ricci curvature.

Finally, the sectional curvature of $M$ is influenced by the warping function as follows (see [8] Section 2):

**Lemma 37**: Let $\mathscr{X},\mathscr{Y} \in \mathfrak{X}(B)$ and $\mathscr{U},\mathscr{V} \in \mathfrak{X}(F)$, and let $K^B$, $K^F$ and $K$ be the sectional curvature of $B$, $F$, and $M$ respectively. Then:

i. $K_{\mathscr{X}\mathscr{Y}} = K^B_{\mathscr{X}\mathscr{Y}}$

ii. $K_{\mathscr{X}\mathscr{V}} = -\dfrac{H(f)(\mathscr{X},\mathscr{X})}{f\,\|\mathscr{X}\|^2}$

iii. $K_{\mathscr{U}\mathscr{V}} = \dfrac{K^F(\mathscr{U},\mathscr{V}) - \|\mathrm{grad}\, f\|^2}{f^2}$ ∎

As before, the sectional curvature along the leafs is not modified (list item (i)). The sectional curvature of subplanes with mixed directions (spanned by one vector tangential to the fibre, the other vector tangential to the leaf) is determined by the warping function only and the direction tangential to the fibre is irrelevant (list item (ii)). The sectional curvature along a fibre is influenced by the warping function (list item (iii); even if the fibre has constant vanishing sectional curvature (i.e. $K^F = 0$), a non-vanishing gradient of the warping function creates an overall manifold whose sectional curvature is not vanishing (i.e. even a flat space may result in a curved spacetime).

### 8.3. Robertson-Walker Spacetime

A Robertson-Walker spacetime is a special warped product: its basis is a time interval and its fibre is a Riemannian manifold of constant sectional curvature.

**Definition 58:** Let $I \subseteq_{\mathrm{open}} \mathbb{R}$ and $g_I := -\,dt \otimes dt$ the pseudo-metric on $I$, let $(S, g_S)$ be a Riemannian manifold of constant sectional curvature. Then, the warped product $M = I \times_f S$ is a Lorentzian manifold called *Robertson-Walker Spacetime.* The metric $-dt \otimes dt + (f \circ \pi)^2 g_S$ is also called *Friedmann–Lemaître–Robertson–Walker (FLRW) metric.* □

In this context, the fibre is often called *space* and the basis is called *time*. Note that for $\dim S = n$ it is $S$ isometric to $\mathbb{R}^n$, $\mathbb{S}^n(r)$, or $\mathbb{H}^n_{n,1}(r)$ (for some $r > 0$) according to Theorem 9. The FLRW metric is the most general metric for describing an expanding, homogeneous, isotropic universe. These corresponding manifolds are used to discuss big bang, cosmological redshift, expansion of the universe etc..

As before, the warping function $f$ lets the basis $I$ effect the geometry of the fiber $S$. In general relativity $I$ is the time dimension and the fiber $S$ is the space dimension. The warping function $f$ describes the expansion or contraction of the universe over time (in this context also referred to as "scaling factor”).

Astronomic observations suggest that our universe (i.e. the space) is isotropic and homogeneous: indicators for this are, for example, the cosmic microwave background and the distribution of galaxies.

*Isotropy* means that the universe looks the same in all directions. Especially, space has no preferred direction. I.e. for each $p \in S$ and for any two $v, w \in T_pS$ there is a vector space isometry w.r.t. $g_S$ that transforms $v$ to $w$.

*Homogeneity* means that the universe looks the same at each point. Especially, the space is locally isometric. I.e. any two points $p, q \in S$ have neighborhoods $U_p, U_q$ that are isometric w.r.t. $g_S$.

This results in a space $S$ with constant sectional curvature, i.e. it is isometric to $\mathbb{R}^n$, $\mathbb{S}^n(r)$, or $\mathbb{H}^n_{n,1}(r)$. Over time, the space may change the absolute value of its curvature i.e. the overall spacetime is a *foliation* of spaces of constant but varying sectional curvature (see Figure 34): the space $S(t)$ at each time $t \in I$ is a leaf of the foliation (note the confusing terminology: the fibers of the warped product are called "leafs" in the context of foliations). The sectional curvature of each leaf (i.e. the fiber) is constant at a given time but the magnitude of the curvature of each leaf (i.e. the fibre) at different times may change. Because the change of the sectional curvature $K_{S(t)}$ is continuous (see next Lemma 38) all leafs are either positive or negative curved, or not curved at all. The warping function determines this dynamics of the universe, i.e. the evolution of the leafs (i.e. the fibers).

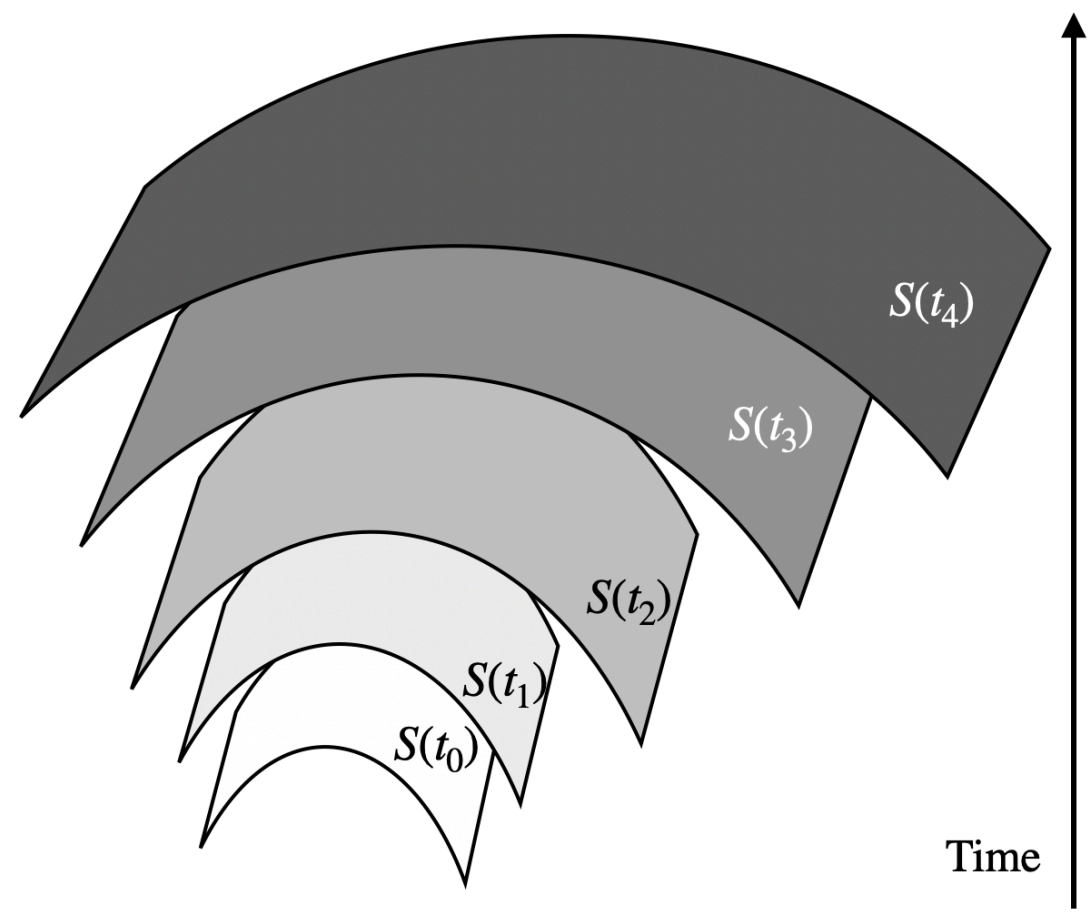

**Fig. 34**. Spacetime as Foliated Manifold

The proof of the following lemma is given in ([30] Ch.12, Prop. 6):

**Lemma 38**: Let $M = I \times_f S$ be a Robertson-Walker spacetime and let $K_S$ be the constant sectional curvature of $S$. Then for each $t \in I$ the fibre $S(t) = \pi^{-1}(t)$ has constant sectional curvature $K_{S(t)} = \dfrac{K_S}{f(t)^2}$. ∎

Ricci curvature, Ricci tensor, and sectional curvature are completely determined by the warping function $f$ (see [3] Rem. 1.38, and [30] Ch.12, Cor. 10):

**Lemma 39**: Let $p \in S(t)$, $\dim S = n$, $v \in T_pI$ with $\|v\| = 1$, and $\mathcal{X}, \mathcal{Y} \in T_pS(t)$. Then the curvatures of $M$ are as follows:

i. $\mathrm{Ric}_p(v, v) = -n\dfrac{f''(t)}{f(t)}$

ii. $\mathrm{Ric}_p(\mathcal{X}, v) = 0$

iii. $\mathrm{Ric}_p(\mathcal{X},\mathcal{Y}) = \left(\dfrac{f''(t)}{f(t)} + \dfrac{(n-1)(K_S + f'(t)^2)}{f(t)^2}\right) g_p(\mathcal{X},\mathcal{Y})$

iv. $\mathrm{Scal}_p = n\left(\dfrac{2\,f''(t)}{f(t)} + \dfrac{(n-1)(K_S + f'(t)^2)}{f(t)^2}\right)$ ∎

The equation of list item (i) provides the curvature in the direction of the time. The Ricci tensor vanishes whenever one argument points in the direction of time (item (ii)). List item (iii) shows that the Ricci tensor in all other cases equates the metric scaled by the warping function and its derivatives. Finally, the scalar curvature is determined by the sectional curvature of the fibre as well as the the warping function and its derivatives.

### 8.4. de Sitter and Anti-de Sitter

Further specializations of Robertson-Walker spacetimes result in important solutions of the Einstein vacuum equation (Equation 70): de Sitter and Anti-de Sitter spaces (see [13]). These spaces play a fundamental role to understand the dynamics of the universe.

Topologically, a de Sitter space is diffeomorphic to the sphere $\mathbb{S}^n_{n,1}(r)$ in Minkowski space $\mathbb{R}^{n+1}_{n,1}$ (i.e. it is a one-sheeted hyperboloid - see Equation 75 and Figure 30). Thus, $\mathbb{S}^n_{n,1}(r)$ is diffeomorphic to $\mathbb{R} \times \mathbb{S}^{n-1}(r)$ (see Note 17) but as a de Sitter space it is endowed with different pseudo-metrics:

**Definition 59:** A Robertson-Walker spacetime $dS_n := \mathbb{R} \times_f \mathbb{S}^{n-1}(r)$ of constant positive sectional curvature $K = \frac{1}{r^2}$ is called a *de Sitter space*. □

From a topological point of view $dS_n$ has a Euclidian sphere $\mathbb{S}^{n-1}(r)$ as fibre i.e. as its *space*. However, the sectional curvature $K_S$ of the space dimension of $dS_n$ may be flat, positive, or negative depending on the warping function *f* chosen (see [4] Ch. 39, and [13] Equations 4.14, 4.4, 4.16):

**Lemma 40**: Let $dS_n = \mathbb{R} \times_f \mathbb{S}^{n-1}(r)$ be a de Sitter space of constant sectional curvature $K = \frac{1}{r^2}$. Then, the space dimension has different constant sectional curvature $K_S$ depending on the warping function *f*:

i. $K_S = 0$ for $f(t) = e^{\frac{2t}{r}}$

ii. $K_S > 0$ for $f(t) = r\cosh\dfrac{t}{r}$

iii. $K_S < 0$ for $f(t) = r\sinh\dfrac{t}{r}$ ∎

Surprisingly, while the overall sectional curvature of $dS_n$ is positive by definition its space dimension may have a very different geometry (although its topology is $\mathbb{S}^{n-1}(r)$): besides being spherical it can be flat or even hyperbolic.

In dimension 4, $dS_4$ is a model of the spacetime of our universe with an overall sectional curvature $K = \frac{1}{r^2}$. Lemma 40 results in the corresponding metrics on $dS_4$ as a warped product:

For space dimensions with positive curvature this metric is (see [13] Equation 4.4)

$$\begin{aligned} ds^2 &= -dt^2 + r^2 \cosh^2 \frac{t}{r} \left( d\chi^2 + \sin^2 \chi (d\theta^2 + \sin^2 \theta d\phi^2) \right) \\ &= -dt^2 + r^2 \cosh^2 \frac{t}{r} d\Omega_3^2 \end{aligned} \tag{82}$$

where $d\Omega_3^2$ is the standard metric on the 3-sphere (see Equation 47).

For space dimensions with negative curvature this metric is (see [13] Equation 4.16)

$$\begin{aligned} ds^2 &= -dt^2 + r^2 \sinh^2 \frac{t}{r} \left( d\chi^2 + \sinh^2 \chi (d\theta^2 + \sin^2 \theta d\phi^2) \right) \\ &= -dt^2 + r^2 \sinh^2 \frac{t}{r} d\mathbb{H}_{2,1}^2 \end{aligned} \tag{83}$$

where $d\mathbb{H}_{2,1}^2$ is the standard metric on the 2-sheeted hyperboloid, i.e. the hyperbolic 3-space (see [13] Equation B.8 and section B.2).

For space dimensions with vanishing curvature this metric is (see [13] Equation 4.14)

$$ds^2 = -dt^2 + e^{\frac{2t}{r}} \left( dx^2 + dy^2 + dz^2 \right) \tag{84}$$

where the latter is the metric on Euclidian space $\mathbb{R}^3$.

An Anti-de Sitter space is a single sheeted hyperboloid $\mathbb{H}_{n-1,2}^n(r)$ in Minkowski space $\mathbb{R}_{n-1,2}^{n+1}$. It is diffeomorphic to $\mathbb{R}^{n-1} \times \mathbb{S}^1$ (see Note 19) but is endowed with a different pseudo-metric:

**Definition 60:** A Robertson-Walker spacetime $AdS_n := \mathbb{S}^1 \times_f \mathbb{R}^{n-1}$ of constant negative sectional curvature $K = -\frac{1}{r^2}$ is called an *Anti-de Sitter space*. □

$AdS_n$ has the upper sheet $\mathbb{H}_{n-1,1}^{n-1}(r)$ of a two-sheeted hyperboloid as fibre a.k.a. *space*. For space dimensions with negative curvature the metric on $AdS_4$ is (see [13] Equation 5.18)

$$\begin{aligned} ds^2 &= -dt^2 + r^2 \cos^2 \frac{t}{r} \left( d\chi^2 + \sinh^2 \chi (d\theta^2 + \sin^2 \theta d\phi^2) \right) \\ &= -dt^2 + r^2 \cos^2 \frac{t}{r} \, d\mathbb{H}_{2,1}^2 \end{aligned} \tag{85}$$

where $d\mathbb{H}_{2,1}^2$ is the metric on the 2-sheeted hyperboloid, i.e. the hyperbolic 3-space (see [13] Equation B.8 and section B.2). There are several coordinate transformations of this metric on $AdS_4$ resulting in metrics that are specific to study different usage of $AdS_4$ as a spacetime model (see [13] Section 5, or [4] Section 39.3). Also, there is a lemma similar to Lemma 40 for $AdS_n$ (see [4] Equations 39.126 and 39.127):

**Lemma 41**: Let $AdS_n := \mathbb{S}^1 \times_f \mathbb{R}^{n-1}$ be an Anti-de Sitter space of constant sectional curvature $K = -\frac{1}{r^2}$. Then, the space dimension has different constant sectional curvature $K_S$ depending on the warping function *f*:

i. $K_S = 0$ for $f(\rho) = e^{\pm\rho}$
ii. $K_S = -1$ for $f(\rho) = \cosh\rho$
iii. $K_S = +1$ for $f(\rho) = \sinh\rho$ ∎

Note the parameter $\rho$ of the warping function $f$: it represents $\mathbb{S}^1$ via polar coordinates. The metrics on $AdS_4$ w.r.t. these functions are in analogy to the ones in equations 82 - 84 (see [4] Equations 39.126 and 39.127). Furthermore, $AdS_4$ can be topologically perceived as a solid cylinder $\mathbb{B}^3 \times \mathbb{R}$ with boundary $\mathbb{S}^2 \times \mathbb{R}$ (see [4] Equations 39.85 and 39.86).

To understand the importance of de Sitter and Anti-de Sitter spaces the Einstein vacuum equation 70 has to be revisited: as before the universe is a 4-dimensional pseudo-Riemannian manifold called *spacetime*. Its geometry and dynamics is governed by the *Einstein field equation* (which is the general equation considering mass):

$$\mathrm{Ric} - \frac{1}{2}\mathrm{Scal} \cdot g + \Lambda g = \kappa T \tag{86}$$

In this equation, $g$ is a pseudo-metric of signature $(3,1)$ reflecting the validity of special relativity at each point, i.e. the tangent spaces are Minkowski spaces. Ric and Scal are the corresponding Ricci tensor and scalar curvature. $\Lambda$ is the so-called *cosmological constant* reflecting the energy density of the vacuum (“dark energy”). $T$ is the *energy-momentum* (also called *stress-energy*) *tensor* which describes the density and flux of mass and momentum; $T = 0$ means absent of masses or being far away from any mass. $\kappa$ is the *Einstein constant*.

Now, $dS_n$ and $AdS_n$ are pseudo-Riemannian manifolds with constant sectional curvature $K$. I.e. $\mathrm{Ric} = (n-1)Kg$ and $\mathrm{Scal} = n(n-1)K$ (see Theorem 8). Thus:

$$\kappa T = \mathrm{Ric} - \frac{1}{2}\mathrm{Scal} \cdot g + \Lambda g \Leftrightarrow \kappa T = (n-1)Kg - \frac{1}{2}n(n-1)Kg + \Lambda g$$

Far away from any masses (i.e. in a vacuum) it is $T = 0$. Consequently:

$$0 = (n-1)Kg - \frac{1}{2}n(n-1)Kg + \Lambda g \Leftrightarrow 0 = (n-1)K(1 - \frac{1}{2}n) + \Lambda$$

which means that

$$\Lambda = K\frac{(n-1)(n-2)}{2}$$

For $n = 4$ it is $\Lambda = 3K$ thus $K = \frac{\Lambda}{3}$. Thus, which of the spacetimes is a solution of the Einstein vacuum equation depends on the cosmological constant $\Lambda$:

**Lemma 42**: The Einstein field equation for the vacuum $\text{Ric} - \frac{1}{2}\text{Scal} \cdot g + \Lambda g = 0$ has the following solutions:

i. $\Lambda > 0 \Rightarrow$ de Sitter space $dS_4$ is a solution
ii. $\Lambda < 0 \Rightarrow$ Anti-de Sitter space $AdS_4$ is a solution
iii. $\Lambda = 0 \Rightarrow$ Minkowski space $\mathbb{R}^4_{3,1}$ is a solution. ∎

Especially, all variants of curvatures $K_S$ of the space dimension $S$ (Lemma 40 and Lemma 41) are allowed in these solutions.

# 9. Exterior Derivatives

In this section vector fields are extended towards tensor fields, and the Lie derivative is generalized to apply to tensor fields also. Differential forms as special tensor fields are defined and their algebraic properties (de Rham algebra) is revealed. The fundamental concept of external derivative is introduced and its tight relation to topological properties (de Rham cohomology) is discussed.

## 9.1. Tensors

Tensors are generalizations of linear maps from a vector space to the real numbers. Such maps build a vector space themselves:

**Definition 61:** Let $V$ be a vector space over $\mathbb{R}$ with $\dim V < \infty$. The *dual space* $V^*$ is the set of all linear maps $\varphi : V \to \mathbb{R}$. $\varphi \in V^*$ is called a *covector* or a *linear form.* □

As one can easily prove, $V^*$ is a vector space over $\mathbb{R}$ by means of point-wise addition and multiplication:

i. $(\varphi + \psi)(x) := \varphi(x) + \psi(x)$
ii. $(a\varphi)(x) := a(\varphi(x))$

The proof of the following theorem can be found in ([2] Th. 8.4):

**Theorem 10 (Riesz)**: $\forall \varphi \in V^* \, \exists ! v \in V : \varphi(x) = \langle v, x \rangle$, i.e. linear forms are induced by a scalar product and $v \mapsto \varphi$ induces an isomorphism $V^* \cong V$. ∎

The covector $\varphi$ from Riesz' theorem is called the *dual vector* of $v$, denoted by $\varphi = v^*$. The matrix representation of $v^*$ is a row vector.

Let $\{b_1, \ldots, b_n\}$ be a basis of $V$. Let $\alpha_1, \ldots, \alpha_n \in V^*$ such that $\alpha_i(b_j) = \delta_{ij}$, i.e. $\alpha_i$ is the projection onto the i-th coordinate of a vector w.r.t. the basis $\{b_1, \ldots, b_n\}$. Note, that $\alpha_1, \ldots, \alpha_n \in V^*$ with $\alpha_i(b_j) = \delta_{ij}$ exist because linear maps are defined by their values on a basis. Then (see [34] Prop. 3.1):

**Lemma 43**: $\alpha_1, \ldots, \alpha_n \in V^*$ is a basis of $V^*$. Especially, $\dim V = \dim V^*$. ∎

The basis $\{\alpha_1, \ldots, \alpha_n\}$ is called *dual basis* (more precise:. *dual* to $\{b_1, \ldots, b_n\}$ and vice versa) and has the property $\alpha_i(b_j) = \delta_{ij}$.

**Definition 62:** A multi-linear function $f : \bigtimes_{i=1}^{k} V_i \to \mathbb{R}$ (with $V_i = V$) is called *k-tensor* on $V$.

Hereby, $f$ is *multi-linear* $:\Leftrightarrow f(\ldots, av + bw, \ldots) = af(\ldots, v, \ldots) + bf(\ldots, w, \ldots)$ for each argument of $f$.

$f$ is called *symmetric* $:\Leftrightarrow f\left(v_{\sigma(1)}, \ldots, v_{\sigma(k)}\right) = f(v_1, \ldots, v_k)$ for each permutation $\sigma \in S_k$.

$f$ is called *alternating* $:\Leftrightarrow f\left(v_{\sigma(1)}, \ldots, v_{\sigma(k)}\right) = (\mathrm{sgn}\sigma) f(v_1, \ldots, v_k)$ for each permutation $\sigma \in S_k$. □

Examples: A scalar product $\langle v_1, v_2 \rangle$ on $\mathbb{R}^n$ is a symmetric 2-tensor. The determinant $\det[v_1, \ldots, v_n]$ on $\mathbb{R}^n$ is an alternating n-tensor.

Verifying whether a tensor is alternating or not can be done in an easier manner than considering all permutations by checking that the sign of the tensor is changing whenever two indices are exchanged (see [22] Ex. 12.17):

**Note 21**: $f$ is alternating $\Leftrightarrow f(\ldots, v_i, \ldots, v_j, \ldots) = -f(\ldots, v_j, \ldots, v_i, \ldots)$ for any $i \neq j$. ∎

Via point-wise addition $(f + g)(v_1, \ldots, v_k) := f(v_1, \ldots, v_k) + g(v_1, \ldots, v_k)$ and point-wise multiplication by a scalar $(af)(v_1, \ldots, v_k) := a(f(v_1, \ldots, v_k))$ it can be verified that the k-tensors form a vector space. Of special importance are alternating tensors which form a subspace (obvious by using Note 21:

**Note 22**: The set of all alternating k-tensors $A_k(V)$ is a vector space over $\mathbb{R}$. Also, the notation $\Lambda^k(V^*)$ is used as an alternative for $A_k(V)$ (see the explanation after Note 27). ∎

Let $\sigma \in S_k$ be a permutation and let $f$ be a k-tensor. Then, by means of $(\sigma f)(v_1, \ldots, v_k) := f\left(v_{\sigma(1)}, \ldots, v_{\sigma(1)}\right)$ a k-tensor $\sigma f$ is defined. This tensor inherits the following properties from $f$ (see [34] Sec. 3.4 first paragraph):

**Note 23**: $f$ is symmetric $\Leftrightarrow \sigma f = f$ for all $\sigma \in S_k$, and $f$ is alternating $\Leftrightarrow \sigma f = (\mathrm{sgn}\sigma) f$ for all $\sigma \in S_k$. ∎

This allows to turn each tensor into a symmetric one or an alternating one:

**Definition 63:** $(Sf)(v_1, \ldots, v_k) := \sum_{\sigma \in S_k} \sigma f$ is called *symmetrization* of $f$, and $(Af)(v_1, \ldots, v_k) := \sum_{\sigma \in S_k} (\mathrm{sgn}\sigma) \sigma f$ is called *alternation* of $f$. □

The name of $Sf$ and $Af$ is justified by the following (see [34] Prop. 3.12):

**Note 24**: $Sf$ is symmetric and $Af$ is alternating. ∎

In case a tensor is already alternating, the following applies (see [34] Prop. 3.14):

**Note 25**: If $f$ is alternating, then $Af = (k!) f$. ∎

Tensors form not only a vector space but they can also be multiplied:

**Definition 64:** Let $f$ be a k-tensor and $g$ be an m-tensor on $V$, and define $(f \otimes g)(v_1, \ldots, v_{k+m}) := f(v_1, \ldots, v_k) g(v_{k+1}, \ldots, v_{k+m})$. Then $f \otimes g$ is called the *tensor product* of $f$ and $g$. □

Obviously, this can be extended to a tensor product of more than two tensors. With this terminology a metric can be described as a 2-tensor (see [34] Sec. 3.6):

**Note 26:** Let $\{b_1, \ldots, b_n\}$ be a basis of $V$ and $\{\alpha_1, \ldots, \alpha_n\}$ be the dual basis and let $\langle \cdot, \cdot \rangle$ be a bilinear form on $V$, $g_{ij} := \langle b_i, b_j \rangle$ the corresponding Gram matrix, $v = \sum v_i b_i$ and $w = \sum w_i b_i$. Then it is $\langle v, w \rangle = \sum_{i,j} g_{ij} \cdot (\alpha_i \otimes \alpha_j)(v, w)$. □

Of special interest in differential topology and differential geometry is the following product of alternating tensors:

**Definition 65:** Let $f \in A_k(V)$ and $g \in A_m(V)$. Define the *wedge product* or *exterior product* $f \wedge g$ of $f$ and $g$ as

$$\begin{aligned}(f \wedge g)(v_1, \ldots, v_{k+m}) &:= \frac{1}{k!m!} A(f \otimes g)(v_1, \ldots, v_{k+m}) \\ &= \frac{1}{k!m!} \sum_{\sigma \in S_{k+m}} (\operatorname{sgn} \sigma) f(v_{\sigma(1), \ldots, \sigma(k)}) g(v_{\sigma(k+1), \ldots, \sigma(k+m)})\end{aligned} \tag{87}$$

$f \wedge g$ is an alternating $(k+m)$-tensor. □

The wedge product has the typical properties of a product (associativity, distributivity and its extension as bilinearity), which justified its name, but also the property of being anticommutativ (see [22] Prop 14.11):

**Lemma 44**: The map $\wedge : A_k(V) \times A_m(V) \to A_{k+m}(V)$ has the following properties:

i. Associativity: $(f \wedge g) \wedge h = f \wedge (g \wedge h)$
ii. Bilinearity: $(af + bg) \wedge h = af \wedge h + bg \wedge h$
iii. Anticommutativity: $f \wedge g = (-1)^{km} g \wedge f$ ∎

Note, that $c \in \mathbb{R}$ is considered a 0-tensor; then:

$$c \wedge g = cg \tag{88}$$

Very useful in concrete computations is the following (see [34] Prop. 3.27, and [22] Prop 14.9):

**Lemma 45**: Let $\beta_1, \ldots, \beta_k \in V^*$ be arbitrary covectors and $v_1, \ldots, v_k \in V$ be arbitrary vectors. Then:

i. $\beta_1 \wedge \ldots \wedge \beta_k(v_1, \ldots, v_k) = \det \left( \beta_i(v_j) \right)_{i,j}$

Let $\dim V = n$, $\omega \in A_n(V)$ and $v_1, \ldots, v_n \in V$. If $A : V \to V$ is linear, then:

ii. $\omega(Av_1, \ldots, Av_n) = (\det A)\, \omega(v_1, \ldots, v_n)$ ∎

The wedge product provides a means to give a basis for the vector space $A_k(V)$ (see [34] Prop. 3.29):

**Note 27:** Let $\{\alpha_1, \ldots, \alpha_n\} \subset V^*$ be the basis dual to $\{b_1, \ldots, b_n\} \subset V$, $\dim V = n$. Then: $\{\alpha_{i_1} \wedge \ldots \wedge \alpha_{i_k} \mid 1 \leq i_1 < \ldots < i_k \leq n\} \subset A_k(V)$ is a basis of $A_k(V)$ (also denoted by $\Lambda^k(V^*)$) . ∎

The last notion originates from the fact that the k-fold wedge product of linear forms from $V^*$ build a basis of $A_k(V)$. Because of this the notation $\Lambda^k(V^*)$ is often used instead of $A_k(V)$ (also $\bigwedge^k(V^*)$ is used). An important property of alternating k-tensors is that they vanish for $k > \dim V$ (see [11] Lemma 2.1.12):

**Lemma 46**: $k > \dim V \Rightarrow \Lambda^k(V^*) = 0$ and $\dim \Lambda^n(V^*) = 1$. ∎

The construction of the basis of $\Lambda^k(V^*)$ its dimension can be easily computed (see [34] Cor. 3.30):

**Note 28:** $\dim \Lambda^k(V^*) = \binom{n}{k}$ ∎

### 9.2. Tensor fields

In analogy to tangent spaces and tangent bundles, the following is defined:

**Definition 66:** The set of all covectors of $T_xM$ is called *cotangent space* and is denoted by $T_x^*M$. The cotangent space $T_x^*M$ is a vector space with $\dim T_x^*M = \dim M$. $T^*M := \bigcup_{x \in M} \{x\} \times T_x^*M$ is called *cotangent bundle* of $M$. □

Similar to the tangent bundle, the cotangent bundle is a manifold (see [22] Prop 11.9):

**Lemma 47**: $T^*M$ is a differentiable manifold with $\dim T^*M = 2 \dim M$. ∎

Based on the definition of differentiable maps between manifolds (Definition 10) the following can be defined:

**Definition 67:** Let $\pi : T^*M \to M$, $(x, \alpha) \mapsto x$ be the *canonical projection* of the cotangent bundle of $M$. A differentiable map $\omega : M \to T^*M$ with $\pi \circ \omega = \mathrm{id}_M$ is called *(differential) 1-form* (or *covector field*) on $M$. The set of all 1-forms is denoted by $\Omega^1(M)$. □

The basis of $T_p^*M$ dual to $\{\partial/\partial x_1, \ldots, \partial/\partial x_n\}$ is denoted by $\{dx_1, \ldots, dx_n\}$. Thus, it is

$$dx_i(\partial/\partial x_j) = \delta_{ij} \tag{89}$$

and for $v \in T_pM$ it is

$$dx_i(v) = dx_i\left(\sum v_j \partial_j\right) = v_j\left(\sum dx_i(\partial/\partial x_j)\right) = v_i \tag{90}$$

Just like vector fields (Lemma 8), 1-forms have a local representation via differentiable functions (see [34] Prop. 17.6):

**Lemma 48**: $\omega \in \Omega^1(M) \Leftrightarrow \omega_p = \sum_i a_i(p)dx_i$ with $a_i \in C^\infty(U)$ for a chart $(U, \varphi)$ around $p$. The function $a_i$ are called *component functions* of $\omega$. ∎

For $\omega \in \Omega^1(M)$ and $\mathcal{X} \in \mathfrak{X}(M)$ it is $\omega(\mathcal{X})_p := \omega_p(\mathcal{X}_p) \in \mathbb{R}$ (for a $p \in M$). This defines a differentiable function on $M$ (see [34] Prop. 17.9):

**Note 29**: It is $\omega(\mathcal{X}) \in C^\infty(M)$. ∎

The following linearity of 1-forms over smooth functions is also often used (see [34] Prop. 17.8):

**Note 30**: For $f \in C^\infty(M)$ it is $\omega(f\mathcal{X}) = f\omega(\mathcal{X})$. ∎

According to Lemma 9, for each $v \in T_pM$ there exists a $\mathcal{V} \in \mathfrak{X}(M)$ such that $\mathcal{V}(p) = v$. And according to Lemma 8, for $\mathcal{V} \in \mathfrak{X}(M)$ exists a $V_i \in C^\infty(M)$ such that $\mathcal{V} = \sum_i V_i \frac{\partial}{\partial x_i}$. Thus, it is $\mathcal{V}(p) = \sum_i V_i(p)\frac{\partial}{\partial x_i} = v = \sum_i v_i \frac{\partial}{\partial x_i}$ which means that $V_i(p) = v_i$. Now, let $f \in C^\infty(M)$ be a differentiable function; then it is (Equation 24) $\mathcal{V}f = \sum_i V_i \frac{\partial f}{\partial x_i}$, i.e. $(\mathcal{V}f)(p) = \sum_i V_i(p)\frac{\partial f}{\partial x_i}(p) = \sum_i v_i \frac{\partial f}{\partial x_i}(p)$. Now, we define (using Equation 90)

$$df\,|_p(v) := \sum_i v_i \frac{\partial f}{\partial x_i}(p) = \sum_i \frac{\partial f}{\partial x_i}(p)dx_i(v) \tag{91}$$

Then it is $df\,|_p \in T_p^*M$. This results in

**Definition 68:** With $df := \sum \frac{\partial f}{\partial x_i} dx_i$ it is $df \in \Omega^1(M)$. $df \in \Omega^1(M)$ is called the *differential* of $f$. □

The tensor product (Definition 64) can be used to "aggregate" cotangent bundles: the k-times tensor product $T^*M \otimes \ldots \otimes T^*M$ is a differentiable manifold called *k-tensor bundle* of $M$ denoted by $T^kT^*M$:

$$T^kT^*M := \bigcup_{x \in M} \{x\} \times \overbrace{T_x^*M \otimes \ldots \otimes T_x^*M}^{k \text{ times}} \tag{92}$$

The proof on the next lemma can be found in [22] Ex. 12.18:

**Lemma 49**: $T^kT^*M$ is a differentiable manifold with $\dim T^kT^*M = n + n^k$ (where $n = \dim M$). ∎

As before, the map $\pi : T^kT^*M \to M,\ (x, f) \mapsto x$ denotes the canonical projection. This allows to extend differential 1-forms into arbitrary k-tensor fields:

**Definition 69:** A differentiable map $A : M \to T^kT^*M$ with $\pi \circ A = \mathrm{id}_M$ is called *(differential) k-tensor field* on $M$. The set of all differentiable k-tensor fields on $M$ is denoted as $\mathcal{T}^k(M)$. □

Let $A \in \mathcal{T}^k(M)$ be a k-tensor field and $\mathcal{X}_1, \ldots, \mathcal{X}_k \in \mathfrak{X}(M)$ be vector fields. By defining

$$A(\mathcal{X}_1, \ldots, \mathcal{X}_k)(p) := A\,|_p(\mathcal{X}_1(p), \ldots, \mathcal{X}_k(p)) \tag{93}$$

a differentiable function results (see [22], details of the proof of Lemma 12.24; and [30] Ch. 2, Prop. 2; and [11] Lemma 2.2.7):

**Note 31**: It is $A(\mathcal{X}_1, \ldots, \mathcal{X}_k) \in C^\infty(M)$. Furthermore, the value $A(\mathcal{X}_1, \ldots, \mathcal{X}_k)(p)$ only depends on the values $\mathcal{X}_i(p) =: v_i$ but not on the field $\mathcal{X}_i$. ∎

As before, differentiability of a tensor field is locally equivalent to the differentiability of its component functions (see [22] Prop 12.19):

**Lemma 50**: $A \in \mathcal{T}^k(M) \quad \Leftrightarrow \quad A_p = \sum a_{i_1 \cdots i_k}(p)\, dx_{i_1} \otimes \cdots \otimes dx_{i_k}$ with differentiable functions $a_{i_1 \cdots i_k} \in C^\infty(U)$ for a chart $(U, \varphi)$ around $p$. The $a_{i_1 \cdots i_k}$ are called *component functions* of $A$. ∎

Finally, a pseudo-metric is revealed a 2-tensor field (see [15] Lemma 2.1.1; and [30] Ch. 3, Def. 1):

**Lemma 51**: Let $(M, g)$ be a (pseudo-) Riemannian manifold. Then, $g \in \mathcal{T}^2(M)$. ∎

The Lie derivative can also be defined for tensor fields:

**Definition 70:** Let $A \in \mathcal{T}^k(M)$, $\mathcal{V} \in \mathfrak{X}(M)$, and $\gamma_t := \Phi^{(p)}_{\mathcal{V},\, t}$ be the corresponding diffeomorphism (Lemma 10). Then, the *Lie derivative of* $A$ w.r.t. $\mathcal{V}$ at $p$ is:

$$(\mathcal{L}_{\mathcal{V}} A)_p := \frac{d}{dt}\bigg|_{t=0} (\gamma_t^* A)_p \tag{94}$$ □

Just like the Lie derivative of a vector field is again a vector field (Lemma 12), similar hold for tensor fields (see [22] Prop 12.30):

**Lemma 52**: $\mathcal{L}_{\mathcal{V}} A \in \mathcal{T}^k(M)$. ∎

The Lie derivative can be computed as follows (see [22] Cor 12.33):

**Lemma 53**: With $\mathcal{X}_1, \ldots, \mathcal{X}_k \in \mathfrak{X}(M)$ and $A \in \mathcal{T}^k(M)$ it is:

$$(\mathcal{L}_{\mathcal{V}} A)(\mathcal{X}_1, \ldots, \mathcal{X}_k) = \mathcal{V}(A(\mathcal{X}_1, \ldots, \mathcal{X}_k)) - \sum_{i=1}^k A(\mathcal{X}_1, \ldots, [\mathcal{V}, \mathcal{X}_i], \ldots, \mathcal{X}_k) \tag{95}$$ ∎

Herein, $A(\mathcal{X}_1, \ldots, \mathcal{X}_k)$ is function, i.e. $\mathcal{V}$ acts as a derivation on this function. The Lie bracket $[\mathcal{V}, \mathcal{X}_i]$ can be computed by means of Lemma 16. Because $\mathcal{L}_{\mathcal{V}} \mathcal{W} = [\mathcal{V}, \mathcal{W}]$ according to Lemma 15 the following is valid:

**Corollary 1**: With $\mathcal{X}_1, \ldots, \mathcal{X}_k \in \mathfrak{X}(M)$ and $A \in \mathcal{T}^k(M)$ it is:

$$(\mathcal{L}_{\mathcal{V}} A)(\mathcal{X}_1, \ldots, \mathcal{X}_k) = \mathcal{V}(A(\mathcal{X}_1, \ldots, \mathcal{X}_k)) - \sum_{i=1}^k A(\mathcal{X}_1, \ldots, \mathcal{L}_{\mathcal{V}} \mathcal{X}_i, \ldots, \mathcal{X}_k) \tag{96}$$ ∎

Herein, $\mathscr{L}_{\mathscr{V}}\mathscr{X}_i$ can be computed based on Equation 26.

### 9.3. Differential k-Forms

According to Note 27 and Equation 89 the following is obvious:

**Note 32**: $\{dx_{i_1} \wedge \ldots \wedge dx_{i_k} \mid 1 \leq i_1 < \ldots < i_k \leq n\}$ is a basis of $\Lambda^k(T_p^*M)$ . ∎

Together with Lemma 50 this proves:

**Note 33**: $\omega_p \in \Lambda^k(T_p^*M) \quad \Rightarrow \quad \omega_p = \sum_{1 \leq i_1 < \ldots < i_k \leq n} a_{i_1,\ldots,i_k} dx_{i_1} \wedge \ldots \wedge dx_{i_k}$ for $a_{i_1,\ldots,i_k} \in \mathbb{R}$ . ∎

While the tensor product is used to construct the tensor bundle $T^kT^*M$, the bundle of alternating tensors can be build by means of the wedge product:

**Note 34**: $\Lambda^kT^*M := \bigcup_{x \in M}\{x\} \times \Lambda^k(T_x^*M)$ is a differentiable manifold. ∎

As usual, $\pi : \Lambda^kT^*M \to M$ id the canonical projection - which supports

**Definition 71:** A differentiable map $\omega : M \to \Lambda^kT^*M$ with $\pi \circ \omega = \mathrm{id}_M$ is called *(differential) k- form* on $M$. The set of all differential k-forms is denoted by $\Omega^k(M)$. □

The convention is $\Omega^0(M) := C^\infty(M)$. As implied by Note 32 (for the claim about being a module see the remark after Definition 72 next):

**Note 35**: $\Omega^k(M)$ is vector space over $\mathbb{R}$ (even a module over $C^\infty(M)$). ∎

In analogy to Lemma 50, k-forms have a similar local representation ( see [22] Prop 10.22):

**Lemma 54**: $\omega \in \Omega^k(M) \quad \Leftrightarrow \quad \omega_p = \sum_{1 \leq i_1 < \ldots < i_k \leq n} a_{i_1,\ldots,i_k}(p)\, dx_{i_1} \wedge \ldots \wedge dx_{i_k}$ with $a_{i_1,\ldots,i_k} \in C^\infty(U)$ for a chart $(U, \varphi)$ around $p$. ∎

The wedge product is extended to alternating tensor fields by taking the product point-wise:

**Definition 72:** The *wedge product* (or *exterior product*) of k-forms is defined by $\wedge : \Omega^k(M) \times \Omega^m(M) \to \Omega^{k+m}(M)$ with $(\omega \wedge \eta)_p := \omega_p \wedge \eta_p$ for $p \in M$. □

With $f \in C^\infty(M) = \Omega^0(M)$ and $\omega \in \Omega^k(M)$ it is $f \wedge \omega \in \Omega^{0+k}(M) = \Omega^k(M)$, and with $f \wedge \omega = f\omega$ it is $f\omega \in \Omega^k(M)$ for $f \in C^\infty(M)$ and $\omega \in \Omega^k(M)$. Consequently, $\Omega^k(M)$ is a module over $C^\infty(M)$.

Let $k$ be odd; with Lemma 44(iii) it is $\omega \wedge \omega = (-1)^{k^2}\omega \wedge \omega = -\omega \wedge \omega$, thus $2\omega \wedge \omega = 0$ — this proves

**Note 36**: $\omega \in \Omega^k(M)$ and let k be odd. Then $\omega \wedge \omega = 0$. Thus, $dx_i \wedge dx_i = 0$. ∎

In general, it is $\omega \wedge \omega \neq 0$: For $\omega = dx_1 \wedge dx_2 + dx_3 \wedge dx_4 \in \Omega^4(M)$ using Note 35 and Lemma 44(ii) results in $\omega \wedge \omega = dx_1 \wedge dx_2 \wedge dx_3 \wedge dx_4 \neq 0$.

According to Lemma 48 it is $A_k(T_xM) = 0$ for $k > \dim T_xM = \dim M$. Thus:

**Note 37**: Let $\omega \in \Omega^k(M)$, $k > \dim M = n$; then $\omega = 0$. Especially, $\Omega^k(M) = 0$ for $k > \dim M$. ∎

$\Omega^k(M)$ is a vector space and the wedge product is an additional operation on this vector space resulting in a new algebraic structure. Let $V_1, \ldots, V_m$ be vector spaces; the *direct sum* $\bigoplus_{k=1}^m V_k$ is the cartesian product $\times_{k=1}^{m} V_k$ endowed with the component-wise addition and scalar multiplication. This turns $\bigoplus_{k=1}^m V_k$ into a vector space. A vector space $V$ with a bilinear operation ("multiplication") $\star : V \times V \to V$ is called an *algebra*.

Define $\Omega^*(M) := \bigoplus_{k=0}^n \Omega^k(M)$ and let $\wedge : \Omega^*(M) \times \Omega^*(M) \to \Omega^*(M)$ be the wedge product. Then:

**Lemma 55**: $\left(\Omega^*(M), \wedge\right)$ is an algebra, called *de Rham algebra*. ∎

Beside the differential forms $\Omega^*(M)$ the following are algebras: the set of vector fields $\mathfrak{X}(M)$ is a vector space (see comments before Lemma 7, and the Lie bracket $[\cdot,\cdot] : \mathfrak{X}(M) \times \mathfrak{X}(M) \to \mathfrak{X}(M)$ (Lemma 13) results in the algebra $\left(\mathfrak{X}(M), [,]\right)$. Also, $\left(\mathrm{GL}(n, \mathbb{K}), \cdot\right)$ is an algebra with the matrix multiplication.

Finally, if there is a differential map $F : M \to N$, differential forms on $N$ result in differential forms on $M$:

**Definition 73:** Let $F : M \to N$ be a differential map. The *pullback* of differential forms $F^* : \Omega^k(N) \to \Omega^k(M)$ is defined as follows:

$$(F^*\omega)_p(v_1, \ldots, v_k) := \omega_{f(p)}(dF(p)v_1, \ldots, dF(p)v_k) \qquad (97)$$ □

The pullback $F^*$ has the following properties that are often used in concrete computations (see [22] Lemma 14.16 and [T ] Prop. 19.5):

**Lemma 56**: Let $F^* : \Omega^k(N) \to \Omega^k(M)$ be the pullback of differential forms. Then:

i. $F^*$ is $\mathbb{R}$-linear
ii. $F^*(\omega \wedge \eta) = F^*(\omega) \wedge F^*(\eta)$
iii. In any chart it is

$$F^*\left(\sum_{1 \le i_1 < \ldots < i_k \le n} a_{i_1,\ldots,i_k} dx_{i_1} \wedge \ldots \wedge dx_{i_k}\right) = \sum_{1 \le i_1 < \ldots < i_k \le n} (a_{i_1,\ldots,i_k} \circ F)\, d(x_{i_1} \circ F) \wedge \ldots \wedge d(x_{i_k} \circ F)$$

iv. $F^*(d\omega) = d(F^*\omega)$ ∎

### 9.4. Closed and Exact Forms

As seen before, a differentiable function, i.e. a 0-form $f \in C^\infty(M) = \Omega^0(M)$ can be differentiated (Definition 68) resulting in a 1-form $df \in \Omega^1(M)$. This can be generalized:

**Definition 74:** The map $d : \Omega^k(M) \to \Omega^{k+1}(M)$ with

$$d\omega = \sum_{1 \le i_1 < \ldots < i_k \le n} \left( \sum_{j=1}^{n} \frac{\partial a_{i_1,\ldots,i_k}}{\partial x_j} dx_j \right) \wedge dx_{i_1} \wedge \ldots \wedge dx_{i_k}$$

is called *exterior derivative*. □

The exterior derivative has the following properties (see [11] Th. 2.3.1):

**Lemma 57**: Let $d : \Omega^k(M) \to \Omega^{k+1}(M)$ be the exterior derivative. Then:

i. $d$ is $\mathbb{R}$-linear
ii. $\omega \in \Omega^k(M), \eta \in \Omega^m(M) \Rightarrow d(\omega \wedge \eta) = d\omega \wedge \eta + (-1)^k \omega \wedge d\eta$
iii. $d \circ d \equiv 0$
iv. $f \in \Omega^0(M) \Rightarrow df \in \Omega^1(M)$ is the differential of $f$. ∎

These properties characterize the exterior derivative completely (see [22] Th. 14.24):

**Theorem 11 (Uniqueness of the Exterior Derivative)**: If $\delta : \Omega^k(M) \to \Omega^{k+1}(M)$ is a map with the properties (i) - (iv) above then $\delta = d$. ∎

Of special importance are forms the external derivative of which vanish, and forms that are already external derivatives:

**Definition 75:** $\omega \in \Omega^k(M)$ is called *closed* $:\Leftrightarrow d\omega = 0$. $\omega \in \Omega^k(M)$ is called *exact* $:\Leftrightarrow \exists \eta \in \Omega^{k-1}(M) : d\eta = \omega$. □

Since the exterior derivative is a linear map (Lemma 57(i)) its kernel $\ker d$ can be considered. With $\dim M = n$ it is $\Omega^{n+1}(M) = 0$ (Note 37), thus. $\ker d^n = \Omega^n(M)$. This implies:

**Note 38**: Let $\dim M = n$. Then: $\omega \in \Omega^n(M)$ is closed.. ∎

Because of $d \circ d \equiv 0$ (Lemma 57(iii)): every exact form is closed

**Lemma 58**: Every exact form is closed. ∎

The reverse of Lemma 58 is not true as shown in the next section.

### 9.5. Intermezzo: A Closed Form That is Not Exact

Let $(r, \alpha)$ be polar coordinates on $\mathbb{R}^2 \setminus \{0\}$. For a point with Cartesian coordinates $(x, y)$ it is $\tan \alpha = \frac{y}{x}$. With

$$\frac{d}{d\alpha} \tan \alpha = \frac{d}{d\alpha} \left( \frac{\sin \alpha}{\cos \alpha} \right) = \frac{\cos^2 \alpha + \sin^2 \alpha}{\cos^2 \alpha} = 1 + \frac{\sin^2 \alpha}{\cos^2 \alpha} = 1 + \tan^2 \alpha$$

i.e. it is

$$d \tan \alpha = (1 + \tan^2 \alpha) d\alpha = (1 + \frac{y^2}{x^2}) d\alpha,$$

and similarly

$$d\left(\frac{y}{x}\right) = -\frac{y}{x^2}dx + \frac{1}{x}dy\,.$$

With $\tan\alpha = \frac{y}{x}$ it is $d\,\tan\alpha = d\left(\frac{y}{x}\right)$, thus

$$(1 + \frac{y^2}{x^2})d\alpha = -\frac{y}{x^2}dx + \frac{1}{x}dy\,.$$

This equation is transformed into

$$d\alpha = \left(\frac{x^2}{x^2 + y^2}\right)\left(-\frac{y}{x^2}dx + \frac{1}{x}dy\right) = \frac{-y\,dx + x\,dy}{x^2 + y^2} =: \tau$$

resulting in a 1-form $\tau \in \Omega^1(\mathbb{R}^2\backslash\{0\})$. Because of Lemma 57(iii) $d\tau = d^2\alpha = 0$, i.e. $\tau$ is closed. Now, assume $\tau$ is exact, i.e. there exists a 0-form (i.e. a function) $\mu \in \Omega^0(\mathbb{R}^2\backslash\{0\}) = C^\infty(\mathbb{R}^2\backslash\{0\}$ with $d\mu = \tau$. Then,

$$\int_{\mathbb{S}^1} d\mu = \int_{\mathbb{S}^1} \tau = \int_{\mathbb{S}^1} d\alpha = \int_0^{2\pi} d\alpha = \alpha\,|_0^{2\pi} = 2\pi\,.$$

In polar coordinates it is $\mu = \mu(r, \alpha)$, i.e.

$$\int_{\mathbb{S}^1} d\mu \overset{(*)}{=} \int_{\mathbb{S}^1} \frac{\partial\mu}{\partial\alpha}d\alpha = \int_0^{2\pi} \frac{\partial\mu}{\partial\alpha}d\alpha = \mu(1{,}2\pi) - \mu(1{,}0) \overset{(**)}{=} 0$$

where (*) is valid because according to Definition 68 it is $d\mu = \frac{\partial\mu}{\partial r}dr + \frac{\partial\mu}{\partial\alpha}d\alpha$ and with $r \equiv 1$ for $\mathbb{S}^1$ it is $\frac{\partial\mu}{\partial r} = 0$ on $\mathbb{S}^1$. (**) is valid because $(1{,}2\pi)$ and $(1{,}0)$ are the same points on $\mathbb{S}^1$. This is a contradiction because $0 \neq 2\pi$. Consequently, $\tau \in \Omega^1(\mathbb{R}^2\backslash\{0\})$ is closed but not exact.

But locally, each form is exact (which is a consequence of the Poincaré Lemma Theorem 12). Thus, certain global (topological) properties are obstructions to global exactness of forms. This is the subject of section 9.6.

## 9.6. de Rham Cohomology

We use a more precise notation where appropriate: $d^k : \Omega^k(M) \to \Omega^{k+1}(M)$ will denote the exterior derivative of k-forms. Figure 35 depicts the resulting chain of linear maps (Lemma 57(i)) called the *de Rham Complex*. On the left side the vector space consisting of a function constant zero is considered to be embedded by $d^{-1}$ in $\Omega^0(M) = C^\infty(M)$ (see the convention after Definition 71). This chain is finite, more precisely of length $n = \dim M$, because $\Omega^k(M) = 0$ for $k > \dim M$ (Note 37).

$$0 \overset{d^{-1}}{\to} \overbrace{\Omega^0(M)}^{C^\infty(M)} \overset{d^0}{\to} \Omega^1(M) \overset{d^1}{\to} \Omega^2(M) \overset{d^2}{\to} \dots \overset{d^{k-1}}{\to} \Omega^k(M) \overset{d^k}{\to} \Omega^{k+1}(M) \overset{d^{k+1}}{\to} \dots \overset{d^{n-1}}{\to} \Omega^n(M) \to 0$$

**Fig. 35**. De Rham Complex

Both, kernel and image of a linear map are subspaces (see [24] Lemma 10.7) and $d$ is a linear map (Lemma 57(i)). Thus, the set of all closed k-forms is the subspace $\ker d^k \subseteq \Omega^k(M)$ (the kernel of $d^k$), and the set of all exact (k+1)-forms is the subspace im $d^k \subseteq \Omega^{k+1}(M)$ (the image of $d^k$).

Because of $d^k \circ d^{k-1} \equiv 0$, it is $\mathrm{im}\ d^{k-1} \subseteq \ker d^k$. Consequently, the quotient space

$$H^k(M) := \ker d^k / \mathrm{im}\ d^{k-1} \tag{99}$$

can be build and is again a vector space:

**Definition 76:** $H^k(M)$ is called *k-th de Rham cohomology group* and dim $H^k(M)$ is called *k-th Betti Number* $b_k(M)$ of $M$. □

While $H^k(M)$ is a vector space, considering its group structure (w.r.t. to the vector addition) is only needed in many applications; thus, $H^k(M)$ is called a “group”. The Betti number $b_k(M)$ measures the obstruction of closed forms being exact.

If $\omega \in \Omega^k(M)$ is closed then $\omega + d\tau$ is closed also for each $\tau \in \Omega^{k-1}(M)$: this is because $d\omega = 0$ implies $d(\omega + d\tau) \overset{(*)}{=} d\omega + dd\tau = 0$ where (*) is valid because $d$ is a linear map.

If $\omega \in \Omega^k(M)$ be exact; then $\omega + d\tau$ is exact for every $\tau \in \Omega^{k-1}(M)$: assume $\omega = d\mu$ for a $\mu \in \Omega^{k-1}(M)$; then, $\omega + d\tau = d\mu + d\tau = d(\mu + \tau)$ with $\mu + \tau \in \Omega^{k-1}(M)$.

Vice versa, let $\omega \in \Omega^k(M)$ be not exact; then $\omega + d\tau$ is also not exact for every $\tau \in \Omega^{k-1}(M)$: $\omega \in \Omega^k(M)$ is not exact, i.e. $\omega \neq d\mu$ for each $\mu \in \Omega^{k-1}(M)$. Assume $\omega + d\eta$ is exact for some $\eta \in \Omega^{k-1}(M)$; thus, there is a $\zeta \in \Omega^{k-1}(M)$ such that $\omega + d\eta = d\zeta$. This implies $\omega = d\zeta - d\eta = d(\zeta - \eta)$ with $\zeta - \eta \in \Omega^{k-1}(M)$ which is a contradiction.

Together this shows:

**Note 39**: Let $\omega \in \Omega^k(M)$. Then:

i. $\omega \in \ker d^k \Rightarrow \omega + d\tau \in \ker d^k$ for each $\tau \in \Omega^{k-1}(M)$
ii. $\omega \in \mathrm{im}\ d^{k-1} \Rightarrow \omega + d\tau \in \mathrm{im}\ d^{k-1}$ for every $\tau \in \Omega^{k-1}(M)$
iii. $\omega \notin \mathrm{im}\ d^{k-1} \Rightarrow \omega + d\tau \notin \mathrm{im}\ d^{k-1}$ for every $\tau \in \Omega^{k-1}(M)$ ∎

Thus, for $\omega \in \ker d^k$ the set $\omega + \mathrm{im}\ d^{k-1} =: [\omega] \in H^k(M) = \ker d^k / \mathrm{im}\ d^{k-1}$ is an "equivalence class" w.r.t. being exact or non-exact.

**Definition 77:** $[\omega] \in H^k(M)$ is called the *cohomology class* of $\omega$. □

By definition, for a form $\hat{\omega} \in [\omega]$ it is $\hat{\omega} = \omega + d\tau$, i.e. $\hat{\omega}, \omega$ differ by an exact form. Such two forms get a name:

**Definition 78:** Two forms $\hat{\omega}, \tilde{\omega} \in [\omega]$ are called *cohomologous*. □

Remember that $\Omega^p(M) = 0$ for $p >$ dim $M$ (Note 37); thus (see [34] Prop. 24.2):

**Lemma 59**: If dim $M = n$ then $H^p(M) = 0$ for each $p > n$. ∎

The de Rham cohomology groups of homeomorphic manifolds are isomorphic (see [22] Cor. 17.12):

**Lemma 60**: Let $M, N$ be homeomorphic. Then $H^p(M) \cong H^p(N)$ for all $p$. ∎

This an astonishing result because the cohomology groups are defined based on the differential structure of a manifold. Since every diffeomorphism is especially a homeomorphism Lemma 60 is also valid for diffeomorphic manifolds.

The next lemma gives a first hint that cohomology groups somehow reflect the topology of the underlying manifolds (see [34] Prop. 24.1):

**Lemma 61**: If $M$ has $q$ connected components then $H^0(M) = \mathbb{R}^q$. ∎

The shape of manifold plays an important role in cohomology:

**Definition 79:** Let $U \subseteq \mathbb{R}^n$ or $U \subseteq \mathbb{H}^n$. $U$ is called *star-shaped* :⇔ there is a $p \in U$ such that for each $q \in U$ the line segment from $p$ to $q$ is contained in $U$, i.e. $\{p + t(q - p) \,|\, 0 \leq t \leq 1\} \subset U$. □

Figure 36 depicts a set that is star-shaped. The fundamental theorem in this domain is now (see [22] Th. 17.14):

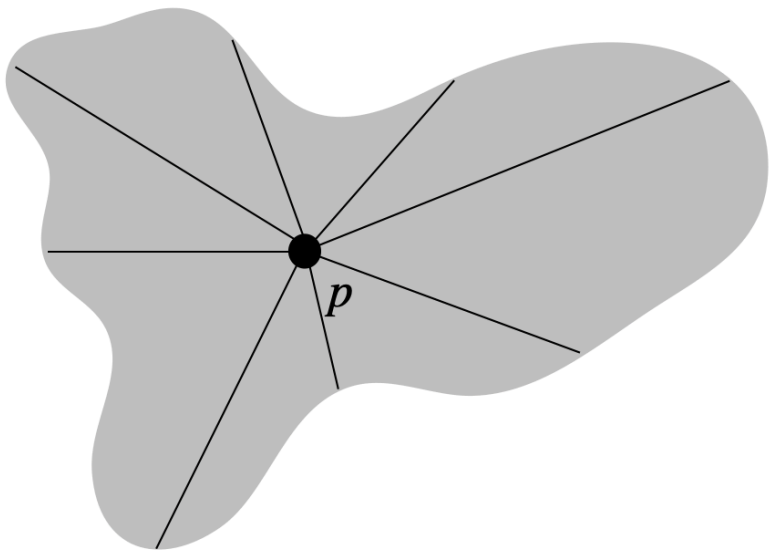

**Fig. 36**. A Star-Shaped Set

**Theorem 12 (Poincaré Lemma)**: Let $U \subseteq_{\text{open}} \mathbb{R}^n$ or $U \subseteq_{\text{open}} \mathbb{H}^n$ be star-shaped. Then every closed $p$-form on $U$ is exact ($p \geq 1$). ∎

Since every point of a manifold has a neighborhood that is diffeomorphic to an open ball in $\mathbb{R}^n$ or $\mathbb{H}^n$ and such a ball is obviously star-shaped, and Theorem 12 prove the following (see also [22] Cor. 17.15):

**Corollary 2 (Local Exactness of Closed Forms)**: Every point on a manifold $M$ has a neighborhood on which every closed form is exact. ∎

Remember that the example after Lemma 58 shows that this is in general not the case globally: if a form $\omega$ is closed on a set $U \subseteq M$ it is not necessarily exact on $U$, i.e. in general there is no form $\tau$ on $U$ with $\omega = d\tau$.

The form $\tau \in \Omega^1(\mathbb{R}^2 \backslash \{0\})$ (as in the example after Lemma 58) is closed but not exact. With $\mathbb{H}^2 := \{(x, y) \mid x > 0\}$, this form $\tau \in \Omega^1(\mathbb{H}^2)$ is in fact both, closed and exact because $\mathbb{H}^2$ is star-shaped and the Poincaré Lemma applies. Thus, the property of a form of being exact or not depends on properties of the domain of the form: the same form maybe exact on one domain and not exact on another domain.

Such type of domain properties are discussed next.

**Definition 80:** Let $f, g : X \to Y$ be two continuous maps between topological spaces. A continuous map $H : X \times [0,1] \to Y$ is called a *homotopy* between $f$ and $g$ :⇔ $H(x,0) = f(x)$ and $H(x,1) = g(x)$ for all $x \in X$. $f$ and $g$ are called *homotopic*, in symbols $f \sim g$. □

$H$ can be considered as a parameterized family of maps $H_t(x) := H(x,t)$ with $H_0 = f$ and $H_1 = g$.

**Definition 81:** Let $f : X \to Y$ and $g : Y \to X$. If $g \circ f \sim \mathrm{id}_X$ and $f \circ g \sim \mathrm{id}_Y$ then the pair $f, g$ is called a *homotopy equivalence*, and $X$ and $Y$ are called *homotopy equivalent*, in symbols $X \simeq Y$. If $X \simeq Y$ then $X$ and $Y$ are said to have the same *homotopy type*. $X$ is called *contractible* :⇔ $X \simeq p$ for a $p \in X$. □

Note that every homeomorphism is a homotopy equivalence.

**Definition 82:** For $R \subseteq X$ a continuous map $r : X \to R$ with $r|_R = \mathrm{id}_R$ is called a *retraction*. $R \subseteq X$ is called *deformation retract* (of $X$) :⇔ the inclusion $\iota_R : R \hookrightarrow X$ and a retraction $r$ build a homotopy equivalence; i.e. $X \simeq R$. □

These definitions become better comprehensible by examples:

i. Let $U \subseteq \mathbb{R}^n$ be star-shaped around $p \in U$. The $U$ is contractable.
The map $H(x,t) := p + t(x - p)$ is a corresponding homotopy (called *straight-line homotopy*) between the inclusion $H_0 = \iota_{\{p\}} : \{p\} \hookrightarrow U$ and the identity $H_1 = \mathrm{id}_U : U \to U$.

ii. $\mathbb{R}^n$ is contractable: $\mathbb{R}^n \simeq \{0\}$
$\mathbb{R}^n$ is star-shaped, i.e. the claim follows from (i). Otherwise, the straight-line homotopy $H(x,t) = t \cdot 0 + (1-t)x$ is the homotopy between $H_0 = \mathrm{id}_{\mathbb{R}^n}$ and $H_1 = \iota_{\{0\}}$.

iii. $\mathbb{S}^{n-1}$ is a deformation retract of $\mathbb{R}^n \backslash \{0\}$, i.e. $\mathbb{R}^n \backslash \{0\} \simeq \mathbb{S}^{n-1}$.
It is $H(x,t) = (1-t)x + t\dfrac{x}{\|x\|}$ the homotopy equivalence between $H_0 = \mathrm{id}_{\mathbb{R}^n}$ and $H_1 = \iota_{S^{n-1}}$. Figure 37(a) depicts the homotopy.

iv. Let $\mathscr{M}$ be the Möbius strip. Then: $\mathscr{M} \simeq \mathbb{S}^1$.
Refer to ([32] page 10) for a formal definition of $\mathscr{M}$. Figure 37(b) indicates the homotopy. The central circle $\mathbb{S}^1$ is a deformation retract of the Möbius strip.

v. Let $\mathscr{Z} := \mathbb{S}^1 \times [0,1]$ a cylinder. Then: $\mathscr{Z} \simeq \mathbb{S}^1$.
The straight-line homotopy along z-axis along the z-axis is the homotopy. I.e. the circle $\mathbb{S}^1$ is a deformation retract of the cylinder $\mathscr{Z}$.

vi. $\mathscr{M} \simeq \mathscr{Z}$
Homotopic $\sim$ is an equivalence relation (see [22] Lemma. 6.28). Thus, homotopy equivalence is especially transitivity; the claim follows from (iv) and (v). The Möbius strip and the cylinder are not homeomorphic, i.e. homotopy equivalence does not imply homeomorphism.

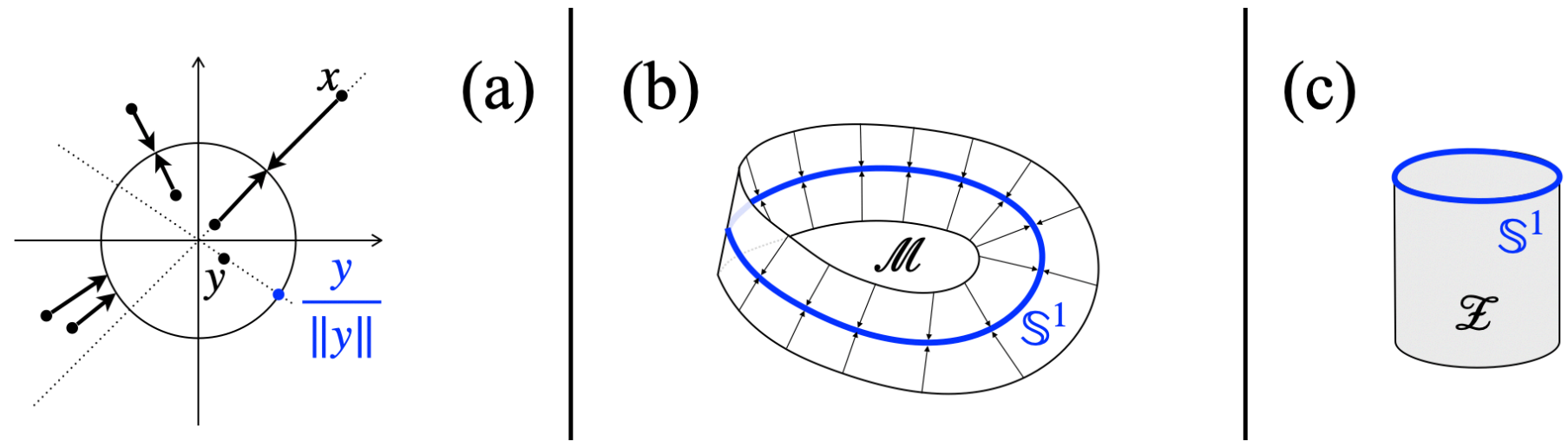

**Fig. 37**. Three Sample Homotopy Types

With these definitions the following can be proved (see [22] Th. 17.11):

**Theorem 13 (Homotopy Invariance of Cohomology)**: If $M$ and $N$ are homotopy equivalent ($M \simeq N$) then $H^p(M) \cong H^p(N)$ for each $p$. ∎

A manifold $M = \{x\}$ consisting of a single point $x$ only has dimension $\dim M = 0$. Lemma 59 implies that $H^p(\{x\}) = 0$ for each $p$. Together with Theorem 13 this proves:

**Lemma 62**: Let $M$ be a contractible manifold. Then, every closed p-form is exact, i.e. $H^p(M) = 0$, for every $p \geq 1$. ∎

We can now continue giving de Rham cohomology groups:

i. Let $M$ be a connected manifold (as always: with or without boundary). Then: $H^0(M) = \mathbb{R}$.
$d^{-1} \equiv 0$, i.e.. im $d^{-1} = 0$. Furthermore, $f \in \Omega^0(M)$ is closed iff $df = 0$, i.e. $f \equiv c$ for $c \in \mathbb{R}$. Thus, $\ker d^0 \equiv \mathbb{R}$. Together this results in $H^0(M) = \ker d^0 / \mathrm{im}\ d^{-1} = \mathbb{R}/0 = \mathbb{R}$.

ii. Let $U \subseteq \mathbb{R}^n$ be star-shaped. Then: $H^p(U) = 0$ for $p \geq 1$.
Every closed form (i.e. $\ker d^p$) is exact on a star-shaped domain, i.e. it is contained in im $d^{p-1}$: $\ker d^p \subseteq \mathrm{im}\ d^{p-1}$. But exact forms are always closed, i.e. $\mathrm{im}\ d^{p-1} \subseteq \ker d^p$. Together: $\ker d^p = \mathrm{im}\ d^{p-1}$, thus $H^p(U) = \ker d^p / \mathrm{im}\ d^{p-1} = 0$.

iii. $H^1(\mathbb{R}^2 \backslash \{\varnothing\}) \neq 0$, thus with $\mathbb{R}^n \backslash \{0\} \simeq \mathbb{S}^{n-1}$ it is $H^1(\mathbb{S}^1) \neq 0$.
The form $\tau \in \Omega^1(\mathbb{R}^2 \backslash \{\varnothing\})$ as discussed in the example after Lemma 58 is closed but not exact, i.e.$\tau \in \ker d^1$ but $\tau \notin \mathrm{im}\ d^0$. Thus, $\mathrm{im}\ d^0 \subsetneqq \ker d^1$, i.e. $H^1(\mathbb{R}^2 \backslash \{\varnothing\}) = \ker d^1 / \mathrm{im}\ d^0 \neq 0$. The second claim follows from Theorem 13.

iv. $H^1(\mathbb{S}^1) = \mathbb{R}$.
The proof is non-trivial and can be found in [34] Sec. 26.2.

v. $H^p(\mathbb{S}^n) \neq 0$ for $p = 0$ and $p = n$, $H^p(\mathbb{S}^n) = 0$ otherwise.
The proof can be found in [11] Prop. 4.4.16.

vi. $H^k(T^n) = \mathbb{R}^{\binom{n}{k}}$ for $T^n := \underbrace{\mathbb{S}^1 \times \cdots \times \mathbb{S}^1}_{\text{n times}}$.

The proof can be found in [28] Example 7.1.49.

### 9.7. Antiderivation

While the exterior derivate increases the degree of a differential form, i.e. it maps a k-form to a (k+1)-form, a map the decreases the degree is useful. This is introduced for alternating tensors first:

**Definition 83:** Let $v \in V$. The *interior multiplication* by $v$ is the map $\iota_v : \Lambda^k(V^*) \to \Lambda^{k-1}(V^*)$ with $\iota_v\omega(v_1, \ldots, v_{k-1}) := \omega(v, v_1, \ldots, v_{k-1})$ (also the notation $v \lrcorner \omega = i_v\omega$ is used). □

Note that sometimes the interior multiplication $\iota_v$ is also called *contraction* by $v$. This definition can be extended to differential forms:

**Definition 84:** Let $\mathscr{V} \in \mathfrak{X}(M)$ be a vector field. Then $\iota_{\mathscr{V}} : \Omega^k(M) \to \Omega^{k-1}(M)$ is defined as $\iota_{\mathscr{V}}\omega(\mathscr{V}_1, \ldots, \mathscr{V}_{k-1}) = \omega(\mathscr{V}, \mathscr{V}_1, \ldots, \mathscr{V}_{k-1})$ with $\iota_{\mathscr{V}}\omega := 0$ for $\omega \in \Omega^0(M)$. □

The interior multiplication has the following important properties which are in analogy to the exterior derivative (see [22] Lemma 14.13):

**Note 40**: $\iota_{\mathscr{V}}$ has the following properties:

i. Let $\omega \in \Omega^k(M)$and $\eta \in \Omega^m(M)$. Then:
$\iota_{\mathscr{V}}(\omega \wedge \eta) = (\iota_{\mathscr{V}}\omega) \wedge \eta + (-1)^k \omega \wedge (\iota_{\mathscr{V}}\eta)$

ii. $\iota_{\mathscr{V}} \circ \iota_{\mathscr{V}} = 0$ ∎

Both, $\iota_{\mathscr{V}}$ as well as $d$ have the same properties (i.e. Note 40 (i) and (ii) as well as Lemma 56 (ii) and (iii)) which results in the following generalization:

**Definition 85:** A map $\aleph : \Omega^k(M) \to \Omega^{k+q}(M)$ is called an *antiderivation of degree q* $:\Leftrightarrow \omega \in \Omega^k(M), \eta \in \Omega^m(M) \Rightarrow \aleph(\omega \wedge \eta) = \aleph\omega \wedge \eta + (-1)^k \omega \wedge \aleph\eta$. □

$\iota_{\mathscr{V}}$ and $d$ are antiderivations whose square is zero. Especially:

**Note 41**: $d : \Omega^*(M) \to \Omega^*(M)$ is an antiderivation of degree $q = +1$, and $\iota_{\mathscr{V}} : \Omega^*(M) \to \Omega^*(M)$ is an antiderivation of degree $q = -1$. ∎

Finally, there is a relation between the Lie derivative of a differential form and the exterior derivative easing the computation of Lie derivatives of differential forms:

**Theorem 14 (Cartan's Magic Formula)**: Let $\mathscr{V} \in \mathfrak{X}(M)$ and $\omega \in \Omega^*(M)$. Then:

$$\mathscr{L}_{\mathscr{V}}\omega = \iota_{\mathscr{V}} d\omega + d(\iota_{\mathscr{V}}\omega)$$ ∎

## 10. Orientations

This section reminds that integration of real functions $f : [a,b] \to \mathbb{R}$ depends on the orientation of the interval $[a,b]$, i.e. the domain of integration. Since we want to extend the apparatus of integration to manifolds in the next section, the notion of orientation of a manifold is defined.

### 10.1. Orientation of Vector Spaces

With $g' = f$ it is $\int_a^b f(x)dx \overset{(i)}{=} g(b) - g(a) = -(g(a) - g(b)) = -\int_b^a f(x)dx$, where (i) is valid because of the fundamental theorem of calculus (see [33] Th, 3.3.1). Thus, it is $\int_a^b f(x)dx = -\int_b^a f(x)dx$, which means that. the direction of the interval matters.

Figure 38 depicts two orientations of the real link: $\mathcal{O}$ is the "orientation" given by increasing numbers, and $\widehat{\mathcal{O}}$ is the "orientation" given by decreasing numbers. Thus, properly thinking about integration needs to consider "orientation"

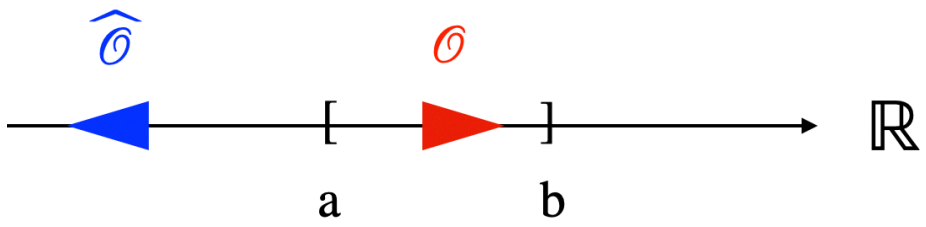

**Fig. 38**. Orientations of the Real Line

We begin by defining "orientation" of vector spaces.

**Definition 86:** Let $(v_1, \ldots, v_n)$ and $(w_1, \ldots, w_n)$ be ordered bases of a vector space $V$, $\dim V = n$. $(v_1, \ldots, v_n)$, $(w_1, \ldots, w_n)$ are called *consistently oriented* (in symbol: $(v_1, \ldots, v_n) \sim (w_1, \ldots, w_n)$) $:\Leftrightarrow \exists f : V \to V$ linear with $f(v_i) = w_i$ and $\det f > 1$ (a linear map $f$ with $\det f > 1$ is called *orientation preserving*). □

Obviously, "~" is an equivalence relation on the set of all ordered bases with exactly two equivalence classes referred to as $\mathcal{O}_1$ and $\mathcal{O}_2$.

**Definition 87:** Each of the two equivalence classes $\mathcal{O}_1$ and $\mathcal{O}_2$ is called an *orientation* of $V$. The equivalence class of the standard basis $(e_1, \ldots, e_n)$ is called *positive* orientation. The other equivalence class is called *negativ* orientation. □

Figure 39 depicts the positive orientation of $\mathbb{R}$, $\mathbb{R}^2$, and $\mathbb{R}^3$: the positive orientation of $\mathbb{R}$ is given by the standard vector $e_1$ pointing "to the right" (part (a)). In $\mathbb{R}^2$ the positive orientation as represented by $e_1, e_2$ and is called "counter-clockwise" indicating that $e_1$ has to be rotated counter-clockwise to reach $e_2$ (part (b)). In part (c) the positive orientation is represented by $e_1, e_2, e_3$; the orientation is called "right-handed" because using a right hand with the thumb pointing in $e_3$-direction, $e_1$ is the forefinger and $e_2$ is the middle finger and a counter-clockwise rotation around the "thumb axis" turns $e_1$ into $e_2$.

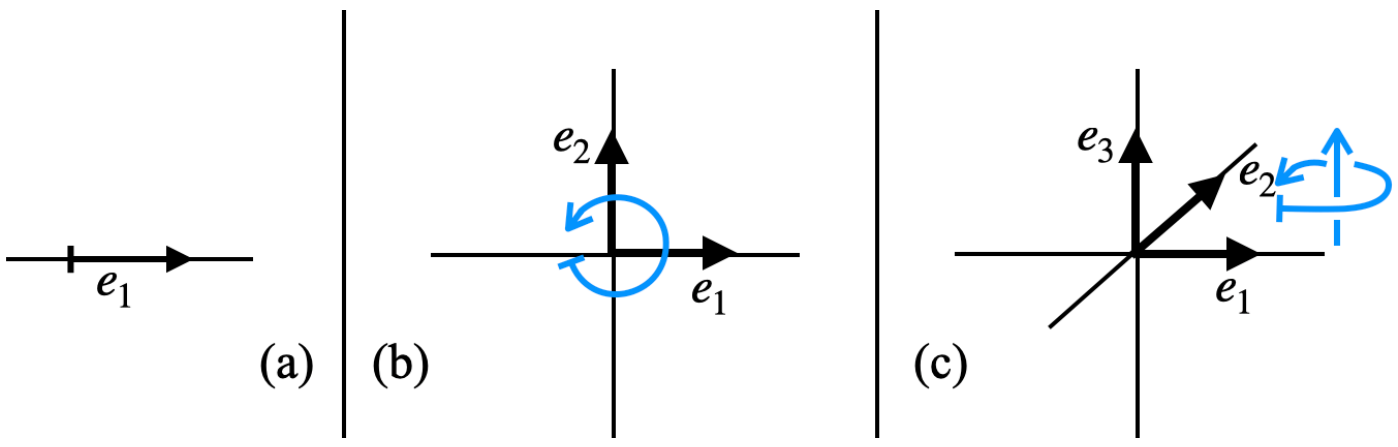

**Fig. 39**. Positive Orientations in Euclidian Spaces

### 10.2. Orientation on Manifolds

Let $M$ be a manifold (with or without a boundary) and $U \subseteq_{\text{open}} M$, $\dim M = n$.

**Definition 88:** The tuple $(\mathscr{X}_1, \ldots, \mathscr{X}_n) \in \mathfrak{X}(U)$ is called a *(local) frame* $:\Leftrightarrow$ $(\mathscr{X}_1|_p, \ldots, \mathscr{X}_n|_p)$ is an ordered basis for $T_pM$ for each $p \in U$. $(\mathscr{X}_1, \ldots, \mathscr{X}_n)$ is called a *global frame* in case $U = M$. □

A manifold with a global frame is called *parallelizable.* The existence of a global frame on $M$ is equivalent to $TM \cong M \times \mathbb{R}^n$. i.e. the definition of "parallelizable" as introduced in section 3.3 after Lemma 6 is the same as the Definition 88 (see [22] Cor. 10.20). Not every manifold is parallelizable (see examples below).

Let $\mathscr{Y}_1, \ldots, \mathscr{Y}_n \in \mathfrak{X}(U)$ be another frame on $U$. The equivalence relation "$\sim$" of being consistently oriented is now defined point-wise:

**Definition 89:** Two frames $(\mathscr{X}_1, \ldots, \mathscr{X}_n)$ and $(\mathscr{Y}_1, \ldots, \mathscr{Y}_n)$ on U are called *consistently oriented* on $U$ (in symbol $(\mathscr{X}_1, \ldots, \mathscr{X}_n) \sim (\mathscr{Y}_1, \ldots, \mathscr{Y}_n)$ ) $:\Leftrightarrow$ $(\mathscr{X}_1|_p, \ldots, \mathscr{X}_n|_p) \sim (\mathscr{Y}_1|_p, \ldots, \mathscr{Y}_n|_p)$ for each $p \in U$. □

Thus, the orientation defined by $(\mathscr{X}_1|_p, \ldots, \mathscr{X}_n|_p)$ on $T_pM$ is the same on $T_xM$ for each $x \in U$.

**Definition 90:** The $\sim$-equivalence class $[(\mathscr{X}_1, \ldots, \mathscr{X}_n)]$ is called a *continuous orientation* on $U$. The orientation induced on $T_xM$ is denoted by $\mathcal{O}_x := [(\mathscr{X}_1|_x, \ldots, \mathscr{X}_n|_x)]$. □

The important situation is when the orientations $\mathcal{O}_x$ are identical on all of $M$, i.e. in case a global frame induces all orientations on the tangent spaces of $M$:

**Definition 91:** A manifold $M$ with a continuous orientation on $M$ is called *orientable.* $\mathcal{O} := \{\mathcal{O}_p\}$ is called an *orientation* on $M$, the pair $(M, \mathcal{O})$ is called an *oriented manifold.* □

Since a manifold is locally diffeomorphic to an open subset of $\mathbb{R}^n$ or $\mathbb{H}^n$ the corresponding charts $(U_i, \varphi_i)$ "inherit" an orientation from Euclidian space: the orientation defined by the standard basis $\left(e_1, \ldots, e_n\right)$ is "lifted" to $(U_i, \varphi_i)$:

**Definition 92:** Let $\mathfrak{A} = \{(U_i, \varphi_i)\}_{i\in I}$ be an atlas for $M$. A chart $(U_i, \varphi_i)$ is called (*positively*) *oriented* :⇔ $[\{\partial/\partial x_1, \ldots, \partial/\partial x_n\}] =: \mathcal{O}_p$ for each $p \in U_i$. An atlas $\mathfrak{A}$ is called *consistently oriented* :⇔ For each $i, j \in I$ the transition map $\varphi_i \circ \varphi_j^{-1}$ has positive Jacobian determinant: $\det(d(\varphi_i \circ \varphi_j^{-1})) > 0$. □

The nice result is that defining an orientation via a frame or a consistently oriented atlas is equivalent (see [22] Prop. 15.6):

**Lemma 63**: If $\mathfrak{A} = \{(U_i, \varphi_i)\}_{i\in I}$ is a consistently oriented atlas for $M$, then there is a unique orientation $\mathcal{O}$ for $M$ such that $[\{\partial/\partial x_1|_p, \ldots, \partial/\partial x_n|_p\}] = \mathcal{O}_p$. If $M$ is oriented ($\partial M = \emptyset$ or $\dim M > 1$), then the collection of all oriented charts is a consistently oriented atlas for $M$. ∎

Yet another possibility is to use a differential n-form to get an orientation on $M$. This is first observed locally: Choose any $\omega \in \Lambda^n(V^*)$, $\omega \neq 0$. Let $(v_1, \ldots, v_n)$ and $(w_1, \ldots, w_n)$ be two ordered bases of $V$. Let $A$ be the change-of-basis matrix between these bases, i.e. $w_j = Av_j$. Then (see Lemma 45(iii))

$$\omega(w_1, \ldots, w_n) = \omega(Av_1, \ldots, Av_n) = (\det A)\ \omega(v_1, \ldots, v_n)$$

Let $(v_1, \ldots, v_n) \sim (w_1, \ldots, w_n)$, i.e. $\det A > 0$. Then $\omega(v_1, \ldots, v_n) > 0$ ⇔ $\omega(w_1, \ldots, w_n) > 0$, i.e. $\omega$ defines an orientation ${}^{\omega}\mathcal{O}$ by defining two bases $(v_1, \ldots, v_n)$ and $(w_1, \ldots, w_n)$ as equivalent for which $\omega(v_1, \ldots, v_n) > 0$ and $\omega(w_1, \ldots, w_n) > 0$.

Vice versa, it is $(v_1, \ldots, v_n), (w_1, \ldots, w_n) \in {}^{\omega}\mathcal{O}$ if and only if $\omega(v_1, \ldots, v_n) > 0$ and $\omega(w_1, \ldots, w_n) > 0$. Thus, for the change-of-basis matrix $A$, i.e. $w_j = Av_j$, it is (see Lemma 45(iii)) $\det A = \omega(w_1, \ldots, w_n)/\omega(v_1, \ldots, v_n) > 0$; i.e. it is $\omega(v_1, \ldots, v_n) > 0$ and $\omega(w_1, \ldots, w_n) > 0$, thus $(v_1, \ldots, v_n) \sim (w_1, \ldots, w_n)$. Thus, picking an $\omega \in \Lambda^n(V^*)$, $\omega \neq 0$ means to choose the orientation ${}^{\omega}\mathcal{O}$ on $V$. This proves (see also [22] Prop. 15.3):

**Lemma 64**: For $n \geq 1$, each $\omega \in \Lambda^n(V^*)$, $\omega \neq 0$ determines an orientation ${}^{\omega}\mathcal{O}$ : ${}^{\omega}\mathcal{O}$ is the set of ordered bases $(v_1, \ldots, v_n)$ such that $\omega(v_1, \ldots, v_n) > 0$. ${}^{\omega}\mathcal{O}$ is called an *orientation form*. ∎

This point-wise construction can be extended globally (se [22] Prop. 15.5):

**Lemma 65**: For $\dim M = n$ choose $\omega \in \Omega^n(M)$ with $\omega|_p \neq 0$ for each $p \in M$. ∎ Then, $\omega$ defines a unique orientation $\mathcal{O}$ for $M$ such that $\mathcal{O}|_p = {}^{\omega|_p}\mathcal{O}$. If $\mathcal{O}$ is an orientation for $M$ there exists an $\omega \in \Omega^n(M)$ with $\omega|_p \neq 0$ for each $p \in M$ such that $\mathcal{O}|_p = {}^{\omega|_p}\mathcal{O}$. ∎

Together Lemma 63 an Lemma 65 show that defining an orientation via a global frame or a consistently oriented atlas or an orientation form are equivalent.

The following examples and facts are quite useful:

i. Let $M_1, \ldots, M_k$ be oriented manifolds. Then, $M_1 \times \ldots \times M_k$ is an orientable manifold (see [22] Prop. 15.7).
ii. Each parallelizable manifold is orientable (see [22] Prop. 15.17).
iii. Spheres and tori $\mathbb{S}^n$, $\mathbb{T}^n$ are orientable (see [22] Ex. 15.22).
iv. The projective space $\mathbb{R}P^n$ is orientable $\Leftrightarrow$ $n$ is odd (see [22] Ex. 15.37).
v. The Mobius strip is not orientable (see [22] Ex. 15.38).
vi. Let $f : \mathbb{R}^n \to \mathbb{R}$ be smooth, $M := f^{-1}(x) \neq \emptyset$ for an $x \in \mathbb{R}$ and $f|_M$ submersive (i.e. $M$ is a regular level set of $f$). Then $M$ is orientable (see [22] Cor. 15.23).

The boundary of a manifold is again a manifold (Lemma 2). In case the enclosing manifold is oriented the boundary can be turned into an oriented manifold too. This is discussed next.

**Definition 93:** Let $p \in \partial M$ and $v \in T_pM \setminus T_p\partial M$. $v$ is called *inward-pointing* :$\Leftrightarrow$ It exists $\varepsilon > 0$ and $\gamma : [0,\varepsilon[ \to M$ with $\gamma(0) = p$ and $\gamma'(0) = v$. If $\gamma : ]-\varepsilon,0] \to M$ then $v$ is called *outward-pointing*. □

Figure 40 depicts an inward-pointing tangent vector $v$ at point $p$, and an outward-pointing vector $w$ at point $q$. The direction of the corresponding curves $\gamma$ and $\delta$ are indicated by solid arrow heads: $\gamma$ is followed from 0 to $\varepsilon$, $\delta$ is followed from $-\varepsilon$ to 0.

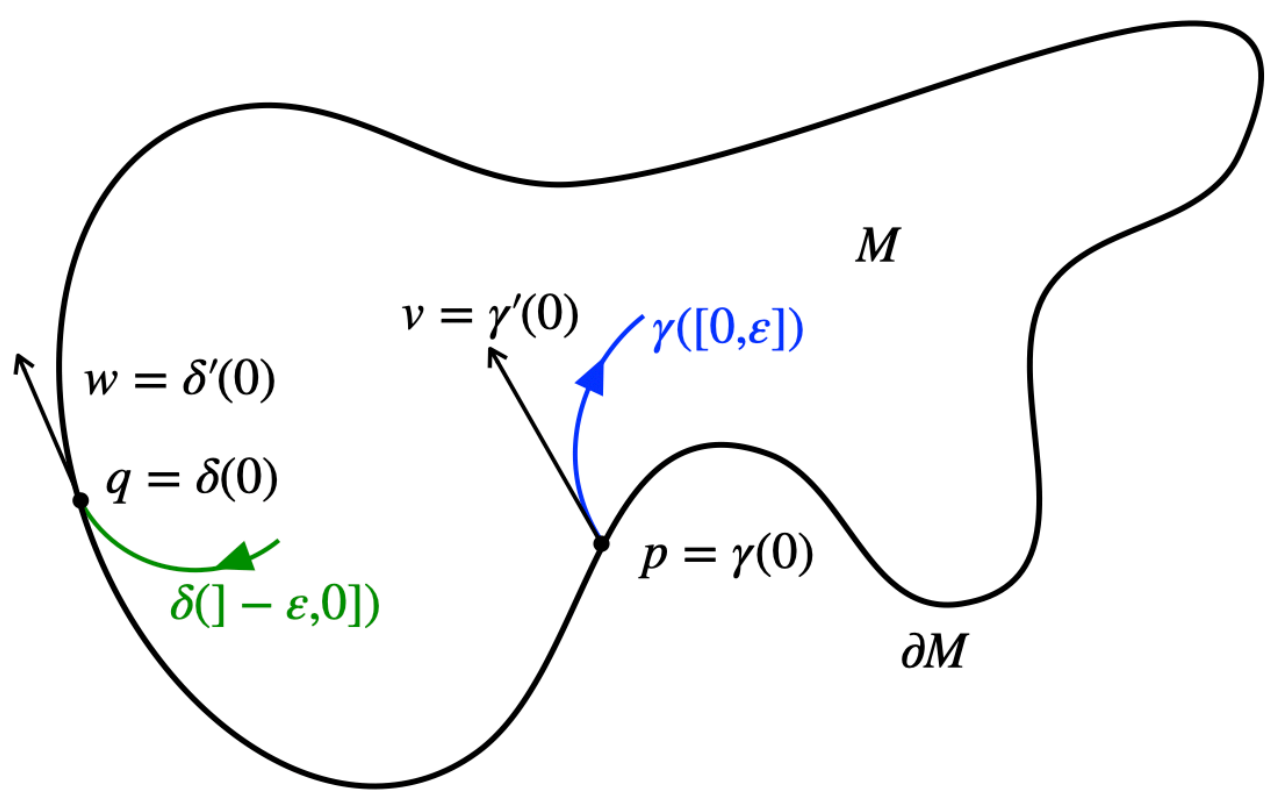

**Fig. 40**. Inward-Pointing and Outward-Pointing Tangent Vectors

A *vector field along* $\partial M$ is a differentiable map $\mathcal{N} : \partial M \to TM$ (not $T\partial M$), i.e. a vector field along $\partial M$ associates with each point of $\partial M$ a vector from possibly all of the tangent space of $M$. Let $\mathfrak{X}(\partial M)$ denote the set of all such vector fields. Then (see [34] Prop. 22.10):

**Lemma 66**: For each manifold $M$ with boundary there exists an outward-pointing $\mathcal{N} \in \mathfrak{X}(\partial M)$. ∎

If $\mathcal{N}$ is outward-pointing, $(-\mathcal{N}) \in \mathfrak{X}(\partial M)$ is obviously inward-pointing. An $\mathcal{N} \in \mathfrak{X}(\partial M)$ is called *nowhere tangent* to $\partial M$ if and only if $\mathcal{N}(p) \notin T_p\partial M$ for each $p \in \partial M$: Simply choose $\mathcal{N} \in \mathfrak{X}(\partial M)$ to be outward-pointing or inward-pointing.

Let $M$ be an oriented manifold with boundary, $\dim M = n$. Let $\omega \in \Omega^n(M)$ be an orientation form on $M$ (Lemma 65). Let $\mathcal{N} \in \mathfrak{X}(\partial M)$ be an outward-pointing vector field (Lemma 66). Then see [34] Prop. 22.11):

**Lemma 67**: The contraction $\iota_{\mathcal{N}}\omega$ of $\omega$ by $\mathcal{N}$ is an orientation form on $\partial M$. Hence, $\partial M$ is orientable. ∎

Finally, for compact manifolds the cohomology is also related to being orientable or not (see [32] Ch. 8 - Th. 9, Th. 10. and Th. 11):

**Lemma 68**: Let $M$ be a connected manifold with $\dim M = n$. Then:

i. $M$ compact and oriented $\Rightarrow$ $H^n(M) = \mathbb{R}$
ii. $M$ compact and non-oriented $\Rightarrow$ $H^n(M) = 0$
iii. $M$ not compact (oriented or not) $\Rightarrow$ $H^n(M) = 0$ ∎

### 10.3. Riemannian Volume Form

Riemannian manifolds that are oriented admit special volume form (see [22] Prop. 15.29):

**Lemma 69**: Let $(M, g)$ be an oriented Riemannian manifold, $\dim M \geq 1$. There exists a unique orientation form $\omega_g \in \Omega^n(M)$ with $\omega_g(\mathscr{E}_1, \ldots, \mathscr{E}_n) = 1$ for each local, oriented, orthonormal frame $(\mathscr{E}_1, \ldots, \mathscr{E}_n)$ for $M$. ∎

This unique orientation form is of utmost importance for integration on manifolds and gets a name:

**Definition 94:** $\omega_g$ is called *Riemannian volume form*, sometimes denoted by $dV_g$. □

Caution: the prefix "*d*" in $dV_g$ is not supposed to denote an exact form. It should remind the notation $\int f(x)\, dx$ for the purpose of integration (see next chapter). For concrete computations the following is used (see [22] Prop. 15.31):

**Lemma 70**: In each oriented chart the Riemannian volume form $\omega_g$ has the following local coordinate expression:

$$\omega_g = \sqrt{\det(g_{ij})}\, dx_1 \wedge \ldots \wedge dx_n$$

where $g_{ij}$ are the components of $g$ in these coordinates. ∎

## 11. Integration

This section introduces the concept of integration on manifolds. Stoke's theorem is central which is a generalization of the fundamental theorem of calculus to manifolds.

Oriented Riemannian manifolds allow to compute their volumes by means of integration. Interestingly, integration provides a criterion about the exactness of 1-forms, i.e. about solvability of corresponding differential equations. Finally, variational calculus is used (which exploits infinite dimensional vector spaces and the notion of Gâteaux derivative in such spaces) to indicate how the Einstein Field Equation can be derived.

### 11.1. Line Integrals

An *antiderivative* of a function $f : [a, b] \to \mathbb{R}$ is a differentiable function $F : [a, b] \to \mathbb{R}$ such that $F' = f$ . The Fundamental Theorem of Calculus (see [33] Th. 3.3.2) teaches that if $f : [a, b] \to \mathbb{R}$ is continuous then $F : ]a, b[ \to \mathbb{R}$ with $F(x) := \int_a^x f(t)dt$ is an antiderivative of $f$.

If $f$ is even differentiable then $f\,dt \in \Omega^1(]a, b])$, $F \in \Omega^0(]a, b])$ and $dF = f$. Thus, there is close relation between antiderivatives and exact 1-forms on $\mathbb{R}$: every 1-form $f\,dt \in \Omega^1(]a, b])$ is exact.

This is not true in higher dimensions, i.e. for 1-forms on (open subsets of) $\mathbb{R}^n$ as the example after Lemma 58 already indicates. But by using integrals a sufficient condition for exactness of 1-forms can be given. This requires the notion of a line integral:

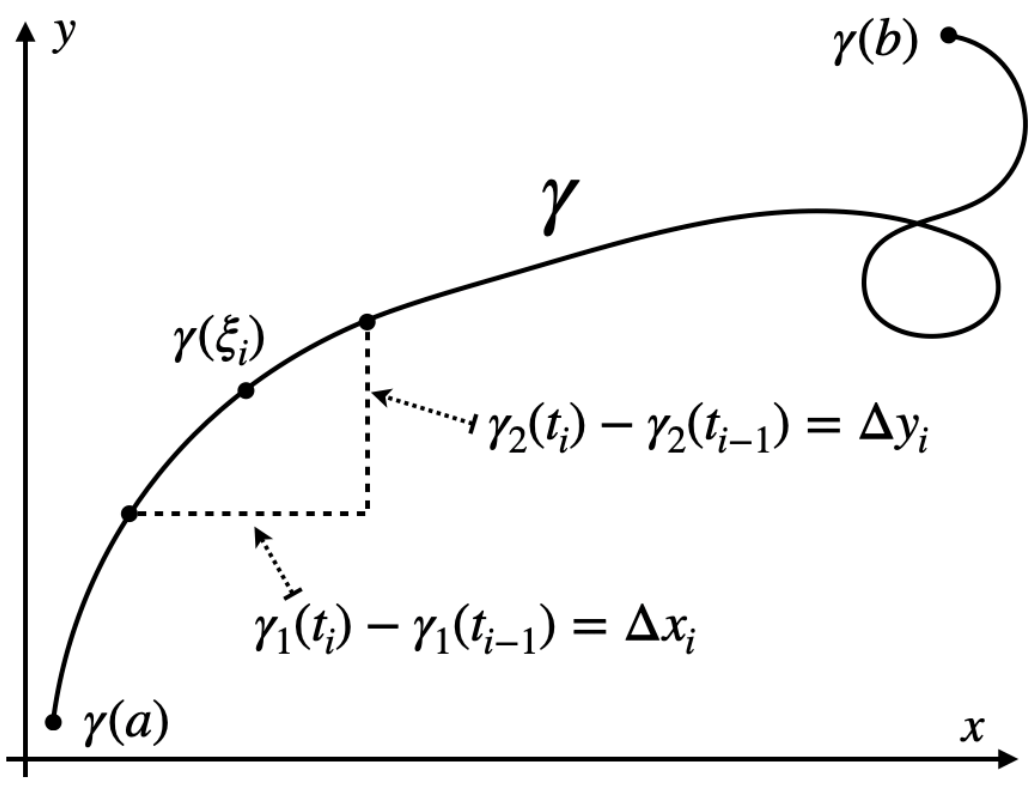

**Fig. 41**. Ingredients of a Line Integral

Figure 41 depicts the image of a smooth curve $\gamma = (\gamma_1, \gamma_2) : [a, b] \to \mathbb{R}^2$, and let $\omega = f dx + g dy \in \Omega^1(\mathbb{R}^2)$. We define the line integral of $\omega$ along $\gamma$ as follows by an approximation. For this purpose, let $a = t_0 < t_1 < \cdots < t_{n-1} < t_n = b$ define a partition $P = \{t_o, \ldots, t_n\}$ of $[a, b]$. Choose $\Xi = \{\xi_1, \ldots, \xi_n\}$ with $\xi_i \in [t_{i-1}, t_i]$. Define

$$S(P, \Xi) := \sum_{i=1}^n f(\gamma(\xi_i))\Delta x_i + g(\gamma(\xi_i))\Delta y_i \tag{100}$$

with $\gamma_1(t_i) - \gamma_1(t_{i-1}) = \Delta x_i$ and $\gamma_2(t_i) - \gamma_2(t_{i-1}) = \Delta y_i$. The width of the partition $P$ is defined as the maximum distance of two consecutive supporting points, i.e.

$\|P\| := \max_i(t_i - t_{i-1})$. If $S(P, \Xi)$ converges for $\|P\| \to 0$ (i.e. the number of supporting points increases), then this limit is defined as the line integral of $\omega$ along $\gamma$:

$$\lim_{\|P\|\to 0} S(P, \Xi) =: \int_\gamma f dx + g dy \tag{101}$$

More precisely, the existence of the limit means that for each $\varepsilon > 0$ there is a $\delta > 0$ such that for all $P$ with $\|P\| < \delta$ and all corresponding choices of $\Xi$ it is

$$\left| S(P, \Xi) - \int_\gamma f dx + g dy \right| < \varepsilon.$$

According to the mean value theorem (see [33] Th. 2.3.11), it is

$$\Delta x_i = \gamma_1(t_i) - \gamma_1(t_{i-1}) = \frac{d\gamma_1}{dt}(\alpha_i)(t_i - t_{i-1})$$

for an $\alpha_i \in [t_i - t_{i-1}]$ as well as

$$\Delta y_i = \gamma_2(t_i) - \gamma_2(t_{i-1}) = \frac{d\gamma_2}{dt}(\beta i)(t_i - t_{i-1})$$

for a $\beta_i \in [t_i - t_{i-1}]$. Substituting the latter two expression in Equation 100 results in

$$S(P, \Xi) = \sum_{i=1}^n \left( f(\gamma(\xi_i)) \frac{d\gamma_1}{dt}(\alpha_i) + g(\gamma(\xi_i)) \frac{d\gamma_2}{dt}(\beta i) \right)(t_i - t_{i-1}).$$

Shrinking the width of the partition results in

$$\lim_{\|P\|\to 0} S(P, \Xi) = \int_a^b \left( f(\gamma(t)) \frac{d\gamma_1}{dt}(t) + g(\gamma(t)) \frac{d\gamma_2}{dt}(t) \right) dt$$

By comparing Equation 101 with the latter equation implies

$$\begin{aligned} \int_\gamma \left( f dx + g dy \right) &= \int_a^b \left( f(\gamma(t)) \frac{d\gamma_1}{dt}(t) + g(\gamma(t)) \frac{d\gamma_2}{dt}(t) \right) dt \\ &= \int_a^b \left\langle \omega(\gamma(t), \frac{d\gamma}{dt}(t) \right\rangle dt \end{aligned} \tag{102}$$

Finally we get

$$\int_\gamma \omega = \int_a^b \langle \omega(\gamma(t), \gamma'(t) \rangle dt \tag{103}$$

This can be extended to higher dimensions. Let $\gamma = (\gamma_1, \ldots, \gamma_n) : [a, b] \to \mathbb{R}^n$ be smooth curve and $\omega = \sum_i a_i \, dx_i \in \Omega^1(\mathbb{R}^n)$. Then: $\int_\gamma \omega := \int_a^b \langle \omega(\gamma(t), \gamma'(t) \rangle dt$.

Using the notion of pullbacks of differential forms (Definition 73) and Lemma 56(iii) shows that

$$F^*\left(\sum_{1\le i_1<\ldots<i_k\le n} a_{i_1,\ldots,i_k} dx_{i_1}\wedge\ldots\wedge dx_{i_k}\right) = \sum_{1\le i_1<\ldots<i_k\le n}(a_{i_1,\ldots,i_k}\circ F)\, d(x_{i_1}\circ F)\wedge\ldots\wedge d(x_{i_k}\circ F)$$

with $\omega = \sum_i a_i\, dx_i \in \Omega^1(\mathbb{R}^n)$ it is $\gamma^*\left(\sum_{i=1}^n a_i\, dx_i\right) = \sum_{i=1}^n (a_i\circ\gamma)\; d(x_i\circ\gamma)$. But $d(x_i\circ\gamma) = d\gamma_i = \gamma_i' dt$, thus,

$$\begin{aligned}\gamma^*\omega &= \sum_{i=1}^n (a_i\circ\gamma)\;\gamma_i' dt \\ &= \left\langle \begin{pmatrix} a_1\circ\gamma \\ \vdots \\ a_n\circ\gamma\end{pmatrix}, \begin{pmatrix}\gamma_1' dt \\ \vdots \\ \gamma_n' dt\end{pmatrix}\right\rangle \\ &= \left\langle \begin{pmatrix} a_1\circ\gamma \\ \vdots \\ a_n\circ\gamma\end{pmatrix}, \begin{pmatrix}\gamma_1' \\ \vdots \\ \gamma_n'\end{pmatrix}\right\rangle dt \\ &= \langle\omega(\gamma), \gamma'\rangle\, dt\end{aligned}$$

Comparing this with Equation 103 gives

$$\int_\gamma \omega = \int_{[a,b]} \gamma^*\omega \tag{104}$$

In Figure 41 only the image of $\gamma$ is shown, i.e. the parameterization of the curve has been ignored. Similarly, Equation 104 ignores such a parameterization. The question is whether $\int_\gamma \omega$ is sensitive to a special parameterization. The answer is:

**Lemma 71**: Let $\chi : [c,d] \to [a,b]$ be an orientation-preserving *parameter transformation*, i.e. $\chi$ is a smooth map with $\chi(c) = a$ and $\chi(d) = b$. Then:

$$\int_{\gamma\circ\chi} \omega = \int_\gamma \omega \tag{105}$$

Proof: Applying the chain rule (see [33] Th. 2.3.5) gets $(\gamma\circ\chi)'(u) = \gamma'(\chi(u))\chi'(u)$. Define $\tilde\gamma := \gamma\circ\chi$ and $t = \chi(u)$. Then, $dt = \chi'(u)du$ (Definition 68) and it is

$$\langle\omega(\tilde\gamma(u), \tilde\gamma'(u)\rangle = \langle\omega(\gamma(\chi(u))), \gamma'(\chi(u))\chi'(u)\rangle = \langle\omega(\gamma(\chi(u))), \gamma'(\chi(u))\rangle\cdot\chi'(u)$$

Then (Equation 103):

$$\int_{\gamma\circ\chi}\omega = \int_{\tilde{\gamma}}\omega = \int_c^d \langle\omega(\tilde{\gamma}(u)), \tilde{\gamma}'(u)\rangle du$$

$$= \int_c^d \langle\omega(\gamma(\chi(u))), \gamma'(\chi(u))\rangle \cdot \chi'(u) du$$

$$= \int_a^b \langle\omega(\gamma(t)), \gamma'(t)\rangle dt = \int_\gamma \omega$$

∎

Thus, a line integral is independent of a concrete (orientation preserving) parameter transformation. In case the parameter transformation changes the orientation the sign of the integral changes:

**Note 42**: If $\chi$ is orientation-reversing (i.e. $\chi(c) = b, \chi(d) = a$), then $\int_{\gamma\circ\chi}\omega = -\int_\gamma \omega$.

<u>Proof</u>: $\chi : [c,d] \to [a,b]$ can be seen as being composed of an orientation-preserving parameter transformation $\xi : [c,d] \to [a,b]$ and an orientation-reversing $\rho : [c,d] \to [c,d]$ with $\rho(c) = d$ and $\rho(d) = c$, i.e. it is $\chi = \xi \circ \rho$. With $v = \chi(u) = \xi(\rho(u))$ it is $dv = \xi'(\rho(u))\rho'(u)du$. Then (Equation 103):

$$\int_{\gamma\circ\chi}\omega = \int_c^d \langle\omega(\gamma(\chi(u))), \gamma'(\chi(u))\rangle \cdot \chi'(u) du$$

$$= \int_c^d \langle\omega(\gamma(\xi(\rho(u)))), \gamma'(\xi(\rho(u)))\rangle \cdot \xi'(\rho(u))\rho'(u) du$$

$$\stackrel{(a)}{=} \int_{\rho(c)}^{\rho(d)} \langle\omega(\gamma(\xi(u))), \gamma'(\xi(u))\rangle \cdot \xi'(u) du$$

$$= \int_d^c \langle\omega(\gamma(\xi(u))), \gamma'(\xi(u))\rangle \cdot \xi'(u) du$$

$$= -\int_c^d \langle\omega(\gamma(\xi(u))), \gamma'(\xi(u))\rangle \cdot \xi'(u) du$$

$$\stackrel{(b)}{=} -\int_{\gamma\circ\xi}\omega$$

$$\stackrel{(c)}{=} -\int_\gamma \omega$$

(a) is making use of "integration by substitution" (see [33] Th. 3.4.5). (b) is Equation 103. (c) is Lemma 71 because $\xi$ is orientation-preserving. ∎

Thus, reversing the orientation reverses the sign of the integral - which is a generalization of the behavior of integrals of real functions (see the introduction of section 10.1).

Line integrals are parameter independent as long as the orientation is preserved (otherwise the sign is changed). Even more, line integrals are path-independent, i.e. the line integral of a differential only depends on the endpoints of the curve. I.e. different curves with the same endpoints (Figure 42) result in the same line integral.

**Theorem 15 (Fundamental Theorem for Line Integrals)**: Let $U \subseteq_{\text{open}} \mathbb{R}^n$ and $f : U \to \mathbb{R}$ be smooth; i.e. $df \in \Omega^1(U)$. Let $\gamma = (\gamma_1, \ldots, \gamma_n) : [a, b] \to U$ be a smooth curve. Then:

$$\int_\gamma df = f(\gamma(b)) - f(\gamma(a)) \tag{106}$$

Proof: It is $\langle df(\gamma(t)), \gamma'(t)\rangle = \sum \frac{\partial f_i}{\partial x_i} \gamma_i'(t) = \frac{d}{dt} f(\gamma(t))$ where the last equation is the chain rule (see [33] Th. 2.3.5). By Equation 103 it is $\int_\gamma df = \int_a^b \langle df(\gamma(t), \gamma'(t)\rangle dt$, i.e.

$$\int_\gamma df = \int_a^b \frac{d}{dt} f(\gamma(t))\, dt \overset{(*)}{=} f(\gamma(t))\big|_a^b = f(\gamma(b)) - f(\gamma(a))$$

where (*) is the fundamental theorem of calculus (see [33] Th. 3.3.1). ∎

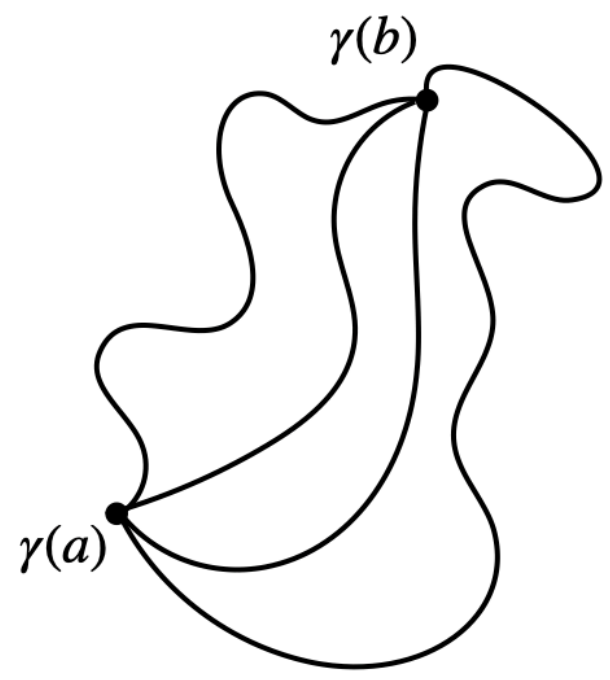

**Fig. 42**. Paths With the Same Endpoints

Theorem 15 has an important consequence:

**Corollary 3**: If $\gamma$ is closed, i.e. $\gamma(a) = \gamma(b)$, it is $\int_\gamma df = 0$ ∎

Path-independence can be more crisply formulated:

**Note 43**: Let $\gamma, \hat{\gamma} : [a, b] \to U$ be two curves with $\gamma(a) = \hat{\gamma}(a)$, $\gamma(b) = \hat{\gamma}(b)$. Then $\int_\gamma df = \int_{\hat{\gamma}} df$. ∎

If $\omega \in \Omega^1(U)$ is an exact form, then $\omega = df$ for some $f \in C^\infty(U)$). Thus:

**Note 44**: $\int_\gamma \omega = \int_{\hat{\gamma}} \omega$ for every exact form $\omega \in \Omega^1(U)$ and any two curves with the same endpoints. ∎

This means that the line integral of every exact 1-form is path-independent. Also, the reverse is true (see [22] Th. 11.42):

**Theorem 16**: Let $\omega \in \Omega^1(U)$. If $\int_\gamma \omega = \int_{\hat{\gamma}} \omega$ for any two curves $\gamma, \hat{\gamma} : [a,b] \to U$ with the same start point and the same endpoint, then $\omega$ is exact. ∎

Thus, path-independence of a 1-form means the existence of its antiderivative. Exact 1-forms are key in physics: A function $f \in C^\infty(U)$ such that $\omega = df$ for $\omega \in \Omega^1(U)$ is called a *potential* for $\omega$. In a force field $\omega$ that has a potential, the total work done by moving from one point in the field to another is independent of the path taken. A force field with a potential is called *conservative*.

### 11.2. Stoke's Theorem

The concept of line integrals, i.e. integrals of 1-forms along 1-dimensional curves, can be extended to integrals of n-forms on n-dimensional manifolds. First, this requires to define the domains on which integrals can be properly constructed: $D \subseteq \mathbb{R}^n$ is called a *domain of integration* if $D$ is a bounded set whose boundary has measure zero (note, that we are not defining what "measure zero" means here but intuitively this should be clear). Let $\omega = f\, dx_1 \wedge \ldots \wedge dx_n \in \Omega^n(\bar{D})$ ($\bar{D}$ is the topological closure of $D$; note that since $D$ is bounded, $\bar{D}$ is compact according to the theorem of Heine-Borel - see [16] Th. 5.14). The integral of $\omega$ over $D$ is defined as

$$\int_D \omega = \int_D f\, dx_1 \wedge \ldots \wedge dx_n := \int_D f\, dx_1 \ldots dx_n \qquad (107)$$

where $\int_D f\, dx_1 \ldots dx_n$ is the well-known Riemann integral (see [20]Chapter 6, for example).

Integration of $\omega \in \Omega^n(M)$ on a compact oriented manifold $M$ with $\dim M = n$ is defined via an oriented atlas $\{(U_i, \varphi_i)\}$ of $M$ with a subordinated partition of unity $\{\psi_i\}$ (see [22] Th. 2.23):

$$\int_M \omega := \sum_i \int_M \psi_i \cdot \omega \qquad (108)$$

With $\operatorname{supp} \psi_i \subseteq U_i$ it is

$$\int_M \psi_i \cdot \omega = \int_{U_i} \psi_i \cdot \omega := \int_{\varphi_i(U_i)} (\varphi_i^{-1})^*(\psi_i \cdot \omega) \qquad (109)$$

where the last integral in Equation 109 is computed via Equation 107. Thus, integrals are taken "piece-wise" via charts and are then summed up. The integral on a chart is computed via the pullback. Note that if non-compact manifolds are considered, $\omega$ must have a compact support (see [22] pp. 405), i.e. $\omega$ must vanish outside a compact subset of $M$.

Integrals of n-forms on manifolds have the following properties which generalize properties known from Riemann integrals (see [22] Prop. 16.6):

**Lemma 72**: Let $\chi : [c, d] \to [a, b]$ be an orientation-preserving *parameter transformation*, i.e. $\chi$ is a smooth map with $\chi(c) = a$ and $\chi(d) = b$. Then:

i. If $a, b \in \mathbb{R}$ then: $\int_M a\omega + b\eta = a\int_M \omega + b\int_M \eta$

ii. If $-M$ denotes the manifold with the opposite orientation than $M$, then: $\int_{-M} \omega = -\int_M \omega$

iii. If $\omega$ is an orientation form of $(M, \mathcal{O})$, then: $\int_M \omega > 0$

iv. If $F : N \to M$ is an orientation preserving diffeomorphism, then: $\int_M \omega = \int_N F^*\omega$ ∎

The next theorem is the main result about integration on manifolds (see [22] Th. 16.11):

**Theorem 17 (Stoke's Theorem)**: Let $M$ be a compact oriented manifold, $\dim M \geq 1$, and let $\omega \in \Omega^{n-1}(M)$. Then:

$$\int_M d\omega = \int_{\partial M} \omega \tag{110}$$ ∎

Stoke's theorem is a (tremendous) generalization of the Fundamental Theorem of Calculus: Let $F : [a, b] \to \mathbb{R}$ be smooth, i.e. $dF = F'dt \in \Omega^1([a, b])$. Then, $\int_a^b F'\,dt = F(b) - F(a)$ (see [33] Th, 3.3.1). With $\int_a^b F'\,dt = \int_{[a,b]} dF$ it is $\int_{[a,b]} dF = \int_a^b F'\,dt = F(b) - F(a) = \int_{\{a,b\}} F = \int_{\partial[a,b]} F$ which is Stoke's theorem applied to the compact oriented 1-manifold $[a, b]$ with boundary $\partial[a, b] = \{a, b\}$ and the 0-form $F \in C^0([a, b]) = \Omega^0([a, b])$.

Similarly, let $\gamma : [a, b] \to \mathbb{R}^n$ be smooth, $\gamma' \neq 0$ and let $\gamma$ be a homeomorphism onto $\gamma([a, b])$, i.e. $\gamma$ is an embedding and $\gamma([a, b]) \subset \mathbb{R}^n$ is a 1-manifold with $\partial\gamma([a, b]) = \{\gamma(a), \gamma(b)\}$. Then

$$\int_\gamma df = f(\gamma(b)) - f(\gamma(a)) = \int_{\{\gamma(a),\gamma(b)\}} f = \int_{\partial\gamma([a,b])} f$$

which is the fundamental theorem for line integrals (Theorem 15). Thus, Stoke's theorem is a generalization of this theorem too.

The next lemma which generalizes the integration by parts rule from calculus of one variable (see [33] Th. 3.4.4) is helpful in many situation:

**Lemma 73 (Integration by Parts)**: Let $u \in \Omega^k(M)$ and $v \in \Omega^l(M)$ with $\dim M = k + l + 1$. Then:

$$\int_M du \wedge v = \int_{\partial M} u \wedge v - (-1)^k \int_M u \wedge dv$$

<u>Proof</u>: With $d(u \wedge v) = du \wedge v + (-1)^k u \wedge dv$ we get

$$\int_M d(u \wedge v) = \int_M (du \wedge v + (-1)^k u \wedge dv)$$
$$= \int_M du \wedge v + \int_M (-1)^k u \wedge dv$$

and thus

$$\int_M du \wedge v = \int_M d(u \wedge v) - (-1)^k \int_M u \wedge dv$$
$$= \int_{\partial M} u \wedge v - (-1)^k \int_M u \wedge dv$$

∎

For $u, v \in \Omega^0(M) = C^\infty(M)$ and $M = [a,b] \subset \mathbb{R}$ (i.e. $\dim M = 1$) it is $d(u \wedge v) = d(uv) = u'v dx + uv'dx$, i.e. $\int_a^b u'v\ dx = [uv]_a^b - \int_a^b uv'\ dx$ - which is the rule of integration by parts.

The next two corollaries of Stoke's theorem are also very useful in many situations. If $M$ is a n-manifold without boundary then the integral of any $\omega \in \Omega^{n-1}(M)$ over the empty set vanishes, i.e. $\int_{\partial M} \omega = 0$. Thus:

**Corollary 4**: If $\partial M = \varnothing$, then $\int_M d\omega = 0$. ∎

If $\omega \in \Omega^{n-1}(M)$ is closed, i.e. $d\omega = 0$, then $\int_M d\omega = 0$. Thus:

**Corollary 5**: If $\omega$ is closed, then $\int_{\partial M} \omega = 0$ ∎

The following criterion on exactness is also helpful in many situations:

**Lemma 74**: Let $S \subseteq M$ be an oriented compact submanifold with $\dim S = k$ and $\partial S = \varnothing$; let $\omega \in \Omega^k(M)$ be closed. If $\int_S \omega \neq 0$, then $\omega$ is not exact on M .

Proof: Assume $\omega$ is exact, i.e. $\omega = d\eta$ for $\eta \in \Omega^{k-1}(M)$. Then $\int_S \omega = \int_S d\eta = \int_{\partial S} \eta = 0$ (Corollary 4) - contradiction. ∎

The example from section 9.5 is an application of this Lemma 74: the 1-form $\tau \in \Omega^1(\mathbb{R}^2 \backslash \{0\})$ with

$$\tau = \frac{-y dx + x dy}{x^2 + y^2}$$

is a closed 1-form, and $\mathbb{S}^1$ is an oriented compact submanifold of dimension one, $\mathbb{S}^1 \subset \mathbb{R}^2 \backslash \{0\} = M$, $\partial \mathbb{S}^1 = \varnothing$. With $\int_{\mathbb{S}^1} \tau = 2\pi \neq 0$ the lemma shows that $\tau$ is not exact.

### 11.3. Volumes

The length of a curve $\gamma = (\gamma_1, \ldots, \gamma_n)$ in $\mathbb{R}^n$ can be approximated by a series of secants $s_i$ (see Figure 43). The length $L(\gamma)$ is then roughly the sum of the length $L(s_i)$ of the secants: $L(\gamma) \approx \sum L(s_i)$.

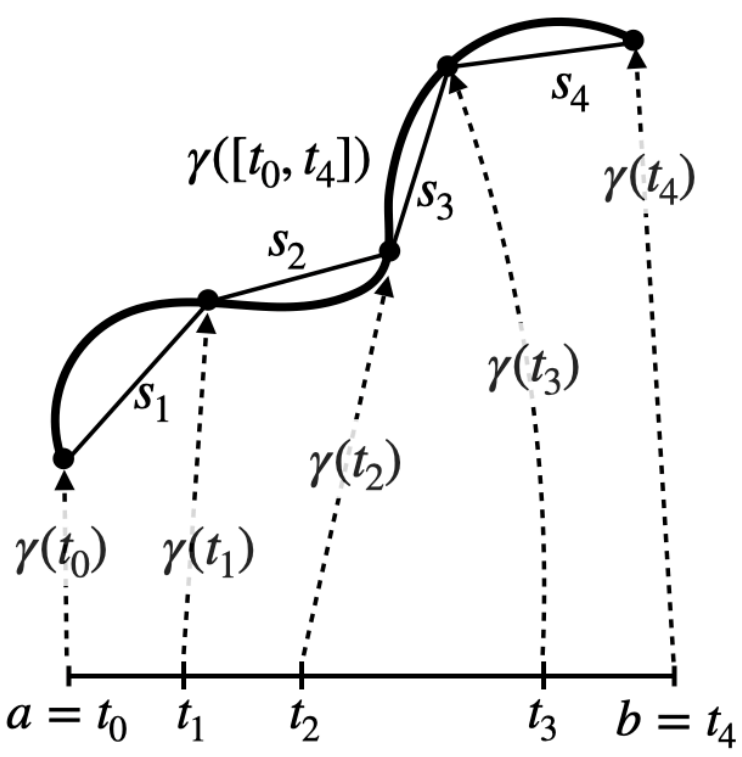

**Fig. 43**. Approximating Curve Lenghts

It is

$$L(s_i) = \|\gamma(t_i) - \gamma(t_{i-1})\| = \sqrt{\sum_j (\gamma_j(t_i) - \gamma_j(t_{i-1}))^2}.$$

By the mean-value theorem

$$\exists \xi_j \in [t_{i-1}, t_i] : \gamma_j(t_i) - \gamma_j(t_{i-1}) = \gamma_j'(\xi_j)(t_i - t_{i-1}).$$

Thus,

$$L(s_i) = \sqrt{\sum_j (\gamma_j'(\xi_j)(t_i - t_{i-1})^2} = \sqrt{\sum_j \gamma_j'(\xi_j)^2}(t_i - t_{i-1}).$$

This implies

$$L(\gamma) \approx \sum_i L(s_i) = \sum_i \sqrt{\sum_j \gamma_j'(\xi_j)^2}(t_i - t_{i-1}).$$

Increasing the number of secants $n$ means $t_i \to t_{i-1}$, thus $\xi_j \to t_i$ , i.e.

$$L(s_i) \to ds_i = \sqrt{\sum_j \gamma_j'(t_i)^2} dt_i = \sqrt{\langle \gamma'(t_i), \gamma'(t_i) \rangle} dt_i = \|\gamma'(t_i)\| dt_i$$

where $ds_i$ is the length of an infinitesimal line segment. For $n \to \infty$:

$$L(\gamma) = \int_a^b \|\gamma'(t)\| dt \tag{111}$$

For a Riemannian manifold $(M, g)$ and a curve $\gamma : I \to M$ on $M$ it is

$$\|\gamma'(t)\| = \sqrt{g_{\gamma(t)}(\gamma'(t), \gamma'(t))}$$

and $\|\gamma'(t)\| dt = \sqrt{g_{\gamma(t)}(\gamma'(t), \gamma'(t))} dt$ is an infinitesimal line segment. Thus, in this general case:

$$L(\gamma) = \int_I \sqrt{g_{\gamma(t)}(\gamma'(t), \gamma'(t))}\, dt \tag{112}$$

In order to generalize this further to measure volumes on Riemannian manifolds some facts from elementary geometry is needed. The volume of a parallelogram (i.e. its area) is its basis times its height, i.e. $b \times h$ (see Figure 44(a)).

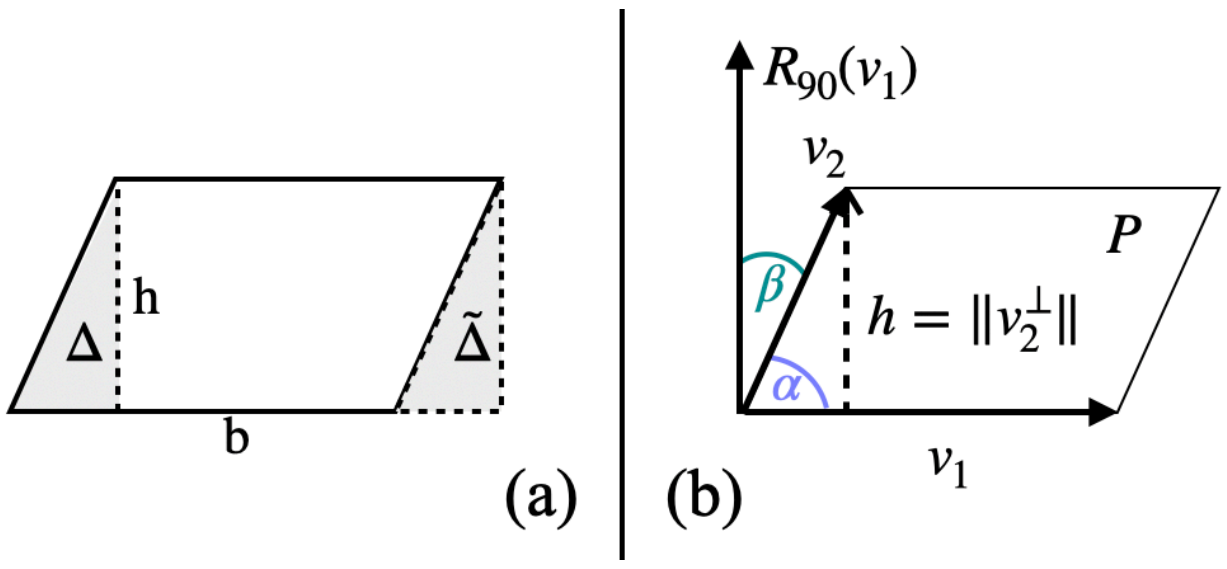

**Fig. 44**. Area of a Parallelogram

This can be expressed in terms of a determinant:

**Note 45**: Let $P$ be the parallelogram spanned by $v_1 = \begin{pmatrix} v_{11} \\ v_{12} \end{pmatrix}$ and $v_2 = \begin{pmatrix} v_{21} \\ v_{22} \end{pmatrix}$. Then $\mathrm{vol} P = \det \begin{pmatrix} v_{11} & v_{21} \\ v_{12} & v_{22} \end{pmatrix}$.

<u>Proof</u>: (see Figure 44(b)) Let $P$ be spanned by $v_1, v_2 \in \mathbb{R}^2$. Then, $h = |v_2| \cdot \sin\alpha = |v_2| \cdot \cos(90° - \alpha) = |v_2| \cdot \cos\beta$. This implies $\mathrm{vol} P = |v_1| \cdot |v_2| \cdot \cos\beta$. Let $R_{90} = \begin{pmatrix} 0 & -1 \\ 1 & 0 \end{pmatrix}$ be the rotation by 90° counter-clockwise, i.e. $|R_{90}(v_1)| = |v_1|$. Thus, $\mathrm{vol} P = |R_{90}(v_1)| \cdot |v_2| \cdot \cos\beta$. But

$$|R_{90}(v_1)| \cdot |v_2| \cdot \cos\beta = \langle R_{90}(v_1), v_2 \rangle = \langle \left(\begin{smallmatrix} -v_{12} \\ v_{11} \end{smallmatrix}\right), \left(\begin{smallmatrix} v_{21} \\ v_{22} \end{smallmatrix}\right) \rangle = v_{11}v_{22} - v_{12}v_{21}$$

Thus, $\mathrm{vol} P = v_{11}v_{22} - v_{12}v_{21} = \det \begin{pmatrix} v_{11} & v_{21} \\ v_{12} & v_{22} \end{pmatrix}$. ∎

This can be generalized to compute the volume of an n-dimensional parallelotope in $\mathbb{R}^k$ $(k \geq n)$:

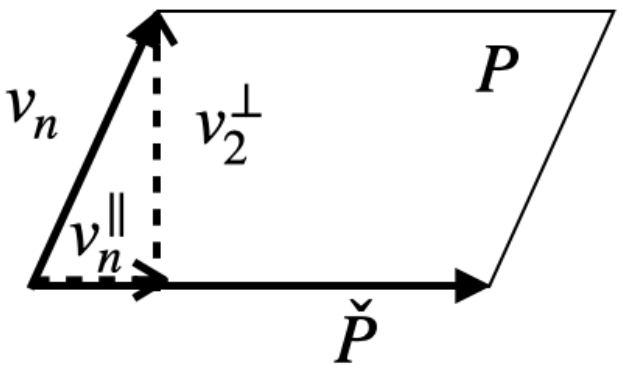

**Fig. 45**. Volume of a Parallelotope

**Note 46**: Let $P := \{ \sum_{i=1}^{n} a_i v_i | 0 \leq a_1, a_2, \ldots, a_n \leq 1 \} \subseteq \mathbb{R}^k$ be an n- dimensional parallelotope. With $V = \left(v_1 v_2 \ldots v_n\right)$ it is $(\mathrm{vol}\ P)^2 = \det\left(V V^T\right)$.

Proof: (by induction) $n = 2$ has been shown before (Note 45). Let $n > 2$. Let $P$ be spanned by the ordered list of vectors $\{v_1, \ldots, v_n\} \subseteq \mathbb{R}^n$. The basis of $P$ is the parallelotope $\check{P} = \{ \sum_{i=1}^{n-1} a_i v_i \,|\, 0 \leq a_1, a_2, \ldots, a_{n-1} \leq 1\}$ (see Figure 45). As before, vol $P$ is defined as the volume of its basis times its heights: vol $P = (\text{vol } \check{P}) \cdot \|v_n^\perp\|$. $v_n$ is decomposed as $v_n = v_n^\perp + v_n^\parallel$ with $v_n^\parallel = \sum_{i=1}^{n-1} c_i v_i \in \text{span}\{v_1, \ldots, v_{n-1}\}$ and $v_n^\perp \in \text{span}\{v_1, \ldots, v_{n-1}\}^\perp$, i.e. $\langle v_n^\perp, v_i \rangle = 0$ for $1 \leq i \leq n-1$. Let $V = \left(v_1 v_2 \ldots v_n\right) =: \left(Q\, v_n\right)$, i.e. $Q = \left(v_1 v_2 \ldots v_{n-1}\right)$ is the matrix consisting of the first $n-1$ columns of $V$. Next, elementary column operations are applied to $V$: the summands $-c_i v_i$ of $v_n^\parallel$ is added to the column $v_n$ for $1 \leq i \leq n-1$. Each of these operations corresponds to a so-called elementary matrix $G_i$ (see [24] Sec. 5.1). Because $v_n^\perp = v_n - v_n^\parallel = v_n - \sum_{i=1}^{n-1} c_i v_i$, these operations transform $V$ into $\tilde{V} = \left(Q\, v_n^\perp\right)$. It is $\tilde{V} = G_{n-1} \ldots G_1 V$. Applying $G_i$ or $G_i^T$ to another matrix doesn't change the latter matrix' determinant (see [24] Lemma 7.13). Thus, $\det(\tilde{V}^T \tilde{V}) = \det(V^T G_1^T \ldots G_{n-1}^T G_{n-1} \ldots G_1 V) = \det(V^T V)$. It is

$$\tilde{V}^T \tilde{V} = \begin{pmatrix} Q^T \\ (v_n^\perp)^T \end{pmatrix} \begin{pmatrix} Q\, v_n^\perp \end{pmatrix} = \begin{pmatrix} Q^T Q & Q^T v_n^\perp \\ (v_n^\perp)^T Q & (v_n^\perp)^T v_n^\perp \end{pmatrix}$$

Now it is

$$Q^T v_n^\perp = \begin{pmatrix} v_1^T \\ \vdots \\ v_{n-1}^T \end{pmatrix} v_n^\perp = \begin{pmatrix} \langle v_1, v_n^\perp \rangle \\ \vdots \\ \langle v_{n-1}^T, v_n^\perp \rangle \end{pmatrix} = \begin{pmatrix} 0 \\ \vdots \\ 0 \end{pmatrix}$$

and similarly $(v_n^\perp)^T Q = \begin{pmatrix} 0 \\ \vdots \\ 0 \end{pmatrix}$. This implies $\tilde{V}^T \tilde{V} = \begin{pmatrix} Q^T Q & 0 \\ 0 & \|v_n^\perp|^2 \end{pmatrix}$. Thus $\det(\tilde{V}^T \tilde{V}) = \det(Q^T Q) \cdot \|v_n^\perp\|^2$ ("Laplace expansion", see [24] Cor. 7.22). With $\det(\tilde{V}^T \tilde{V}) = \det(V^T V)$ it is $\det(V^T V) = \|v_n^\perp\|^2 \cdot \det(Q^T Q)$. By induction hypothesis it is $\det(Q Q^T) = (\text{vol } \check{P})^2$, i.e. $\det(V V^T) = \|v_n^\perp\|^2 \cdot (\text{vol } \check{P})^2 =$ $= \left(\|v_n^\perp\| \cdot (\text{vol } \check{P})\right)^2 = (\text{vol } P)^2$. ∎

For $n = k$ (i.e. $P$ is an n-dimensional parallelotope in $\mathbb{R}^n$) it is $\det(V V^T) = (\det V) \cdot (\det V^T) = (\det V) \cdot (\det V) = (\det V)^2$, i.e. $(\text{vol } P)^2 = (\det V)^2$, thus vol $P = |\det V|$. This results in:

**Note 47**: Let $P = \{ \sum_{i=1}^{n} a_i v_i \,|\, 0 \leq a_1, a_2, \ldots, a_n \leq 1\} \subseteq \mathbb{R}^n$ be an n- dimensional parallelotope. With $V = \left(v_1 v_2 \ldots v_n\right)$ it is vol $P = |\det V|$. ∎

Before, curves (i.e. 1-manifolds) have been linearly approximated by "small" secants (i.e. 1-parallelotopes). Similarly, n-manifolds can be linearly approximated by "small" n-parallelotopes (see Figure 46).

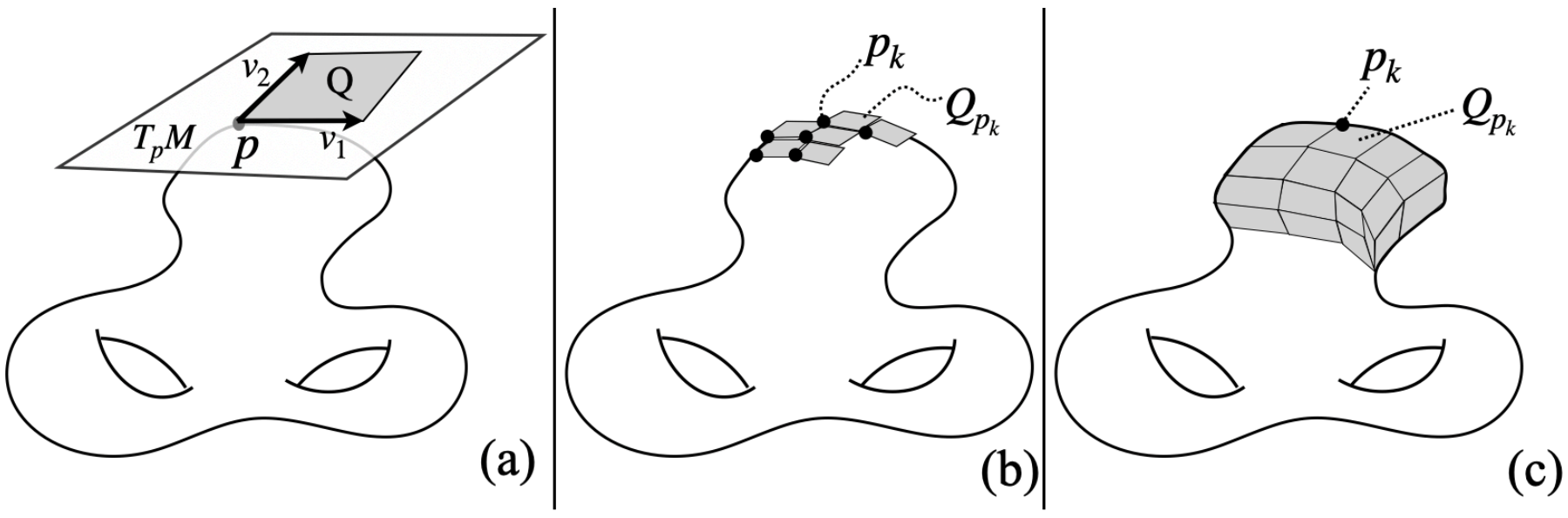

**Fig. 46**. Approximating Volumes

For this purpose, take $n = \dim M$ "small" linear independent vectors $v_1, \ldots, v_n$ in $T_pM$. They span a parallelotope $Q$ of volume $\text{vol } Q = |\det\left(v_1 v_2 \ldots v_n\right)|$ (see Figure 46(a)). This is repeated for points that are "below" the vertices of the parallelotope (see Figure 46(b)). "Glueing" these parallelotope together gives an approximation of the Riemannian manifold $M$ (see Figure 46(c)). Thus, the volume of $M$ is approximated by the sum of the volumes of these parallelotopes: $\text{vol } M \approx \sum_i \text{vol } Q_{p_k}$.

According to Note 46 it is $\text{vol } P = \sqrt{\det\left(V V^T\right)}$, but $\det(V^T V) = \det(\langle v_i, v_j \rangle)$. Thus, $\text{vol } Q_{p_k} = \sqrt{\det(\langle v_i, v_j \rangle)}$. For a Riemannian manifold $(M, g)$ the scalar product is $g = \langle , \rangle$. Thus, $\text{vol } Q_{p_k} = \sqrt{\det(g_{p_k}(v_i, v_j)}$. This implies (for $M$ compact and oriented)

$$\text{vol } M \approx \sum_k \text{vol } Q_{p_k} = \sum_k \sqrt{\det(g_{p_k}(v_i, v_j)}$$

Pick $v_r = \partial_r$ $(1 \leq r \leq n)$ as $v_1, \ldots, v_n \in T_pM$. With $g_{ij}|_p = g_{ij}(\partial_i|_p, \partial_j|_p)$ it is it is $\text{vol } M \approx \sum_k \sqrt{\det(g_{ij}|_{p_k})}$. With the volume form $\omega_g = \sqrt{\det(g_{ij})} dx_1 \wedge \ldots \wedge dx_n$ it is $\text{vol } M := \int_M \omega_g$. Often, $dV_g$ is used instead of $\omega_g$. Thus:

$$\text{vol } M = \int_M \sqrt{\det(g_{ij})} dx_1 \wedge \ldots \wedge dx_n = \int_M dV_g \tag{113}$$

Finally, this allows to define the integral of a function $f \in C^\infty(M)$ over a compact oriented Riemannian manifold. Note, that it is $f\omega_g \in \Omega^n(M)$ (Note 35). Then the integral of $f$ over $M$ is:

$$\int_M f\, dV_g = \int_M f\omega_g = \int_M f\sqrt{\det(g_{ij})} dx_1 \wedge \ldots \wedge dx_n \tag{114}$$

### 11.4. Variational Calculus and Einstein's Field Equation

Variational calculus provides a means to derive maps that are extremal values of (problem-specific) conditions. E.g. it allows to compute shortest paths between two points, or to determine the equations of motion of masses. In this context, infinite dimensional vector spaces appear (i.e. sets of maps that form the domain for conditions that are to be made extremal) as well as a notion of derivative on such vector spaces (note that all spaces on which derivatives had been defined have been finite dimensional).

The concept of a basis is central for studying vector spaces:

**Definition 95:** Let $V$ be a vector space over $\mathbb{K}$ and $S \subset V$. The (linear) *span* of $S$ is the set $\text{span}(S) := \{a_1 v_1 + \ldots + a_n v_n \mid n \in \mathbb{N} \wedge a_1, \ldots, a_n \in \mathbb{K}\}$ i.e. $\text{span}(S)$ is the set of all finite linear combinations of vectors from $S$.

A subset $B \subset V$ is *linear independent* :⇔ For each finite subset $\{v_1, \ldots, v_n\} \in B$ and $a_1 v_1 + \ldots + a_n v_n = 0$ for $a_1, \ldots, a_n \in \mathbb{K}$ it is $a_1 = \ldots = a_n = 0$.

A subset $B \subset V$ is a *basis* of $V$ :⇔ $B$ is linear independent and $\text{span}(B) = V$.

A vector space $V$ is *infinite dimensional* :⇔ For each finite set $S \subset V$ it is $\text{span}(S) \neq V$, in symbols: $\dim_{\mathbb{K}} V = \infty$. □

Note, that the field $\mathbb{K}$ is a subset of the dim symbol because $\mathbb{K}$ has an influence on the dimension of $V$ (see a bit later). The following is a key fact (see [2] Th. 2.1):

**Note 48**: Every vector space has a basis. ∎

Especially, all bases of a vector space have the same cardinality, and this cardinality is called the dimension of the vector space. Especially, (see [2] Th. 2.2):

**Note 49**: $\dim V = \infty \Leftrightarrow$ For each basis $B$ it is $\text{card} B = \infty$. ∎

For infinite dimensional vector spaces the defined concept of a basis, which is the usual one in linear algebra, is also called a *Hamel* basis. This is because another definition of a basis (*Schauder* basis) allows linear combinations of infinitely many basis vectors as converging series (which assumes that the vector space is a topological vector space) - see [2] Ch. 8.3. Note, that the existence of a Schauder basis is not guaranteed ([2] Ch. 8.3) but the existence of a Hamel basis is (Note 48).

Examples for infinite dimensional vector spaces:

1. $\mathbb{R}$ is a vector space over $\mathbb{Q}$, $\dim_{\mathbb{Q}} \mathbb{R} = \infty$
   - Let $\{b_1, \ldots, b_n\} \subset \mathbb{R}$ be a finite basis. $\mathbb{Q}$ is countable thus, the $\text{span}_{\mathbb{Q}}\{b_1, \ldots, b_n\} = \{q_1 b_1 + \ldots + q_n b_n \mid q_i \in \mathbb{Q}\} \subset \mathbb{R}$ is countable also. But $\mathbb{R}$ is uncountable thus the $\text{spann}_{\mathbb{Q}}\{b_1, \ldots, b_n\}$ is not all of $\mathbb{R}$, i.e. $\text{spann}_{\mathbb{Q}}\{b_1, \ldots, b_n\} \subsetneq \mathbb{R}$. Consequently, for each $n \in \mathbb{N}$, $\{b_1, \ldots, b_n\} \subset \mathbb{R}$ is not a basis: Each basis must be uncountable, i.e. the dimension of $\mathbb{R}$ over $\mathbb{Q}$ is infinite (Note 49).
   - $\mathbb{R}$ is a vector space over $\mathbb{R}$ with $\dim_{\mathbb{R}} \mathbb{R} = 1$.
2. Let $\mathbb{P} := \{a_n x^n + \ldots + a_1 x + a_0 \mid n \in \mathbb{N} \wedge a_i \in \mathbb{R} \wedge a_n \neq 0\}$ be the set of polynomials. $\mathbb{P}$ is an infinite dimensional vector space (over $\mathbb{R}$ or $\mathbb{C}$) with basis $\{x^i\}_{i \in \mathbb{N}}$.
3. Let $M$ be a differentiable manifold, $\dim M \geq 1$. The set $C^k(M)$ with $k \geq 0$ is an infinite dimensional vector space (over $\mathbb{R}$)

The vector space $C^k(M)$ can be generalized: Let $\mathscr{F}$ be a vector space of functions (e.g. $\mathscr{F} = C^\infty(M)$). A map $\mathscr{F} \to \mathbb{R}$ is called a *functional*; different from this is a map $\mathscr{F} \to \mathscr{F}$ called an *operator. Variational calculus* deals with finding extremal points (often, minima) of functionals as follows: let $y \in C^\infty([a,b])$ and let $x \in [a,b]$ be the parameter $y$ depends on. Furthermore, let $F \in C^\infty([a,b,] \times \mathbb{R} \times \mathbb{R})$. Define

$$\mathscr{I}(y) = \int_a^b F(x, y(x), y'(x))\, dx \tag{115}$$

Then, $\mathscr{I} : C^\infty([a,b]) \to \mathbb{R}$ is a functional. Assume that $y \in C^\infty([a,b])$ is given and the problem is to verify that $y$ minimizes the functional $\mathscr{I}$. The idea of variational calculus is to vary $y$ a little bit by "disturbing" it with an arbitrary "perturbation function" $\varphi \in C^\infty([a,b])$ satisfying $\varphi(a) = 0$ and $\varphi(b) = 0$. Thus, with $\varepsilon \in \mathbb{R}$ as parameter $\bar{y}(x) = y(x) + \varepsilon\varphi(x)$ is a one-parameter family of functions satisfying the condition $\bar{y}(a) = y(a), \bar{y}(b) = y(b)$ ( see Figure 47).

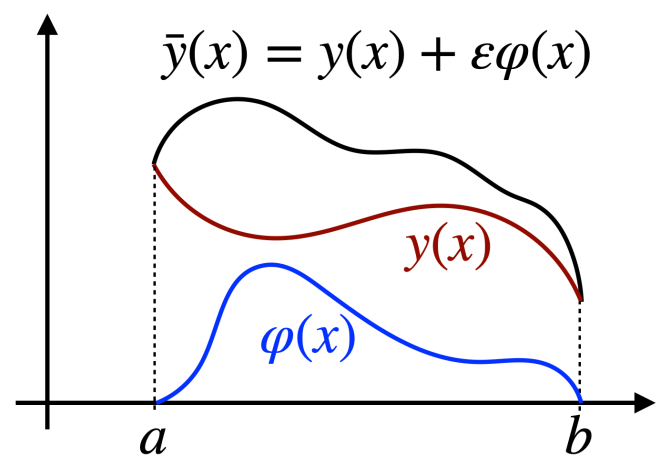

**Fig. 47**. One-Parameter Family $\bar{y}$ of Functions Varying $y$

Define $\Phi \in C^\infty(\mathbb{R})$ with $\Phi(\varepsilon) = \mathscr{I}(y + \varepsilon\varphi)$, i.e.

$$\Phi(\varepsilon) = \int_a^b F(x, y(x) + \varepsilon\varphi(x), y'(x) + \varepsilon\varphi'(x))\, dx \tag{116}$$

The problem of determining an extremal $y$ is now reduced to the one-variable calculus problem of determining whether 0 is an extremal value of $\Phi$ , i.e. to verify that $\Phi'(0) = 0$. Now, it is (Definition 1)

$$\Phi'(0) = \lim_{\varepsilon \to 0} \frac{\Phi(\varepsilon) - \Phi(0)}{\varepsilon} = \lim_{\varepsilon \to 0} \frac{\mathscr{I}(y + \varepsilon\varphi) - \mathscr{I}(y)}{\varepsilon} = \frac{\partial}{\partial\varepsilon}(\mathscr{I}(y + \varepsilon\varphi))\Big|_{\varepsilon=0} \tag{117}$$

which is called the *variation of y in direction of* $\varphi$. This can be generalized as follows:

**Definition 96:** Let $\mathscr{J} : \mathscr{F} \to \mathbb{R}$ be a functional. If the following limit exists

$$d\mathscr{I}(y, \varphi) := \lim_{\varepsilon \to 0} \frac{\mathscr{I}(y + \varepsilon\varphi) - \mathscr{I}(y)}{\varepsilon},$$

then $\mathscr{J}$ is called *Gâteaux differentiable* in $y$ in direction of $\varphi$ and $d\mathscr{I}(y, \varphi)$ is called *Gâteaux derivative* of $\mathscr{I}$. □

In general, $d\mathcal{J}(y,\varphi)$ is neither continuous nor linear in $\varphi$ (see [17] example on pp. 4). For $d\mathcal{J}(y,\varphi)$ being linear in $\varphi$:

**Definition 97:** If $d\mathcal{J}(y,\varphi)$ is linear in $\varphi$, then the notation $\delta\mathcal{J}(y)\varphi$ is used instead and is called *first variation of y in direction of $\varphi$.* □

Note, that the Gâteaux derivative $\delta$ is generalization of the directional derivative (Definition 7) on infinite dimensional vector spaces. The following is often used in computations:

**Lemma 75**: The *Gâteaux derivative* $\delta$ satisfies the Leibniz rule (a.k.a. product rule):

$$\delta(\mathcal{J}\mathcal{G})(y)\varphi = \mathcal{J}(y)\,\delta\mathcal{G}(y)\varphi + \mathcal{G}(y)\,\delta\mathcal{J}(y)\varphi \qquad \blacksquare$$

Verifying that $y$ minimizes $\Phi(\varepsilon) = \mathcal{J}(y+\varepsilon\varphi)$ means to compute $\Phi'(0) = \left.\frac{\partial\Phi}{\partial\varepsilon}\right|_{\varepsilon=0}$, thus we compute

$$\frac{\partial\Phi}{\partial\varepsilon} = \frac{\partial}{\partial\varepsilon}\int_a^b F(x,\bar{y}(x),\bar{y}'(x))\,dx \overset{(a)}{=} \int_a^b \frac{\partial}{\partial\varepsilon}F(x,\bar{y}(x),\bar{y}'(x))\,dx$$

Herein, $(a)$ valid is because integration and differentiation may be interchanged if variables differ (see [20]Th. 6.22). Next, the derivative $\frac{\partial F}{\partial\varepsilon}$ is computed by applying the chain rule (see [20]Th. 3.8):

$$\frac{\partial}{\partial\varepsilon}F(x,\bar{y}(x),\bar{y}'(x)) = \frac{\partial F}{\partial x}\overbrace{\frac{\partial x}{\partial\varepsilon}}^{=0} + \frac{\partial F}{\partial\bar{y}}\overbrace{\frac{\partial\bar{y}}{\partial\varepsilon}}^{=\varphi} + \frac{\partial F}{\partial\bar{y}'}\overbrace{\frac{\partial\bar{y}'}{\partial\varepsilon}}^{=\varphi'}$$

Note that $x$ is independent of $\varepsilon$, i.e. $\frac{\partial x}{\partial\varepsilon} = 0$. With $\bar{y}(x) = y(x) + \varepsilon\varphi(x)$ it is $\frac{\partial\bar{y}}{\partial\varepsilon} = \varphi$ (because $y$ is independent of $\varepsilon$), and $\bar{y}'(x) = y'(x) + \varepsilon\varphi'(x)$ implies $\frac{\partial\bar{y}'}{\partial\varepsilon} = \varphi'$. Together, this results in

$$\frac{\partial\Phi}{\partial\varepsilon} = \int_a^b\left(\frac{\partial F}{\partial\bar{y}}\varphi + \frac{\partial F}{\partial\bar{y}'}\varphi'\right)dx = \int_a^b\frac{\partial F}{\partial\bar{y}}\varphi\,dx + \int_a^b\frac{\partial F}{\partial\bar{y}'}\varphi'\,dx \qquad (118)$$

Next, integration by parts (Lemma 73) is used for the last summand, and $\varphi(a) = 0$ and $\varphi(b) = 0$ is applied:

$$\int_a^b\frac{\partial F}{\partial\bar{y}'}\varphi'\,dx = \overbrace{\left[\frac{\partial F}{\partial\bar{y}'}\varphi(x)\right]_a^b}^{=0} - \int_a^b\frac{d}{dx}\left(\frac{\partial F}{\partial\bar{y}'}\right)\varphi'\,dx$$

$$= -\int_a^b\frac{d}{dx}\left(\frac{\partial F}{\partial\bar{y}'}\right)\varphi\,dx$$

Substituting this result in Equation 118 in (considering that $\bar{y} = y$ for $\varepsilon = 0$)

$$\left.\frac{\partial \Phi}{\partial \varepsilon}\right|_{\varepsilon=0} = \int_a^b \left(\frac{\partial F}{\partial y}\varphi - \frac{d}{dx}\left(\frac{\partial F}{\partial y'}\right)\varphi\right) dx = \int_a^b \left(\frac{\partial F}{\partial y} - \frac{d}{dx}\left(\frac{\partial F}{\partial y'}\right)\right)\varphi \, dx \quad (119)$$

Finally, the fundamental lemma of variational calculus (i.e. $\int_a^b fg\,dx = 0$ for all $g$ implies $f = 0$ - see [26] Lemma 3.1) proves:

**Theorem 18 (Euler-Lagrange Equation)**: A minimizing $y \in C^\infty([a,b])$ of the functional $\mathcal{J}(y) = \int_a^b F(x, y(x), y'(x))\, dx$ satisfies:

$$\frac{\partial F}{\partial y} - \frac{d}{dx}\left(\frac{\partial F}{\partial y'}\right) = 0 \quad (120) \blacksquare$$

This equation can be generalized:

Let $F \in C^\infty([a,b] \times \mathbb{R}^n \times \mathbb{R}^n)$, $y : [a,b] \to \mathbb{R}^n$ with $y_i \in C^\infty([a,b])$ and $\mathcal{J}(y) = \int_a^b F(x, y(x), y'(x))\, dx$ ($F$ is called *Lagrange function*). With $y_i' := \dfrac{dy_i}{dx}$ the Euler-Lagrange Equation becomes ($1 \le i \le n$)

$$\frac{\partial F}{\partial y_i} - \frac{d}{dx}\left(\frac{\partial F}{\partial y_i'}\right) = 0 \quad (121)$$

i.e. the Euler-Lagrange equation is a system of $2n$ partial differential equations.

If $M$ is a manifold, $y : [a,b] \to M$ and $F : TM \to \mathbb{R}$ be both smooth, furthermore $\mathcal{J}(y) = \int_a^b F(x, y(x), y'(x))\, dx$, the Euler-Lagrange Equation is ($1 \le i \le n$)

$$\frac{\partial F}{\partial x_i}(t, y(t), y'(t)) - \frac{d}{dt}\left(\frac{\partial F}{\partial v_i}(t, y(t), y'(t))\right) = 0 \quad (122)$$

if $(y(t), y'(t)) = (x_1(t), \ldots, x_n(t), v_1(t), \ldots, v_n(t)) \in TM$ in proper coordinates $(x, v)$ of $TM$.

Lagrange functions and the Euler-Lagrange Equation enables a fundamental principle of physics: the *Principle of Least Action*. This principle states that a system always changes in a way such that a certain kind of physical quantity (describing the system) called *action* is minimal. Here, an *action* is the integral of a Langrange function over the time and the Langrange function is specific to the physical quantity of interest.

As an example, let $x : [a,b] \to \mathbb{R}^3$ be the trajectory of a point mass $m$. Its kinetic energy is $T(x') = \frac{1}{2}m(x_1'^2 + x_2'^2 + x_3'^2)$ and $V(x)$ is its potential energy. $F := T - V$ is a Langrange function and $\mathcal{J}(x) = \int_a^b \left(T(x') - V(x)\right) dt$ the action in this system. According to Equation 121 the Euler-Langrange equation is then:

$$\frac{\partial F}{\partial x_i} - \frac{d}{dt}\left(\frac{\partial F}{\partial x_i'}\right) = 0\,, \; i = 1,2,3$$

i.e. $\frac{\partial F}{\partial x_i} = -\frac{\partial V}{\partial x_i}$ because $T$ is independent of $x_i$ and with $V$ independent of $x_i'$ it is $\frac{d}{dt}\left(\frac{\partial F}{\partial x_i'}\right) = \frac{d}{dt}(\frac{1}{2}m\,x_i'2) = m\,x_i''$. Together, it is $-\frac{\partial V}{\partial x_i} = \frac{d}{dt}\left(m\,x_i'\right)$ for $i = 1{,}2{,}3$, thus $-\text{grad}V = m\,x''$ which is the well-know equation of motions of point masses.

Finally, we indicate (very superficially) how Einstein's Field Equation follows from the principle of least action (for the details see [4] Ch. 20, and [10] pp. 114 ff). Thus, the goal is to derive the equation of motion of a point mass in a spacetime $(M, g)$, i.e. a pseudo-Riemannian manifold. To apply the principle of least action the Lagrange function $F_H(x) = \sqrt{g}\,\text{Scal}$ is used and yields the (gravitational) action $S_H = \int_M \sqrt{g}\,\text{Scal}\,d^4x$ (called *Hilbert action* or *Einstein-Hilbert action*) with the scalar curvature Scal (Definition 50) $\sqrt{g} := \sqrt{\det(g_{ij})}$ and $d^4x := d\,x_1 \wedge \ldots \wedge d\,x_4$. Applying the variational calculus results in $\text{Ric} - \frac{1}{2}\text{Scal}\,g = 0$, i.e. in Einstein's vacuum field equation. Similarly, using $S_H = \int_M \sqrt{g}(\text{Scal} - 2\Lambda)\,d^4x$ results in Einstein's vacuum field with cosmological constant $\Lambda$, i.e. $\text{Ric} - \frac{1}{2}\text{Scal}\,g + \Lambda g = 0$. Finally, adding a summand $F_M$ reflecting the mass of the spacetime to $F_H$, i.e. the Langrange function becomes $F_H(x) = \sqrt{g}(\text{Scal} - 2\Lambda) + F_M$, results in the full Einstein field equation

$$\text{Ric} - \frac{1}{2}\text{Scal}\,g + \Lambda g = \kappa T \tag{123}$$

## 12. Outlook

Entanglement is one of the fundamental phenomena in quantum physics. Systems of entangled entities are strongly correlated even across huge spacial distances implying that measuring one entity reveals immediate information about the state of the others. This seems contradictory for example to special relativity limiting communication to the speed of light.

Einstein postulated the existence of hidden variables which are set when entities become entangled. The corresponding values prescribe measurement results. The Bell test and its experimental verification proved that such hidden variables do not exist. Thus other explanations how entanglement is enacted are looked for.

One explanation suggested recently is that entanglement can be described by wormholes. Such a wormhole is a connection between blackholes in spacetime. Wormholes have been described in an article by Einstein and Rosen ("ER"), while entanglement has been published in a contribution bei Einstein, Podolski and Rosen ("EPR") - both in the same year 1935. Because of the analogy between wormholes and entanglement the corresponding explanation is referred to as "ER=EPR". Understanding wormholes, black holes, and spacetime require the underlying knowledge presented in this contribution: based on this, the authors are preparing a follow-up article describing ER=EPR.

Furthermore, analyzing black holes and entropy implies that all information about the internals of a black hole is encoded as qubits on its "surface" (the so-called event horizon). This information encoded on a two-dimensional surface about a three-dimensional entity is like a hologram of the entity. Going beyond this, spacetime itself can be considered as such a hologram. Even further, spacetime itself may not be a fundamental concept of physics but a fabric consisting of entangled qubits. Thus, "it" (i.e. the spacetime) is created from qubits by entanglement: this is referred to as "it from qubit". The follow-up article of the authors will describe this also.

**Author Contributions**: Writing – original draft, F.L.; Writing – review & editing, J.B. All authors have read and agreed to the published version of the manuscript.

(Links were last followed on May 25, 2026)